\documentclass{mythesis}
\usepackage[utf8]{inputenc}

\newcommand{\intinf}{\int_{-\infty}^\infty}
\newcommand{\R}{\mathbb{R}}
\newcommand{\mathd}{\mathrm{d}}
\newcommand{\mathe}{\mathrm{e}}
\newcommand{\mathi}{\mathrm{i}}
\newcommand{\der}[2][]{\frac{\mathd#1}{\mathd#2}}
\newcommand{\pder}[2][]{\frac{\partial#1}{\partial#2}}
\newcommand{\argmin}{\operatornamewithlimits{argmin}}
\newcommand{\res}{\operatornamewithlimits{Res}}
\DeclareMathOperator\lip{Lip}
\DeclareMathOperator{\supp}{supp}
\newcommand{\opnorm}[1]{{\left\vert\kern-0.25ex\left\vert\kern-0.25ex\left\vert #1 
    \right\vert\kern-0.25ex\right\vert\kern-0.25ex\right\vert}}

\newcommand{\barh}{\bar{h}}
\newcommand{\surften}{\gamma}
\newcommand{\lagrange}{\ell}
\newcommand{\Lagrange}{\mathcal{L}}
\newcommand{\Energy}{E}

\newcommand{\vis}{\mu}
\newcommand{\mob}{m}
\newcommand{\Rey}{\mathrm{Re}}
\newcommand{\Ca}{\mathrm{Ca}}
\newcommand{\Bo}{\mathrm{Bo}}
\newcommand{\Ma}{\mathrm{Ma}}
\newcommand{\Bi}{\mathrm{Bi}}

\newcommand{\markup}[1]{#1}
\definecolor{aquamarine}{HTML}{008A73}
\newcommand{\amend}[1]{#1}

\title{Applications and Novel Regularization of the Thin-Film Equation}
\author{Khang Ee Pang}
\date{\today}

\begin{document}

% \maketitle
\mytitle

\tolerance=400

\pagenumbering{roman}
\newpage
\tableofcontents

% \listoffigures
% \listoftables

\newpage
% \addcontentsline{toc}{chapter}{Abstract}
\thispagestyle{plain}
\topskip0pt
\vspace*{\fill}
\begin{center}
\textbf{Abstract}
\end{center} 
The classical no-slip boundary condition of the Navier-Stokes equations fails to describe the spreading motion of a droplet on a substrate due to the missing small-scale physics near the contact line. In this thesis, we introduce a novel regularization of the thin-film equation to model droplet spreading. The solution of the regularized thin-film equation -- the Geometric Thin-Film Equation is studied and characterized. 
Two robust numerical solvers are discussed, notably, a fast and mesh-free numerical scheme for simulating thin-film flows in two and three spatial dimensions. Moreover, we prove the regularity and convergence of the numerical solutions. The existence and uniqueness of the solution of the Geometric Thin-Film Equation with respect to a wide range of measure-valued initial conditions are also discussed. 

\vspace*{\fill}

\newpage
\thispagestyle{plain}
\topskip0pt
\vspace*{\fill}
\begin{center}
\textbf{Statement of Original Authorship}
\end{center} 
I hereby certify that the submitted work is my own work, was completed while registered as a candidate for the degree stated on the Title Page, and I have not obtained a degree elsewhere on the basis of the research presented in this submitted work.
\vspace*{\fill}

\newpage
% \addcontentsline{toc}{chapter}{Collaborations}
\thispagestyle{plain}
\topskip0pt
\vspace*{\fill}
\begin{center}
\textbf{Collaborations}
\end{center}  
\textbf{Chapter 2:} This work was done in collaboration with Yutaku Kita, Charles Cuvillier, and Lennon \'O N\'araigh. 
Kita provided experimental data and expertise on the point-heated droplet. The base-state solution was first obtained by \'O N\'araigh. The derivations of the perturbed equation were derived independently by Cuvillier and I. All other aspects of this chapter were derived and implemented by me, with the help of \'O N\'araigh. 

\textbf{Chapter 3 \& 5:} The work in these chapters was done in collaboration with Lennon \'O N\'araigh. The models were formulated around the geometric diffuse-interface method developed by \'O N\'araigh. The numerical solvers and optimizations were developed by me, with the help of \'O N\'araigh. 

\textbf{Chapter 4:} This work was done in collaboration with Richard Smith and Lennon \'O N\'araigh. The global existence theorem and a partial convergence of the particle solution were obtained by me, with some help from \'O N\'araigh. Smith proposed an alternative framework and developed the necessary tools for proving the convergence theorem and the regularity of the solution for the Camassa-Holm equation. These were then adapted for the Geometric-Thin Film Equation by me, with the help of Smith. Any numerical computations were performed by me. 

\vspace*{\fill}

\newpage
% \addcontentsline{toc}{chapter}{Acknowledgement}
\thispagestyle{plain}
\topskip0pt
\vspace*{\fill}
\begin{center}
\textbf{Acknowledgement}
\end{center}  
Firstly, I would like to thank my examination committee, Dr James Herterich for chairing and organizing my Viva Voce, Professor Uwe Thiele for acting as the external examiner, and Professor Fr\'ed\'eric Dias for acting as the internal examiner. 

I would like to express my deepest appreciation to my supervisor Dr Lennon \'O N\'araigh, for the guidance he has given me. I feel fortunate to have such an excellent supervisor. 
I am also thankful to Dr Richard Smith for his help with the necessary functional analysis to tackle the well-posedness problem. 
I thank Professor Prashant Valluri for hosting me at the University of Edinburgh during my research visit. 
Additionally, I’d like to recognize Charles Cuvillier for his work on the point-heated droplet before me and Dr Yutaku Kita for his feedback on our manuscript. 

I acknowledge the funding I received from Science Foundation Ireland (Grant Number 18/CRT/6049). 
I am also grateful to the CRT directors and cohort members, especially Tiziana Comito, for their support. 

Lastly, I would be remiss in not mentioning my family and my other half, Shu Min, for being the inspiration, constant support, and encouragement without which I would not have reached this height.  

% This publication has emanated from research supported in part by a Grant from Science Foundation Ireland under Grant Number 18/CRT/6049. 
\vspace*{\fill}

% \newpage
% \thispagestyle{plain}
% \topskip0pt
% \vspace*{\fill}
% \begin{center}
% \textbf{Dedication}
% \end{center}
% % I dedicate this thesis to my significant other Shu Min, for being the inspiration and the constant support and encouragement without which I would not reach this height. 
% \vspace*{\fill}

\newpage
\thispagestyle{plain}
\topskip0pt
\vspace*{\fill}
\begin{center}
\textbf{Dedication} 

To Shu Min.
\end{center} 
\vspace*{\fill}

\newpage
\thispagestyle{plain}
\begin{center}
\textbf{List of Publications}
\end{center}
\begin{enumerate}
    \item \textbf{Khang Ee Pang} and Lennon \'O N\'araigh. ``A mathematical model and mesh-free numerical method for contact-line motion in lubrication theory''. \textit{Environ Fluid Mech} \textbf{22}, 301–336 (2022). DOI: 10.1007/s10652-021-09827-0
    \item Lennon \'O N\'araigh, \textbf{Khang Ee Pang}, and Richard J. Smith. ``A new convergence analysis of the particle method for the Camassa-Holm equation''. Submitted.
    arXiv:2112.10589    
    \item  Lennon \'O N\'araigh, \textbf{Khang Ee Pang}, and Richard J. Smith. ``Convergence Analysis of the Geometric Thin-Film Equation''. Submitted. 
    arXiv:2207.14175
    \item \textbf{Khang Ee Pang}, Lennon \'O N\'araigh, Charles Cuvillier, and Yutaku Kita. ``Symmetry Breaking in Point-Heated Droplets''. Submitted. \\ 
    arXiv:2307.09310
    % \item Lennon \'O N\'araigh, \textbf{Khang Ee Pang}, and Richard J. Smith. ``Well-Posedness of the Geometric Thin-Film Equation''. Unpublished.
\end{enumerate}
Some passages and theorems in the thesis have been quoted verbatim from the above sources.

\hspace{1em}

\begin{center}
\textbf{Contributed Talks}
\end{center}
\begin{enumerate}
    \item 9th International Symposium on Bifurcations and Instabilities in Fluid Dynamics, University of Groningen. 
    \item 3rd ThermaSMART Workshop, York University.
    \item 9th Annual Irish SIAM Student Chapter Conference, University of Galway.
    \item UCD Applied \& Computational Mathematics Seminar Series, University College Dublin. 
\end{enumerate}

\newpage
\pagenumbering{arabic}

\chapter{Introduction} \label{ch:1}

\section{Background}

% \commentkp{Application} The deposition of droplets on a substrate followed by the wetting of the substrate occurs in many 

When a droplet of fluid (surrounded by a gaseous atmosphere) is deposited on a dry solid substrate, the solid, fluid, and gas form a triple-phase contact line at the points where the three phases meet. The contact angle $\theta_{cl}$ is then measured as the angle between the liquid-gas interface and the solid surface (conventionally, through the liquid). The droplet will then either spread or recede, depending on whichever is more energetically favourable, until it reaches an equilibrium configuration. Droplet spreading then describes the dynamic phase before the attainment of this equilibrium.

\begin{figure}[htb]
    \centering
    \begin{tikzpicture}[scale=1.4]
        % axis
        \draw [->, line width=0.3mm] (-1.7,0.75-0.607) -- (-1.7,1.7);
        \draw (-1.7,1.7) node[left] {$z$};
        % plane @height=-0.607
        \draw [line width=0.3mm] (1.1,0.75-0.607) -- (4.3,0.75-0.607) -- (2.8,-0.75-0.607) -- (-3.2,-0.75-0.607) -- (-1.7,0.75-0.607) -- (1.1,0.75-0.607) -- (-0.4,-0.75-0.607);
        \draw (-3.2,-0.75-0.607) node[left] {$x$};
        \draw (4.3,0.75-0.607) node[above] {$y$};
        % sphere @camera sin30=0.5 @intersection cos235= 
        % \draw (0, 0) circle (1);
        \draw [orange, line width=0.2mm] (0-1,1) arc [start angle=90, end angle=250, x radius=0.5cm, y radius=1cm];
        \draw [orange, line width=0.3mm, dotted] (0-1,1) arc [start angle=90, end angle=-20, x radius=0.5cm, y radius=1cm];
        \draw [color=orange, line width=0.2mm] (-1-1,0) arc [start angle=180, end angle=360, x radius=1.0cm, y radius=0.5cm];
        \draw [color=orange, line width=0.3mm, dotted] (1-1,0) arc [start angle=0, end angle=180, x radius=1.0cm, y radius=0.5cm];
        % base @intersect h=pi/4 @r2=cos(h) @h'=hcos30
        % \draw (0.707,-0.607) arc [start angle=0, end angle=360, x radius=0.707cm, y radius=0.5*0.707cm];
        \draw [color=orange, line width=0.7mm] (-0.612-1,-0.607-0.177) arc [start angle=210, end angle=330, x radius=0.707cm, y radius=0.5*0.707cm];
        \draw [color=orange, line width=0.7mm, dashed] (0.612-1,-0.607-0.177) arc [start angle=-30, end angle=210, x radius=0.707cm, y radius=0.5*0.707cm];
        % spherical cap
        \draw [color=orange, line width=0.7mm] (0.588-1,-0.809) arc [start angle=-54, end angle=234, x radius=1cm, y radius=1cm];
        % angle
        \draw [line width=0.3mm, dashed] (-1,-0.607) -- (-1+0.588,-0.809) -- (0.2,-0.3);
        \draw [line width=0.7mm, domain=40:160] plot ({0.3*cos(\x)-1+0.588}, {0.3*sin(\x)-0.809});
        % interior contour
        \draw [orange, line width=0.2mm] plot[domain=1.36:2,smooth] (\x,{-4*(\x-2)^2+0.8*1.2^2-0.607});
        \draw [orange, line width=0.3mm, dotted] plot[domain=2:2.4,smooth] (\x,{-4*(\x-2)^2+0.8*1.2^2-0.607});
        % \draw (0.87+2,-0.13) arc [start angle=0, end angle=360, x radius=0.87cm, y radius=0.5*0.87cm];
        \draw [orange, line width=0.2mm] (-0.87+2,-0.13) arc [start angle=180, end angle=360, x radius=0.87cm, y radius=0.5*0.87cm];
        \draw [orange, line width=0.3mm, dotted] (0.87+2,-0.13) arc [start angle=0, end angle=180, x radius=0.87cm, y radius=0.5*0.87cm];
        % base
        % \draw (1.16+2,-0.607) arc [start angle=0, end angle=360, x radius=1.16cm, y radius=0.5*1.16cm];
        \draw [color=orange, line width=0.7mm] (-1.142+2,-0.607+0.1) arc [start angle=170, end angle=370, x radius=1.16cm, y radius=0.5*1.16cm];
        \draw [color=orange, line width=0.7mm, dashed] (1.142+2,-0.607+0.1) arc [start angle=10, end angle=170, x radius=1.16cm, y radius=0.5*1.16cm];
        % parabolic cap
        \draw[orange, line width=0.7mm] plot[domain=0.84:3.16,smooth] (\x,{0.8*(-(\x-2)^2+1.2^2)-0.607});
        % % angle
        \draw [line width=0.3mm, dashed] (2,-0.607) -- (3.16,-0.53) -- (2.7,0.5);
        \draw [line width=0.7mm, domain=115:185] plot ({0.3*cos(\x)+3.16}, {0.3*sin(\x)-0.53});
        % arrow
        \draw [line width=0.2mm] (-0.8,-1.8) -- (-1,-0.607-0.5*0.707);
        \draw [line width=0.2mm] (-0.8,-1.8) -- (1.6,-0.607-0.5*1.16+0.03);
        \filldraw (-1,-0.607-0.5*0.707) circle (0.03);
        \filldraw (1.6,-0.607-0.5*1.16+0.03) circle (0.03);
        \draw (-0.8,-1.8) node[below] {Contact line};
        % contact angle
        \draw (-1+0.588,-0.85) node[below] {$\theta_{cl}$};
        \draw (3.16,-0.33) node[right] {$\theta_{cl}$};
        % \draw [->, line width=0.4mm] (1,1) -- (-0.5,-0.5);
        % \draw [->, line width=0.4mm] (1,1) -- (2.9,-0.4);
        % \draw (1,1) node[above] {contact angle};
    \end{tikzpicture} 
    \caption{Radially symmetrical droplets on a horizontal surface. The liquid interface is marked in orange. (left) The droplet on the left has a contact angle greater than $90^\circ$, whereas (right) the droplet on the right has a contact angle less than $90^\circ$.}
    \label{fig:sketch_3d}
\end{figure}
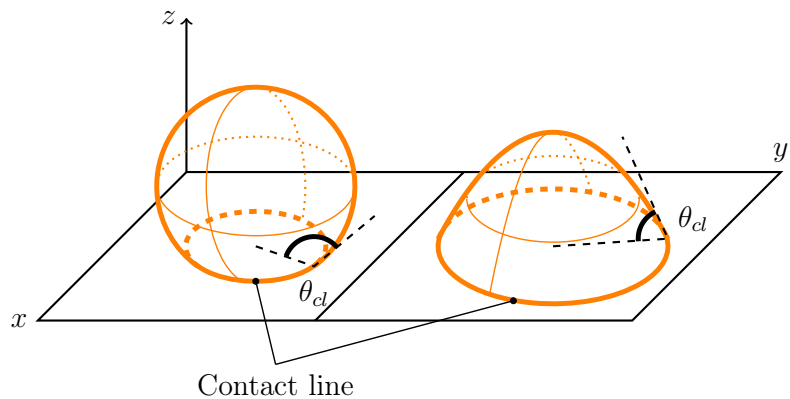 

The balance of the surface tension forces between the three phases at the contact line maintains the equilibrium contact angle. This can be computed via Young's equation, given by
\begin{equation}
    \gamma_{sg} = \gamma_{sl} + \gamma_{lg}\cos(\theta_{eq}),
\end{equation}
where $\gamma_{sg}$, $\gamma_{sl}$, and $\gamma_{lg}$ are the solid-gas, solid-liquid phases, and liquid-gas surface tensions, respectively. Three regimes are possible depending on the surface tensions:
\begin{enumerate}[label=(\roman*)]
    \item Weak wetting. If $\gamma_{sg}-\gamma_{sl}\leq0$, then $\theta_{eq}$ is greater than $90^\circ$ and the droplet beads up on the surface (\Cref{fig:sketch_3d} left).
    \item Strong wetting. If $\gamma_{lg}>\gamma_{sg}-\gamma_{sl}>0$, then $\theta_{eq}$ is less than $90^\circ$, and the droplet tends to spread out on the surface (\Cref{fig:sketch_3d} right).
    \item Complete wetting. If $\gamma_{sg}\geq\gamma_{sl}+\gamma_{lg}$, then $\theta_{eq}=0$ and the droplet will spread indefinitely, coating the whole surface. %, at the least, until the continuum theory fails. 
\end{enumerate}
% Evidently, if $\gamma_{sg}-\gamma_{sl}<0$, then $\theta_{eq}>90^\circ$, the system corresponds to the weak-wetting regime and the droplet beads up on the surface (Figure~\ref{fig:sketch_3d}a). On the contrary, if $\gamma_{sg}-\gamma_{sl}>0$, then $\theta_{eq}<90^\circ$, the system falls into the strong-wetting regime, and the droplet tends to spread out on the surface (Figure~\ref{fig:sketch_3d}b). Finally, in the case of $\gamma_{sg}>\gamma_{sl}+\gamma_{lg}$, then $\theta_{eq}=0$, the system is deemed complete wetting, and the droplet will spread indefinitely, coating the whole surface. 
Cases (i) and (ii) are also commonly referred to together as partial wetting in the literature. If the liquid is water, the weak wetting and strong wetting regimes correspond to the surface being hydrophobic and hydrophilic, respectively. Moreover, the surface is also called super-hydrophilic if $\theta_{eq}<10^\circ$. 

An ambiguity of the value of $\theta_{eq}$, known as the contact-angle hysteresis, may occur if the surface contains imperfections or chemical inhomogeneity. This causes pinning of the contact line \textit{before} the actual equilibrium contact angle is reached. Thus $\theta_{eq}$ varies depending on whether the contact line is advancing or receding. The contact-angle hysteresis can be described by modelling the contact line as a stick-slip system in which the sticking results in the early pinning of the contact line \cite{shanahan1994simple}. However, this work will not consider contact-angle hysteresis, and we assume there is no ambiguity for the value of $\theta_{eq}$.

% \commentkp{Introduce droplet spreading and moving contact line} During the initial deposition, the droplet is often not in equilibrium. The dynamic phase of attaining the equilibrium is known as droplet spreading. This introduces a complication in the model due to the moving contact line. Moving contact line is ubiquitous in multi-phase flow besides droplet spreading. The rising of meniscus in a capillary, the sinking of a solid, the pouring of liquid from a reservoir, the condensation and evaporation of droplets, they all exhibit a moving contact line, but the droplet spreading is one of the most well studied 

\section{\markup{Motivation and Overview of Thesis}}

The general modelling of viscous fluids is a difficult one. The Navier-Stokes equations can be used to describe the motion of the fluid, resulting in a set of nonlinear partial differential equations (PDEs) that can be difficult to solve. The lubrication theory provides an invaluable tool for reducing the Navier-Stokes equations to a more mathematically attractable equation under the assumption that the length scale in one of the spatial dimensions is much smaller than the rest. This assumption corresponds to the near-complete-wetting regime where the droplet assumes a thin shape. The resulting thin-film equation is a fourth-order parabolic PDE that describes the evolution of the height of the liquid-gas interface. A review of the lubrication approximation for the thin liquid film is given in \Cref{sec:review_classic}. 

However, the lubrication theory itself is insufficient to describe droplet spreading. In droplet spreading (or receding), the contact line is in motion relative to the solid substrate.  And yet this contradicts the classical no-slip assumption in viscous fluid flow, which stipulates that there should be no relative motion between a substrate in contact with a fluid. In the frame of reference of the solid substrate, the no-slip boundary condition is imposed by requiring that the velocity field vanishes at the liquid-solid interface. The moving contact line paradox was first discussed by Huh and Scriven~\cite{huh1971hydrodynamic} and has since been subjected to numerous mathematical modellings and experimental investigations (review papers include \cite{de1985wetting,dussan1974on,oron1997long,bonn2009wetting}). Nevertheless, the dynamics and mechanism of the moving contact line are still under much debate. In \Cref{sec:intro_motivation}, we review two well-studied droplet-spreading models rooted in the lubrication theory -- the slip-length model and the precursor-film model. 

In Chapter~\ref{sec:point_heating}, we look at a case study of applying the thin-film equation by investigating the dynamics of a point-heated droplet using the slip-length and precursor-film models. The stability of the Marangoni convection in a sessile drop heated from below with a point source is studied. We see that either model is not able to describe the system completely; hence a mixed approach is required. In particular, the slip-length model is inadequate for simulating the evolution of complex droplet profiles, while the precursor-film model is unsuitable for linear stability analysis. This is not ideal, as different models rely on different underlying assumptions and mechanisms. Thus, we seek an alternative model for describing droplet spreading that is robust to mathematical and numerical treatments. 

Our model is based on a recent work of Holm, \'O N\'araigh, and Tronci \cite{holm2020gdim} for a novel regularization method of the thin-film equation. The geometric diffuse-interface method (G-DIM) and the regularized thin-film equation -- the Geometric Thin-Film Equation are introduced in Chapter~\ref{sec:theory}. The main objective of this thesis is to develop the regularization method further. The departure of our work from Holm et al. is highlighted. Two robust numerical solvers, a finite-difference solver and a particle method solver, are developed for the exposition of numerical solutions of the Geometric Thin-Film Equation. Furthermore, we discussed optimizations for the particle method to reduce the computational time greatly. The numerical solvers' convergence rate and computational complexity are also studied and compared. 

The primary motivation of the Geometric Diffuse-Interface method, as introduced in \cite{holm2020gdim}, is to preserve the gradient-flow structure of the thin-film equation. The resulting Geometric Thin-Film Equation then admits a family of singular solutions known as particle solutions. The particle solution transforms the PDE to a system of finite dimensional ODEs, which can be solved efficiently. The finite-dimensional system also allows us to apply standard dynamical system theory to study the behaviour of the solution. As we shall see in Chapter~\ref{sec:theory_wellpose}, the particle solutions provide a basis for the exposition of the regularity of the solution of the Geometric Thin-Film Equation for a wide class of measured-value initial conditions. The well-posedness of the Geometric Thin-Film Equation is also discussed. 

% In this thesis, we discuss a novel regularization method for the thin-film equation which build upon the work by Holm, \'O N\'araigh, and Tronci \cite{holm2020gdim} dubbed the Geometric Diffuse-Interface Method (G-DIM). The key idea of the G-DIM is to regularize the energy functional of the classical thin-film equation in such a way to produce a smooth energy gradient. This preserves the gradient flow structure of the thin-film equation which allows for the application of gradient flow theory to the droplet spreading problem. The G-DIM regularization of the thin-film equation is discussed in Chapter~\ref{sec:theory}. The resulting equation, the so-called Geometric Thin-Film Equation, exhibits some advantages compared to existing models, including the existence of a family of weak solutions of the form $h(x,t)=\sum_{i=1}^Nw_i\delta(x-x_i(t))$. This give rise of a Lagrangian numerical algorithm known as the particle method for solving the PDE. The particle method will be discussed in Chapter~\ref{sec:particle} and the analysis of the weak solution of the Geometric Thin-Film Equation is discussed in Chapter~\ref{sec:weak}.

To this point, the Geometric Thin-Film Equation only describes droplets in the complete wetting regime. An opposing force to the liquid-gas surface tension must be introduced to model partial wetting.  In \Cref{sec:partial}, the surface tension energy between the liquid and the solid substrate is modelled in the G-DIM framework and incorporated into the Geometrical Thin-Film Equation. An analytical description of the equilibrium droplet profile for the partial wetting model is obtained. We also explored the evolution of the solution numerically using methods developed in Chapter~\ref{sec:theory}. 

% Chapter~\ref{sec:3d} explores numerical methods for solving thin-film flow in the G-DIM framework in three spatial dimensions. This corresponds to solving the Geometric Thin-Film Equation with two-spatial variables (the height of the interface occupies the third dimension). Extending the finite-difference method to solve the Geometric Thin-Film Equation in higher dimensions is straightforward. The solution of the Geometric Thin-Film Equation with radial symmetry (rotational symmetry along the $z$-axis) is briefly discussed.  On the other hand, the optimization for the particle method of 2D droplets no longer applies in higher dimensions. Instead, we investigate a fast multipole method (FMM) for improving the numerical solver. 

\section{Review of the Classical Thin-Film Equation} \label{sec:review_classic}

\begin{figure}[thb]
    \centering
    \subfloat[]{
    \begin{tikzpicture}
        \fill [black!10] (0,0) rectangle (10,-0.6);
        % axes
        \draw [->, line width=0.3mm] (0,0) -- (10,0);
        \draw [->, line width=0.3mm] (0,0) -- (0,3);
        \draw (10,0) node[right] {$x$};
        \draw (0,3) node[above] {$z$};
        % length scale
        \draw [<->, line width=0.2mm] (3.14,0.8) -- (3*3.14,0.8);
        \draw (2*3.14,0.8) node[above] {$\lambda_0$};
        \draw [line width=0.2mm, dashed] (0,1.5) -- (10,1.5);
        \draw (0,1.5) node[left] {$h_0$};
        % interface
        \draw[orange,line width=0.7mm] plot[domain=0:10,smooth] (\x,{0.3*cos(180*\x/3.14)+1.5});
        \draw (9,2.2) node {$z=h(x,t)$};
        \draw (1.2,2.2) node {gas};
        \draw (1.2,0.8) node {liquid};
        \draw (1.2,-0.3) node {solid};
        % gravity
        \draw [<-, line width=0.2mm] (3.14,2.3) -- (3.14,2.9);
        \draw (3.14,2.6) node[right] {$g$};;
    \end{tikzpicture}}

    \subfloat[]{
    \begin{tikzpicture}
        \fill [black!10] (0,0) rectangle (10,-0.6);
        % axes
        \draw [->, line width=0.3mm] (0,0) -- (10,0);
        \draw [->, line width=0.3mm] (0,0) -- (0,3);
        \draw (10,0) node[right] {$x$};
        \draw (0,3) node[above] {$z$};
        % length scale
        % \draw [<->, line width=0.2mm] (0,-0.1) -- (2*3.14,-0.1);
        \draw (2*3.14,0) node[below] {$\lambda_0$};
        \draw (0,1.8) node[left] {$h_0$};
        % interface
        \draw[orange,line width=0.7mm] plot[domain=0:6.28,smooth] (\x,{1.8*(1-(\x/6.28)^2)});
        \draw (4,1.8) node {$z=h(x,t)$};
        \draw (1.2,2.2) node {gas};
        \draw (1.2,0.8) node {liquid};
        \draw (1.2,-0.3) node {solid};
        % gravity
        \draw [<-, line width=0.2mm] (8,1.3) -- (8,1.9);
        \draw (8,1.6) node[right] {$g$};
        % contact angle
        \draw [dashed] (6.28,0) -- (5.3,0.65);
        \draw [line width=0.3mm] (6.28-0.7,0) arc (180:147:0.7);
        \draw (5.5,0.3) node[left] {$\theta_{cl}(t)$};
        % contact line velocity
        \draw [->, line width=0.3mm] (6.28,0.02) -- (7,0.02) node[above] {$\dot{x}_{cl}(t)$};
        \fill [orange] (6.28,0) circle (0.08);
    \end{tikzpicture}}
    \caption{Schematic of a two-dimensional (a) thin-film flow and (b) droplet spreading setup. The orange line indicates the liquid-gas interface, and $h(x,t)$ describes the height of the interface measured from the substrate. $h_0$ and $\lambda_0$ are the typical vertical and horizontal length scales, respectively.}
    \label{fig:sketch_thinfilm}
\end{figure}
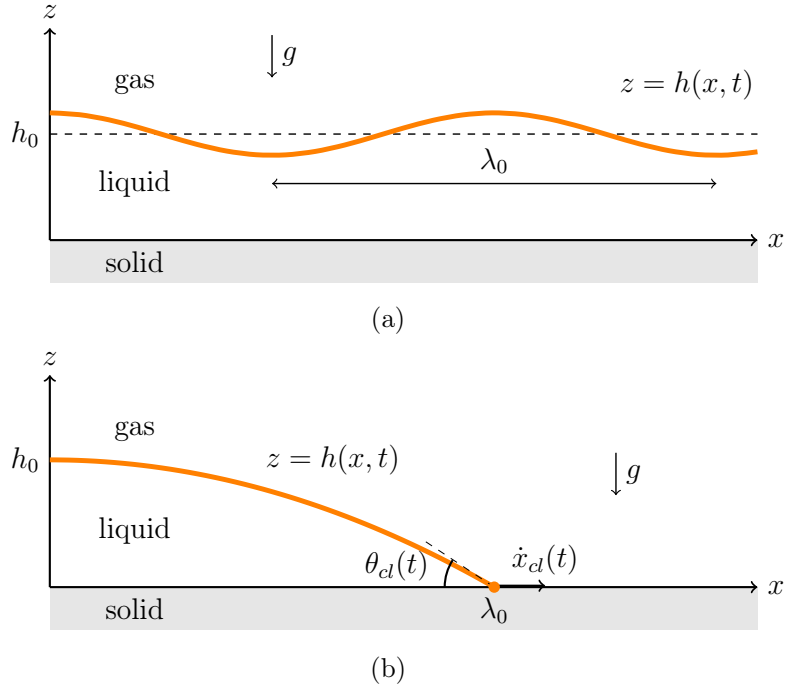

As with many of the droplet spreading models in the literature, the starting point for modelling droplet spreading is the thin-film equation. The thin-film equation arises when the horizontal length scale of the system is much larger than the vertical length scale ($\epsilon:=h_0/\lambda_0\ll1$, \cref{fig:sketch_thinfilm}). In such a limit (so-called lubrication limit), 
%the vertical component of the velocity field is assumed to be negligible, and the dynamics of the fluid flow can be drastically reduced. 
the Navier-Stokes equations then reduce to a single fourth-order parabolic partial differential equation (PDE) for the interface height $h(x,t)$, known as the thin-film equation. For two-dimensional surface-tension-driven thin-film flow, the thin-film equation is given by
\begin{equation} \label{eq:ctfe}
    \pder[h]{t} = -\pder{x}\left(\frac{\surften}{3\vis}h^3\pder[^3h]{x^3}\right), \qquad (x,t)\in\Omega\times[0,T],
\end{equation}
where $\gamma=\gamma_{lg}$ is the liquid-gas surface tension and $\vis$ is the constant dynamic viscosity of the fluid. % The derivation of the classical thin-film equation is given in Subsection~\ref{sec:classical}. 

The lubrication limit applies for thin droplets where $\partial_xh(x,t)\ll1$, which in turn corresponds to the near-complete-wetting regime ($\partial_xh(x_{cl},t)\approx\theta_{eq}\ll1$, where $x_{cl}$ is the contact line position). 
% We also assume Stokes flow for the flow within the film such that the capillary force dominates over the inertia forces. 
To discuss the relevant variables at play and the different regularization models, we show that \cref{eq:ctfe} arises from the Navier-Stokes equations in the lubrication limit. We refer the reader to \cite{oron1997long} for an in-depth discussion of the derivation.

\subsection{Lubrication Approximation}

We begin with the Navier-Stokes equations in two dimensions with gravity acting in the $-z$ direction
\begin{subequations}
\begin{equation} \label{eq:ns2d}
    \begin{cases}
        \rho \left(\partial_tu + u\partial_xu + w\partial_zu\right) = -\partial_xp + \mu\nabla^2u, \\
        \rho \left(\partial_tw + u\partial_xw + w\partial_zw\right) = -\partial_zp + \mu\nabla^2w - \rho g, 
    \end{cases}
\end{equation}
with the incompressible condition
\begin{equation} \label{eq:ns_incompressible}
    \partial_xu + \partial_zw = 0.
\end{equation}
Furthermore, we impose the no-slip boundary condition at the bottom wall and stress-free boundary condition at the interface
\begin{align} 
    u = w = 0 \qquad \text{at $z=0$}, \label{eq:ns_noslip} \\
    \partial_zu = 0 \qquad \text{at $z=h$}, \label{eq:ns_nostress}
\end{align}
as well as the kinematic boundary conditions at the free surface, which ensures that ``particles at the interface stay at the interface'', 
\begin{equation} \label{eq:ns_kinematic}
    w = \partial_th + u\partial_xh \qquad \text{at $z=h$}.
\end{equation}
In particular, we note that at equilibrium (i.e. $\partial_t h=0$), the velocity field at the interface is tangential to the interface $(u,w)\cdot(-\partial_x h,1) = 0$. 
Finally, we supplement the equation with the Young-Laplace equation for the pressure at the interface such that the pressure is proportional to the surface curvature $\kappa$:
\begin{equation} \label{eq:ns_pressure}
    p = -\gamma\kappa = -\gamma\frac{\partial_x^2h}{(1+(\partial_xh)^2)^{3/2}} \qquad \text{at $z=h$}.
\end{equation}
\end{subequations}
Next, we rescale the axes such that $\tilde{x},\tilde{z}=O(1)$, recalling that the ratio between the length scales is $\epsilon=h_0/\lambda_0$. %and nondimensionalize the rest of the equations as usual. 
As such, the non-dimensional variables are given by
\begin{gather}
    \tilde{x} = \frac{\epsilon x}{h_0}, \qquad \tilde{z} = \frac{z}{h_0}, \qquad \tilde{h} = \frac{h}{h_0}, \qquad
    \tilde{u} = \frac{u}{U_0}, \qquad \tilde{w} = \frac{w}{\epsilon U_0}.
    % \\ \tilde{t} = \frac{\epsilon U_0 t}{h_0}, \qquad \tilde{p} = \frac{\epsilon h_0}{\mu U_0}p.
\end{gather}

Applying the transformation, \cref{eq:ns2d} becomes
\begin{align*} 
    % \begin{dcases}
        \frac{\rho U_0^2}{\lambda_0} \left(\frac{\lambda_0}{U_0}\pder[\tilde{u}]{t} + \tilde{u}\pder[\tilde{u}]{\tilde{x}} + \tilde{w}\pder[\tilde{u}]{\tilde{z}}\right) &= -\frac{1}{\lambda_0}\pder[p]{\tilde{x}} + \frac{\mu U_0}{\lambda_0^2}\pder[^2\tilde{u}]{\tilde{x}^2} + \frac{\mu U_0}{h_0^2}\pder[^2\tilde{u}]{\tilde{z}^2}, \\
        \epsilon^2\frac{\rho U_0^2}{h_0} \left(\frac{\lambda_0}{U_0}\pder[\tilde{w}]{t} + \tilde{u}\pder[\tilde{w}]{\tilde{x}} + \tilde{w}\pder[\tilde{w}]{\tilde{z}}\right) &= -\frac{1}{h_0}\pder[p]{\tilde{z}} + \epsilon^3\frac{\mu U_0}{h_0^2}\pder[^2\tilde{w}]{\tilde{x}^2} + \epsilon\frac{\mu U_0}{h_0^2}\pder[^2\tilde{w}]{\tilde{z}^2} - \rho g.
    % \end{dcases}
\end{align*}
Next we divide by $\mu U_0/h_0^2$ and $\mu U_0/\epsilon h_0^2$ respectively
\begin{align*} 
    % \begin{dcases}
        \epsilon\frac{\rho U_0h_0}{\mu} \left(\frac{\lambda_0}{U_0}\pder[\tilde{u}]{t} + \tilde{u}\pder[\tilde{u}]{\tilde{x}} + \tilde{w}\pder[\tilde{u}]{\tilde{z}}\right) &= -\frac{\epsilon h_0}{\mu U_0}\pder[p]{\tilde{x}} + \epsilon^2\pder[^2\tilde{u}]{\tilde{x}^2} + \pder[^2\tilde{u}]{\tilde{z}^2}, \\
        \epsilon^3 \frac{\rho U_0h_0}{\mu} \left(\frac{\lambda_0}{U_0}\pder[\tilde{w}]{t} + \tilde{u}\pder[\tilde{w}]{\tilde{x}} + \tilde{w}\pder[\tilde{w}]{\tilde{z}}\right) &= -\frac{\epsilon h_0}{\mu U_0}\pder[p]{\tilde{z}} + \epsilon^4\pder[^2\tilde{w}]{\tilde{x}^2} + \epsilon^2\pder[^2\tilde{w}]{\tilde{z}^2} - \frac{\epsilon h_0^2\rho g}{\mu U_0}.
    % \end{dcases}
\end{align*}
Thus we are motivated to scale the time and pressure by
\begin{gather}
    \tilde{t} = \frac{U_0}{\lambda_0}t, \qquad \tilde{p} = \frac{\epsilon h_0}{\mu U_0}p.
\end{gather}
Furthermore, we identify the dimensionless Reynolds number and Gravity number to be
\begin{equation}
    \Rey=\frac{\rho U_0h_0}{\mu}, \qquad G=\frac{\epsilon h_0^2\rho g}{\mu U_0}.
\end{equation}
Finally (with the tildes dropped), we obtain
\begin{subequations}
\begin{equation} \label{eq:ns2d2}
    \begin{cases}
        \epsilon\Rey \left(\partial_tu + u\partial_xu + w\partial_zu\right) = -\partial_xp + \epsilon^2\partial_x^2u + \partial_z^2u, \\
        \epsilon^3 \Rey \left(\partial_tw + u\partial_xw + w\partial_zw\right) = -\partial_zp + \epsilon^4\partial_x^2w + \epsilon^2\partial_z^2w - G.
    \end{cases}
\end{equation}
Meanwhile, \cref{eq:ns_incompressible,eq:ns_noslip,eq:ns_nostress,eq:ns_kinematic} are invariant under the rescaling transformation, and \cref{eq:ns_pressure} becomes
\begin{equation} \tag{1.7f} \label{eq:ns_pressure2}
    p = \frac{1}{\Ca}\frac{\partial_x^2h}{(1+\epsilon^2(\partial_xh)^2)^{3/2}} \qquad \text{at $z=h$}, 
\end{equation}
\end{subequations}
where $\Ca$ is the (scaled) capillary number given by
\begin{equation*} 
    \qquad \Ca=\frac{U_0\mu}{\epsilon^3\gamma}.
\end{equation*}
In the lubrication limit, we take $\epsilon\ll1$ while asserting that $\Rey,\Ca=O(1)$. Such assertion implies that $\gamma$ is large relative to $U_0\mu$, i.e. the flow is surface tension driven. By dropping the $\epsilon$ terms, the leading-order approximation of \cref{eq:ns2d2} is given by
\begin{equation} \label{eq:ns2d3}
    \begin{cases}
        \partial_z^2u - \partial_xp = 0, \\
        \partial_zp + G = 0, 
    \end{cases}
\end{equation}
with \cref{eq:ns_pressure2} becoming
\begin{equation} \label{eq:tfe_derp}
    p=\frac{1}{\Ca}\partial_x^2h, \qquad \text{at $z=h$}. 
\end{equation}
In the lubrication theory, the flow is assumed to be laminar, and the inertia of the fluid is neglected. We note that it is possible to include the inertia effect by taking the next order approximation and that $\epsilon\Rey=O(1)$. Hence, the left-hand side of the momentum equation in the $x$-direction is kept. This procedure gives rise to the K\'arm\'an-Pohlhausen approximation~\cite{prokopiou1991}. 

Next, we integrate the second equation in \cref{eq:ns2d3} with respect to $z$ to obtain
\begin{equation}
    p + Gz = f(x,t),
\end{equation}
where $f(x,t)$ is some constant of integration, which can be determined by substituting \cref{eq:tfe_derp}, giving
\begin{equation}
    f(x,t) = \frac{1}{\Ca}\partial_x^2h + Gh.
\end{equation}
Hence, the pressure within the film is given by 
\begin{equation}
    p = \frac{1}{\Ca}\partial_x^2h + G(h-z),
\end{equation}
and 
\begin{equation} \label{eq:tfe_derp2}
    \partial_xp = \partial_x\left(\frac{1}{\Ca}\partial_x^2h + Gh\right).
\end{equation}
Since we have that $\partial_xp$ is independent of $z$, integrating the first equation in \cref{eq:ns2d3} from $h$ to $z$ and apply the no-stress boundary condition gives
\begin{equation}
    \partial_zu = \partial_xp(z-h).
\end{equation}
Integrating once more, this time from $0$ to $z$ and applying the no-slip condition gives
\begin{equation} \label{eq:tfe_deru}
    u = \partial_xp \left(\frac{1}{2}z^2-hz\right).
\end{equation}

On the other hand, integrating \cref{eq:ns_incompressible} from $0$ to $h$ with respect to $z$, we get
\begin{equation*}
    w|_{z=h}-w|_{z=0} = -\int_0^h \partial_xu \,\mathd z.
\end{equation*}
Next, we substitute in the boundary conditions for $w$ from \cref{eq:ns_noslip,eq:ns_kinematic} to obtain
\begin{equation} \label{eq:tfe_der1}
    \partial_th + u|_{z=h}\partial_xh = -\int_0^h \partial_xu \,\mathd z.
\end{equation}
By the Leibniz integral rule, we have
\begin{equation*}
    \partial_x\int_0^h u\,\mathd z = u|_{z=h}\partial_xh + \int_0^h \partial_xu\,\mathd z.
\end{equation*}
So \cref{eq:tfe_der1} becomes
\begin{align} 
    \partial_th = -\partial_x\int_0^h u\,\mathd z = \partial_x\left(\frac{1}{3}h^3\partial_xp\right).
\end{align}
Finally, using \cref{eq:tfe_derp2} we obtain
\begin{equation} \label{eq:hdef}
    \partial_th = \partial_x\left\{\frac{1}{3}h^3\partial_x\left(\frac{1}{\Ca}\partial_x^2h + Gh\right)\right\}. 
\end{equation}
% We will refer back to this derivation in later section when we extend the model to account for different external effect. 

% \begin{table}[h]
%     \centering
%     \begin{tabular}{|c|c|c|}
%         \hline 
%         Variable & Description & Dimension \\
%         \hline \hline
%         $u, v$ & Velocity field in the $x$ and $y$-direction & $LT^{-1}$ \\
%         $\rho$ & Density of liquid & $ML^{-3}$ \\
%         $\mu$ & Dynamic viscosity & $ML^{-1}T^{-1}$ \\
%         $p$ & Pressure & $ML^{-1}T^{-2}$ \\ 
%         $f_x, f_y$ & External force in the $x$ and $y$-direction & $MLT^{-2}$ \\
%         $\Rey$ & Reynold's number & $1$ \\
%         $\Ca$ & Capillary number & $1$ \\
%         \hline
%     \end{tabular}
%     \caption{Caption}
%     \label{tab:my_label}
% \end{table}

It is worthwhile to remark that the gravity number can be rewritten as the ratio between the Bond number and the capillary number,
\begin{equation}
    G = \frac{\Bo}{\Ca},
\end{equation}
where $\Bo$ is given by
\begin{equation}
    \Bo = \frac{\lambda_0^2\rho g}{\gamma}.
\end{equation}
We observe that $\Bo$ is tied directly to the size of the droplet through $\lambda_0$. In particular, if one wishes to neglect the gravitational effect (i.e. $G\ll1$ regime), the Earth's gravity limits the droplet's size. For water droplets, this limit is no more than a few millimetres. We note that the limit can be circumvented by conducting experiments in a micro-gravity environment, which is what McCraney et al. \cite{mccraney2022} did on board the International Space Station. This allows them to study the contact-line dynamics and vibration modes of droplets up to a few centimetres in radius. 

% \commentkp{Expand this paragraph} 
In the capillary-dominant case, we can rewrite \cref{eq:hdef} as 
\begin{equation}
    \Ca\partial_th = \partial_x\left\{\frac{1}{3}h^3\partial_x\left(\partial_x^2h + \Bo h\right)\right\},  
\end{equation}
which suggests a new, much shorter timescale for the system
\begin{equation}
    \tau = \frac{1}{\Ca}\tilde{t} = \frac{\epsilon^3\gamma}{\lambda_0\mu}t.
\end{equation}
% This limit is used by \cite{hocking1981sliding,hocking1982the,holm2020gdim}
In the other extreme, when $\Ca\ll1$, the time derivative can be dropped completely. This corresponds to a quasi-steady spreading state regime where the solution can be approximated by a parabolic profile and spreading is modelled explicitly \cite{greenspan1978}.

\subsection{Similarity Solution}

In the context of droplet spreading, it is desirable to propose a similarity solution to Equation~\eqref{eq:hdef}, corresponding to a self-similar droplet that retains some overall structural properties even as the base of the droplet spreads out. In this section, we consider the equation 
% derive the similarity solution for a slight generalization of the mobility with $h^n$:
\begin{equation}
    \frac{\partial h}{\partial t}+\frac{\partial}{\partial x}\left(h^n\frac{\partial^3 h}{\partial x^3}\right)=0.
    \label{eq:hdefn}
\end{equation}
When $n=3$, we recover the classical thin-film equation, and $n=1$ describes the flow in a Hele-Shaw cell \cite{constantin1993}. 
We start by considering the transformation 
\begin{equation}
    z=\varepsilon^ax, \qquad s=\varepsilon^bt, \qquad \psi(z,s)=\varepsilon^ch(\varepsilon^{-a}z,\varepsilon^{-b}s).
\end{equation}
Under the transformation, Equation~\eqref{eq:hdefn} becomes 
\begin{equation}
    \varepsilon^{b-c}\pder[\psi]{s} = \varepsilon^{4a-(n+1)c}\pder{z}\left(\psi^n\pder[^3\psi]{z^3}\right).
    \label{eq:simtr}
\end{equation}
For Equation~\eqref{eq:simtr} to be invariant under the transformation, we require $c = (4a-b)/n$. Next, we look for solutions of the form
\begin{equation}
    h(x,t)=t^{c/b}f(\eta), \qquad \eta=xt^{-a/b}.
    \label{eq:simtr2}
\end{equation}
Then Equation~\eqref{eq:hdefn} becomes
\begin{equation}
    t^{c/b-1}\left(\frac{c}{b}f-\frac{a}{b}\eta f'+\der{\eta}(f^nf''')\right)=0.
\end{equation}
If we set $a=-c$, then the expression within the bracket becomes a total derivative of $\eta$
\begin{equation}
    t^{c/b-1}\der{\eta}\left(\frac{c}{b}\eta f+f^nf'''\right)=0.
\end{equation}
So the similarity solution is given by
\begin{equation} \label{eq:similarity}
    h(x,t)=t^{-\alpha}f(\eta), \qquad \eta=xt^{-\alpha} , \qquad \alpha=\frac{1}{n+4},
\end{equation}
and Equation~\eqref{eq:hdefn} reduces to
\begin{equation} \label{eq:simtr3}
    f^nf'''=\alpha\eta f.
\end{equation}

An appropriate droplet spreading boundary conditions for Equation~\eqref{eq:simtr3} is
\begin{align} \label{eq:simbc}
\begin{split}
    f(0)=1, \qquad f'(0)=0, \qquad \\
    f=f'=0, \qquad \text{at $\eta=\eta_0>0$},
\end{split}
\end{align}
where $\eta_0$ corresponds to the outermost extent of the droplet. In other words, we require the solution to be compactly supported and touch down at $\eta=\eta_0$ in a smooth manner. 

\begin{figure}
    \centering
    \includegraphics[width=0.7\textwidth]{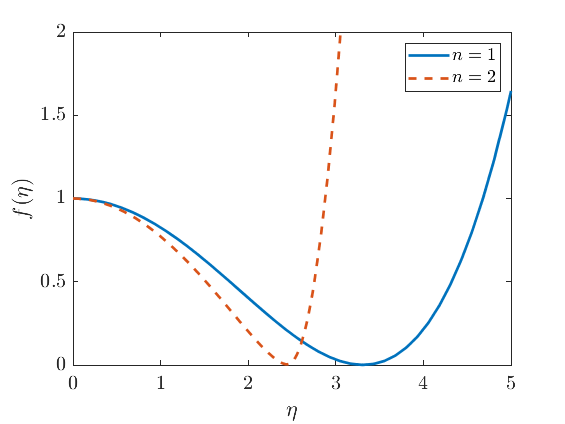}
    \caption{Similarity solution using the shooting method for case $n=1$ and $n=2$.}
    \label{fig:intro_sim}
\end{figure}

Such one-dimensional boundary-value problems can be solved via the shooting met\-hod \cite{hulshof2001some}, where we fixed the left ($\eta=0$) boundary condition and vary the value of $f''(0)$ such that the right boundary condition at $\eta=\eta_0$ is satisfied (e.g. Figure~\ref{fig:intro_sim}). Bernis et al. \cite{bernis1992} prove that Equation~\eqref{eq:simtr3} with boundary condition Equation~\eqref{eq:simbc} has no non-trivial solution for $n\geq3$, and that there only exists a unique even and nonnegative solution when $0<n<3$. At the critical value $n=3$, which corresponds to the no-slip assumption at $h=0$, $f(\eta)$ degenerates into a Dirac delta function centred at $\eta=0$, and the droplet does not spread \cite{bernis1999}.  

\subsection{Contact Line Singularity}
Tanner pioneered empirical measurement of droplet spreading \cite{tanner1979} by means of measuring the optical lensing caused by the refraction from the droplet. In the complete-wetting regime, it is found that the relation between the contact-line velocity and the contact angle is $\dot{x}_{cl}\sim\theta_{cl}^3$. Here we have used the dot to represent the derivative with respect to $t$. This relation is now known as Tanner's law. Assuming a parabolic droplet profile of the form $h(x,t)=A(x_{cl}(t)^2-x^2)$, then for fixed droplet volume $V$, the contact angle is completely determined from the contact line position:
\begin{align}
    % h(x,t) &= \frac{3V}{4x_{cl}^3}(x_{cl}^2-x^2) \\
    % h(r,t) &= \frac{2V}{\pi r_{cl}^4}(r_{cl}^2-r^2)
    \partial_xh(x_{cl}(t),t) = \frac{3V}{2x_{cl}(t)^2}, \qquad \text{in 2D}, \\ 
    \partial_rh(r_{cl}(t),t) = \frac{4V}{\pi r_{cl}(t)^3}, \qquad \text{in 3D}.
\end{align}
In the spirit of lubrication theory, taking $\partial_xh(x_{cl},t)=\theta_{cl}$, then applying Tanner's law and integrating once yield the relation $x_{cl}\sim t^{1/7}$ for two-dimensional droplets and $r_{cl}\sim t^{1/10}$ for three-dimensional droplets. The parabolic-profile assumption is not special, and the same conclusion can be reached with other profile shapes. We remark that the $1/7$ exponent agrees with the similarity variable \cref{eq:similarity} with $n=3$. This gives us a first hint that the classical thin-film equation is still relevant in describing droplet spreading. 

The reason for the failure of the classical thin-film equation in describing a moving contact line owes to the fact that \cref{eq:hdef} inherits the no-slip assumption from the Navier-Stokes equations. Since the contact line is a point at the boundary, in contact with the substrate, a moving contact line is incompatible with the no-slip assumption. This manifests as a pressure field that behaves like $1/r$ as the contact line is approached \cite{huh1971hydrodynamic}. The force exerted on the substrate over a patch of length $r$ is, therefore, $F\propto\log r$ and thus unbounded. As Huh and Scriven originally put it:

\begin{quote}
    Not even Hercules could sink a solid if the physical model were entirely valid.
\end{quote}

The resolution of the moving-contact-line paradox is that there is missing physics and that on a sufficiently small scale, there is slip, the dynamics of which are governed by the interactions between the fluid molecules and the substrate molecules~\cite{dussan1979on}.  These molecular-level interactions can be incorporated into a macroscopic fluid model via a so-called regularization technique.  
In practice, we regularize the classical thin-film equation in such a way as to resolve the singularity at the contact line while keeping the desired physical properties of the original model. In general, the regularization procedure introduces a small regularization parameter $\varepsilon\ll1$ such that the regularization effect dominates when $h(x,t)=O(\varepsilon)$ and the thin-film flow dynamics is recovered when $h(x,t)\gg\varepsilon$. Various regularisation methods have been proposed in the literature. In this work, we will focus our attention on two well-studied regularization techniques rooted in the lubrication theory, namely the slip-length model~\cite{greenspan1978,hocking1981sliding,hocking1983spreading} and the precursor film model~\cite{hervet1984,de1985wetting,oron1997long}. 
Both regularization models give accurate and consistent descriptions of the contact-line spreading~\cite{eggers2004characteristic,savva2011dynamics} and yield the same qualitative and quantitative results when used to model droplet dynamics. This consistency between the different approaches gives a solid justification for the general approach of model regularization. 

% Using the theory of matched asymptotic expansions, it has been shown~\cite{hocking1982the} that the solution of Equation~\eqref{eq:hdefslip} consists of an outer solution and an inner solution.  The outer solution resembles the similarity solution~\eqref{eq:similarity1} and is valid on large scales, far from the contact line.  The inner solution is valid on small scales close to the contact line, and provides for contact-line motion.  Indeed, by matching the inner and outer solutions across an intermediate matching zone, the contact-line $a(t)$ is shown to satisfy the so-called \textit{Tanner's Law},
% %
% \begin{equation}
% \frac{\mathd a}{\mathd t}=\frac{\surften\theta_0^3R^6}{3\vis}\epsilon\left[1+\epsilon\left(2-\ln\frac{2a}{R}\right)\right]a^{-6}.
% \end{equation} 
% Where $\theta_0$ is the initial contact angle. Thus, $a(t)\sim t^{1/7}$, which is the scaling that would be expected if the similarity solution could be made to extend down to the microscopic scale.

% The slip-length model therefore provides for a resolution of the contact-line singularity.  However, the stress $\surften h_{xx}$ remains singular at the contact line.  For these reasons, an alternative regularization of the Thin-Film Equation has been proposed, namely the Precursor-Film model~\cite{de1985wetting,bonn2009wetting}.    Following Reference~\cite{holm2020gdim}, in this work we present the Geometric Thin-Film Equation as an alternative regularization, the advantage of this approach as we reveal in subsequent sections is the remarkable simplicity of the numerical solutions produced by this model.

\section{Classical Droplet Spreading Models}
\label{sec:intro_motivation}

% The resolution of the moving contact line paradox is that there is missing physics, and that on a sufficiently small scale, there is slip, the dynamics of which are governed by the interactions between the fluid molecules and the substrate molecules~\cite{dussan1979on}.  These molecular-level interactions can be incorporated into a macroscopic fluid model via a so-called regularization technique.  In this work, we focus our attention on two regularization techniques in the literature -- the slip length~\cite{hocking1981sliding} and the precursor film~\cite{de1985wetting}.  Although methodologically distinct, these yield the same qualitative and quantitative results when used to model droplet dynamics.  This consistency between the different approaches gives a solid justification for the general approach of model regularization. 

% In this thesis, a novel regularization method that does not fall into any of the methods above is discussed. The theory is based upon a previous work by Holm, \'O N\'araigh, and Tronci \cite{holm2020gdim} known as the geometric diffuse-interface method (G-DIM), which aims to address some disadvantages of the previous models. In this section, we briefly discuss the key ideas of the three regularization methods and to highlight the role of the G-DIM in modelling droplet spreading. 

\subsection{Slip-length Model} 
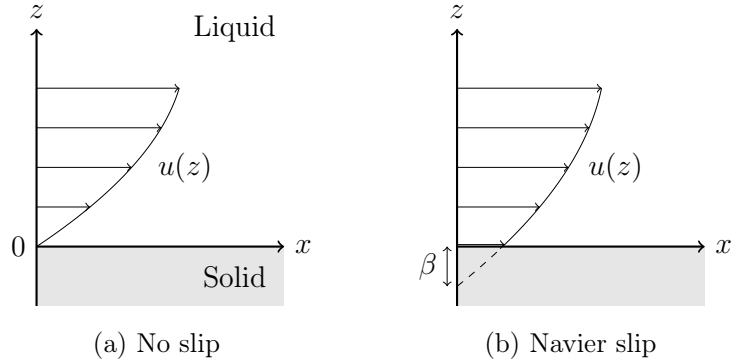
\begin{figure}
    \centering
    \subfloat[No slip]{
    \begin{tikzpicture}[scale=1.4]
        \fill [black!10] (0,0) rectangle (2.5,-0.6);
        \draw (2,-0.3) node {Solid};
        \draw (2,2.2) node {Liquid};
        \draw [->,line width=0.3mm] (0,0) node[left] {$0$} -- (2.5,0) node[right] {$x$};
        \draw [->,line width=0.3mm] (0,-0.6) -- (0,2.2) node[above] {$z$};
        \draw (1.5,0.8) node {$u(z)$};
        % interface
        % \draw [orange,line width=0.7mm] plot[domain=0:2.5,smooth] (\x,{0.3*cos(180*\x/3.14)+1.3});
        % velocity profile
        \draw plot[domain=-2:-0.4,smooth] ({0.375*\x^2+1.5},\x+2);
        \foreach \y in {-0.4,-0.8,-1.2,-1.6}
        {\draw [->] (0,\y+2) -- ({0.375*\y^2+1.5},\y+2);}
    \end{tikzpicture}
    }
    \hspace{2em}
    \subfloat[Navier slip]{
    \begin{tikzpicture}[scale=1.4]
        \fill [black!10] (0,0) rectangle (2.5,-0.6);
        \draw [->,line width=0.3mm] (0,0) -- (2.5,0) node[right] {$x$};
        \draw [->,line width=0.3mm] (0,-0.6) -- (0,2.2) node[above] {$z$};
        \draw (1.6,0.8) node {$u(z)$};
        % interface
        % \draw [orange,line width=0.7mm] plot[domain=0:2.5,smooth] (\x,{0.3*cos(180*\x/3.14)+1.3});
        % velocity profile
        \draw plot[domain=-2:-0.4,smooth] ({0.260*\x^2+1.5},\x+2);
        \draw [dashed] plot[domain=-2.4:-2,smooth] ({0.260*\x^2+1.5},\x+2);
        \foreach \y in {-0.4,-0.8,-1.2,-1.6,-1.98}
        {\draw [->] (0,\y+2) -- ({0.260*\y^2+1.5},\y+2);}
        % slip length
        \draw [<->] (-0.1,0) -- (-0.1,-0.4);
        \draw (-0.1,-0.2) node[left] {$\beta$};
        \end{tikzpicture}
    }
    \caption{Geometrical interpretation of the boundary conditions. The slip-length parameter $\beta$ can be thought of as the extrapolation length of the velocity profile into the substrate. }
    \label{fig:sketch_slip}
\end{figure}

In the slip-length model, the Navier-slip condition replaces the no-slip boundary condition. For two-dimensional droplets, this is given by
\begin{equation}
    u = \beta\pder[u]{z}, \qquad w = 0, \qquad \text{at $z=0$},
\end{equation}
where $0<\beta\ll1$ is the slip-length parameter. Thus, the fluid is allowed slip parallel to the substrate, and the velocity is directly proportional to the shear stress at the boundary (e.g. \Cref{fig:sketch_slip}). Using the Navier-slip condition, \cref{eq:tfe_deru} becomes
\begin{equation}
    u = \partial_xp\left(\frac{1}{2}z^2-hz-\tilde{\beta} h\right),
\end{equation}
where $\tilde{\beta}=\beta/h_0$ is the nondimensionalized slip length coefficient. The updated thin-film equation with slip length is now given by
\begin{align} \label{eq:hdefslip}
    \pder[h]{t}+\frac{1}{\Ca}\pder{x}\left\{\left(\frac{1}{3}h^3+\tilde{\beta}h^2\right) \pder{x}\left(\pder[^2h]{x^2} + \Bo h\right)\right\}=0.
    % m(h) = \frac{1}{3}h^3+\tilde{\beta} h^2,
\end{align}
Note that $\tilde{\beta}$ is the regularization parameter for the slip-length model and that \cref{eq:hdef} is recovered when $h\gg\tilde{\beta}$. Furthermore, the boundary conditions at the contact line of a compactly supported droplet are given by
\begin{gather}
    % \partial_xh(0,t) = \partial_{xxx}h(0,t) = 0, \\
    h(x_{cl}(t),t)=0, \qquad \partial_xh(x_{cl}(t),t) = -\theta_{cl}(t).
\end{gather}

The second piece of the slip-length model is the equation of motion of the contact line. Often, this is modelled \textit{a posteriori}. An example would be basing the model on Tanner's law \cite{oron1997long}, giving an equation of the form $\dot{x}_{cl}(t) = K(\theta_{cl}(t)-\theta_{eq})^3$, for some constant $K$. An alternative approach involves a procedure of matched asymptotic expansion (derivation given in Appendix A), which gives a contact line velocity of
\begin{equation} \label{eq:velhocking}
    \dot{x}_{cl}(t) = \frac{1}{3}\varepsilon\left(x_{cl}(t)^{-6}-\theta_{eq}^3\right)\left(1+\varepsilon(2-\ln(2x_{cl}(t)))\right),
\end{equation}
for another small parameter $\varepsilon=1/\ln(\beta^{-1})$. Tanner's law emerges as an $\varepsilon^1$-approxi\-ma\-tion of \cref{eq:velhocking}.

Due to the requirement of explicit modelling of contact-line motion, the slip-length model is usually only practical for 2D droplets or 3D axisymmetrical droplets, where the contact line is simple. The motion becomes difficult to model for more general systems, where the contact line exhibits complex shapes. Numerical methods for the slip-length model also need to pay extra attention to the contact line where only the one-sided derivatives are defined. We look to the precursor film model for implicit modelling of the contact line dynamics.

\subsection{Precursor Film Model}

\begin{figure}[htb]
    \centering
    \begin{tikzpicture}
        \fill [black!10] (0,0) rectangle (10,-0.6);
        % axes
        \draw [->, line width=0.3mm] (0,0) -- (10,0) node[right] {$x$};
        \draw [->, line width=0.3mm] (0,0) -- (0,3) node[above] {$z$};
        % interface
        \draw[orange,line width=0.7mm] plot[domain=0:5,smooth] (\x,{1.5*(1-(\x/5)^2)+0.3});
        \draw [orange, line width=0.7mm] (5,0.3) arc [start angle=240, end angle=270, x radius=1, y radius=1];
        \draw[orange,line width=0.7mm] (5.5,0.17) -- (9,0.17);
        \draw (3.4,1.7) node {$h(x,t)$};
        \draw [orange, line width=0.7mm] (9,0.16) arc [start angle=90, end angle=0, x radius=0.5, y radius=0.17];
        % film thickness
        \draw [->] (8,-0.3) -- (8,0);
        \draw [->] (8,0.47) -- (8,0.17);
        \draw (8,0.47) node[above] {Precursor film $h_\infty$};
        % contact line region
        \draw (5,0.4) rectangle (5.5,0);
        \draw (5.5,0.4) -- (6.5,2) node[above] {Contact region};
        \draw (1.2,2.2) node {gas};
        \draw (1.2,0.8) node {liquid};
        \draw (1.2,-0.3) node {solid};
    \end{tikzpicture}
    \caption{Sketch of the droplet flow geometry with a precursor film of thickness $h_\infty$ in beyond the contact region. }
    \label{fig:sketch_precursor}
\end{figure}
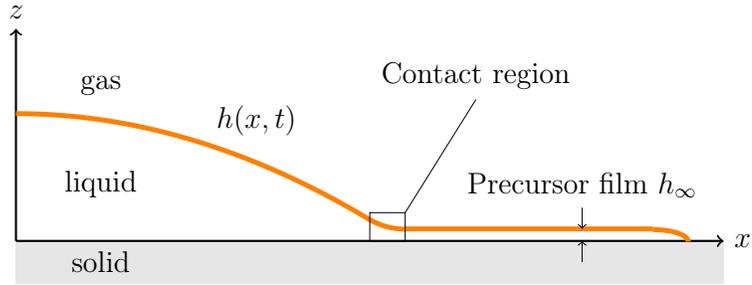

The precursor film first proposed by Hervet and de Gennes \cite{hervet1984} is a microscopic film preceding the droplet core, prewetting the substrate in advance of the contact line. A sketch of the system is shown in Figure~\ref{fig:sketch_precursor}. This spreading on the prewetted surface essentially alleviates the triple point and the contact line singularity. However, the transition from the droplet core to the precursor film is a smooth one. This means the contact line is no longer well-defined, and a contact region, also called the droplet foot, is discussed instead.  

\begin{figure}
    \centering
    \begin{tikzpicture}[scale=1.5]
        % axes
        \draw [->, line width=0.3mm] (0,-1.5) -- (5,-1.5) node[right] {$h$};
        \draw [black!30] (5,0) -- (0,0) node[left] {$0$};
        \draw (0,0) node[left] {$0$};
        \draw [->, line width=0.3mm] (0,-1.5) -- (0,1.6);
        % disjoining pressure
        \pgfmathsetmacro{\eps}{0.1};
        \draw [orange,line width=0.7mm] plot[domain=0.092:0.5,samples=100] (\x/0.1,{10*(-(\eps/\x)^3+(\eps/\x)^4)});
        \draw (4,-0.2) node[below] {$\Phi(h)$};
        % regions
        \draw [dashed] (1.32,-1.5) node[below] {$h_\infty$} -- (1.32,1.5);
        \draw (0.66,1.4) node {Repulsive};
        \draw (3.16,1.4) node {Attractive};
        % epsilon
        \draw [<->] (0,0) -- (1,0);
        \draw (0.5,0) node[above] {$\varepsilon$};
    \end{tikzpicture}
    \caption{Plot of the two-term potential and the corresponding repulsive and attractive lengthscales.}
    \label{fig:sketch_vdw}
\end{figure}
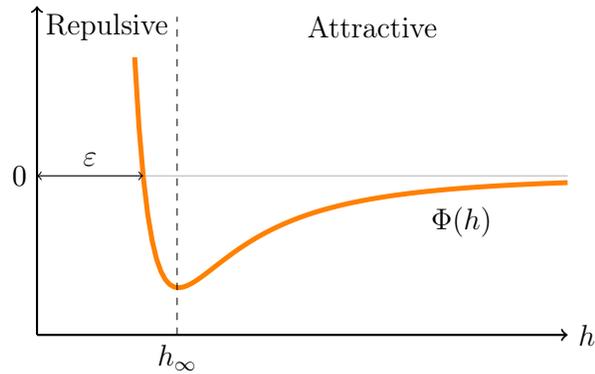

Although physically, such a precursor film does exist, it has a very small thickness, typically around $10-100\,\mathrm{nm}$~\cite{bonn2009wetting}. In such small-length scales, the precursor is maintained by the van der Waals forces between the solid and the liquid. This is often modelled using a two-term disjoining pressure, corresponding to the short-range repulsive and long-range attractive forces, respectively, of the form
\begin{equation} \label{eq:vdwdef}
    \Phi(h) = \mathcal{A}\left[\left(\frac{\varepsilon}{h}\right)^n-\left(\frac{\varepsilon}{h}\right)^m\right].
\end{equation}
for some $n>m>1$, $\mathcal{A}$ is the Hamaker constant, usually treated as a system parameter. $\varepsilon$ is the small parameter of the precursor model, which defines the thickness of the precursor film $h_\infty=O(\varepsilon)$. A plot of the disjoining pressure is shown in Figure~\ref{fig:sketch_vdw}. The long-range attractive term creates a potential well, so the ``dry'' region is energetically favourable. In contrast, the short-range repulsive term prevents the rupturing of the liquid film, which would lead to a finite-time blowup. With this, the thin-film equation with van der Waals force is given by
\begin{align} \label{eq:hdefprecursor}
    \pder[h]{t}+\frac{1}{\Ca}\pder{x}\left\{\frac{1}{3}h^3\pder{x}\left(\pder[^2h]{x^2} + \Bo h + \Phi(h)\right)\right\}=0.
\end{align}
Again, when $h\gg\varepsilon$, the disjoining pressure is very close to zero, and the original thin-film equation is recovered. 
Even though the contact line is not well defined, it is still possible to compute and relate the equilibrium contact angle to the precursor film parameters as shown by \cite{schwartz1998hysteretic}. Using a force balance equivalence in the contact region and taking the $\theta_{eq}$ to be the angle at the inner edge of the contact region, in the lubrication limit, one obtains the relation 
\begin{equation}
    \theta_{eq} = \sqrt{\frac{2h_\infty\mathcal{A}(n-m)}{(n-1)(m-1)}}.
\end{equation}
The energy associated with the van der Waals potential can be shown to be
\begin{equation}
    E_{vdw}[h] = -\int_{-\infty}^\infty \mathcal{A}h\left[\frac{1}{-n+1}\left(\frac{\varepsilon}{h}\right)^n-\frac{1}{-m+1}\left(\frac{\varepsilon}{h}\right)^m\right]\,\mathd x,
\end{equation}
which competes with the surface-tension energy. The balance of the energies decides the final equilibrium shape of the droplet. This implicitly determines the contact line dynamics, in contrast to the explicit modelling of the contact line motion in the slip-length model. 

Despite the robustness of the precursor film model, the thickness of the precursor film means that such a small scale must be resolved in the model: in particular, the numerical grid size must be at least as small as the precursor-film thickness~\cite{schwartz1998hysteretic}.  Although this approach is just about feasible for millimetre-scale droplets, it may not be possible for larger ones. Beyond the millimetre scale, an unphysically large precursor-film thickness can be used in numerical investigations (and the results checked for robustness to changes in the value of the precursor-film thickness).   

\section{\markup{Preliminaries}}

In this section, we briefly introduce some preliminaries and definitions that will be helpful in later chapters for the novel regularization of the thin-film equation and the classification of the solutions of the regularized thin-film equation. 

\subsection{\markup{Gradient Flows}}

The thin-film equation is formally a class of partial differential equations with a gradient flow structure \cite{lisini2012}. This places the thin-film equation with many well-known equations such as the Holm-Putkaradze aggregation equation for chemotaxis \cite{holm2005aggregation}, the Cahn-Hilliard equation for the separation of binary fluids, and the Fokker–Planck equation for generalized diffusion. The gradient flow equation has the form 
\begin{equation} \label{eq:intro_gf}
    \pder[h]{t} = \pder{x}\left( m(h)\pder{x}\frac{\delta E}{\delta h} \right),
\end{equation}
where $m(h)$ is a non-negative function known as mobility. The functional $E:H^1_0(\mathbb{R})\rightarrow\mathbb{R}$ is called the free-energy of the system
\begin{equation}
    E[h] = \int_{-\infty}^\infty L(x,h,\partial_x h) \,\mathd x.
\end{equation}
In Equation \eqref{eq:intro_gf}, the functional derivative of $E$ is taken. This can be computed using the Euler-Lagrange equation as
\begin{equation}
    \frac{\delta E}{\delta h} = \pder[L]{h} - \pder{x}\pder[L]{(\partial_x h)}. 
\end{equation}
An important aspect of the gradient flow equation is that it is mass preserving, and the energy $E(t)$ decreases monotonically. More precisely, at every time $t$, the solution $h(x,t)$ evolves in a way as to minimize $E[h](t)$ locally at $h$ until a minimum is reached. Therefore, the system evolves in the direction of the negative of the gradient of $E$ with respect to $h$, hence the name gradient flow. 
% Since many physical phenomena are conservative dynamics

For the thin-film equation, the surface tension seeks to minimize the surface area of the fluid interface, and the associated surface tension energy of the system is given by
\begin{equation}
    E_{lg}[h] = \int_{-\infty}^\infty \frac{1}{2}\surften(\partial_xh)^2 \,\mathd x + E_0,
\end{equation}
where $E_0$ is some constant reference energy, giving the gradient
\begin{equation}
    \frac{\delta E_{lg}}{\delta h} = -\gamma\pder[^2h]{x^2},
\end{equation}
and the mobility is given by $m(h) = \tfrac{1}{3\mu}h^3$. 

Many of the gradient flow equations, such as the Holm-Putkaradze aggregation equation, admit particle solutions of the form
\begin{equation} \label{eq:intro_particle}
    h^N(x,t) = \sum_{i=1}^N w_i\delta(x-x_i(t)).
\end{equation}
This is a class of weak solutions where the $h$ is given by a linear combination of weighted Dirac measures located at $x_i$. The particle solutions provide a few desirable properties, including positivity preserving and mass conserving. However, the thin-film equation is incompatible with the particle solutions due to the hyper-diffusivity, where the point-like particles get smoothed out and do not remain particles. 
%
% \begin{equation}
%     \dot{x}_i = \left[f(h)\pder{x}\frac{\delta E}{\delta h}\right]_{x=x_i}
% \end{equation}
%
In a recent work by Carrillo et al. \cite{carrillo2019}, they developed a deterministic particle method for diffusive processes, where particles remain particles. Inspired by their work, we seek a regularization of the thin-film equation in a similar manner that admits particle solutions. 

\subsection{\markup{Function Spaces}}

Consider a general fourth-order Cauchy problem
\begin{subequations} \label{eq:cauchy_prob}
\begin{equation}
    \pder[h]{t} = F\left( t,h,\partial_xh,\partial_x^2h,\partial_x^3h,\partial_x^4h \right), \qquad \text{in $\Omega\times (0,T)$},
\end{equation}
satisfying an initial condition and a boundary condition
\begin{align}
    h &= h_0,  \qquad \text{on $\Omega\times\{0\}$}, \\
    h &= 0,  \qquad \text{on $\partial\Omega\times[0,T]$}. 
\end{align}
\end{subequations}
Intuitively, a solution of Equation~\eqref{eq:cauchy_prob} is one that evolves continuously over time. For this, we introduce the following definition. 

\begin{definition}
Let $X$ be a Banach space with norm $\|\cdot\|$. The space $C^k([0,T];X)$ comprises all $C^k$ functions $u:[0,T]\rightarrow X$ with
\begin{equation}
    \|u\|_{C^k([0,T];X)} := \sum_{i=0}^k \max_{0\leq t\leq T}\|u^{(i)}(t)\| \leq \infty.
\end{equation}
With slight abuse of notation, we sometimes write $u(\cdot,t)\in X$ to denote $u(t)$ for $t\in[0,T]$. 
\end{definition}

A classical solution of Equation~\eqref{eq:cauchy_prob} is a function $h\in C^1((0,T);C^4(\Omega))$ where the solution is continuously differentiable. However, a differential equation may have solutions which are not differentiable in the classical sense but nevertheless satisfy a variational form of the PDE:
\begin{equation}
    -\int_0^T\int_{\partial\Omega} h\pder[\phi]{t}\,\mathd t\,\mathd x = \int_0^T\int_{\partial\Omega} F\left( t,h,\partial_xh,\partial_x^2h,\partial_x^3h,\partial_x^4h \right)\phi \,\mathd t\,\mathd x,
\end{equation}
for all $\phi\in C_c^\infty(\Omega\times(0,T))$ compactly supported and infinitely differentiable test functions. 
The particle solutions are one such family of solutions. To define the space of weak solutions, we introduce a few more notations. 

Given a Banach space $X$ with norm $\|\cdot\|$ we denote by $B_X:=\{x\in X : \|x\|\leq 1 \}$ its closed unit ball.
Let $C_0(\mathbb{R})$ denote the space of continuous real-valued functions that vanish at infinity, equipped with the supremum norm $\|\cdot\|_\infty$. The space of (signed) Radon measures $\mathcal{M}(\mathbb{R})$ is then the dual space of $C_0(\mathbb{R})$. That is, $\mathcal{M}(\mathbb{R})$ is the space of continuous linear functionals $\mu:C_0(\mathbb{R})\rightarrow\mathbb{R}$ and we write
\begin{equation}
    \mu(f) := \int_{-\infty}^\infty f(x) \,\mathd\mu(x), \qquad f\in C_0(\mathbb{R}). 
\end{equation}
The Radon measures space is endowed with the total variation norm $\|\cdot\|_1$ given by
\begin{equation}
    \|\mu\|_1 = \sup\left\{\mu(f):f\in B_{C_0(\mathbb{R})}\right\}. 
\end{equation}
Furthermore, we require the solutions of the thin-film equation to be nonnegative, and for that, we consider only the space of positive Radon measures
\begin{equation}
    \mathcal{M}^+(\mathbb{R}) := \{ \mu\in\mathcal{M}(\mathbb{R}):\|\mu\|>0 \}.
\end{equation}
Note that the particle solutions \eqref{eq:intro_particle} are contained in $\mathcal{M}^+(\mathbb{R})$ for each $t$. 

To compute numerically and visualize the solutions, the solution should be at least weakly differentiable. We define the Hilbert space
\begin{equation}
    H^k(\mathbb{R}) = \left\{ u\in L^2(\mathbb{R}): \begin{array}{l}
    \text{for all $1\leq i\leq k$, $\partial_{x}^i u$ exists in the} \\
    \text{weak sense and $\partial_{x}^i u\in L^2(\mathbb{R})$}.
  \end{array}\right\},
\end{equation}
with norm induced by the inner product
\begin{equation}
    \langle f,g\rangle_{H^k(\mathbb{R})} = \left(\sum_{i=0}^k \int_{-\infty}^\infty f^{(i)}(x)\,g^{(i)}(x)\,\mathd x\right)^{1/2}.
\end{equation}
To recover a weakly differentiable solution, we will construct an operator $T^*:\mathcal{M}^+(\mathbb{R})\rightarrow H^3(\mathbb{R})$ via convolution with a kernel to obtain a solution in the space $C(\mathbb{R}^+; H^3(\mathbb{R}))$. 

% In the theory of functional analysis, an important tool for obtaining convergence given a sequence of approximate solutions is proving a sequential compactness result. 

\chapter{Point Heated Droplets}
\label{sec:point_heating}

\section{Overview}

In this chapter, we investigate the stability of thermocapillary convection within droplets when heated by a point source from below. A description of thin films on an inhomogeneously heated substrate in the context of the lubrication theory is introduced. Linear stability analysis shows that symmetry breaking does not occur when the heating is perfectly centred. However, transient simulations in three spatial dimensions reveal that vortices develop within the droplet when the heating is slightly off-centred to the centre of the droplet. The vortices persist even when the contact line is pinned. 

In Section~\ref{sec:pbg}, the aim of the chapter is introduced in the context of describing the mechanism of the onset of symmetry breaking in a point-heated droplet. A literature review on thermocapillary flow is also presented.  
In Section~\ref{sec:phtheory}, we present the theoretical model along with the key assumptions. A slightly different derivation and nondimensionalization of the thin-film equation from the previous chapter is introduced. 
In Section~\ref{sec:paxi}, we look at the radially symmetric base state (equilibrium solution) and its stability when the solution is perturbed slightly. When the heating is centred, the radially symmetric base state is stable to such perturbations. 
In Section~\ref{sec:poff}, we look at a second possible mechanism for the generation of Marangoni currents in the azimuthal direction -- namely, a small perturbation of the heating point source away from the droplet centre.  We show such currents persist only in the case of pinned droplets.  
In Section~\ref{sec:pother}, other forms of instabilities are documented. In particular, we observed rupturing of the droplet for high-powered heating. This is explored numerically with a transient simulation.
% The implications of our findings for the experimental knowledge already in the literature is discussed and concluding remarks given in Section~\ref{sec:conc}.

\section{Background} \label{sec:pbg}

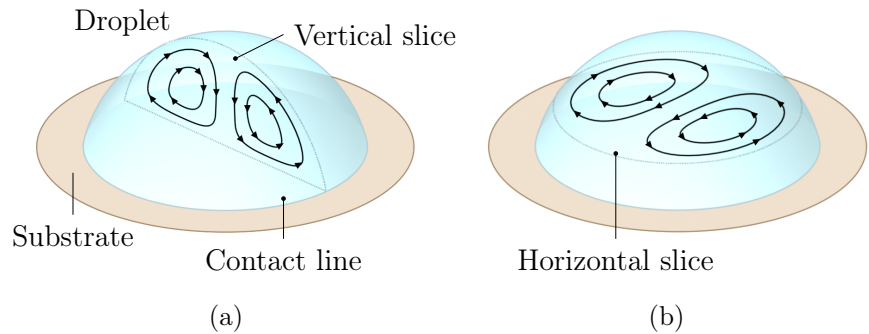
\begin{figure}
\centering
    \subfloat[\label{fig:sketch_marangoni_flow}]{
    \begin{tikzpicture}[scale=0.85]
    \clip (-3.5,2.2) rectangle (3.9,-2.5);
        \node[inner sep=0pt] at (0,0) {\includegraphics[width=0.4\textwidth]{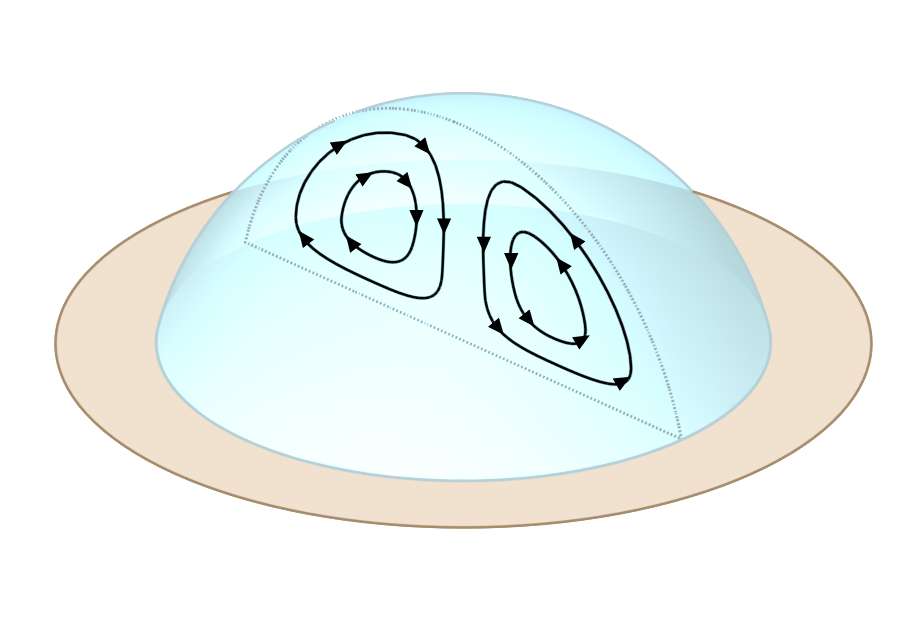}};
        % arrows
        \draw (-2.5,-0.7) -- (-2.5, -1.4);
        \draw (-2.5, -1.4) node[below] {Substrate};
        \draw (1,-1.1) -- (1, -1.8);
        \filldraw (1,-1.13) circle (0.03);
        \draw (1, -1.8) node[below] {Contact line};
        \draw (-0.7, 1.8) node[left] {Droplet};
        \draw (0.2, 1.2) -- (1,1.6); 
        \filldraw (0.2, 1.2) circle (0.03);
        \draw (1,1.6) node[right] {Vertical slice};
    \end{tikzpicture}
}
\subfloat[\label{fig:sketch_perpendicular_flow}]{
    \begin{tikzpicture}[scale=0.85]
    \clip (-3.3,2.2) rectangle (3.3,-2.5);
        \node[inner sep=0pt] at (0,0) {\includegraphics[width=0.4\textwidth]{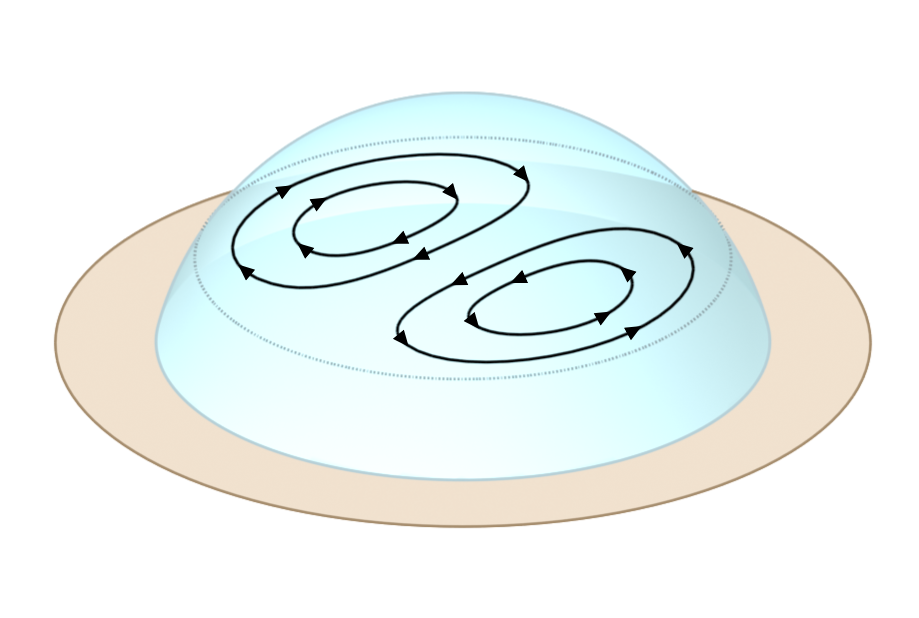}};
        % \draw (0.2, 1.22) -- (1,1.6); 
        % \filldraw (0.2, 1.22) circle (0.03);
        % \draw (1,1.6) node[right] {Horizontal slice};
        \draw (-1, -0.38) -- (-1, -1.8);
        \filldraw (-1, -0.38) circle (0.03);
        \draw (-1, -1.8) node[below] {Horizontal slice};
    \end{tikzpicture}
}
\caption{Two types of flows inside a droplet. (a) Axisymmetric Marangoni convection for homogeneously heated substrate. (b) Twin vortices perpendicular to the substrate. } 
\label{fig:sketch_flow}
\end{figure}

When a droplet is heated from below relative to the surroundings, the temperature gradient causes a difference in the surface tension on the liquid-gas interface of the droplet. This surface tension gradient, in turn, drives a flow from regions of low surface tension to high surface tension, known as the Marangoni current. For a homogeneously heated substrate, the convection flow is from the edge to the top of the droplet, where it then flows down at the centre and then radially outward to the edge of the droplet \cite{ehrhard1991}. A sketch of the Marangoni convection is shown in Figure~\ref{fig:sketch_marangoni_flow}.
This chapter is concerned with the theoretical modelling of the flow inside a sessile droplet heated at the substrate by a point source. Such local heating causes a difference in the surface tension on the droplet surface, which can drive a fluid flow from regions of low surface tension to high surface tension.
Thermocapillary flows have been observed experimentally in point-heated, millimetre-sized water droplets~\cite{yutaku2016,yutaku2017}.  In particular, when such droplets are heated from below by a point heat source at the droplet centre,  a vortex pair perpendicular to the substrate is observed, and the system is no longer axisymmetric. A sketch of this type of flow is shown in Figure~\ref{fig:sketch_perpendicular_flow}.  This chapter aims to obtain a theoretical understanding to explain the onset of such vortices.

These experimental results concern hydrophobic substrates, where the equilibrium contact angle is around $110^\circ$.  This particular setup is difficult to model analytically.  Therefore, as a first attempt to understand theoretically the origin of the vortex pair, we investigate hydrophilic substrates, where the equilibrium contact angle is small and where lubrication theory can be used for the analytical modelling.  As such, the present chapter aims to gain qualitative insights into the formation of the vortex pair in the point-heated droplet rather than precise quantitative insights.  In particular, we seek to determine if, given a radially-symmetric equilibrium solution for the point-heated droplet in lubrication theory, can linear stability analysis explain the onset of thermocapillary flows in the azimuthal direction?  In other words, is the radially-symmetric equilibrium solution susceptible to symmetry-breaking via linear instability?  The answer to this question -- at least in the lubrication theory -- turns out to be in the negative.  Hence, in this work, we also investigate other mechanisms to break the radial symmetry of the equilibrium base state.

% Lubrication theory is a key tool in analysing thermocapillary flows in thin films and droplets -- provided the latter possesses a sufficiently small equilibrium contact line.   Lubrication theory refers to a particular limiting geometry where the lengthscale of flow variations in the film (or droplet) in the lateral direction greatly exceeds that in the vertical direction.  In such a scenario, there is a natural small parameter, being the ratio of these two lengthscales, which enables an expansion of the Navier--Stokes equations~\cite{oron1997long}.  At lowest order in the expansion, one obtains a single equation for the height of the film (or droplet) as a function of the lateral variations and time.  The flow inside the film (or droplet) is Stokes flow, for which analytical expressions can be obtained.
Lubrication theory, as introduced in \Cref{sec:review_classic}, will be a vital tool in analysing thermocapillary flows. As such, we impose the lubrication assumption that the droplet possesses a sufficiently small equilibrium contact angle. 
Following this approach, Ehrhard and Davis~\cite{ehrhard1991} have studied the spreading of   3D axisymmetric droplets on a homogeneously heated substrate.  They found that the Marangoni current impede the spreading of the droplets. Increasing the Marangoni number results in a lower equilibrium droplet radius for fixed droplet volume and contact angle.  The theoretical predictions agreed well with the experiments.  Similarly, Tan et al.~\cite{tan1990}, and Van Hook et al.~\cite{vanhook1997} studied the rupturing of 2D and 3D thin films, respectively, on a substrate subjected to a spatially periodic heat source. An attractive van-der Waals potential $\phi=Ah^{-3}$ is used to model the dewetting. Local dewetting of the film is observed in high-temperature regions. A critical Marangoni number is found to which the deformed steady-state becomes unstable, and rupturing occurs.  Gravity is found to be stabilizing and delays the onset of rupture. Bostwick~\cite{bostwick2013} extended the work in Reference~\cite{ehrhard1991} to include thermal variations in the radial direction. He found that multiple stable equilibrium droplet solutions exist when the droplet's core is cooled relative to the surroundings. On the other hand, when the droplet is heated at the centre, they do not find bi-stability. 

Besides the lubrication theory, direct numerical simulation (DNS) is often employed to study droplet dynamics on heated substrates. 
S\'aenz et al. \cite{saenz2015} simulated 3D droplets on a homogeneously heated substrate with a non-circular contact line. A vortex pair perpendicular to the substrate was observed for a highly asymmetrical contact line shape. Shi et al. \cite{shi2017} investigate thin droplets on a homogeneously heated substrate with a spherical cap interface. They observed the development of multiple hexagonal B\'enard-Marangoni convection cells above a critical Marangoni number. Lu et al. \cite{lu2011} study evaporating droplets in an axisymmetrical setting. The free surface is modelled as a spherical cap with a constant radius and decreasing volume depending on the evaporation flux. They found that the Marangoni convection dominates the natural convection by about three orders of magnitudes for millimetre-sized droplets. Lee et al. \cite{Lee2022} used a similar method to study the effect of localised heating. When the droplet is heated at the centre, they observed a reversal of the convection flow compared to the homogeneously heated droplet, where the fluid rises at the centre of the droplet. 

% Our own results in the case of point heating also show rupture in regions of high heating, and for which a critical Marangoni number can be found.  However, a main focus of our work is to identify mechanisms for the generation of azimuthal flows in the case of point heating.  The work is organized as follows.

\section{Model Formulation} \label{sec:phtheory}

\subsection{Lubrication Theory} 

We use standard lubrication theory in three spatial dimensions \cite{oron1997long} and already described in Chapter 1. We start with dimensional variables; dimensionless variables are introduced at appropriate points in the exposition. The starting point is the incompressibility condition: 
\begin{equation} \label{eq:pht1}
    \pder[u]{x} + \pder[v]{y} + \pder[w]{z} = 0.
\end{equation}
We integrate Equation~\eqref{eq:pht1} once with respect to $z$ to produce:
\begin{equation} \label{eq:pht2}
    w|_{z=h} - w|_{z=0} = - \int_0^h \left(\pder[u]{x}+\pder[v]{y}\right) \,\mathd z.
\end{equation}
The interface is located at $z=h(x,y,t)$. We apply the no-slip boundary condition $\bm{u}|_{z=0}=0$ and the free-surface kinematic condition
\begin{equation}
    w = \pder[h]{t} + u\pder[h]{x} + v\pder[h]{y}, \qquad \text{at $z=h$.}
\end{equation}
Hence, Equation~\eqref{eq:pht2} becomes
\begin{equation} \label{eq:pht3}
    \pder[h]{t} + \pder{x}(\langle u\rangle h) + \pder{y}(\langle v\rangle h) = 0,
\end{equation}
where $\langle u\rangle$ and $\langle v\rangle$ are the depth-average velocities, 
\begin{equation}
    \langle u\rangle = \frac{1}{h}\int_0^h u \,\mathd z, \qquad \langle v\rangle = \frac{1}{h}\int_0^h v \,\mathd z.
\end{equation}
As in Chapter 1, we constitute the depth-averaged velocity by assuming Stokes flow in the liquid film, with spatial variations given by Lubrication Theory. As such, the following momentum balance conditions are obtained:
\begin{subequations}
\begin{align}
    -\pder[p]{x} + \mu\pder[^2u]{z^2} - \pder[\Phi]{x} &= 0, \label{eq:phstoke1} \\
    -\pder[p]{y} + \mu\pder[^2v]{z^2} - \pder[\Phi]{y} &= 0, \label{eq:phstoke2} \\
    -\pder[p]{z}-\pder[\Phi]{z} &= 0. \label{eq:phstoke3}
\end{align}
\end{subequations}
Here, $p$ is the fluid pressure and $\Phi$ is the potential of conservative body forces acting on the fluid. 

We integrate Equation~\eqref{eq:phstoke3} once to obtain 
\begin{equation}
    p + \Phi = f(x,y,t), 
\end{equation}
valid for all $z$. We use the standard Laplace-Young condition in the long-wave limit \cite{oron1997long} to compute:
\begin{equation}
    p|_z = -\gamma_0\nabla^2h,
\end{equation}
where $\nabla=(\partial_x, \partial_y)$ and $\nabla^2=\partial_x^2 + \partial_y^2$ are the gradient and Laplacian operator in the $xy$-plane. $\gamma_0$ is a constant reference value of the surface tension. Hence,
\begin{equation}
    p + \Phi = -\gamma_0\nabla^2h + \Phi|_h.
\end{equation}
Referring back to Equation~\eqref{eq:phstoke1}-\eqref{eq:phstoke2}, we have:
\begin{equation}
    \mu\pder[^2]{z^2}(u,v) = \nabla(p+\Phi) = \nabla(-\gamma_0\nabla^2h + \Phi|_h).
\end{equation}
The term $\mu\partial_z^2(u,v)|_{z=h}$ is the tangential stress at the interface. This is matched by surface-tension gradients at the interface:
\begin{equation}
    \mu\pder{z}(u,v) = \nabla\gamma, \qquad \text{at $z=h$},
\end{equation}
where $\nabla\gamma$ is the surface-tension gradient (independent of $z$). We will prescribe the surface-tension gradients later in what follows. In the meantime, we substitute Equation (11) into Equation (10) to obtain:
\begin{equation}
    \mu\pder{z}(u,v) = \nabla\gamma + (z-h)\nabla(-\gamma_0\nabla^2h + \Phi|_h).
\end{equation}
We integrate Equation (12) from $z=0$ to $z$ to obtain:
\begin{equation} \label{eq:phtuv}
    \mu(u,v) = z\nabla\gamma + \left(\frac{1}{2}z^2-hz\right)\nabla(-\gamma_0\nabla^2h + \Phi|_h).
\end{equation}
Thus, the depth-average velocities $\langle u\rangle$ and $\langle v\rangle$ are given by:
\begin{align*}
    \mu\langle u\rangle &= \frac{1}{2}h\pder[\gamma]{x} - \frac{1}{3}h^2\pder{x}(-\gamma_0\nabla^2h + \Phi|_h), \\
    \mu\langle v\rangle &= \frac{1}{2}h\pder[\gamma]{y} - \frac{1}{3}h^2\pder{y}(-\gamma_0\nabla^2h + \Phi|_h).
\end{align*}
Equation~\eqref{eq:pht3} now becomes:
\begin{equation} \label{eq:pht4}
    \mu\pder[h]{t} + \nabla\cdot\left\{\frac{1}{2}h^2\nabla\gamma - \frac{1}{3}h^3\nabla\left(-\gamma_0\nabla^2h + \Phi|_h\right)\right\} = 0.
\end{equation}

\subsection{Temperature Profile}

In standard lubrication theory, the advection-diffusion equation for the temperature in the film is diffusion-dominated and reads:
\begin{equation} \label{eq:pttemp1}
    \pder[^2T]{z^2} = 0. 
\end{equation}
The solution to \eqref{eq:pttemp1} is:
\begin{equation}
    T = A(x,y,t)z + B(x,y,t),
\end{equation}
where $A$ and $B$ are determined from boundary conditions. 

First, we address the boundary condition at the substrate at $z=0$. We assume that the substrate is heated in an inhomogeneous fashion, such that the substrate temperature $T_s$ is given by:
\begin{equation}
    T_s(x,y) = \langle T_s\rangle + (\Delta T_s)\tilde{T}_s(x,y),
\end{equation}
where $\langle T_s\rangle$ denotes the mean temperature and $\Delta T_s=\max T_s - \min T_s$ is the maximum temperature difference across the substrate. Thus $\tilde{T}_s$ is a dimensionless temperature fluctuation. We similarly re-write the temperature inside the film as:
\begin{equation} \label{eq:pttemp2}
    T(x,y,z) = \langle T_s\rangle + \Delta T_s\left(\tilde{T}_s(x,y)+\tilde{T}(x,y,z)\right).
\end{equation}
Continuity of temperature at the interface between the liquid film and the substrate means that $T=T_s$ at $z=0$, hence $\tilde{T}=0$ at $z=0$. We furthermore assume that the film temperature satisfies a Robin boundary condition at $z=h(x,y,t)$; this corresponds to the application of Newton's Law of Cooling at the interface:
\begin{equation} \label{eq:pttemp3}
    -k\pder[T]{z} = \alpha(T-T_g), \qquad z=h(x,y,t).
\end{equation}
Here, $k$ is the thermal conductivity of the film, $\alpha$ is the heat-transfer coefficient, and $T_g$ is the temperature of the gas surrounding the film. Substituting Equation~\eqref{eq:pttemp2} into Equation~\eqref{eq:pttemp3} gives:
\begin{equation}
    -k\pder[\tilde{T}]{z} = \alpha\left(\tilde{T}+\tilde{T}_s(x,y) + \frac{\langle T_s\rangle - T_g}{\Delta T_s}\right),
\end{equation}
at $z=h$. Rearranging, this reads:
\begin{equation} \label{eq:pttemp4}
    -\pder[\tilde{T}]{z} = \frac{\Bi}{h_0}\left(\tilde{T}+\tilde{T}_s(x,y)+\Theta\right), \qquad z=h,
\end{equation}
where $\Theta=(\tilde{T}_s-T_g)/\Delta T_s$ is the scaled temperature difference between the substrate and the surrounding gas, $h_0$ is the vertical length scale of the system, and $\Bi=\alpha h_0/k$ is the Biot number. 

By linearity, the fluctuation $\tilde{T}$ also satisfies the advection-diffusion equation~\eqref{eq:pttemp1}. Hence, $\tilde{T}$ also has the form $\tilde{A}z+\tilde{B}$. Applying the boundary condition $\tilde{T}=0$ at $z=0$, the temperature profile $\tilde{T}$ becomes:
\begin{equation}
    \tilde{T} = \tilde{A}(x,y,t)z.
\end{equation}
Applying the boundary condition~\eqref{eq:pttemp4}, we obtain:
\begin{equation}
    A = -\frac{\Bi\left(\tilde{T}_s(x,y)+\Theta\right)}{1+\Bi\tilde{h}}\frac{1}{h_0},
\end{equation}
hence
\begin{equation} 
    \tilde{T}(x,y,z,t;h) = -\frac{\Bi\left(\tilde{T}_s(x,y)+\Theta\right)}{1+\Bi\tilde{h}}\tilde{z},
\end{equation}
where $\tilde{z}=z/h_0$ and $\tilde{h}=h/h_0$ are in their dimensionless form. We also explicitly denote the dependence on the interface height $h$. The complete temperature profile in the film therefore reads:
\begin{equation} \label{eq:pttemp5}
    T(x,y,z,t;h) = \langle T_s\rangle + \Delta T_s\left(\tilde{T}_s(x,y) - \frac{\Bi(\tilde{T}_s(x,y)+\Theta)}{1+\Bi\tilde{h}}\tilde{z}\right).
\end{equation}
The temperature on the surface of the film is, therefore
\begin{equation} \label{eq:pttemp6}
    T|_{z=h} = \langle T_s\rangle + (\Delta T_s)\psi(x,y,t;h),
\end{equation}
where $\psi$ is a nondimensional quantity given by
\begin{equation} 
    \psi(x,y,t;h) = \frac{\tilde{T}_s(x,y)-\Theta\Bi\tilde{h}}{1+\Bi\tilde{h}}.
\end{equation}

We now couple the equations for the free-surface height \eqref{eq:pht4} and \eqref{eq:pttemp6}. We assume that the surface tension is temperature-dependent:
\begin{equation}
    \gamma = \gamma_0 - \frac{\gamma_1}{\Delta T_s}(T-T_{ref}), \qquad z=h(x,y,t),
\end{equation}
where $\gamma_0$ is the reference level of surface tension, $\gamma_1>0$ is a positive constant, and $T_{ref}$ is a reference temperature. Hence,
\begin{equation}
    \nabla\gamma = -\frac{\gamma_1}{\Delta T_s}\nabla T, \qquad z=h(x,y,t).
\end{equation}
Using Equation~\eqref{eq:pttemp6}, this becomes:
\begin{equation}
    \nabla\gamma = -\gamma_1\nabla\psi(x,y,t;h).
\end{equation}
Substitution into Equation~\eqref{eq:pht4} yields:
\begin{equation} \label{eq:pht5}
    \mu\pder[h]{t} + \nabla\cdot\left\{-\frac{1}{2}\gamma_1h^2\nabla\psi - \frac{1}{3}h^3\nabla\left(-\gamma_0\nabla^2h + \Phi|_h\right)\right\} = 0.
\end{equation}

\subsection{Nondimensionalization}

Let $r_0$ and $h_0$ be the horizontal and vertical length scale of the system, respectively. The aspect ratio is defined by $\epsilon=h_0/r_0$ where $\epsilon\ll1$ for the lubrication theory to hold. We non-dimensionalize, with 
\begin{gather*}
    \tilde{x}=\frac{x}{r_0}, \qquad \tilde{y}=\frac{y}{r_0}, \qquad \tilde{z} = \frac{z}{h_0}, \qquad \tilde{h} = \frac{h}{h_0}, \qquad \tilde{\nabla}=r_0\nabla. 
\end{gather*}
Thus Equation~\eqref{eq:pht5} becomes:
\begin{equation}
    \mu h_0\pder[\tilde{h}]{t} + \tilde{\nabla}\cdot\bigg\{-\frac{1}{2}\gamma_1\epsilon^2\tilde{h}^2\tilde{\nabla}\psi - \frac{1}{3}\epsilon^2h_0\tilde{h}^3\tilde{\nabla}\left(-\frac{\gamma_0h_0}{r_0^2}\tilde{\nabla}^2\tilde{h} + \Phi|_h\right)\bigg\} = 0.
\end{equation}
Divide by $\gamma_0\epsilon^4$ to obtain:
\begin{equation}
    \frac{\mu h_0}{\gamma_0\epsilon^4}\pder[\tilde{h}]{t} + \tilde{\nabla}\cdot\bigg\{-\frac{1}{2}\frac{\gamma_1}{\gamma_0\epsilon^2}\tilde{h}^2\tilde{\nabla}\psi - \frac{1}{3}\tilde{h}^3\tilde{\nabla}\left(-\tilde{\nabla}^2\tilde{h} + \frac{h_0}{\gamma_0\epsilon^2}\Phi|_h\right)\bigg\} = 0.
\end{equation}
Thus we are motivated to scale the time and the potential by
\begin{equation} 
    \tilde{t} = \frac{\gamma_0\epsilon^4}{\mu h_0}t, \qquad \tilde{\phi} = \frac{h_0}{\gamma_0\epsilon^2}\Phi|_h,
\end{equation}
and the dimensionless Marangoni number is identified as
\begin{equation*}
    \Ma = \frac{\gamma_1}{\gamma_0\epsilon^2}.
\end{equation*}
% Typical values of $\gamma_1/\gamma_0$ for water is on the order of $10^2$--$10^3$. 
Finally, with the tildes dropped, the dimensionless thin-film equation reads:
\begin{subequations} \label{eq:pht}
\begin{equation} 
    \pder[h]{t} + \nabla\cdot\left\{-\frac{1}{2}\Ma h^2\nabla\psi - \frac{1}{3}h^3\nabla\left(-\nabla^2h + \phi\right)\right\} = 0,
\end{equation}
and $\psi$ is the (non-dimensional) temperature variation at the interface
\begin{equation} 
    \psi(x,y,t;h) = \frac{T_s(x,y)-\Theta\Bi h}{1+\Bi h}.
\end{equation}
\end{subequations}

\section{Axisymmetric Linear Stability Analysis} \label{sec:paxi}

\subsection{Homogeneously heated substrate}

When the substrate is homogeneously heated, the surface temperature is given by
\begin{equation} 
    \psi(x,y,t;h) = -\frac{\Bi h}{1+\Bi h}, %\qquad \nabla\psi(x,y,t;h) = -\frac{\Bi\nabla h}{(1+\Bi h)^2},
\end{equation}
where we have set $\Theta=1$ without lost of generality. Equation~\eqref{eq:pht} then becomes
\begin{equation} \label{eq:phthomo}
    \pder[h]{t} + \nabla\cdot\left\{\frac{1}{2}\Ma h^2\frac{\Bi\nabla h}{(1+\Bi h)^2} + \frac{1}{3}h^3\nabla\nabla^2h\right\} = 0,
\end{equation}
% Equation~\eqref{eq:phthomo} has been used by Reference \cite{ehrhard1991,oron1997long,oron2000,thiele2004} to investigate the dynamics of droplet and thin-film under heating. 
We note that a constant film height $h=const.$ is an equilibrium solution of Equation~\eqref{eq:phthomo}. The stability of such a solution was studied in \cite{oron1992,oron2000,thiele2004}. The Marangoni current is shown to be always destabilizing to long wave perturbation while gravity is stabilizing. 

Equation~\eqref{eq:phthomo} can be expressed in terms of gradient flow \cite{oron1992,thiele2004}
\begin{equation} \label{eq:phtenergy}
    \pder[h]{t} + \nabla\cdot\left\{\mob(h)\nabla\frac{\delta E}{\delta h}\right\} = 0,
\end{equation}
with mobility $\mob(h)=h^3/3$ and free energy of the Marangoni term given by
\begin{equation}
    E_{mar}[h] = \frac{1}{|\Omega|}\int_\Omega \frac{3}{2}\Ma\,\Bi h\log\left(\frac{h}{1+\Bi h}\right) \,\mathd A,
\end{equation}
where we integrate over the domain $\mathd A=\mathd x\,\mathd y$. Equation~\eqref{eq:phtenergy} evolves to minimize the free energy in a gradient decent sense. We note that, in general, for an inhomogeneously heated substrate ($T_s\neq0$), Equation~\eqref{eq:pht} does not admit a closed-form energy equation. % We will leave the discussion of the gradient flow structure of droplet spreading to later chapter.

\subsection{Axisymmetric Base State Solution} 

We now look at the equilibrium solution in an axisymmetric configuration about the $z$-axis by considering surface temperature $T_s=T_s(r)$ depending only on $r$. In this section, we will also set $\phi=0$. In such cases, the thin-film equation reduces to a one-dimensional equation given by
\begin{subequations} \label{eq:phtfe_r}
\begin{equation} 
    \frac{1}{r}\pder{r}\left\{-\frac{1}{2}\Ma h^2r\pder[\psi]{r} + \frac{1}{3}h^3r\pder{r}\left(\frac{1}{r}\pder{r}\left(r\pder[h]{r}\right)\right)\right\} = 0,
\end{equation} 
defined on the domain $r\in[0,r_*]$ where $r_*$ is the contact line position. For now, we will avoid modelling the moving contact line as it introduces unnecessary complications to the system. However, we do assert boundary conditions at equilibrium given by
\begin{align}
    \partial_r h = 0, \qquad \partial_{rrr} h = 0, \qquad &\text{at $r=0$}, \\
    h = 0, \qquad \partial_r h = -\alpha, \qquad &\text{at $r=r_*$} ,
\end{align}
\end{subequations}
where $\alpha$ is the equilibrium contact angle. 

\begin{figure}[ht]
    \centering
    \begin{tikzpicture}[scale=2.5]
        % plane (concentric)
        \draw [gray!30, line width=0.2mm] (0.25,0) arc [start angle=0, end angle=360, x radius=0.25cm, y radius=0.5*0.25cm];
        \draw [gray!30, line width=0.2mm] (0.5,0) arc [start angle=0, end angle=360, x radius=0.5cm, y radius=0.5*0.5cm];
        \draw [gray!30, line width=0.2mm] (0.75,0) arc [start angle=0, end angle=360, x radius=0.75cm, y radius=0.5*0.75cm];
        \draw [gray!30, line width=0.2mm] (1.2,0) arc [start angle=0, end angle=360, x radius=1.2cm, y radius=0.5*1.2cm];
        % plane (radial) projection = [1,0.5;0,0.5]
        \draw [gray!30, line width=0.2mm] (-0.54,-0.54) -- (0.54,0.54);
        \draw [gray!30, line width=0.2mm] (-1.2,0) -- (1.2,0);
        \draw [gray!30, line width=0.2mm] (-0.67*1.5,-0.67*0.5) -- (0.67*1.5,0.67*0.5);
        \draw [gray!30, line width=0.2mm] (-1.06*0.5,1.06*0.5) -- (1.06*0.5,-1.06*0.5);
        \draw [gray!30, line width=0.2mm] (0,0) -- (0,-0.6);
        \draw [gray!30, line width=0.2mm] (-0.67*1.5,0.67*0.5) -- (0.67*1.5,-0.67*0.5);
        % hotspot
        \node[inner sep=0pt] at (0,0) {\includegraphics[width=.25\textwidth]{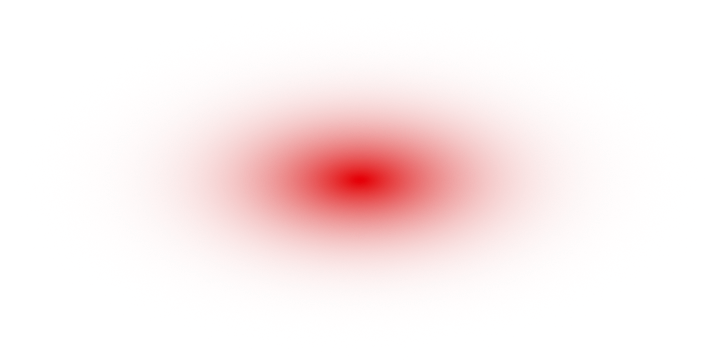}};
        % axis
        \draw [->, line width=0.3mm] (1.2,0) arc [start angle=0, end angle=62, x radius=1.2cm, y radius=0.5*1.2cm];
        \draw (0.67*1.5,0.67*0.5) node[above] {$\varphi$};
        \draw [->, line width=0.3mm] (0,0) -- (1.2,0);
        \draw (0.6,0) node[below] {$r$};
        \draw [->, line width=0.3mm] (0,0) -- (0,1.2);
        \draw (0,1.2) node[left] {$z$};
        % labels
        \draw (1.2,0) node[right, text=gray] {\scriptsize 0};
        \draw (-1.2,0) node[left, text=gray] {\scriptsize $\pi$};
        \draw (0,-0.6) node[below, text=gray] {\scriptsize $\frac{3\pi}{2}$};
        \draw (-1.06*0.5,-1.06*0.5) node[below, text=gray] {\scriptsize $\frac{4\pi}{3}$};
        \draw (-0.67*1.5,0.67*0.5) node[above, text=gray] {\scriptsize $\frac{5\pi}{6}$};
        \draw (-0.67*1.5,-0.67*0.5) node[below, text=gray] {\scriptsize $\frac{7\pi}{6}$};
        \draw (1.06*0.5,-1.06*0.5) node[below, text=gray] {\scriptsize $\frac{5\pi}{3}$};
        \draw (0.67*1.5,-0.67*0.5) node[below, text=gray] {\scriptsize $\frac{11\pi}{6}$};
        % \draw (-1.06*0.5,1.06*0.5) node[above, text=gray] {\scriptsize $2\pi/3$};
        % \draw (-0.67*1.5,0.67*0.5) node[above, text=gray] {\scriptsize $5\pi/6$};
        % \draw (1.06*0.5,1.06*0.5) node[above, text=gray] {\scriptsize $1\pi/3$};
        % \draw (0.67*1.5,0.67*0.5) node[above, text=gray] {\scriptsize $\pi/6$};
        % profile
        \draw[orange, line width=0.6mm] plot[domain=-0.96:0.96,smooth] (\x,{1-abs(\x)^2});
        % base
        \draw [orange, line width=0.6mm] (-0.97,0) arc [start angle=180, end angle=360, x radius=0.97cm, y radius=0.5*0.97cm];
        \draw [orange, line width=0.6mm] (-0.97,0) arc [start angle=180, end angle=170, x radius=0.97cm, y radius=0.5*0.97cm];
        \draw [orange, line width=0.6mm] (0.97,0) arc [start angle=0, end angle=10, x radius=0.97cm, y radius=0.5*0.97cm];
        \draw [orange, line width=0.4mm, dotted] (0.97,0) arc [start angle=0, end angle=180, x radius=0.97cm, y radius=0.5*0.97cm];
        % half
        \draw [orange, line width=0.2mm] (-0.67,0.5) arc [start angle=180, end angle=360, x radius=0.67cm, y radius=0.5*0.67cm];
        \draw [orange, line width=0.3mm, dotted] (0.67,0.5) arc [start angle=0, end angle=180, x radius=0.67cm, y radius=0.5*0.67cm];
        % perpendicular
        \draw [orange, line width=0.3mm, dotted] plot[domain=0:0.44,smooth] (\x,{1-3*abs(\x)^2});
        \draw [orange, line width=0.2mm] plot[domain=-0.45:0,smooth] (\x,{1-7*abs(\x)^2});
    \end{tikzpicture} 
    \caption{Schematic of a point heated droplet in the cylindrical coordinate system $(r,\varphi,z)$ and axisymmetric heating from the bottom.}
\end{figure}
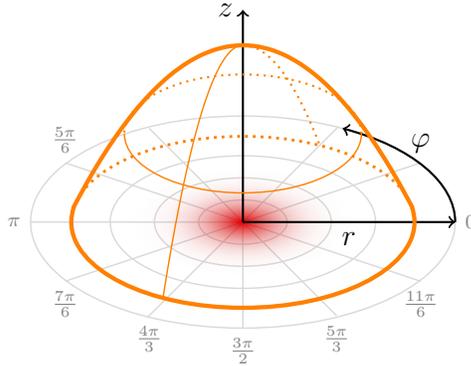

By integrating Equation \eqref{eq:phtfe_r} with respect to $r$ once and asserting $h(r_*)=0$, we obtain a third-order nonlinear ODE
\begin{equation} \label{eq:basestateode}
    h''' = \frac{3}{2}\Ma \frac{\psi'}{h} - \frac{h''}{r} + \frac{h'}{r^2}.
\end{equation}
Here, the primes denote derivatives with respect to $r$. 

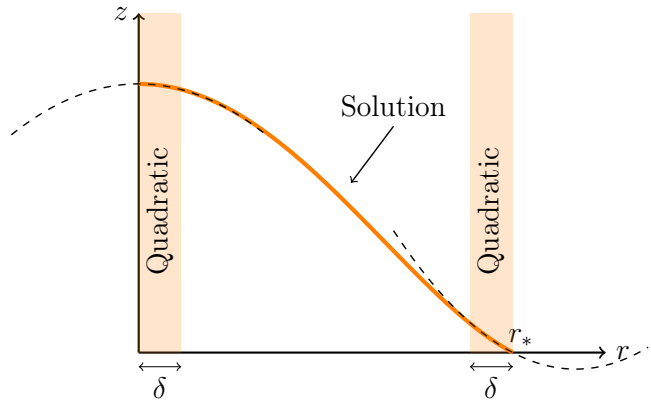
\begin{figure}[ht]
    \centering
    \begin{tikzpicture}[x=6cm, y=4cm]
        % axis
        \draw [->, line width=0.3mm] (0,0) -- (0,1.2);
        \draw (0,1.2) node[left] {$z$};
        \draw [->, line width=0.3mm] (0,0) -- (1.1,0);
        \draw (1.1,0) node[right] {$r$};
        \draw (0.9,0.12) node[below] {$r_*$};
        % profile
        \draw[orange, line width=0.6mm] plot[domain=0:0.8811,smooth] (\x,{(\x^2-1)^2-0.05});
        % quadratic 
        \draw[black, dashed, line width=0.2mm] plot[domain=-0.3:0.3,smooth] (\x,{-2*abs(\x)^2+1-0.05});
        \draw[black, dashed, line width=0.2mm] plot[domain=0.6:1.2,smooth] (\x,{-0.788*(\x-0.881)+2.658*(\x-0.881)^2});
        % arrow
        \draw [->, line width=0.2mm] (0.6,0.8) -- (0.5,0.6);
        \draw (0.6,0.8) node[above] {Solution};
        % regions
        \fill [orange, fill opacity=0.2] (0,0) rectangle (0.1,1.2);
        \fill [orange, fill opacity=0.2] (0.78,0) rectangle (0.8811,1.2);
        \draw [<->] (0,-0.05) -- (0.1,-0.05);
        \draw (0.05,-0.05) node[below] {$\delta$};
        \draw [<->] (0.78,-0.05) -- (0.8811,-0.05);
        \draw (0.83,-0.05) node[below] {$\delta$};
        % more arrows
        \node[rotate=90] at (0.05,0.5) {Quadratic};
        \node[rotate=90] at (0.83,0.5) {Quadratic};
    \end{tikzpicture} 
    \caption{In practice, a small width $\delta$ is chosen such that when $r$ is within $\delta$ away from the problematic points (shaded regions), we switch to solving second order linear ODE instead of Equation \eqref{eq:basestateode}.}
    \label{fig:sketch_shooting}
\end{figure}

We employ a shooting method to solve the ODE: We find the solution of Equation \eqref{eq:basestateode} with the left-hand ``initial'' conditions
\begin{equation}
    h(0) = p, \qquad h'(0) = 0, \qquad h''(0) = q,
\end{equation}
varying $p$ and $q$ such that the right-hand ``final'' conditions
\begin{equation} \label{eq:shootingright}
    h(r_*)=0, \qquad h'(r_*)=-\alpha, 
\end{equation}
are satisfied. We have the freedom to either fix the contact-line position $r_*$ or the equilibrium droplet volume. We choose to fix $r_*$ as it makes computing conditions \eqref{eq:shootingright} easier. There are two problematic points where Equation \eqref{eq:basestateode} is not defined, which are the boundaries $r=0$ and $r=r_*$ (where $h(r_*)=0$). To avoid evaluating Equation \eqref{eq:basestateode} near these points, the solution near the boundaries is approximated with quadratic functions (e.g. Figure~\ref{fig:sketch_shooting}). 

Once the solution $h(r)$ is obtained, the stream function can be computed
\begin{align}
    \Psi(r,z;h) &= \int_0^z u_r(r,s) \,\mathd s, \\
    &= -\frac{1}{2}\Ma z^2\psi' + \left(\frac{1}{2}hz^2-\frac{1}{6}z^3\right)\pder{r}\left(h''+\frac{h'}{r}\right),
\end{align}
valid for $0\leq r\leq r_*$ and $0\leq z \leq h(r)$. 

\begin{figure}[ht]
\centering
\begin{subfigure}{1\textwidth}
    \centering
    \includegraphics[width=1\textwidth]{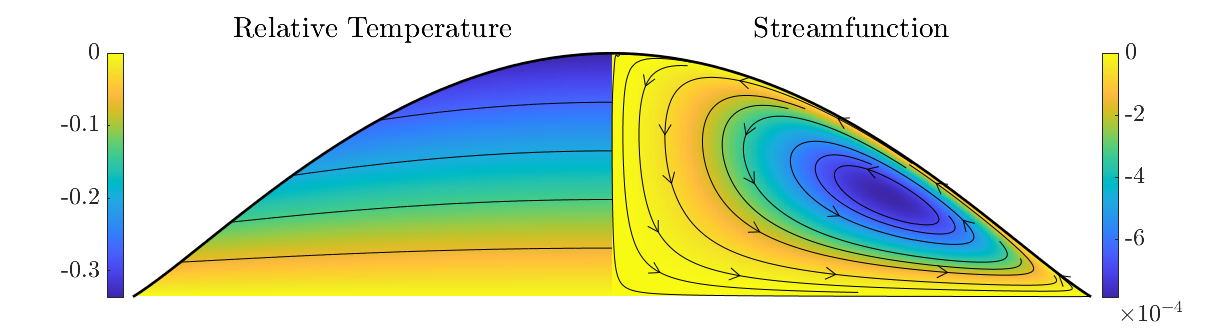} 
    \caption{Homogeneous heating $T_s(r)=0$.}
    \label{fig:bshomogeneous}
\end{subfigure}

\begin{subfigure}{1\textwidth}
    \centering
    \includegraphics[width=1\textwidth]{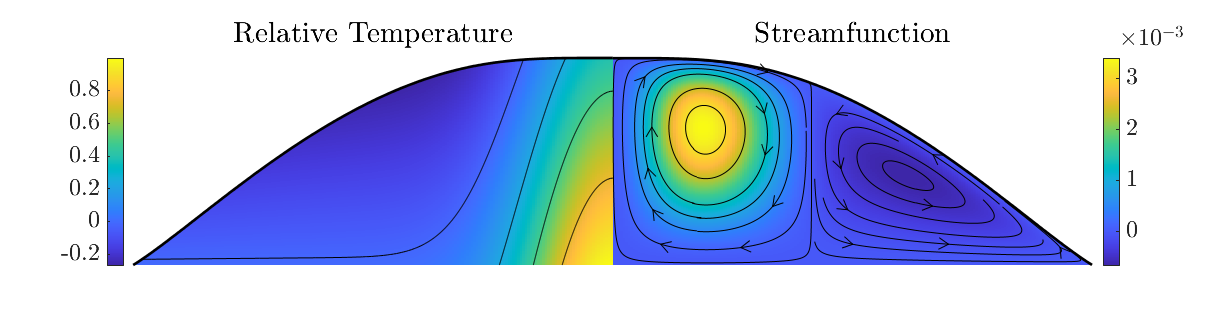}
    \caption{Point heating $T_s(r)=\mathe^{-r^2/0.2^2}$.}
\end{subfigure}
\caption{Equilibrium solutions of different heating profiles. We take $r_*=1$ and $\alpha=0.6$. All other parameters $(\Ma,\Bi,\Theta)$ are taken to be unity. The maximum height of the homogeneously heated droplet is about 0.506, whereas the point heated droplet is shorter at 0.427 due to the reduced surface tension at the centre of the droplet. }
\label{fig:shooting}
\end{figure}

Two profiles of the equilibrium solution are shown in Figure \ref{fig:shooting}, one with homogeneous heating and another with localized heating. The relative temperature $T(r,z;h)$ within the droplet given by Equation~\eqref{eq:pttemp5} (with $\Delta T_s=1$ and the $\langle T_s\rangle$ term dropped) is plotted on the left half of the figure and the stream function $\Psi(r,z;h)$ on the right. Figure \ref{fig:bshomogeneous} shows the equilibrium solution for a homogeneously heated surface. A temperature gradient can be seen where the base of the droplet is hotter at the bottom and cooler towards the top. The stream function on the right shows one convective region flowing in the clockwise direction within the droplet. Figure \ref{fig:bshomogeneous} shows the equilibrium droplet profile on a point-heated substrate. The temperature decreases vertically as well as radially. Two convective regions can be seen from the stream function. The inner convective region has the opposite orientation as the homogeneously heated case, flowing counterclockwise. A second convection cell is observed in the outer region flowing in the clockwise direction at a lower magnitude than the inner region. 

\begin{figure}[ht]
    \centering
    \includegraphics[width=0.7\textwidth]{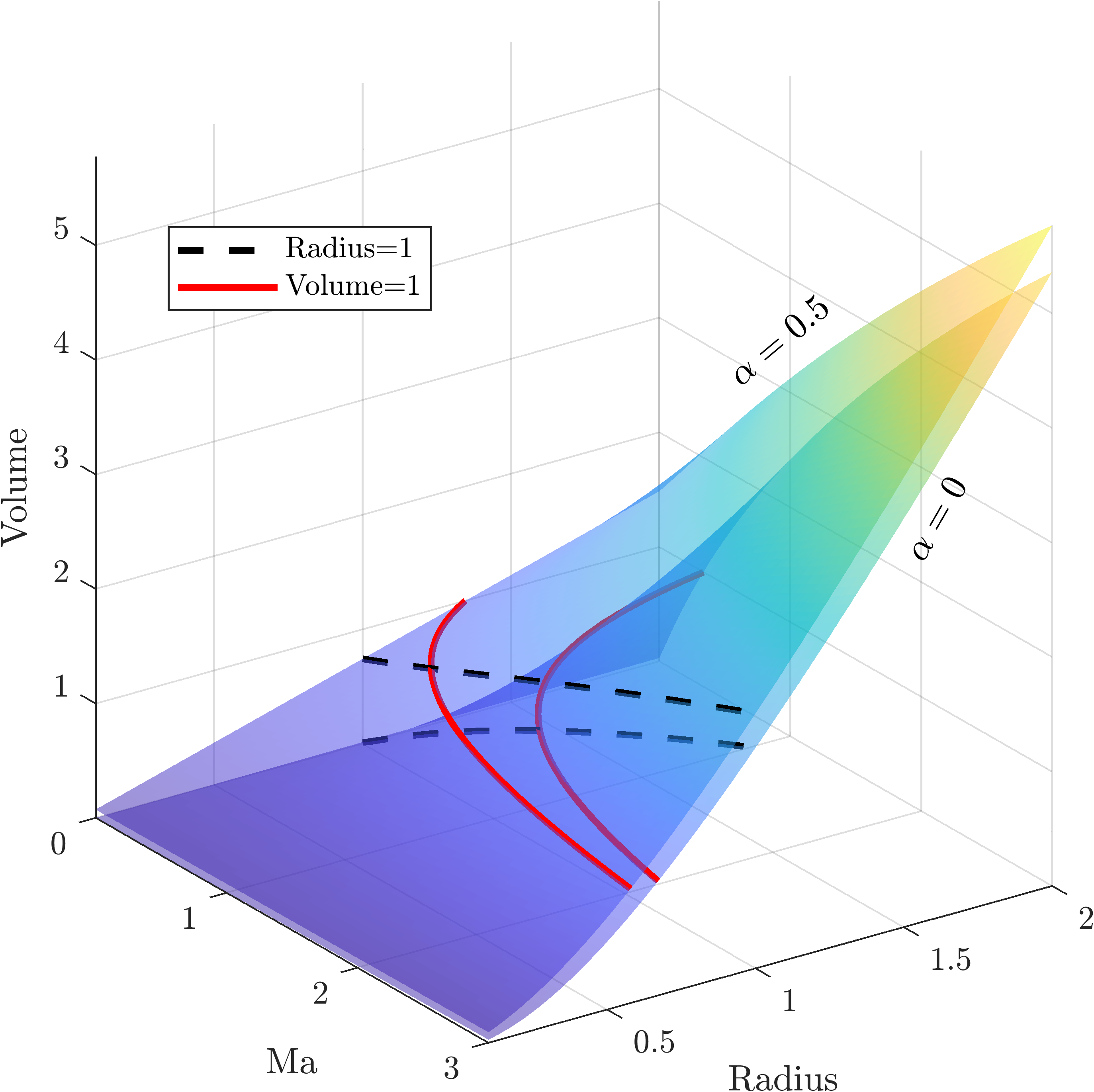}
    \caption{A plot of the base-state droplet volume for homogeneously heated substrate ($T_s=0$) as a function of droplet radius and Marangoni number, and for two values of equilibrium contact angle $\alpha$.  Other parameters are given by $(\Bi,\Theta)=(1,1)$. }
    \label{fig:level_set}
\end{figure}

It is worthwhile to understand qualitatively the properties of the solution as the various parameters in the model are changed.  For that reason, in Figure~\ref{fig:level_set}, we plot the volume of the base-state droplet as a function of droplet radius and Marangoni number.  Such a plot demonstrates the effect of parameter changes on the droplet shape as particular paths through the high-dimensional parameter space are chosen.  In particular, the solid lines in Figure~\ref{fig:level_set} correspond to paths of constant volume and constant equilibrium contact angle; these show that increasing the Marangoni number decreases the equilibrium droplet radius.  This is consistent with earlier findings in Reference~\cite[Figure 8]{ehrhard1991} and \cite[Figure 3]{thiele2004} for homogeneously heated substrates.
In the present work, we investigate instead the dashed lines in the parameter space, corresponding to fixed equilibrium contact angle $\alpha$ and fixed equilibrium droplet radius.  Along such paths, increasing the Marangoni number increases the droplet volume.

\begin{figure}
\centering 
    \subfloat[$T_s(r)=\mathe^{-r^2/0.2^2}$.\label{fig:eq_height}]{
    \includegraphics[width=0.7\textwidth]{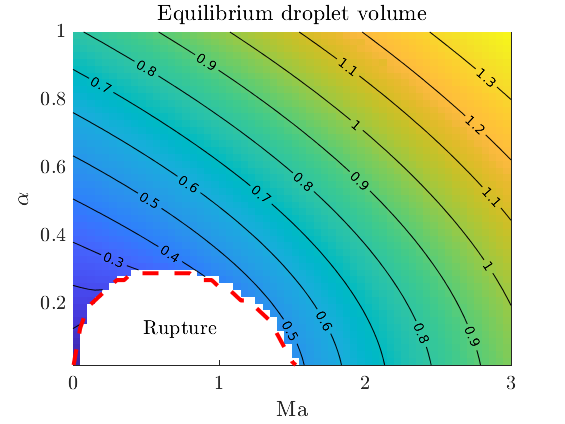} 
}

\subfloat[$T_s(r)=\mathe^{-r^2/0.3^2}$.\label{fig:eq_volume}]{
    \includegraphics[width=0.7\textwidth]{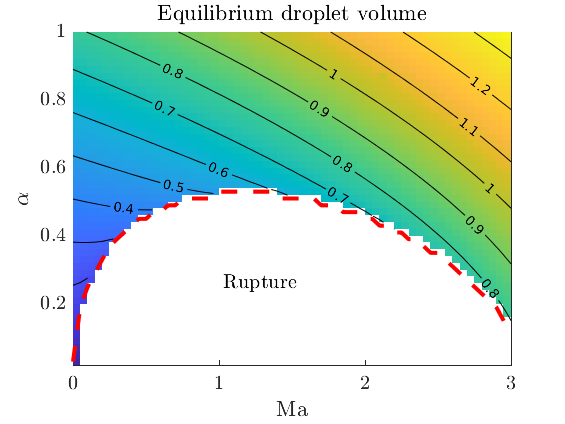} 
}
\caption{Contour plot of the equilibrium droplet volume for two hotspot sizes as a function of the equilibrium contact angle $\alpha$ and Marangoni number $\Ma$. Empty regions correspond to cases where the droplet ruptures. The contact line is fixed at $r_*=1$. Other physical parameters are $(\Bi,\Theta)=(1,1)$.  } 
\label{fig:equibrium_droplet_size}
\end{figure}

Figure~\ref{fig:equibrium_droplet_size} shows the equilibrium droplet volume for varying values of Marangoni number $\Ma$ and equilibrium contact angle $\alpha$. The equilibrium droplet volume and height positively correlate to $\alpha$ and $\Ma$. For small droplet volume and fixed $\Ma$, there is a critical $\alpha$ value where the point heating causes the droplet to rupture and the equilibrium solution ceases to exist. Numerically, the solvers fail to converge to an equilibrium solution, indicating a bifurcation occurring in the equation. The rupturing region is larger for larger hotspot size. This `ring rupture' is explored in more detail below. The plots also show that there is a lower bound for the possible equilibrium droplet size at fixed $\Ma$. 

\subsection{Stability of Base State}

Let $h_0(r)$ be a base state solution. To determine the stability of the base state solution, we consider a small perturbation around the base state of the form 
\begin{equation} \label{eq:bsperturbed}
    h(r,\varphi,t) = h_0(r) + \delta h(r,\varphi)\mathe^{\sigma t}.
\end{equation}
The linearized equation for $\delta h$ is given by
\begin{subequations}
\begin{multline} \label{eq:lbsperturbed}
    \sigma \delta h + \nabla\cdot\bigg\{-\frac{1}{2}\Ma\left(2h_0\delta h\nabla\psi_0+h_0^2\nabla(G\delta h)\right) \\
    +\frac{1}{3}h_0^3\nabla\nabla^2\delta h + h_0^2\delta h\nabla\nabla^2h_0 \bigg\} = 0,
\end{multline} 
where $\psi_0=\psi(h_0)$ and 
\begin{equation}
    G(r) = -\frac{\Bi(\psi_0+\Theta)}{1+\Bi h_0}.
\end{equation}
The boundary condition becomes
\begin{equation}
    \delta h(r_*,\varphi) = \partial_r\delta h(r_*,\varphi) = 0.
\end{equation}
\end{subequations}
This is an eigenvalue problem where the eigenvalue $\sigma$ represents the growth rate: Given a base state, if any of the eigenvalues has a positive real part $\Re(\sigma)>0$, then the corresponding eigenmode $\delta h$ grows exponentially, and the base state is unstable. Otherwise, the base state is stable.  

We seek separable solution of the form $\delta h(r,\varphi) = h_1(r)\mathe^{\mathi k\varphi}$, with wavenumber $k=0,1,2,\dots$. This is sufficient to encode all the possible perturbations of the base state. Equation \eqref{eq:lbsperturbed} then becomes a fourth-order linear ODE of one variable 
\begin{subequations} \label{eq:evproblem}
\begin{align} 
    \mathcal{L}(h_1) &= \sigma h_1, \qquad \mathcal{L} = \sum_{i=0}^4 A_i(r)\pder[^i]{r^i},
\end{align}
where the coefficients are given by
\begin{align}
    A_4(r) &= \frac{1}{3}h_0^3, \\
    A_3(r) &= \frac{2h_0^3}{3r} + h_0^2h_0', \\
    A_2(r) &= -\frac{2k^2+1}{3}\frac{h_0^3}{r^2} + \frac{h_0^2h_0'}{r} - \frac{1}{2}\Ma h_0^2G, \\
    A_1(r) &= \frac{2k^2+1}{3}\frac{h_0^3}{r^3} - (k^2+1)\frac{h_0^2h_0'}{r^2} + h_0^2(\nabla^2h_0)' \nonumber \\
    &\qquad -\frac{1}{2}\Ma\left\{\frac{h_0^2G}{r} + 2h_0^2G' + 2h_0h_0'G + 2h_0\psi_0'\right\}, \\
    A_0(r) &= \frac{k^4-4k^2}{3}\frac{h_0^3}{r^4} + 2k^2\frac{h_0^2h_0'}{r^3} + \nabla\cdot(h_0^2\nabla\nabla^2h_0) \nonumber \\
    &\qquad -\frac{1}{2}\Ma\left\{-k^2\frac{h_0^2G}{r^2} + \nabla\cdot(h_0^2\nabla G + 2h_0\nabla\psi_0)\right\}. 
\end{align}
\end{subequations}
The boundary conditions at $r=1$ are given by
\begin{subequations} \label{eq:chebtaubc}
\begin{equation}
    h_1(1) = h_1'(1) = 0,  \qquad \text{$\forall k$}, 
\end{equation}
and the parity theorem in polar coordinates \cite{boyd2001} dictates the boundary conditions at the pole
\begin{align} 
    h_1'(0) = h_1'''(0) = 0, \qquad &\text{if $k=0$}, \\
    h_1(0) = h_1''(0) = 0, \qquad &\text{if $k=1$}, \\
    h_1(0) = h_1'(0) = 0, \qquad &\text{if $k\geq2$}.
\end{align}
\end{subequations}

A Chebyshev tau method is used to solve the eigenvalue problem \eqref{eq:evproblem}--\eqref{eq:chebtaubc}. The solution is approximated as a truncated series of Chebyshev polynomials $T_n(x)$, $x\in[-1,1]$ with coefficients $a_n$ to be determined
\begin{equation} \label{eq:chebtauh}
    h_1(r) = \sum_{n=0}^{N} a_nT_n\left(2r-1\right).
\end{equation}
However, it is known that the round-off error can be severe when evaluating higher-order derivatives of the Chebyshev polynomials \cite{dongarra1996}. To address this, we introduce a new function $g(r):=h_1''(r)$ and rewrite Equation \eqref{eq:evproblem} as a system of two coupled second order ODE
\begin{equation} \label{eq:chebtau}
    \begin{cases} 
        h_1'' - g = 0, \\
        A_4g'' + A_3g' + A_2g + A_1h_1' + A_0h_1 = \sigma h_1.
    \end{cases}
\end{equation}
This avoids the fourth order derivatives, but we do this at the cost of introducing an additional $N+1$ unknowns, where $g$ can be expressed as 
\begin{equation} \label{eq:chebtaug}
    g(r) = \sum_{n=0}^{N} b_nT_n\left(2r-1\right). 
\end{equation}
Thus, $2N+2$ equations are needed to solve the system. The boundary conditions in Equation \eqref{eq:chebtaubc} give four equations. Further $2N-2$ equations are obtained by evaluating Equation \eqref{eq:chebtau} at the Chebyshev nodes
\begin{equation}
    r_i = \cos(i\Delta x), \qquad \Delta x=\frac{\pi}{N}, \qquad i=1,\dots,N-1. 
\end{equation}
Substituting the expansions \eqref{eq:chebtauh} and \eqref{eq:chebtaug} into Equation \eqref{eq:chebtauh} and \eqref{eq:chebtaubc}, the resulting system of equations has the following structure: 

\begin{center}
\begin{tikzpicture}[scale=0.9]
    \draw (0,0) -- (6,0) -- (6,6) -- (0,6) -- (0,0);
    \draw (3,0) -- (3,6);
    \draw (0,3) -- (6,3);
    \draw (0,0.6) -- (6,0.6);
    \draw (0,3.6) -- (6,3.6);
    \draw (0,2.4) -- (6,2.4);
    \draw (0,5.4) -- (6,5.4);
    \draw (1.5,4.5) node {\scriptsize $T^{(2)}$};
    \draw (4.5,4.5) node {\scriptsize $-T^{(0)}$};
    \draw (1.5,1.5) node {\scriptsize $A_1T^{(1)}+A_0T^{(0)}$};
    \draw (4.5,1.8) node {\scriptsize $A_4T^{(2)}+A_3T^{(1)}$};
    \draw (4.5,1.2) node {\scriptsize $+A_2T^{(0)}$};
    \draw (1.5,3.3) node {\scriptsize $T^{(0)}(1)$};
    \draw (1.5,0.3) node {\scriptsize $T^{(1)}(1)$};
    \draw (1.5,2.7) node {\scriptsize $T^{(1)}(-1)$};
    \draw (1.5,5.7) node {\scriptsize $T^{(0)}(-1)$};
    \draw (4.5,3.3) node {\scriptsize $0$};
    \draw (4.5,0.3) node {\scriptsize $0$};
    \draw (4.5,2.7) node {\scriptsize $0$};
    \draw (4.5,5.7) node {\scriptsize $0$};
\end{tikzpicture}
\begin{tikzpicture}[scale=0.9]
    \draw (0,0) -- (0.6,0) -- (0.6,6) -- (0,6) -- (0,0);
    \draw (0,3) -- (0.6,3);
    \draw (0.3,4.5) node {\scriptsize $\bm{a}$};
    \draw (0.3,1.5) node {\scriptsize $\bm{b}$};
    \draw (0.6,3) node[right] {$=\sigma$};
\end{tikzpicture}
\begin{tikzpicture}[scale=0.9]
    \draw (0,0) -- (6,0) -- (6,6) -- (0,6) -- (0,0);
    \draw (3,0) -- (3,6);
    \draw (0,3) -- (6,3);
    % \draw (0,0.8) -- (6,0.8);
    % \draw (0,3.8) -- (6,3.8);
    \draw (0,0.6) -- (3,0.6);
    \draw (0,2.4) -- (3,2.4);
    \draw (1.5,4.5) node {\scriptsize $0$};
    \draw (4.5,4.5) node {\scriptsize $0$};
    \draw (1.5,1.5) node {\scriptsize $T^{(0)}$};
    \draw (4.5,1.5) node {\scriptsize $0$};
    \draw (1.5,0.3) node {\scriptsize $0$};
    \draw (1.5,2.7) node {\scriptsize $0$};
\end{tikzpicture}
\begin{tikzpicture}[scale=0.9]
    \draw (0,0) -- (0.6,0) -- (0.6,6) -- (0,6) -- (0,0);
    \draw (0,3) -- (0.6,3);
    \draw (0.3,4.5) node {\scriptsize $\bm{a}$};
    \draw (0.3,1.5) node {\scriptsize $\bm{b}$};
\end{tikzpicture}
\end{center}

The generalized eigenvalue problem is then solved using the QZ algorithm with the \hltexttt{eig} function in MATLAB. We verify the solver by finding the eigenvalues for the biharmonic operator $\nabla^4$ with the same boundary conditions. The error of the first ten eigenvalues is shown in Table \ref{tab:biharmerr}. 
\begin{table}[ht]
    \centering
    \begin{tabular}{ |c|c|c|c| } 
        \hline
        $(k,n)$ & Exact & Rel. error ($D^2$) & Rel. error ($D^4$) \\
        \hline
        (0,1) & 104.3631 & $1.131\times10^{-12}$ & $1.417\times10^{-7}$\\ 
        (1,1) & 452.0045 & $5.060\times10^{-13}$ & $1.923\times10^{-9}$\\ 
        (2,1) & 1216.407 & $2.606\times10^{-13}$ & $1.742\times10^{-9}$\\ 
        (0,2) & 1581.744 & $6.140\times10^{-13}$ & $1.629\times10^{-7}$\\ 
        (3,1) & 2604.064 & $2.753\times10^{-12}$ & $9.900\times10^{-10}$\\ 
        (1,2) & 3700.127 & $1.728\times10^{-13}$ & $1.837\times10^{-8}$\\ 
        (4,1) & 4853.327 & $1.573\times10^{-12}$ & $2.492\times10^{-9}$\\ 
        (2,2) & 7154.224 & $6.979\times10^{-13}$ & $1.206\times10^{-9}$\\ 
        (0,3) & 7939.548 & $6.043\times10^{-11}$ & $4.080\times10^{-8}$\\ 
        (5,1) & 8233.563 & $6.030\times10^{-13}$ & $6.003\times10^{-10}$\\
        \hline
    \end{tabular}
    \caption{Error of the first ten eigenvalues of the biharmonic operator with $N=40$. The $D^2$ method solves two second-order ODE, and the $D^4$ method directly solves the fourth-order ODE. }
    \label{tab:biharmerr}
\end{table}

\begin{figure}
\centering
    \subfloat{\includegraphics[width=0.42\textwidth]{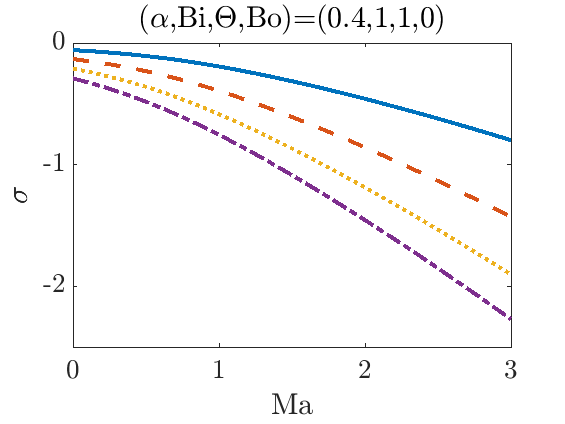}}
    \subfloat{\includegraphics[width=0.42\textwidth]{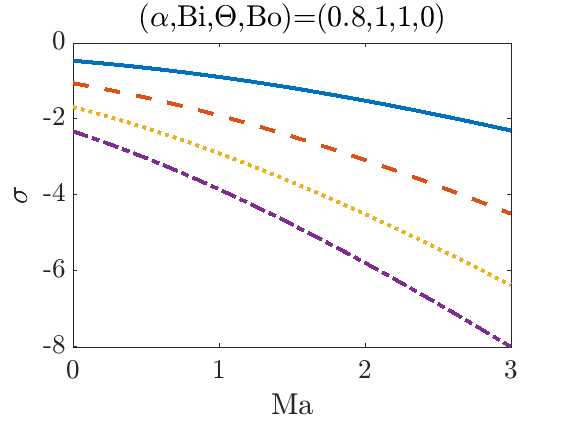}}

    \subfloat{\includegraphics[width=0.42\textwidth]{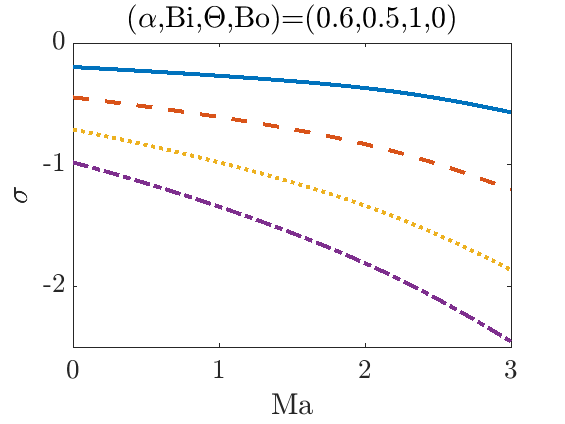}}
    \subfloat{\includegraphics[width=0.42\textwidth]{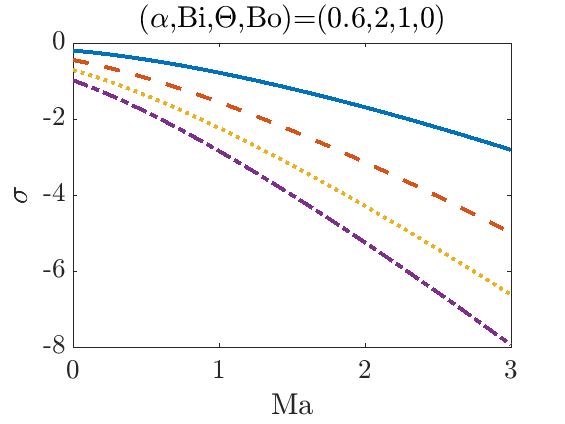}}

    \subfloat{\includegraphics[width=0.42\textwidth]{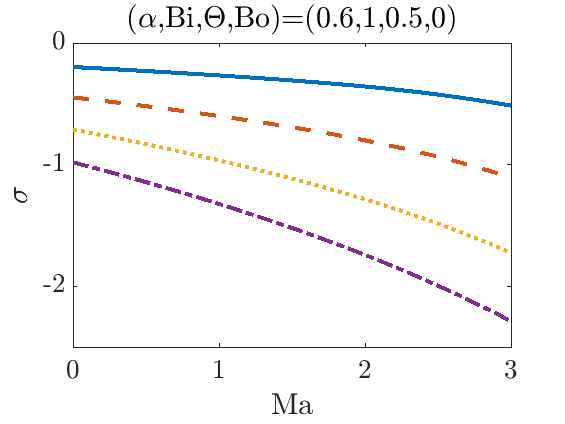}}
    \subfloat{\includegraphics[width=0.42\textwidth]{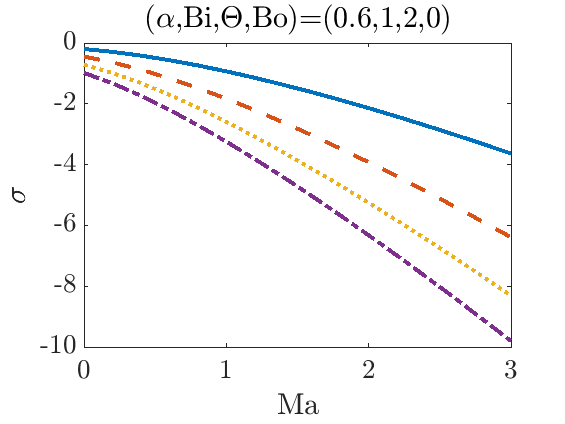}}

    \subfloat{\includegraphics[width=0.42\textwidth]{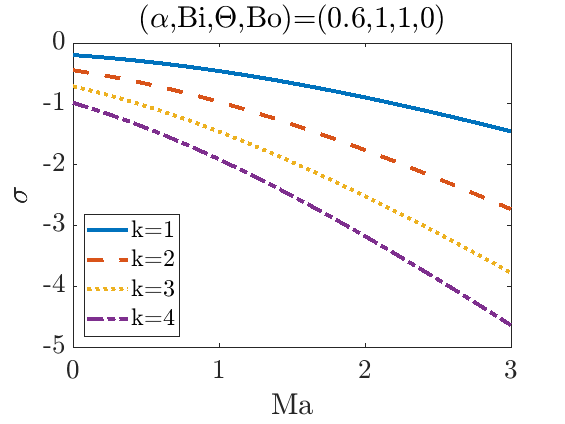}}
    \subfloat{\includegraphics[width=0.42\textwidth]{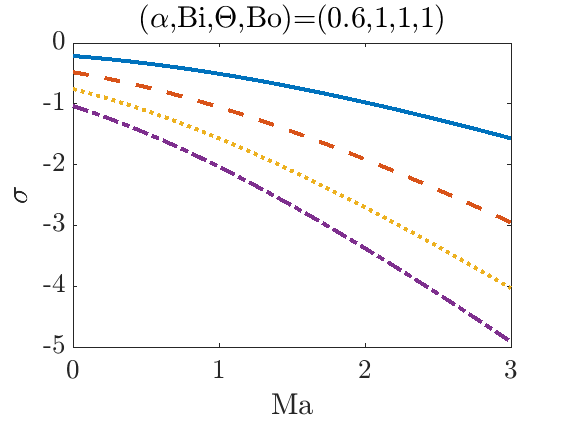}}

    \caption{Dispersion curves for different values of $\alpha$, $\Bi$, $\Theta$, and $\Bo$ for fixed radius droplet $r_*=1$. }
    \label{fig:dispersion}
\end{figure}

\begin{figure}
\centering
    \subfloat{\includegraphics[width=0.42\textwidth]{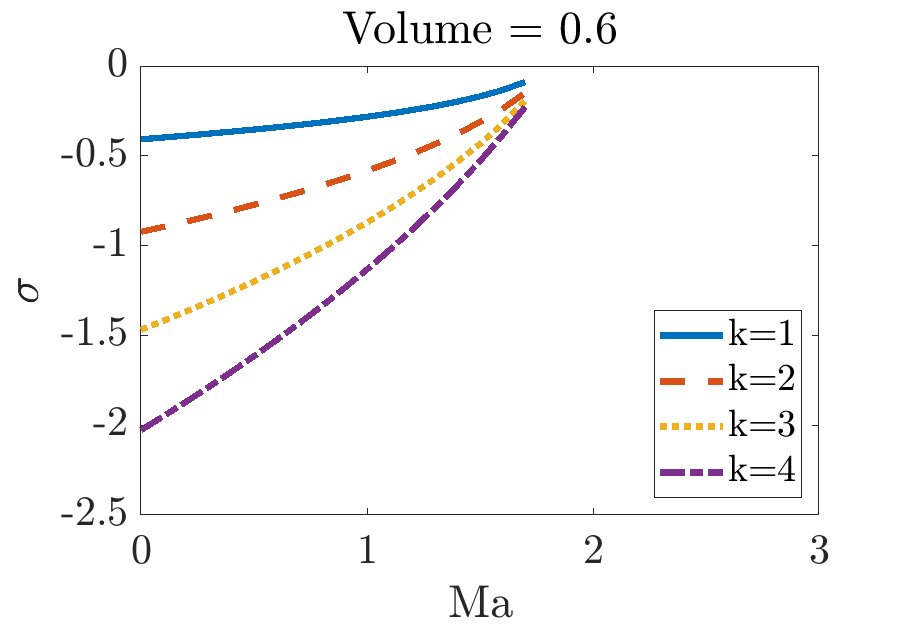}}
    \subfloat{\includegraphics[width=0.42\textwidth]{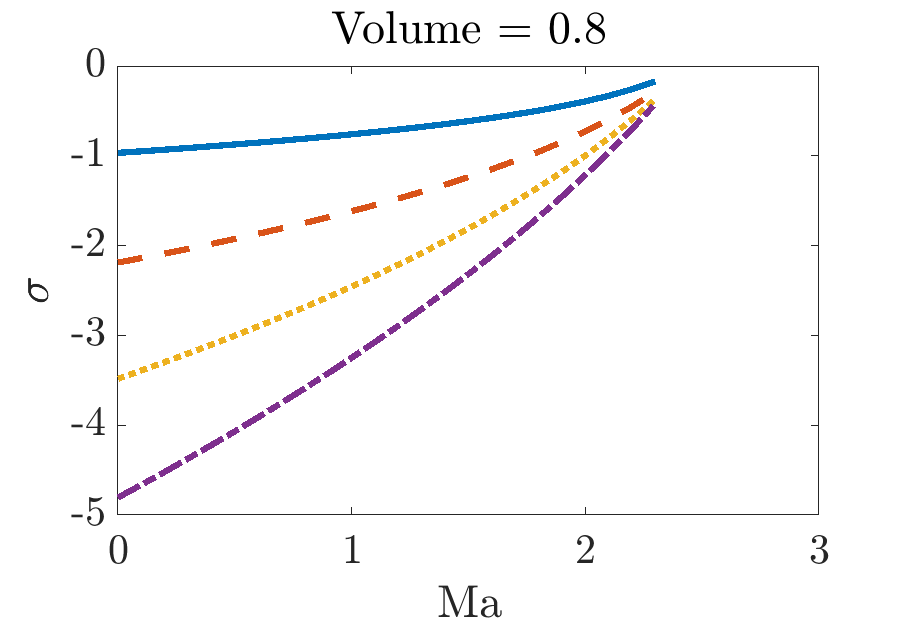}}
    % \subfloat{\includegraphics[width=0.33\textwidth]{images/vol10.png}}
    \caption{Dispersion curves for fix equilibrium droplet volume and radius $r_*=1$, where $\alpha$ is allow to vary. Other parameters are $(\Bi,\Theta,\Bo)=(1,1,0)$. }
    \label{fig:dispersion_vol}
\end{figure}

Results are shown in Figures~\ref{fig:dispersion}--\ref{fig:dispersion_vol}.  Figure \ref{fig:dispersion} shows the largest growth rate $\max_{n}\Re(\sigma_{k,n})$, where $\sigma_{k,n}$ is the $(k,n)$-th eigenvalue, for wavenumber $k=1,2,3,4$ as a function of the Marangoni number.  The figure corresponds to a constant-contact-angle path in parameter space along which an increase in the Marangoni number has a stabilizing effect.  In contrast, 
in Figure~\ref{fig:dispersion_vol}, a constant-volume path in parameter space is taken, along which an increase in the Marangoni number is destabilizing.  There are blank regions in Figure~\ref{fig:dispersion_vol} because the base state solution ceases to exist above a critical Marangoni number.  Summarizing, in both cases, $\max_{n}\Re(\sigma_{k,n})$ remains negative for all considered parameter values.  Hence, the axisymmetric base state is stable with respect to small-amplitude perturbations.

\section{Off-Centered Heating} \label{sec:poff}
\label{sec:offcent}

\subsection{Modelling the Moving Contact Line}  

\begin{figure}[htb]
    \centering
    \includegraphics[width=0.7\linewidth]{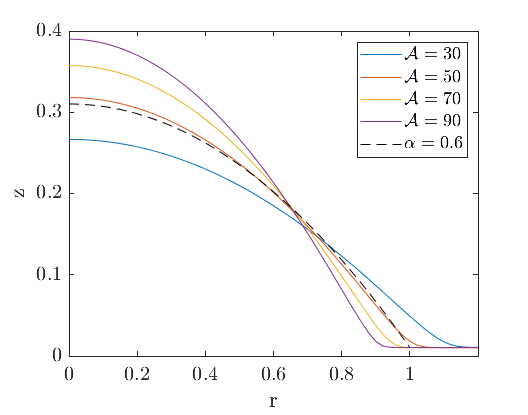}
    \caption{Equilibrium droplet profile for different $\mathcal{A}$ values without the Marangoni effect ($\Ma=0$). The dashed line corresponds to the equilibrium solution without the precursor film with an equilibrium contact angle of $\alpha=0.6$. The droplets are chosen to have the same volume of 0.471. The other parameters are $m=2$, $n=3$, and $\varepsilon=0.01$. }
    \label{fig:precursor_equilibrium}
\end{figure}

In this section, we perform full two-dimensional simulations to investigate the effect when the hotspot is off-centred. We introduce a thin precursor film extending beyond the droplet to model the contact-line motion. The potential is taken to include a two-term Lennard-Jones disjoining pressure of the form 
\begin{equation} \label{eq:ptdisjoint}
    \phi(x,y,t;h) = \mathcal{A}\left[\left(\frac{\varepsilon}{h}\right)^m - \left(\frac{\varepsilon}{h}\right)^n\right], 
\end{equation}
with $0<m<n$. In the precursor model, the parameter $\mathcal{A}$ is related to the equilibrium contact angle via the formula \cite{schwartz1998simulation}
\begin{equation}
    \cos\alpha = 1 - \frac{\varepsilon\mathcal{A}(n-m)}{(n-1)(m-1)}.
\end{equation}
The contact-line region has a scale of $\varepsilon$. Thus to resolve the contact-line region, the simulation resolution should be of the same scale as the precursor-film thickness. With our computing resource, we are limited to $\varepsilon=0.01$, which we will use for the rest of the section. Figure \ref{fig:precursor_equilibrium} shows the equilibrium droplet profile for various $\mathcal{A}$ values with $\varepsilon=0.01$. The profile closest matching an equilibrium contact angle of $\alpha=0.6$ used in the previous section is given by $\mathcal{A}=50$, which we will use for the rest of the section. 

\subsection{Transient Simulations}

As the previous section has not revealed any mechanisms for the breaking of the axisymmetric base state, we explore other non-spontaneous symmetry-breaking mechanisms in this section.  Thus, we  solve Equation~\eqref{eq:pht} with a temperature hotspot that is slightly perturbed from the droplet centre, such that:
\begin{equation}
T_s(x-x_0,y)=\mathrm{e}^{-(x^2+y^2)/s^2},
\end{equation}
where $x_0\ll 1$ is the perturbation, measuring the amount by which the hotspot is off-centre, and $s$ is the size of the hotspot. 
In this scenario, and motivated by References~\cite{yutaku2016,yutaku2017}, we investigate potential symmetry-breaking in the axisymmetric base state by looking at the vorticity inside the droplet.  To calculate the vorticity, we first of all introduce the velocity field:
%From Equation~\eqref{eq:phtuv}, the velocity field in the $xy$-plane is given by 
\begin{align}
    (u,v)(x,y,z,t;h) &= -\Ma z\nabla\psi + \left(\tfrac{1}{2}z^2-hz\right)\nabla(-\nabla^2h + \phi), \\
    w(x,y,z,t;h) &= -\int_0^z \nabla\cdot(u,v) \,\mathd z,
\end{align}
Hence, the $z$-component of the vorticity is given by:
\begin{equation}
    \omega_{z}(x,y,z,t;h) = \pder[v]{x} - \pder[u]{y}.
\end{equation}
% where $u,v,\omega_z$ are defined only within the droplet. 

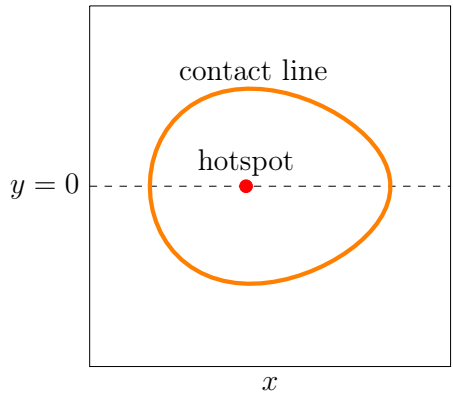
\begin{figure}[tbh]
    \centering
        \begin{tikzpicture}[scale=1.7]
            % axis
            \draw (-1.5,1.5) -- (1.5,1.5) -- (1.5,-1.5) -- (-1.5,-1.5) -- (-1.5,1.5);
            \draw [dashed] (-1.5,0) -- (1.5,0);
            \draw (0,-1.5) node[below] {$x$};
            \draw (-1.5,0) node[left] {$y=0$};
            % curve
            \draw [orange, line width=0.6mm, domain=0:360, samples=60] plot ({-cos(\x)}, {(0.14*cos(\x)+0.8)*sin(\x)});
            \filldraw[red] (-0.2,0) circle (1.5pt);
            \draw (-0.2,0) node[above] {hotspot};
            \draw (-0.14,0.8) node[above] {contact line};
        \end{tikzpicture}
    \caption{The symmetry along the $y$-axis is kept, so we only solve on one side of the domain and reflect the solution to get the full droplet profile. Since Newton's method is at least $O(N^2)$ in computational complexity, with $N$ being the number of grid points, this procedure speeds up the simulation by at least four times. }
    % \label{fig:my_label}
\end{figure}
We solve Equation~\eqref{eq:pht} with the disjoining pressure \eqref{eq:ptdisjoint} using a finite-difference scheme on a rectangular grid. The simulation is performed on the domain $(x,y)\in[-L_x,L_x]\times[-L_y,0]$ with symmetric boundary conditions on all four sides and equally spaced in both directions $\Delta x=\Delta y=0.01$. The second-order centred-difference scheme is used to discretize the space domain, and the Crank–Nicolson method with step size $\Delta t=0.01$ is used to discretize the time domain. This results in a system of nonlinear equations for each time step which is solved using Newton's method~\cite{gtfe2022,witelski2003adi}.   The initial condition is radially symmetrical with the form
\begin{equation}
    h(r,t=-100) = \begin{cases}
        A(1-r^2)^2 + \varepsilon, & \qquad r\leq1, \\
        \varepsilon, & \qquad \text{otherwise.}
    \end{cases}
\end{equation}
We first evolve the system with a homogeneous temperature profile $T_s=0$ for 100 nondimensional time until $t=0$ before turning on the localised heating. The solution is insensitive to varying resolutions, and we conclude that the numerical method has converged. 

\begin{figure}[tbh]
    \centering
    \subfloat{\includegraphics[width=0.5\textwidth]{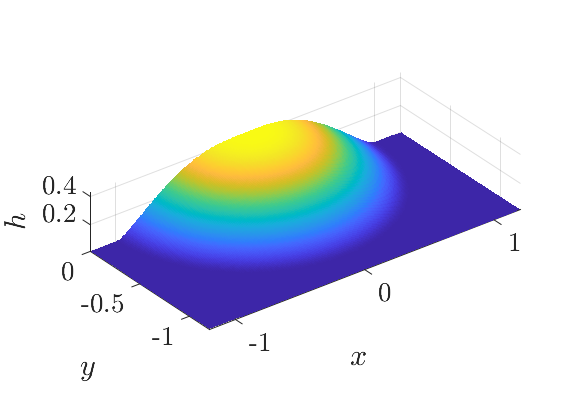}}
    \subfloat{\includegraphics[width=0.5\textwidth]{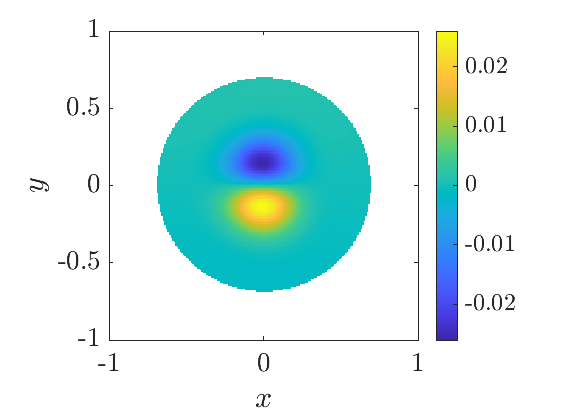}}
    \caption{Solution for an off-centre heated droplet using the precursor film description. (left) Droplet profile $h$ at $t=1$. (right) An $xy$-slice of the $z$-vorticity $\omega_z$. The precursor parameters are $(\mathcal{A}, \varepsilon, m, n)=(50,0.01,2,3)$, heating location at $(x,y)=(-0.01,0)$ and a width of $s=0.2$, and all other parameters $(\Ma,\Bi,\Theta)$ are taken to be unity. }
    \label{fig:precursor_vorticity}
\end{figure}

\begin{figure}[tbh]
    \centering
    \subfloat[]{\includegraphics[width=0.5\textwidth]{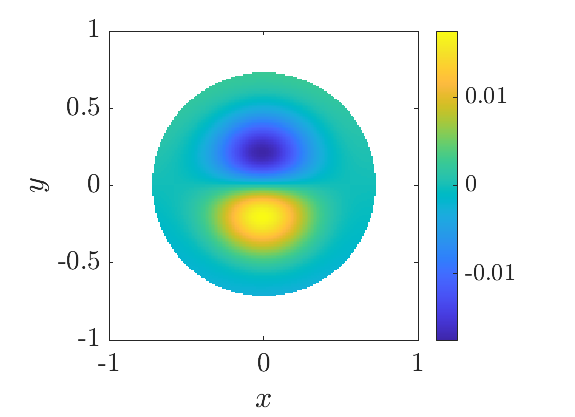}}
    \subfloat[]{\includegraphics[width=0.48\textwidth]{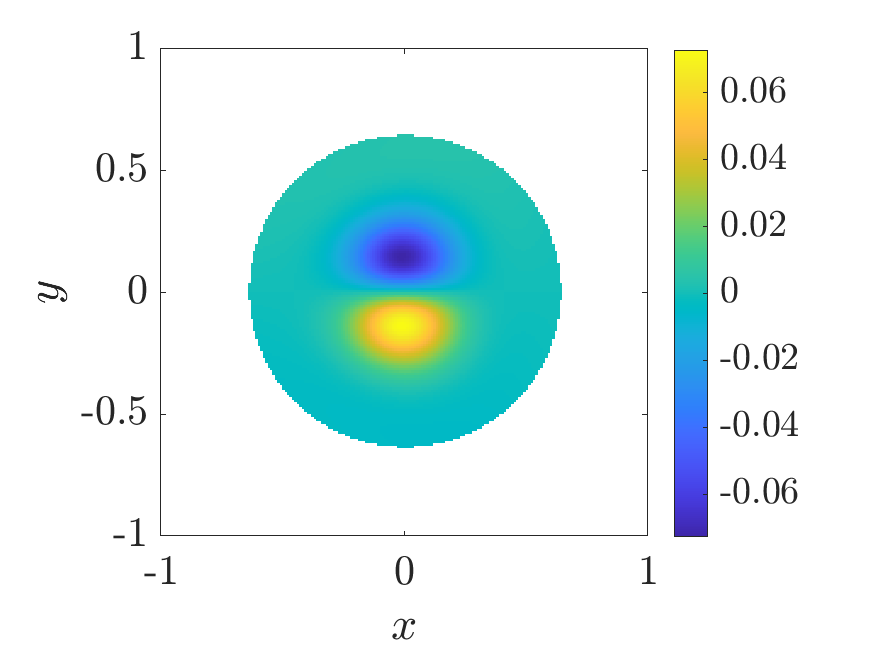}}
    \caption{$\omega_z$ under different heating parameters. (a) Increased hotspot width ($s=0.3$). (b) Increased Marangoni number ($\Ma=2$). All other parameters are the same as Figure~\ref{fig:precursor_vorticity}. }
    \label{fig:precursor_vorticity2}
\end{figure}

A first set of results in Figure~\ref{fig:precursor_vorticity} shows the solution at $t=1$ when the localized heating is located at $(x,y)=(-0.01,0)$.
% Figure \ref{fig:precursor_vorticity} shows $\omega_z$ at the mid-height of the droplet at $t=1$ when the localized heating is located at $(x,y)=(-0.01,0)$.   
The left panel shows the droplet profile $h$, and the second panel on the right shows a plot of $\omega_z$ at the mid-height of the droplet at $z=0.2151$. Two primary vortices in the $z$-direction are observed inside the droplet. 
Hence, the symmetry-breaking and the departure from the axisymmetric base state are a consequence of a slightly off-centre hotspot.  In other words, the observed asymmetric vortex structure is not intrinsic to the studied system but arises from an externally-prescribed asymmetry. Figure~\ref{fig:precursor_vorticity2} shows the effect on the vorticity when the heating power is increased. A larger hotspot size results in an increased vortex size with a lower magnitude, while increasing the Marangoni number results in a larger vorticity strength. 

\begin{figure}[htb]
    \centering
    \subfloat[\label{fig:vorticity_strength}]{\includegraphics[width=0.48\textwidth]{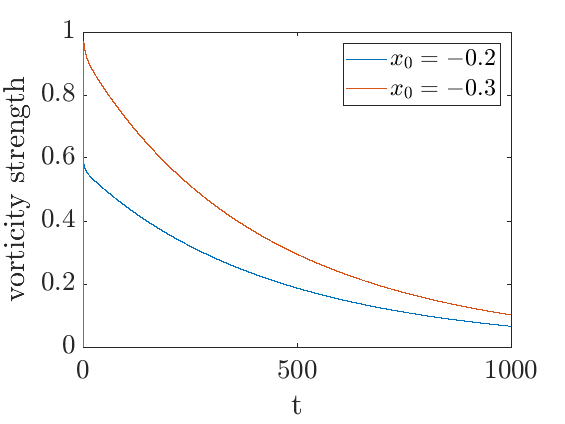}}
    \subfloat[\label{fig:droplet_centre}]{\includegraphics[width=0.48\textwidth]{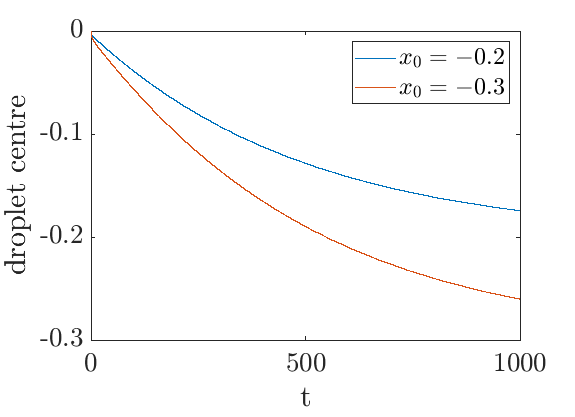}}
    \caption{Evolution of the droplet characteristics for two different heating locations.  (a) Maximum $z$-vorticity at the mid-height of the droplet decays over time. (b) The centres of the droplets travel towards the heating location. Aside from the heating location, all parameters are the same as Figure~\ref{fig:precursor_vorticity}.}
\end{figure}

To further understand this effect, we plotted the vorticity strength over time in Figure~\ref{fig:vorticity_strength}.  The figure shows the maximum vorticity first increasing as the droplet transitions away from a homogeneously-heated configuration (without vorticity, $\omega_z(x,y,z,t=0)=0$). This occurs over a very small timescale of $\sim1$ nondimensional time.  Thereafter, the vorticity strength rises to a maximum before decaying again to zero.  The decay of the vorticity strength corresponds to a `thermotaxis' event where the droplet moves so that its centre coincides with the hotspot centre (Figure~\ref{fig:droplet_centre}), and resumes a radially-symmetric equilibrium shape.  Thus, the asymmetric vortex pair is only a transient event.
% karbalaei2018microfluidic

\subsection{Pinned Contact Line}

The previous transient results reveal that the symmetry breaking is only a transient effect in cases where the droplet contact line can move -- in such a case, the droplet moves via thermotaxis so as to resume a radially symmetric state.  Therefore, in order to investigate a mechanism for persistent symmetry-breaking, we consider numerical solutions of the basic droplet model~\eqref{eq:pht} with slightly off-centred heating and with a pinned contact line.  

For these purposes, we seek an equilibrium solution as $t\rightarrow\infty$ of Equation~\eqref{eq:pht} with a fixed contact line.  Hence,  we again look at the time-independent equation
\begin{equation} \label{eq:phtfe_eq}
    \nabla\cdot\left\{-\frac{1}{2}\Ma h^2\nabla\psi + \frac{1}{3}h^3\nabla\nabla^2h\right\} = 0,
\end{equation} 
with boundary condition
\begin{equation} \label{eq:ociterbc}
    h(r_*,\varphi) = 0, \qquad \partial_rh(r_*,\varphi) = -\alpha. 
\end{equation}
For off-centred heating, the solution $h=h(r,\varphi)$ is no longer axisymmetric.  Hence, Equation~\eqref{eq:phtfe_eq} must be solved on  the disk with $(r,\varphi)\in[0,r_*]\times[0,2\pi)$.   

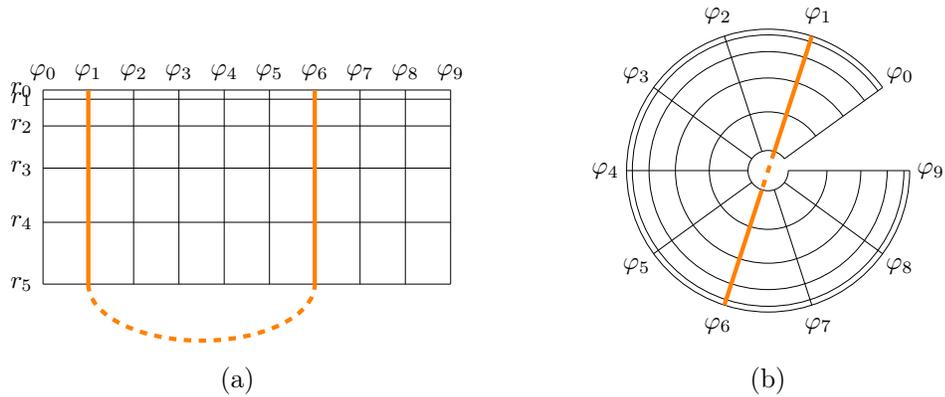
\begin{figure}[tbh]
\centering
\begin{subfigure}{.5\textwidth}
    \centering
    \begin{tikzpicture}[scale=3.2]
        \pgfmathsetmacro{\r}{cos(deg(3.1415*5/11))}
        \pgfmathsetmacro{\dt}{0.2}
        \foreach \i in {0,...,9} {
            \draw (\i*\dt,\r) -- (\i*\dt,1);
            \draw (\i*\dt,1) node[above] {\footnotesize $\varphi_{\i}$};
        }
        \foreach \i in {0,...,5} {
            \pgfmathsetmacro{\r}{cos(deg(3.1415*\i/11))}
            \draw (0,\r) -- (9*\dt,\r);
            \draw (0,\r) node[left] {\footnotesize $r_{\i}$};
        }
        % slice
        \foreach \i in {1,6} {
            \draw [orange, line width=0.6mm] (\i*\dt,\r) -- (\i*\dt,1);
        }
        % \draw [orange, line width=0.6mm, dashed] (2*\dt,\r) -- (7*\dt,\r);
        \draw [orange, line width=0.6mm, dashed] (1*\dt,\r) arc [start angle=180, end angle=360, x radius={2.5*\dt}, y radius=0.25];
    \end{tikzpicture} 
    \caption{}
\end{subfigure}%
\begin{subfigure}{.5\textwidth}
    \centering
    \begin{tikzpicture}[scale=2]
        \foreach \i in {0,...,5} {
            \pgfmathsetmacro{\r}{cos(deg(3.1415*\i/11))}
            \draw (\r,0) arc [start angle=360, end angle=36, x radius=\r, y radius=\r];
        }
        \pgfmathsetmacro{\r}{cos(deg(3.1415*5/11))}
        \foreach \i in {0,...,9} {
            \pgfmathsetmacro{\phi}{deg(2*3.1415*\i/10)}
            \draw ({\r*cos(\phi)}, {\r*sin(\phi))}) -- ({cos(\phi)}, {sin(\phi)});
            \pgfmathsetmacro{\phi}{deg(2*3.1415*(\i+1)/10)}
            \draw ({1.15*cos(\phi)},{1.15*sin(\phi)}) node {\footnotesize $\varphi_{\i}$};
        }
        % slice
        \foreach \i in {2,7} {
            \pgfmathsetmacro{\phi}{deg(2*3.1415*\i/10)}
            \draw [orange, line width=0.6mm] ({\r*cos(\phi)}, {\r*sin(\phi))}) -- ({cos(\phi)}, {sin(\phi)});
        }
        \pgfmathsetmacro{\a}{deg(2*3.1415*2/10)}
        \pgfmathsetmacro{\b}{deg(2*3.1415*7/10)}
        \draw [orange, line width=0.6mm, dashed] ({\r*cos(\a)}, {\r*sin(\a))}) -- ({\r*cos(\b)}, {\r*sin(\b))});
    \end{tikzpicture}
    \caption{}
\end{subfigure}
\caption{Example of a simulation grid in a Cartesian grid arrangement (a) and the mapping onto a polar grid (b). For the boundaries, the pole is avoided by construction, the periodic boundary condition is applied in the azimuthal direction, and the only actual boundary is the $r_0$ row corresponding to the contact line $r=r_*$. It is also crucial that every node has a reflection through the origin for the derivatives to be computed.}
\label{fig:polar_grid}
\end{figure}

A spectral method on the disk is used, which gives a higher resolution solution compared to the previously-introduced finite-difference method. Additionally, using a polar grid removes any spurious four-fold symmetry-breaking that can arise in such simulations due to numerical discretization errors associated with a Cartesian grid. 
The numerical grid is equally spaced in the $\varphi$-direction and has the Chebyshev spacing in the $r$-direction. A sketch of the numerical grid is shown in Figure~\ref{fig:polar_grid}. The equilibrium solution is found using an iterative method with the expression
\begin{equation} \label{eq:ociter}
     \nabla\cdot\left\{\frac{1}{3}(h^n)^3\nabla\nabla^2\right\}h^{n+1} = \nabla\cdot\left\{\frac{1}{2}\Ma (h^n)^2\nabla\psi^n\right\}. 
\end{equation}
Starting with an initial guess $h^0$, we solve for $h^{n+1}$ in Equation~\eqref{eq:ociter} along with boundary conditions \eqref{eq:ociterbc} in a least square sense until the solution converges. The solution is substituted back into Equation~\eqref{eq:phtfe_eq} to ensure the error is small. A limitation of this method is that only droplets with a circular contact line can be expressed. 

Once the equilibrium solution $h$ is found, the velocity field $\bm{u}=u_r\hat{r}+u_\varphi\hat{\varphi}+u_z\hat{z}$ can be computed with 
\begin{align}
    u_r(r,\varphi,z;h) &= -\Ma z\pder[\psi]{r} + \left(hz-\frac{1}{2}z^2\right)\pder{r}\nabla^2h, \\
    u_{\varphi}(r,\varphi,z;h) &= \frac{1}{r}\left\{-\Ma z\pder[\psi]{\varphi} + \left(hz-\frac{1}{2}z^2\right)\pder{\varphi}\nabla^2h\right\}, \\
    u_z(r,\varphi,z;h) &= -\frac{1}{r}\int_0^z \left(\pder{r}(ru_r) + \pder[u_\varphi]{\varphi}\right) \,\mathd z,
\end{align}
and the $z$-vorticity is given by
\begin{equation}
    \omega_z(r,\varphi,z;h) = \frac{1}{r}\left(\pder{r}(ru_\varphi) - \pder{\varphi}u_r\right),
\end{equation}
defined only within the droplet. 

% A slice of the $z$-vorticity at the mid-hight of the droplet is shown in Figure~\ref{fig:pinned_vorticity}. We still see vorticity formation in the droplet even when the symmetry of the contact line is enforced. When the heating is centred, the vorticity vanishes. 

\begin{figure}[htb]
    \centering
    \subfloat[Without interpolation.]{\includegraphics[width=0.42\textwidth]{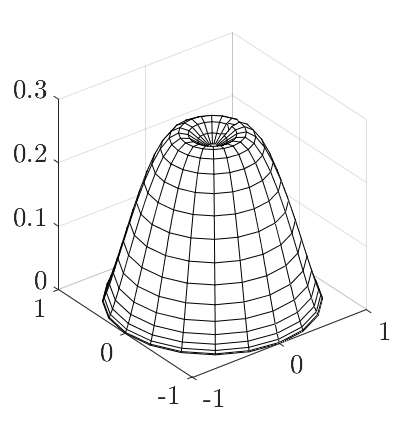}}
    \subfloat[With interpolation.]{\includegraphics[width=0.42\textwidth]{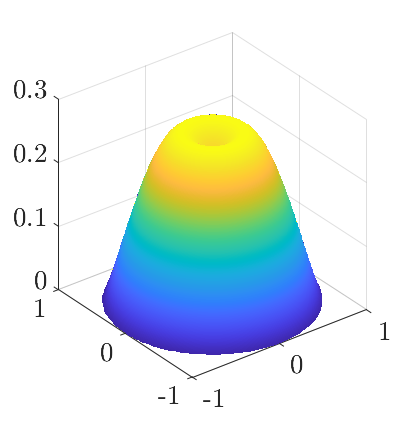}}
    \caption{Equilibrium solution using a spectral method on the disk. Contact angle $\alpha=0.3$, centred heating. All other parameters $(\Ma,\Bi,\Theta)$ are taken to be unity. }
    \label{fig:spectral_solution}
\end{figure}

A solution is shown in Figure~\ref{fig:spectral_solution}. We use $M=20$ points in the azimuthal direction and $N=18$ points in the radial direction. The residue of the iterative scheme is bounded by $\|h^{n+1}-h^n\|_{\infty}<10^{-9}$.
Panel (a) shows the solution evaluated on the spectral grid points. Even though the resolution seems low initially, the spectral solution describes polynomials of degree $N$ in the radial direction and $M$-terms Fourier series in the azimuthal direction \cite{trefethen2000spectral}. The interpolated solution is shown in panel (b). The localized heating results in a depression at the centre of the droplet. 

\begin{figure}[ht]
    \centering
    \subfloat[Off-centered heating]{\includegraphics[width=0.5\textwidth]{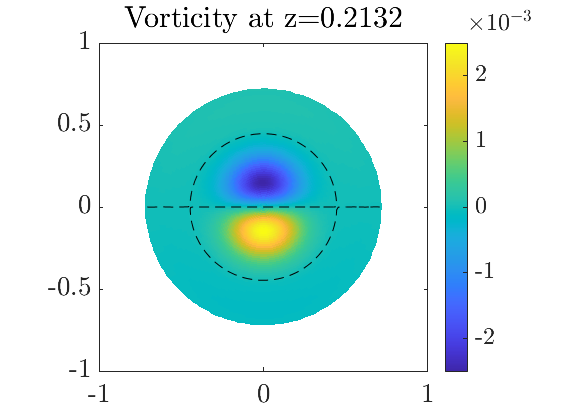}}
    \subfloat[Centered heating]{\includegraphics[width=0.5\textwidth]{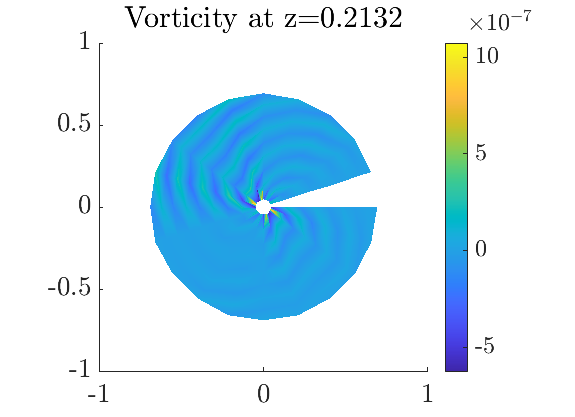}}
    \caption{The $z$-component of the vorticity $\omega_z$ in an off-centre heated droplet with pinned circular contact line at $r_*=1$ and $\alpha=0.6$. Heating location at $(x,y)=(-0.001,0)$ for the off-centered heating. The dashed line represents the level-zero contour, which divides the droplet into four circulation regions. All other parameters $(\Ma,\Bi,\Theta)$ are taken to be unity. }
    \label{fig:pinned_vorticity}
\end{figure}

Figure~\ref{fig:pinned_vorticity} shows the vorticity when the heating is off-centred.  Panel (a) shows the vorticity in the case of off-centred heating.  The vortex pair can be seen clearly.  In contrast, in panel (b), the vorticity in the case of centred heating is shown; this is zero, up to numerical error.  These results establish two necessary criteria for the existence of persistent symmetry-breaking -- off-centred heating and a pinned contact line.
\begin{figure}[htb]
    \centering
    \includegraphics[width=0.5\textwidth]{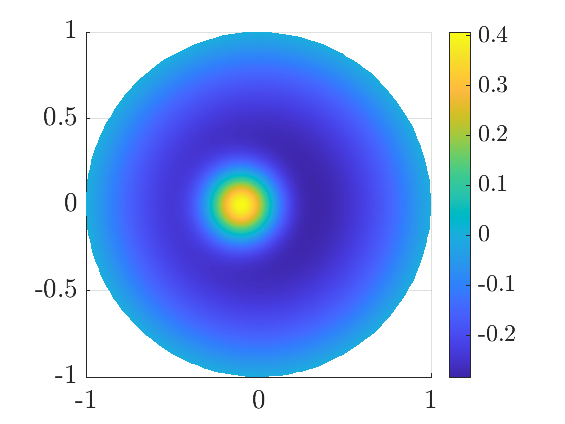}
    \caption{Relative temperature profile at the liquid-gas interface $\psi(x,y;h)$ with heating location at $(x,y)=(-0.1,0)$.}
    \label{fig:surface_temperature}
\end{figure}
We emphasize finally that the vorticity in Figures~\ref{fig:precursor_vorticity}--\ref{fig:pinned_vorticity} is only observed after visualizing the flow inside the droplet.  The temperature at the droplet surface does not exhibit any vorticity signature (e.g. Figure~\ref{fig:surface_temperature}).  This is consistent with the experimental findings~\cite{kita2019}, where the authors observed a vorticity signature in the surface temperature profile at higher contact angle values ($100^\circ$) but not at the lower contact angle values.  

\section{Other Instabilities} \label{sec:pother}

\subsection{Ring Rupture}

There is a different \amend{state to the equilibrium solution for} the point heated droplet, which was alluded to in the previous section as shown in Figure~\ref{fig:equibrium_droplet_size}. When the droplet is sufficiently thin, there exists a critical Marangoni number $\Ma_*$ such that whenever $\Ma>\Ma_*$, the classical equilibrium solution to Equation~\eqref{eq:phtfe_r} ceases to exist. This is due to the finite-time blowup of the solution as $h\downarrow0$ corresponding to the rupturing of the droplet near the heating location. \amend{We emphasise that there is no bifurcation in the sense that the stability of the system does not change during the transition. }

\begin{figure}
\centering
\subfloat{\includegraphics[width=1\textwidth]{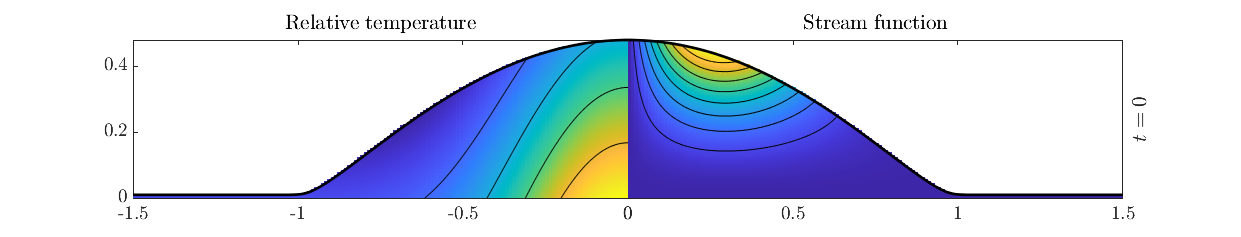}}

\subfloat{\includegraphics[width=1\textwidth]{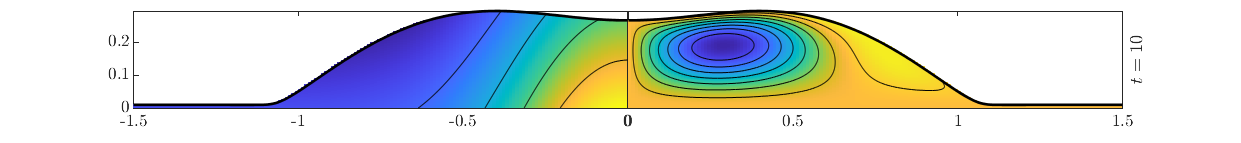}}

\subfloat{\includegraphics[width=1\textwidth]{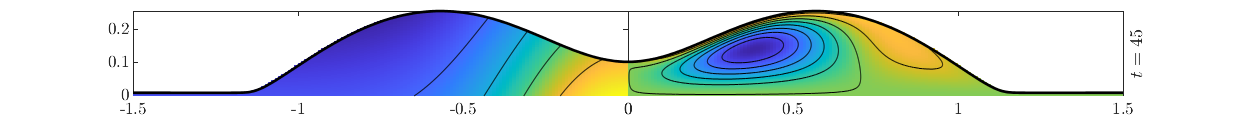}}

\subfloat{\includegraphics[width=1\textwidth]{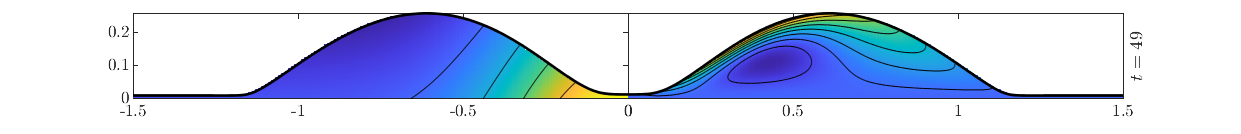}}

\subfloat{\includegraphics[width=1\textwidth]{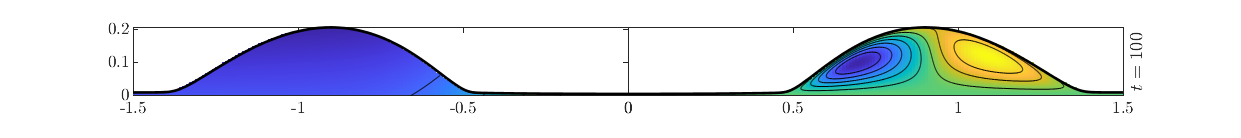}}
\caption{Evolution of the ring rupture process with axisymmetric heating profile $T_s(r)=\mathe^{-r^2/0.4^2}$. Relative temperature is shown on the left, and the stream function is shown on the right. The scale is maintained through the snapshots. The precursor film parameters are $(\mathcal{A}, \varepsilon, m, n)=(50,0.01,2,3)$ and all other parameters $(\Ma,\Bi,\Theta)$ are taken to be unity. }
\label{fig:rupturing_evolution}
\end{figure}

\begin{figure}
    \centering
    \includegraphics[width=0.6\linewidth]{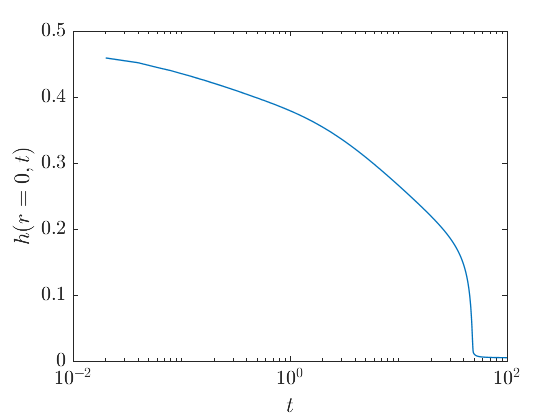}
    \caption{The height of the droplet at $r=0$ for the ring rupture process in Figure \ref{fig:rupturing_evolution}. }
    \label{fig:rupturing_height}
\end{figure}

When a repulsive van der Waals potential such as Equation~\eqref{eq:ptdisjoint} is used, the finite-time blowup is avoided. A transient simulation of the rupturing process of an axisymmetrical droplet is shown in Figure~\ref{fig:rupturing_evolution}. The initial condition is taken to be the equilibrium droplet profile without the hotspot (i.e. homogeneous heating). At $t=0$, the localised heating is turned on, and the droplet is no longer at equilibrium, and we observed the streamlines intersecting with the droplet surface. The localised heating reduces the surface tension at the centre of the droplet, causing a shear force acting from the centre towards the edge of the droplet. This results in a depression at the heated location, eventually leading to the droplet rupturing. The thickness of the droplet at the rupturing point is plotted in Figure~\ref{fig:rupturing_height}. When the system is axisymmetrical, the rupturing occurs at the droplet centre, resulting in a ring-shaped droplet. We note that the ring-shaped droplet is unstable, as one might expect. 


\subsection{\markup{Thermotaxis}}

\begin{figure}
\centering
\subfloat{\includegraphics[width=0.9\textwidth]{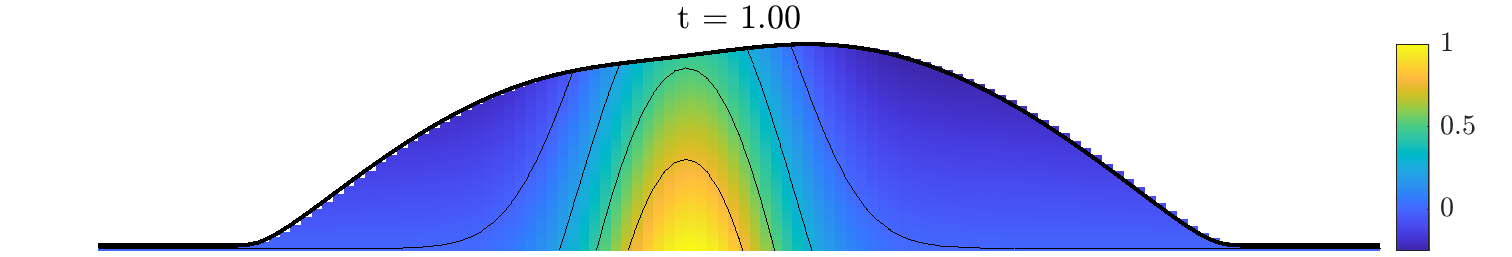}}

\subfloat{\includegraphics[width=0.9\textwidth]{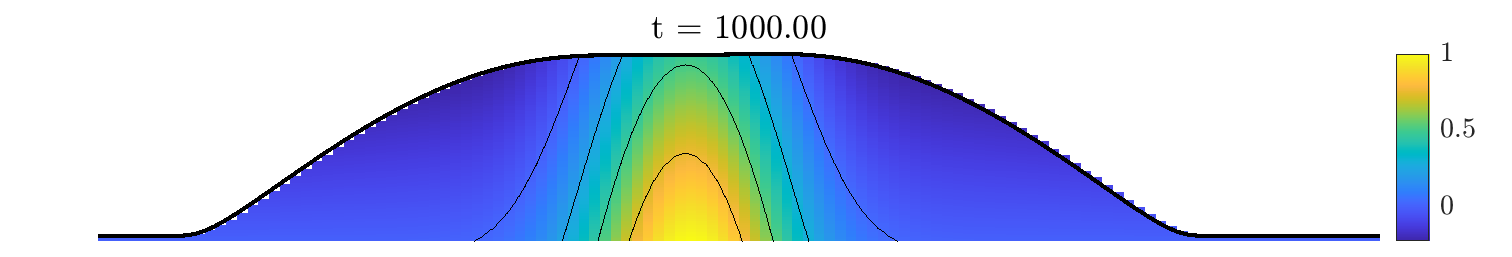}}
\caption{Snapshot of the transient simulation showing the droplet migrating towards the hotspot. The relative temperature in the droplet is shown. }
\label{fig:thermotaxis}
\end{figure}

Thermotaxis is the behaviour where the droplet migrates across a substrate due to a temperature difference on the substrate. This causes an asymmetric Marangoni current which drives the droplet motion. The thermotaxis behaviour is alluded to in \Cref{fig:droplet_centre}, where we observed the droplet moves towards the hotspot. Snapshots of the transient simulation are shown in \Cref{fig:thermotaxis}. Interestingly, this direction is opposite to the behaviour of droplets on a directionally-heated substrate, which moves towards the cooler region of the substrate \cite{karbalaei2016}. The model and numerical solvers developed in this chapter are primed to investigate the thermotaxis behaviour of sessile droplets. We leave this for future work. 

\section{Discussion and Conclusions}

In conclusion, we have developed a descriptive model for generating azimuthal flows in point-heated droplets.  We have drawn inspiration from the experiments in Reference~\cite{yutaku2017}, wherein vortical flows in the azimuthal direction were observed to form spontaneously after applying localized point heating on the substrate.  Throughout, we have emphasized the limited applicability of the present analytical model, focused as it is on small equilibrium contact angles.  However, an advantage of this approach is the resulting simplified mathematical model, which is analytically tractable and does not require large-scale three-dimensional direct numerical simulation.   Despite these limitations, the model does provide some insights into the vortical flows in point-heated droplets.  First, the radially-symmetric base state is revealed to be linearly stable with respect to small-amplitude perturbations in the azimuthal direction.  Thus, linear stability is ruled out as a mechanism for the generation of vortical flows in the azimuthal direction.  

This chapter has explored a second mechanism for the generation of such flows: namely, a small perturbation of the heat source from the droplet centre.  Such perturbations do give rise to azimuthal vortical flows, qualitatively similar to those observed in the experiments in Reference~\cite{yutaku2017}.  Our simulations reveal that such vortices die out in the case of depinned droplets: in this case, the droplet moves to resume a radially-symmetric equilibrium configuration.  Our simulations further reveal that the vortices are persistent when the droplet contact line is pinned.  Thus, a potential mechanism for generating and maintaining such flows is twofold: very slightly off-centred heating combined with contact-line pinning. 

Much work is needed to understand the onset of the twin vortices further. The linear stability analysis used in this chapter is only valid for small perturbations from the base state. A transient-growth analysis can be used to study perturbations of much larger amplitudes and the base state's stability range. The theoretical results are limited to the hydrophilic case where the equilibrium contact angle is small, and DNS methods are needed to study droplets beyond the lubrication limit. We found that the spherical cap assumption for the droplet profile is inadequate to study interfacial dynamics, especially when the system is asymmetric. A free-surface model such as the two-phase level set (TPLS) method \cite{scott2013performance} is better suited to study the Marangoni current in a droplet. 
Evaporation can also be modelled in the lubrication limit to study the effect of the evaporative flux on the droplet stability. 

We have also performed transient simulations. These require separate modelling of the moving contact line. We chose to model the moving contact line using the precursor model. A two-term Lennard-Jones potential is introduced to model the van der Waals force between the fluid and the substrate. The precursor film, however, is incompatible with the Chebyshev tau method used in the linear stability analysis. The reason is twofold, the first being that in the precursor-film model, the droplet bulk transitions smoothly into the precursor film and does not have a well-defined contact line, which is needed to impose the boundary condition for the Chebyshev tau method. One might try to extend the domain beyond the contact-line region. Still, the sharp derivative near the contact-line region also degrades the solution of the Chebyshev tau method, which works best with droplet profiles that do not possess such sharp derivatives. Secondly, even though the precursor film only exists near the droplet, for numerical purposes, the precursor film is usually extended indefinitely to coat the whole substrate to avoid the contact line singularity. Since the film height does not decay, the eigenmodes exhibit oscillations in the precursor film region. This is not physical, as any perturbation should be contained within the bulk of the droplet. The slip length model is also unsuitable in this situation where the transient phase of the droplet evolution has irregular contact line shapes when heated asymmetrically. This poses a challenge to model the contact line motion. For these reasons, we seek an alternative description of droplet spreading which implicitly describes the contact line motion and is compatible with the Chebyshev analysis. So we are motivated in the rest of the thesis to explore - albeit in a preliminary way - other, novel methods to describe moving contact lines.

\chapter{Geometric Thin-Film Equation} \label{sec:theory}

\section{\markup{Overview}}

This chapter discusses a novel regularization method for the thin-film equation using the geometric diffuse-interface method (G-DIM). The G-DIM was first introduced by Holm, \'O N\'araigh, and Tronci in \cite{holm2020gdim}, and we seek to develop G-DIM further.  In \Cref{sec:gdim}, we review the G-DIM theory and discuss the preliminary work by the original authors. 
In \Cref{sec:modelab}, an alternative derivation of the Geometric Thin-Film Equation is proposed. We show that the Geometric Thin-Film Equation arises in certain physical contexts. 
The departure of our model from the model introduced by Holm et al. is highlighted in \Cref{sec:departure}. 
Two robust numerical schemes for solving the Geometric Thin-Film Equation are developed. The finite-difference solver is introduced in \Cref{sec:fdm} and the particle method solver in \Cref{sec:particles}. We demonstrate numerically that the Geometric Thin-Film Equation is able to reproduce spreading, and the spreading rate of the droplet agrees with Tanner's law. Two optimization techniques for the particle method solver are introduced, enabling fast numerical exposition of general thin-film flows. The error and performance of the solvers are compared. 

\section{\markup{Background}}
\label{sec:gdim}

The geometric diffuse-interface method (G-DIM) is inspired by the diffuse-interface method (DIM) for two-phase flows.  The DIM describes the two phases by a scalar `concentration' function $C:\Omega\rightarrow [0,1]$, where $C=0$ indicates the first phase and $C=1$ indicates the second phase.  At the interface, the function $C$ transitions smoothly from one phase to the other, usually with a transition function (such as $\tanh$) over some width $d$. This blurring of the interface means that the interface position is no longer defined precisely, although the level set $C=0.5$ is usually taken to be the interface. When applied to the droplet spreading problem, the smooth transition of the scalar field allows the stress at the contact line to dissipate over the interface region, thereby allowing slip. The G-DIM works similarly, but instead of the whole interface, only the contact line is blurred, allowing the same dissipation to occur across the contact region. The G-DIM has the added benefit of retaining a sharp liquid-gas interface, which allows us to treat the interface as a function. 

In \cite{holm2020gdim}, the model is formulated purely around the gradient-flow theory. The blurring of the contact line is achieved by using a convolution operator with a smoothing kernel $K$. The model now involves a smooth interface height $\barh$, as well as a sharp interface height $h$, interacting via an evolution equation. The evolution equation for $\bar{h}$ is obtained by a regularization of the free-energy functional while preserving the gradient-flow structure of the thin-film equation. The regularized equation was placed in the context of differential geometry as the Lie derivatives on certain one-form -- hence the name \textit{Geometric} Thin-Film Equation. In this chapter, an alternative derivation of the Geometric Thin-Film Equation is proposed. % and we show that the equation arises in certain physical context. These models are introduced in \Cref{sec:modelab}.  

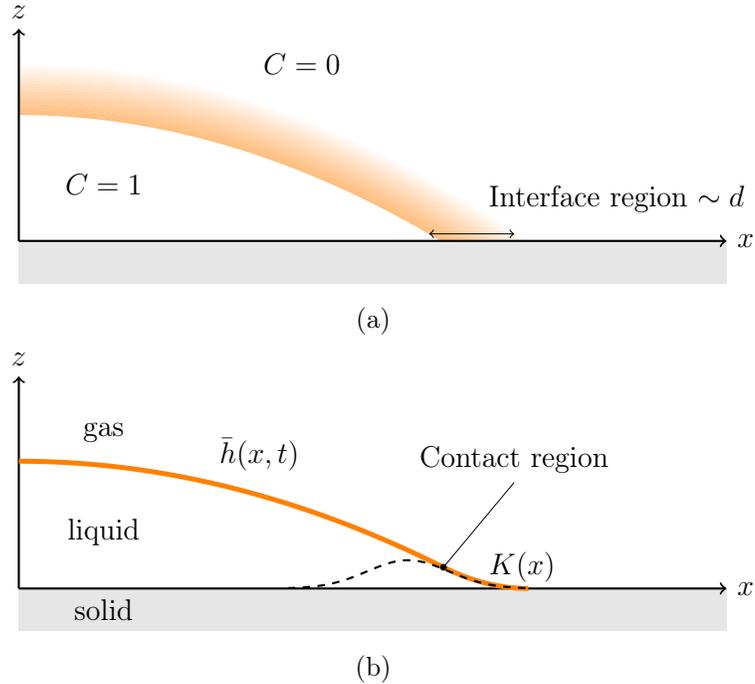
\begin{figure}[tbh]
    \centering
    \subfloat[]{
    \begin{tikzpicture}
        \fill [black!10] (0,0) rectangle (10,-0.6);
        % interface
        \pgfmathsetmacro{\n}{20}
        \foreach \i in {0,...,\n}{
            \pgfmathsetmacro{\f}{\i/\n}
            \draw[orange!50, line width=0.4mm, opacity=1-\f] plot[domain=0:6,smooth] ({(1+0.2*\f)*\x},{(1+0.4*\f)*1.8*(1-(\x/6)^2)});
        };
        % \fill[orange!50] (0,0) -- plot[domain=0:6,smooth] (\x,{1.8*(1-(\x/6)^2)}) --cycle;
        % axes
        \draw [->, line width=0.3mm] (0,0) -- (10,0) node[right] {$x$};
        \draw [->, line width=0.3mm] (0,0) -- (0,3) node[above] {$z$};
        % label
        \draw (4,2.5) node {$C=0$};
        \draw (1.2,0.8) node {$C=1$};
        % contact region
        \draw [<->] (5.8,0.1) -- (7,0.1);
        \draw (6.5,0.6) node[right] {Interface region $\sim d$};
    \end{tikzpicture}}
    
    \subfloat[]{
    \begin{tikzpicture}
        \fill [black!10] (0,0) rectangle (10,-0.6);
        % axes
        \draw [->, line width=0.3mm] (0,0) -- (10,0) node[right] {$x$};
        \draw [->, line width=0.3mm] (0,0) -- (0,3) node[above] {$z$};
        % interface
        \draw[orange,line width=0.7mm] plot[domain=0:6,smooth] (\x,{1.5*(1-(\x/6)^2)+0.3});
        \draw[orange,line width=0.7mm] plot[domain=6:7.2,smooth] (\x,{0.3/(6-7.2)^2*(\x-7.2)^2});
        \draw (3.4,1.9) node {$\bar{h}(x,t)$};
        % contact line region
        \draw[line width=0.3mm,dashed] plot[domain=3.8:7.2,smooth] (\x,{0.4*exp(-1.5*(\x-5.5)^2)});
        \draw (6,.3) -- (7,1.5) node[above] {Contact region};
        \filldraw (6,0.3) circle (0.04);
        \draw (6.5,0.3) node[right] {$K(x)$};
        \draw (1.2,2.2) node {gas};
        \draw (1.2,0.8) node {liquid};
        \draw (1.2,-0.3) node {solid};
    \end{tikzpicture}}
    \caption{Sketch of a droplet in (a) the diffuse-interface method and (b) the geometric diffuse-interface method framework. In DIM, the position of the liquid-gas interface has an uncertainty, while in G-DIM, the interface is sharp, and only the contact line position contains uncertainty, which is defined precisely by the kernel $K(x)$. }
    \label{fig:sketch_gdim}
\end{figure}

Holm et al. numerically studied the two-dimensional droplet spreading solution of the Geometric Thin-Film Equation. 
% An alternative method with improved numerical stability and convergence rate is proposed in \Cref{sec:fdm}. A rigorous error analysis of the finite-difference method is also performed. 
In this chapter, we proposed a new finite-difference solver with improved numerical stability and convergence rate as well as simulating droplet spreading in three spatial dimensions, providing necessary validation of the model for describing droplet spreading. 

Although they were not pursued in detail, Holm et al. noted that the gradient-flow structure of the Geometric Thin-Film Equation admits a family of weak solutions that could potentially be useful for the purposes of developing robust and computationally inexpensive numerical solvers for simulating thin-film flows. We investigate this further, leading to a Lagrangian numerical scheme known as the particle method for solving the Geometric Thin-Film Equation. % The particle method is discussed in \Cref{sec:particles}. 
% Furthermore, a fast summation algorithm is developed for fast evaluation of $\bar{h}$, 
Furthermore, two optimization techniques for the particle method are developed -- a fast summation algorithm for 2D droplets and a fast multipole method for 3D droplets. 
These optimizations result in a significant speed improvement to the particle method solver compared to the finite-difference method solver. Additional analysis tools are needed to study the convergence and regularity of the weak solution of the Geometric Thin-Film Equation, which we delegate to the next chapter. 

\section{Model Formulation} \label{sec:modelab}

In the framework of the Geometric Thin-Film Equation, the starting point is the assumption that there is missing physics on a small scale.  Instead of modelling the missing physics, it is parametrized.  As such, $h(x,t)$ is used to denote the interface location in a crude model with missing physics -- which we call the `noisy' interface location.  The noisy interface location is to be smoothened by a filtering operation to produce a smoothened, more accurate estimate of the interface location, which we denote by $\barh(x,t)$.  The noisy interface location may be different from the true interface location -- for instance, the noisy interface location may be zero, whereas the true interface location may be close to, but different from zero -- as would be the case if the noisy interface location was obtained through an incomplete model with missing small-scale physics.

\subsection{Model A}

To take account of the fact that $\barh$ represents a smoother description of the interface location than $h$, we propose that $h$ and $\barh$ be connected via the expression
\begin{equation}
h=\barh+\eta,
\label{eq:uncertainty}
\end{equation}
where $\eta$ is a fluctuating quantity with mean zero and variance $\sigma^2$.  Then, $\barh(x,t)$ can be made into an accurate estimate of the interface location by minimizing the total interfacial energy given by
\begin{equation}
\Energy[\barh]=\gamma\int_{-S/2}^{S/2} \sqrt{1+|\partial_x\barh|^2}\mathd x,
\label{eq:Adef}
\end{equation}
subject to a fixed-variance-constraint:
\begin{equation}
\int_{-S/2}^{S/2} |h-\barh|^2\mathd x=\sigma^2.
\label{eq:constop}%
\end{equation}%
Here, $S$ is the system size (we take $S\rightarrow\infty$ in what follows), and $\gamma$ is a positive constant representing the surface tension.  The constraint~\eqref{eq:constop} enforces a fixed level of uncertainty between the model with missing physics and the smoothened model.
%, and furthermore enforces the uncertainty model in Equation~\eqref{eq:uncertainty}.
%
In practice, we minimize the surface area in the longwave limit: in this limit, $|\partial_x\barh|$ is small, and the square root in Equation~\eqref{eq:Adef} can be well approximated by 
$1+(1/2)|\partial_x\barh|^2$.  Thus, the energy in Equation~\eqref{eq:Adef} becomes 
\[
\gamma S+\tfrac{1}{2}\surften\int_{-S/2}^{S/2}|\partial_x\barh|^2\,\mathd x.
\]
The constant term here can be ignored, as only energy differences are important.  Furthermore, as the droplet profile rapidly decreases far from the droplet core at $x=0$, we can take $S\rightarrow \infty$ in the foregoing analysis. This then gives the required functional form for the energy in the longwave limit:
\begin{equation}
\Energy[\barh]=\tfrac{1}{2}\surften\int_{-\infty}^{\infty} |\partial_x\barh|^2\mathd x,
\label{eq:Along}
\end{equation}

Equation~\eqref{eq:Along} with constraint~\eqref{eq:constop} (with $S\rightarrow\infty$) is a constrained minimization problem -- to solve it, one would introduce an energy functional with a Lagrange multiplier:
\begin{equation}
\Lagrange[h,\barh]=\Energy[\barh]+\lambda \left[\tfrac{1}{2}\int_{-\infty}^\infty |h-\barh|^2\mathd x-\tfrac{1}{2}\sigma^2\right].
\label{eq:multiplier}
\end{equation}
One would then compute the Euler-Lagrange Equations
\begin{equation}
\frac{\delta \Lagrange}{\delta \barh}=0,\qquad \frac{\delta \Lagrange}{\delta h}=0,
\end{equation}
yielding
\begin{equation}
-\gamma\partial_{xx}\barh-\lambda (h-\barh)=0,\qquad h-\barh=0.
\label{eq:lagrange}
\end{equation}
In practice, solving Equation~\eqref{eq:lagrange} yields inconsistent results, as it implies that $h=\barh$.  But $h$ and $\barh$ live in different function spaces ($h$ is noisy, $\barh$ is smooth), so Equation~\eqref{eq:lagrange} cannot be correct.  
Instead, we can study the dynamics, whereby $\Lagrange[h,\barh]$ in Equation~\eqref{eq:multiplier} reduces over time.  The dynamics are highly conditioned:
\begin{itemize}%[noitemsep]
\item The evolution of $h$ and $\barh$ must be such that $\Lagrange$ tends to a minimum over time;
\item The integrals $\int h(x,t)\mathd x = \int \barh(x,t)\mathd x =: A$ must be conserved quantities, reflecting underlying principles of conservation of fluid mass.
\end{itemize}
\amend{The fact that $h$ and $\bar{h}$ have the same mass should not come as a surprise since $h(x)$ is normally distributed with a mean of $\bar{h}(x)$ for all $x\in\mathbb{R}$. Indeed we have $\int\eta\,\mathd x=0$. }
Under these conditions, the evolution equation for $h$ must be of a conservative-gradient-descent type, hence:
\begin{equation}
\frac{\partial h}{\partial t}=\frac{\partial}{\partial x}\left(h\mob\frac{\partial}{\partial x}\frac{\delta \Lagrange}{\delta h}\right).
\label{eq:evolution_a}
\end{equation}
where $\mob:[0,\infty)\rightarrow[0,\infty)$ is a mobility function to be determined.  The evolution equation for $\barh$ may be similar.  However, for simplicity, we may assume that $\barh$ relaxes instantaneously to a smoothened form of $h$, hence $\delta \Lagrange/\delta\barh=0$, hence
\begin{equation}
-\gamma\partial_{xx}\barh=\lambda(h-\barh),
\label{eq:helmholtz1}
\end{equation}
or
\begin{equation} \label{eq:helmholtz2}
    \barh=\left(1-\frac{\gamma}{\lambda}\partial_{xx}\right)^{-1}h.
\end{equation}
Equation~\eqref{eq:helmholtz2} establishes a natural smoothing operation and hence, smoothing kernel for the formulation, namely, the Helmholtz kernel.

We now use $\delta\Lagrange/\delta h=\lambda(h-\barh)$.  We compare this to Equation~\eqref{eq:helmholtz1}, and get $\delta\Lagrange/\delta h=-\surften\partial_{xx}\barh$.  Substitution of these results into Equation~\eqref{eq:evolution_a} yields:
\begin{equation}
\pder[h]{t} = -\pder{x}\left[h\mob\pder{x}\left(\surften\partial_{xx}\barh\right)\right].
\label{eq:evolution_b}
\end{equation}
The physical model for $h$ is completed by specifying the mobility.  This is done by reference to the classical theory. In \cite{holm2020gdim}, the mobility is explicitly modelled using the kinematic condition on the free surface, to which they obtain 
\begin{equation}
    \bar{\mob}(h,\bar{h}) = \frac{1}{3\vis}\left(\frac{3}{2}h\bar{h} - \frac{1}{2}h^2\right). 
\end{equation}
Instead, we consider mobility purely a function of $\bar{h}$ and use the mobility
\begin{equation}
\mob(\bar{h})=\frac{1}{3\vis}\barh^2;
\label{eq:mobility}
\end{equation}
the reason for using $\barh^2$ in the mobility becomes apparent when we look at particle-like solutions of the regularized model (\cref{sec:particles}). In particular, this aids in the computation of the particle solutions. 
Finally, the value of the Lagrange multiplier $\lambda$ is chosen at each point in time to reflect the model uncertainty:
\begin{equation}
\int_{-\infty}^\infty \left|\left(1-\frac{1}{\lambda}\partial_{xx}\right)^{-1}h-h\right|^2\mathd x=\sigma^2.
\label{eq:lagrange_solve}
\end{equation}
We refer to this model with a fixed level of uncertainty as Model A.

\subsection{Model B}
\label{subsec:modelB}

In practice, recomputing the Lagrange multiplier $\lambda$ at each time $t$ is a difficult task numerically.  However, an equivalent model can be formulated 
by introducing an unconstrained functional,
\begin{equation}
\Lagrange[h,\barh]=\Energy[\barh]+\frac{\gamma}{2\alpha^2}\int_{-\infty}^\infty (h-\barh)^2\mathd x.
\end{equation}
Here, $\alpha$ is a fixed parameter of the model.  The dynamical equation is the same as before (Equation~\eqref{eq:evolution_b}), as is the mobility; however, now $\barh$ is computed as 
\begin{equation}
\barh=\left(1-\alpha^2\partial_{xx}\right)^{-1}h:=K_1*h.
\label{eq:helmholtz3}
\end{equation}
We refer to this model with uncertainty on a small scale $\alpha$ as Model B.  Here,  we have introduced the standard notation for smoothing kernels~\cite{bressan2007global}:
\[
K_1*f(x)=(1-\alpha^2\partial_{xx})^{-1}f(x)=\int_{-\infty}^\infty K_1(x-y)f(y)\mathd y,
\]
and we explicitly use $K_1$ for the Helmholtz kernel, such that
\[
K_1(x)=\frac{1}{2\alpha}\mathe^{-|x|/\alpha}.
\]

Although Model A and Model B are different, there is a one-to-one relationship between them, and they are equivalent -- e.g. $\lambda$ in Equation~\eqref{eq:lagrange_solve} is a $\left[\text{lengthscale}\right]^2$ which depends on time.  We therefore identify
\begin{equation}
\alpha(t)=\left[\lambda(t;\sigma)\right]^{-1/2},
\label{eq:lagrange_solve1}
\end{equation}
and the required uncertainty on a small lengthscale $\alpha$ in the second description of the model is the average value of Equation~\eqref{eq:lagrange_solve1}:
\begin{equation}
\alpha=\lim_{T\rightarrow\infty}\frac{1}{T}\int_0^{\infty}\left[\lambda(t;\sigma)\right]^{-1/2}\mathd t.
\end{equation}
Due to the computational efficiency, Model B is preferred.

The kernel solution~\eqref{eq:helmholtz3} can be substituted back into the expression $\Lagrange[h,\barh]$ to give:
\begin{equation}
\lagrange[h]:=\Lagrange\left[h,\barh=\left(1-\alpha^2\partial_{xx}\right)^{-1}h\right]=\tfrac{1}{2}\gamma\int_{-\infty}^\infty \left[\left(\partial_x\barh\right)^2+\alpha^2\left(\partial_{xx}\barh\right)^2\right]\mathd x.
\label{eq:ell1}
\end{equation}
This can, in turn, be written in several further ways:
\begin{enumerate}
    \item The inner-product pairing of $\partial_x h$ with $\partial_x \barh$:
    \begin{align*} \label{eq:ell2}
        \lagrange[h] =\;& \tfrac{1}{2}\gamma\intinf \left[\left(\partial_x\barh\right)^2+\alpha^2\left(\partial_{xx}\barh\right)^2\right]\mathd x,\tag*{by Eq.~\eqref{eq:ell1}} \\
        =\;& \tfrac{1}{2}\gamma\intinf \partial_x \barh\left(1-\alpha^2\partial_x^2\right) \partial_x\barh\,\mathd x, \tag*{integration by parts} \\
        =\;& \tfrac{1}{2}\gamma\intinf \partial_x \barh\, \partial_x\underbrace{\left(1-\alpha^2\partial_x^2\right)\barh}_{=h}\,\mathd x.
    \end{align*}
    Hence,
    \begin{equation} \label{eq:tfe_engrgy}
        \lagrange[h]=\tfrac{1}{2}\gamma\intinf \partial_x h\, \partial_x\barh\,\mathd x.
    \end{equation}
    \item The weighted inner-product pairing:
    \begin{equation} \label{eq:ell3}
        \lagrange[h] = \tfrac{1}{2}\gamma\langle \partial_x\barh,\partial_x\barh\rangle_{K_1} = \tfrac{1}{2}\gamma\int_{-\infty}^\infty \partial_x \barh\left(1-\alpha^2\partial_x^2\right) \partial_x\barh\,\mathd x.
    \end{equation}
    The pairing $\langle \cdot,\cdot\rangle_K$ defines a Reproducing Kernel Hilbert Space~\cite{evgeniou2000regularization}.
\end{enumerate}
%The pairing $\langle \cdot,\cdot\rangle_K$ defines a Reproducing Kernel Hilbert Space (RKHS), which is an important concept in Machine Learning -- this gives an alternative interpretation of the regularization pursued in this section as one involving a `coarse' model ($h$) and a `smooth' model $\mybar$, where the smooth model `learns' from the coarse model (via Equation~\eqref{eq:
Equation~\eqref{eq:evolution_b} now reads:
\begin{equation} \label{eq:evolution_c}
    \pder[h]{t}=-\pder{x}\left(h\mob\pder{x}\frac{\delta \lagrange}{\delta h}\right),\qquad \barh=K_1*h.
\end{equation}

\subsection{\markup{Higher-Order Smoothing}} \label{sec:tfe_higher}

To generate particle-like solutions of the regularized model (e.g. \cref{sec:particles}), smoothing with the Helmholtz kernel is insufficient due to the thin-film equation's fourth-order derivative.  Therefore, a higher-order smoothing is needed.  We take $\ell[h]$ as before (specifically, Equation~\eqref{eq:ell1}), with evolution equation~\eqref{eq:evolution_c} and smoothing kernel $\barh=K_1*K_1*h$ -- this is a straightforward extension of the basic model. We will use $K_2=K_1*K_1$ to denote the bi-Helmholtz kernel \cite{bi_helmholtz}. This is the Green's function associated with the bi-Helmholtz operator $(1-\alpha^2\partial_{xx})^2K_2=\delta$; specifically, 
\begin{equation} \label{eq:bihelm}
    K_2(x) = \frac{1}{4\alpha^2} (\alpha + |x|) \mathe^{-|x|/\alpha}.
\end{equation}
The derivation of \cref{eq:bihelm} is given in Appendix B. The Geometric Thin-Film Equation is therefore given by
\begin{equation} \label{eq:gtfe}
    \pder[h]{t} = -\frac{\surften}{3\vis}\pder{x}\left(h\barh^2\pder[^3\barh]{x^3}\right), \qquad (1-\alpha^2\partial_{xx})^2\barh = h.
\end{equation}
This is then subjected to an initial condition $h(x,0)=h_0(x)$.

% In higher dimension, this is 
% \begin{equation}
%     \pder[h]{t} = -\nabla\cdot\left(h\mu\nabla(\Delta\bar{h})\right), \qquad (1-\alpha^2\Delta)^2\barh = h.
% \end{equation}

We emphasize that the present higher-order smoothing is introduced to give the required degree of regularity to the numerical solutions (e.g. Chapter~\ref{sec:particles}). We also favour the bi-Helmholtz kernel due to its simplicity and satisfying the bare minimum smoothness requirement for the Geometric Thin-Film Equation to be well defined. Allowing us to study the worst-case scenario of the solution. However, one might consider other higher-order smoothing kernels. The Gaussian kernel is one such candidate. The compactly supported mollifier of the form
\begin{equation}
    \varphi(x) = \begin{dcases}
        A\exp\left(-\frac{1}{\alpha^2-x^2}\right), & |x|\leq\alpha, \\
        0, & \text{otherwise}, 
    \end{dcases}
\end{equation}
can also be used to obtain compactly supported solution $\bar{h}$. By allowing different smoothing kernels, one obtains a whole family of regularized thin-film equations. We will focus on covering the depth of the Geometric Thin-Film Equation, and the breath shall be left for future studies. 

In Holm et al. \cite{holm2020gdim}, another kernel is briefly mentioned, where they modify the Helmholtz problem by considering the operator
\begin{equation} \label{eq:extended_helm}
    \left(1-\alpha^2\partial_{xx} + \tfrac{1}{2}\alpha^4\partial_{xxxx}\right)K_3 = \delta.
\end{equation}
The resulting kernel has been studied alongside the bi-Helmholtz kernel by us, but we found that it is not ideal for describing the droplet spreading problem. In particular, the Green's function of \cref{eq:extended_helm} is given by (derivation given in Appendix B):
\begin{gather}
    K_3(x) = \frac{1}{2}\mathe^{-B|x|}\left\{(A+B)\sin(A|x|) - (A-B)\cos(Ax)\right\}, 
\end{gather}
where
\begin{equation}
    A = \frac{1}{\alpha}\sqrt{\frac{1}{\sqrt{2}}-\frac{1}{2}}, \qquad B=\frac{1}{\alpha}\sqrt{\frac{1}{\sqrt{2}}+\frac{1}{2}},
\end{equation}
which fails the positive preserving property $K_3\ngeq0$. This highlights the importance of choosing an appropriate smoothing kernel for the task. 

\begin{figure}
    \centering
    \includegraphics[width=0.7\linewidth]{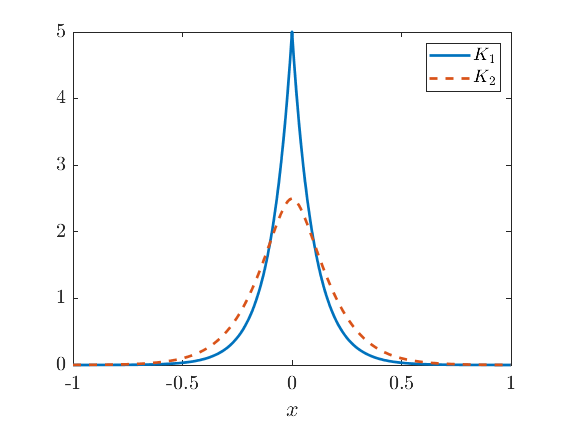}
    \caption{The Helmholtz kernel $K_1(x)$ and bi-Helmholtz kernel $K_2(x)$ for $\alpha=0.1$.}
    \label{fig:kernels}
\end{figure}

% We emphasize that the present higher-order smoothing is introduced to give the required degree of regularity to the numerical solutions (e.g. Chapter~\ref{sec:particles}).  Therefore, the higher-order smoothing is introduced here for mathematical convenience.  However, the higher-order smoothing can be given a physical basis through the construction of appropriate energy functionals, as in Section~\ref{subsec:modelB}, although we do not pursue that approach here.

\section{Departure from Holm et al.}
\label{sec:departure}

We highlight a few departures from the formulation of the Geometric Thin-Film Equation presented in \cite{holm2020gdim}. These changes address some of the issues we encountered when working with the formulation developed by Holm et al. The aim is to simplify the Geometric Thin-Film Equation and regularization of the same sort to emphasize the attractability of the model. 

\subsection{Asymmetrical Regularization}

In the formulation by Holm et al., the regularized free energy functional is given by
\begin{equation}
    E[\bar{h}]=\frac{1}{2}\int_{-\infty}^\infty |\partial_x\bar{h}|^2 \,\mathd x,
\end{equation}
instead of Equation~\eqref{eq:tfe_engrgy}. Using this formulation, with integrand $L=|\partial_x\bar{h}|^2$, the Geometric Thin-Film Equation becomes
\begin{equation} \label{eq:tfe_energy_old}
    \pder[h]{t} = -\pder{x}\left(h\mob(h,\barh)\pder{x}(K_1*\partial_{xx}\bar{h})\right), \qquad \bar{h}=K_1*h,
\end{equation}
where $K_1$ is the Helmholtz kernel discussed in Section \ref{sec:tfe_higher}. Compared to Equation~\eqref{eq:gtfe}, only the energy gradient receives the higher order smoothing in this case. However, this formulation is difficult to extend because of the extra convolution. In particular, if $\delta E/\delta h$ is no longer linear, the convolution $K_1*\delta E/\delta h$ could be difficult to calculate. Alternatively, one could compute the integral numerically, which would require additional and otherwise unnecessary computation, slowing down the simulation. This is akin to the regularization step in Ref \cite{carrillo2019}, although they are only interested in the second-order diffusion operator. Similarly, we favour the pairing $\langle h,\barh\rangle$ over $\|\barh\|_2$ as it introduces one fewer convolution.

\subsection{A Revised Theorem} \label{sec:revise}

As noted by the authors of \cite{holm2020gdim}, the numerical solution of the Geometric Thin-Film Equation exhibits a jump discontinuity at the contact line. In \cite[\nopp Theorem 3.2]{holm2020gdim}, a proof for classifying the regularity of the Geometric Thin-Film Equation in space is given. However, we note that this proof is incomplete, and we wish to address this issue here. We start by stating a few definitions.

\begin{definition}
A moving-contact-line solution of the Geometric Thin-Film Equation is a solution of the form
\begin{equation}
    h(x,t)=\phi(x,x_{cl})\chi_{(-x_{cl},x_{cl})}(x),
\end{equation}
where $x_{cl}=x_{cl}(t)$ is a non-trivial function of time, $\phi(x,x_{cl})$ is a smooth function in the variable $x$ on the interval $(-x_{cl},x_{cl})$, and $\chi_A$ is the indicator function on the set $A$:
\begin{equation}
    \chi_A(x) = \begin{cases}
        1, & x\in A, \\
        0, & x\notin A. 
    \end{cases}
\end{equation}
Furthermore, we assume the droplet is symmetric, such that $h(x,t)$ and hence, $\phi(x,\cdot)$ is an even function of $x$.
\end{definition}

\begin{definition}
A continuous moving contact line solution is one for which $\phi(x_{cl},x_{cl})=0$ for all $t\geq0$. Otherwise, we say that the solution has a jump discontinuity. 
\end{definition}

The following Lemma will be helpful in proving the theorem. 

\begin{lemma} \label{thm:tfe_contact}
If $h(x,t)$ is a moving contact line solution, then the area 
\begin{equation}
    A = \int_{-\infty}^\infty h(x,t) \,\mathd x,
\end{equation}
is conserved if and only if
\begin{equation}
    \int_{-x_{cl}}^{x_{cl}}\pder[\phi]{x_{cl}} + 2\phi(x_{cl},x_{cl}) = 0.
\end{equation}
\end{lemma}
\begin{proof}
We use the Leibniz integral rule 
\begin{align}
    \der[A]{t} &= \der{t}\int_{-\infty}^\infty h(x,t)\,\mathd x, \\
    &= \der{t}\int_{-x_{cl}}^{x_{cl}} \phi(x,x_{cl})\,\mathd x, \\
    &= \dot{x}_{cl}\int_{-x_{cl}}^{x_{cl}} \pder[\phi]{x_{cl}}\,\mathd x + \dot{x}_{cl}\phi(x_{cl},x_{cl}) + \dot{x}_{cl}\phi(-x_{cl},x_{cl}), \\
    &= \dot{x}_{cl}\int_{-x_{cl}}^{x_{cl}} \pder[\phi]{x_{cl}}\,\mathd x + 2\dot{x}_{cl}\phi(x_{cl},x_{cl}).
\end{align}
Requiring $\mathd A/\mathd t=0$ and $\dot{x}_{cl}\neq0$ (moving contact line solution), we deduce that 
\begin{equation}
    \int_{-x_{cl}}^{x_{cl}} \pder[\phi]{x_{cl}}\,\mathd x + 2\phi(x_{cl},x_{cl}) = 0.
\end{equation}
\end{proof}

Here is the rephrased version of \cite[\nopp Theorem 3.2]{holm2020gdim}.

\begin{theorem}
No continuous moving contact line solution $h$ of the Geometric Thin-Film Equation; any contact line solution of the Geometric Thin-Film Equation necessarily has a jump discontinuity at the contact line.
\end{theorem}

That is, the function $h$ is necessary not a function with derivative defined in the classical sense. In the next chapter, we classify the solution space and regularity of $h$ precisely. We also show that the convolution operation gives $\bar{h}$ of sufficient regularity for the exposition of the numerical methods.

\begin{proof}
We take the Geometric Thin-Film Equation in flux conservative form,
\begin{equation}\label{eq:tfe_flux}
    \pder[h]{t} = -\pder{x}(h\mob\partial_{xxx}\bar{h}) =: \pder[J]{x}.
\end{equation}
Because the moving contact line solution is not a classical solution, we consider the weak formulation of Equation~\eqref{eq:tfe_flux}. Let $\psi$ be a test function, then
\begin{equation}
    \left\langle \psi,\pder[h]{t} \right\rangle = -\left\langle\pder[\psi]{x},J \right\rangle,
\end{equation}
where $\langle\cdot,\cdot\rangle$ is the usual inner product on functions. We substitute in for the moving contact line solution $h(x,t)=\phi(x,x_{cl})\chi_{(-x_{cl},x_{cl})}(x)$ and obtain
\begin{multline} \label{eq:tfe_weak1}
    \dot{x}_{cl}\left(\int_{-x_{cl}}^{x_{cl}}\psi(x)\pder[\phi]{x_{cl}}\mathd x + \phi(x_{cl},x_{cl})(\psi(x_{cl})+\psi(-x_{cl}))\right) = \\
    -\int_{-\infty}^{\infty} \pder[\psi]{x}J\,\mathd x.
\end{multline}
By construction, $J=-\phi\chi_{(-x_{cl},x_{cl})}\mob\partial_{xxx}\bar{h}$ is smooth on $(-x_{cl},x_{cl})$. Thus, Equation~\eqref{eq:tfe_weak1} can be re-written as
\begin{multline}
    \dot{x}_{cl}\left(\int_{-x_{cl}}^{x_{cl}}\psi(x)\pder[\phi]{x_{cl}}\mathd x +\phi(x_{cl},x_{cl})(\psi(x_{cl})+\psi(-x_{cl}))\right) = \\ 
    \Big[\psi(x)\phi(x,x_{cl})\mob\partial_{xx}\bar{h}\Big]_{-x_{cl}}^{x_{cl}} - \int_{-x_{cl}}^{x_{cl}} \psi(x)\pder{x}(\phi(x,x_{cl})\mob\partial_{xx}\bar{h})\,\mathd x.
\end{multline}
Re-arranging gives
\begin{multline}
    \int_{-x_{cl}}^{x_{cl}}\psi(x)\left(\dot{x}_{cl}\pder[\phi]{x_{cl}} + \pder{x}(\phi(x,x_{cl})\mob\partial_{xx}\bar{h})\right)\mathd x = \\
    +\left[\Big(\phi(x,x_{cl})\mob\partial_{xx}\bar{h} - \operatorname{sgn}(x)\dot{x}_{cl}\phi(x,x_{cl})\Big)\psi(x)\right]_{-x_{cl}}^{x_{cl}}.
\end{multline}
Since $\psi$ is an arbitrary test function, in particular, if $\operatorname{supp}\psi\subset\subset(-x_{cl},x_{cl})$, then we deduce that 
\begin{equation}
    \dot{x}_{cl}\pder[\phi]{x_{cl}} + \pder{x}(\phi(x,x_{cl})\mob\partial_{xxx}\bar{h}) = 0. 
\end{equation}
Integrate from $-x_{cl}$ to $x_{cl}$ to obtain
\begin{equation} \label{eq:tfe_weak2}
    \dot{x}_{cl}\int_{-x_{cl}}^{x_{cl}} \pder[\phi]{x_{cl}}\mathd x = -2\phi(x_{cl},x_{cl})\left[\mob\partial_{xxx}\bar{h}\right]_{x_{cl}}.
\end{equation}

If $\phi(x_{cl},x_{cl})=0$, then by Lemma~\ref{thm:tfe_contact}, we have $\int_{-x_{cl}}^{x_{cl}}\pder[\phi]{x_{cl}}\mathd x=0$, and Equation~\eqref{eq:tfe_weak2} reduces to the indeterminate form $\dot{x}_{cl}\times0=0$. 

If $\phi(x_{cl},x_{cl})\neq0$, again by Lemma~\ref{thm:tfe_contact}, we have $\int_{-x_{cl}}^{x_{cl}}\pder[\phi]{x_{cl}}\mathd x\neq0$, hence Equation~\eqref{eq:tfe_weak2} gives a non-trivial evolution equation for the moving contact line
\begin{equation} \label{eq:tfe_weak3}
    \dot{x}_{cl} = -\frac{2\phi(x_{cl},x_{cl})\left[\mob\partial_{xxx}\bar{h}\right]_{x_{cl}}}{\int_{-x_{cl}}^{x_{cl}} \partial_{x_{cl}}\phi\,\mathd x} = \left[\mob\partial_{xxx}\bar{h}\right]_{x=x_{cl}}.
\end{equation}
Thus, the Geometric Thin-Film Equation has an explicit built-in law for the motion of the contact line.
\end{proof}

We will see later in \Cref{sec:particles} that in the particle method, the evolution of the particles has a similar equation of motion as Equation~\eqref{eq:tfe_weak3}, and Equation~\eqref{eq:tfe_weak3} can be thought of as the Lagrangian specification of the flow field of the contact line.

\subsection{Numerical Stability}

The Geometric Thin-Film Equation (for $\partial\barh/\partial t$) is a fourth-order parabolic equation, which makes solving the equation numerically stiff \cite{witelski2003adi}. This imposes harsh bounds on the step size $\Delta t$ for certain numerical methods and causes them to be unstable. For explicit methods, the step size required is of the order $\Delta t=O(\Delta x^4)$, with $\Delta x$ being the grid spacing. This is more widely known as the Courant–Friedrichs–Lewy (CFL) condition for numerical stability. 

In \cite{holm2020gdim}, the authors introduced a finite difference scheme with semi-implicit time discretization to solve the Geometric Thin-Film Equation. This discretizes the Geometric Thin-Film Equation into a system of linear equations, which can then be solved using a matrix inversion at each time step. We notice that this method is conditionally stable, with CFL condition that is first-order in $\Delta x$:
\begin{equation} \label{eq:cfl}
    \frac{\surften}{3\vis} \frac{\Delta t}{\Delta x} < 1,
\end{equation}
We demonstrate this in Section~\ref{sec:fdm_err}. The CFL condition means that $\Delta t$ needs to be sufficiently small for the numerical method to remain stable, limiting the size of the numerical simulation. 

This thesis introduces an unconditionally stable numerical scheme for the Geometric Thin-Film Equation using a fully implicit discretization. This decouples the condition for $\Delta t$ with $\Delta x$, allowing for a larger step size to be taken. This, however, makes solving for the solution at each time step a nonlinear problem. We will address solving the nonlinear equation in Chapter~\ref{sec:fdm}, as well as doing a thorough comparison of the stability of different numerical schemes for solving the Geometric Thin-Film Equation. 

Furthermore, we also explore the particle method as an alternative numerical method for solving the Geometric Thin-Film Equation. The particle method utilizes the gradient-flow structure of the Geometric Thin-Film Equation to transform the PDE problem into a system of finite-dimensional ODEs which describe the equation of motion for the particles. The particle method provides a few advantages, including the positivity preserving and mass conserving of the solution.

\section{A Finite-Difference Method for Nonlinear PDEs} \label{sec:fdm}

In this section, we develop a framework to efficiently and accurately solve a wide range of non-linear PDEs. This framework allows us to convert the PDEs into their finite-difference formulation and translate them into code with minimal effort. We will demonstrate the solver by solving the highly nonlinear fourth-order Geometric Thin-Film equation. However, we emphasise that the finite-difference solver is generic and has applicability to a wide range of nonlinear PDEs. We also pay special attention to optimising the convolution and deconvolution operators in the Geometric Thin-Film equation. The Gauss-Newton method for solving nonlinear optimization problems is at the core of the finite-difference solver, which we introduce next.

% In Section~\ref{sec:fdmgn}, we introduce the Gauss-Newton method for non-linear least squares optimization, which underlies our non-linear finite difference solver. In Section~\ref{sec:fdmpw} we describe a pointwise matrix algebra which will be handy in computing the Jacobian analytically of non-linear objective functions. The notation also allows for a straight forward translation to a vectorized implementation of MATLAB code. In Section~\ref{sec:fdm1}, we outline the non-linear finite difference solver for both a backward Euler and a Crank-Nicolson scheme for the discretizing in the time domain. Finally, in Section~\ref{sec:fdm_num}, we solve the Geometric Thin-Film Equation using the finite difference schemes

\subsection{Gauss-Newton method} \label{sec:fdmgn}

Let $\bm{F}=(F_1,\dots,F_n)^T:\R^n\rightarrow\R^n$. Consider the non-linear least squares problem, where we seek to minimize the cost function  
\begin{equation} \label{eq:gn_cost}
    f(\bm{v}) = \frac{1}{2}\|\bm{F}(\bm{v})\|^2 = \frac{1}{2}\sum_{i=1}^n F_i(\bm{v})^2,
\end{equation}
over $\bm{v}\in\R^n$. We start by Taylor expanding $f$ around $\bm{v}$
\begin{align}
    f(\bm{v}+\delta\bm{v}) &= f(\bm{v}) + \sum_{j=1}^n \delta v_j \pder[f]{v_j}(\bm{v}) \nonumber\\
    &\qquad+ \sum_{j=1}^n \sum_{k=1}^n \delta v_j \delta v_k \pder[^2f]{v_j\partial v_k}(\bm{v}) + O(\|\delta\bm{v}\|^3), \\
    &\approx f(\bm{v}) + \sum_{j=1}^n \delta v_j \sum_{i=1}^n F_i(\bm{v})\pder[F_i]{v_j}(\bm{v}) \nonumber\\
    &\qquad+ \sum_{j=1}^n\sum_{k=1}^n \delta v_j \delta v_k \sum_{i=1}^n \pder[F_i]{v_j}(\bm{v})\pder[F_i]{v_k}(\bm{v}) + O(\|\delta\bm{v}\|^2),
\end{align}
where in the last line, we have dropped the second order derivative of $\bm{F}$, incurring an $O(\|\delta\bm{v}\|^2)$ error. This can be written more compactly as 
\begin{equation} \label{eq:gnt}
    f(\bm{v}+\delta\bm{v}) \approx f(\bm{v}) + \delta\bm{v}^TJ(\bm{v})^T\bm{F}(\bm{v}) + \delta\bm{v}^TJ(\bm{v})^TJ(\bm{v})\delta\bm{v},
\end{equation}
where the Jacobian $J(\bm{v})=\nabla_{\bm{v}}\bm{F}(\bm{v})$ is a matrix of size $n\times n$ with entries given by
\begin{equation}
    [J(\bm{v})]_{ij} = \pder[F_i]{v_j}(\bm{v}).
\end{equation}
%We note that $J(\bm{v})^TJ(\bm{v})$ is symmetric and positive semi-definite. 
Suppose that $\bm{v}=\bm{v}^I$ is the current guess. Then Equation~\eqref{eq:gnt} attains its local extremum when
\begin{align}
    \nabla_{\delta\bm{v}}\left[f(\bm{v}^I) + \delta\bm{v}^TJ(\bm{v}^I)^T\bm{F}(\bm{v}^I) + \delta\bm{v}^TJ(\bm{v}^I)^TJ(\bm{v}^I)\delta\bm{v}\right] &= 0, \\
    J(\bm{v}^I)^T\bm{F}(\bm{v}^I) + J(\bm{v}^I)^TJ(\bm{v}^I)\delta\bm{v} &= 0.
\end{align}
So the optimal value of $\delta\bm{v}$ (i.e. the descent direction) is given by
\begin{equation}
    \delta\bm{v}^I = -J(\bm{v}^I)^{-1} \bm{F}(\bm{v}^I),
\end{equation}
and we update the guess with
\begin{equation} \label{eq:fdmgn}
    \bm{v}^{I+1} = \bm{v}^I + \delta\bm{v}^I.
\end{equation}
We can improve upon this by optimizing over $f$ in the direction of $\delta\bm{v}$. This is much cheaper than the computation of the Jacobian, and we can ensure that we do not overshoot the local minima. The step size now is given by
\begin{equation} \label{eq:fdm_linesearch}
    \alpha^I = \argmin_{\alpha\in[0,1]}f(\bm{v}^I+\alpha\delta\bm{v}^I).
\end{equation}
This step is often referred to as line search. Thus, the next iteration of the Gauss-Newton method with line search is given by
\begin{equation}
    \bm{v}^{I+1} = \bm{v}^I + \alpha^I\delta\bm{v}^I.
\end{equation}
We repeatedly improve the guess until the residual $f(\bm{v}^I)$ is smaller than some desired tolerance $\varepsilon$.

\subsection{A Pointwise Matrix Algebra} \label{sec:fdmpw}

\begin{definition}[Pointwise product]
	Let $\bm{x} = (x_1,\dots,x_n)^T$, $\bm{y} = (y_1,\dots,y_n)^T \in\R^n$, and let $A \in M^{n\times n}(\R)$ with entries $[A]_{ij}=a_{ij}$. Define the pointwise product $\odot:\R^n\times \R^n\rightarrow \R^n$ by
	\[\bm{x}\odot\bm{y} = (x_1y_1,\dots,x_ny_n)^T,\]
	and $\odot:\R^n\times M^{n\times n}(\R)\rightarrow M^{n\times n}(\R)$ by
	\[[\bm{x}\odot A]_{ij} = x_ia_{ij}.\]
\end{definition}
We note that this is equivalent to converting $\bm{x}$ and $\bm{y}$ to a diagonal matrix and using the standard matrix multiplication. However, Proposition~\ref{thm:algebra} does not have the standard linear algebra equivalent. 

With a slight abuse of notation, whenever $g$ is a function from $\R$ to $\R$, we write $g(\bm{x})$ to denote the pointwise application of the function $g$, that is
\begin{equation}
    g(\bm{x}) = (g(x_1),\dots,g(x_n))^T.
\end{equation}

\begin{definition}
	Let $\bm{F}:\R^n\rightarrow\R^n, \bm{F}(\bm{x})=(F_1(\bm{x}),\dots,F_n(\bm{x}))^T$, define the Jacobian of $\bm{F}$ to be
	\[[\nabla\bm{F}]_{ij}=\pder[F_i]{x_j}.\]
\end{definition}

The following result describes how to compute the Jacobian of expressions using the pointwise matrix product operator. 

\begin{proposition} \label{thm:algebra}
	Let $f,g\in C^1(\R)$ and let $A\in M^{n\times n}(\R)$, then
	\begin{enumerate}[label={(\roman*)}]
	    \item $\nabla (Af(\bm{x})) = A\nabla f(\bm{x})$
	    \item $\nabla f(A\bm{x}) = f'(A\bm{x})\odot A$
	    \item $\nabla \left(f(\bm{x})\odot g(\bm{x})\right)=g(\bm{x})\odot\nabla f(\bm{x}) + f(\bm{x})\odot\nabla g(\bm{x})$ \\
	\end{enumerate}
\end{proposition}

\begin{proof} 
(i) By the linearity of differentiation, we have
\begin{align}
    [\nabla(Af(\bm{x}))]_{ij} &= \pder{x_j} \sum_{k=1}^n a_{ik}f(x_k), \\
    &= \sum_{k=1}^n a_{ik}\pder{x_j}f(x_k), \\
    &= [A\nabla f(\bm{x})]_{ij}.
\end{align}
(ii) Using the Chain Rule
\begin{align}
    [\nabla f(A\bm{x})]_{ij} &= \pder{x_j}f\left(\sum_{k=1}^n a_{ik}x_k\right), \\
    &= f'\left(\sum_{k=1}^n a_{ik}x_k\right)a_{ij}, \\
    &= [f'(A\bm{x})\odot A]_{ij}.
\end{align}
(iii) Using the Product Rule
\begin{align}
    [\nabla f(\bm{x})\odot g(\bm{x})]_{ij} &= \pder{x_j} f(x_i)g(x_i), \\
    &= g(x_i) \pder{x_j} f(x_i) + f(x_i) \pder{x_j} g(x_i), \\
    &= [g(\bm{x})\odot\nabla f(\bm{x}) + f(\bm{x})\odot\nabla g(\bm{x})]_{ij}.
\end{align}
\end{proof}

Note that $\nabla\bm{x}=I$ where $I\in M^{n\times n}(\R)$ is the identity matrix. Furthermore, from Prop~\ref{thm:algebra}(ii) we deduce that 
\begin{equation}
    \nabla f(\bm{x}) = \nabla f(I\bm{x}) = f'(\bm{x})\odot I.
\end{equation}

The notation of pointwise operations described above is useful for the exposition of the numerical method, as the numerical code is vectorized and makes use of precisely this pointwise multiplication.

\subsection{The Finite-Difference Solver} \label{sec:fdm1}

In this section, we outline a finite difference solver for the geometric thin-film equation. In the literature, this algorithm is often referred to as the Newton method \cite{witelski2003adi,knoll2004} or, sometimes, the implicit finite-difference method \cite{bahadir2003, srivastava2011}. Even then, the Jacobian matrix is often computed numerically to an approximation and only computed analytically when the PDE is relatively simple. % We demonstrate that it is possible to compute the Jacobian matrix analytically even for long expressions. 

Instead of solving Equation~\eqref{eq:gtfe}, we smooth both sides of the same equation and solve the evolution for the smoothened free-surface height $\barh(x,t)$:
\begin{equation} \label{eq:smooth_tfe}
	\pder[\bar{h}]{t} = -\mathcal{L}^{-1}\left[\pder{x}\left(h\barh^2 \pder[^3\bar{h}]{x^3}\right)\right],\qquad \mathcal{L}(\barh)=h.
\end{equation}
Here, we have written $\mathcal{L}=(1-\alpha^2\partial_{xx})^2$ such that $\bar{h}=\mathcal{L}^{-1}h$, reflecting the choice of the inverse-bi-Helmholtz operator as the smoothing kernel in Equation~\eqref{eq:gtfe}.  We also work in dimensionless variables, such that the prefactor $\surften/(3\vis)$, which was present in Equation~\eqref{eq:gtfe}, is rescaled to one.

\paragraph{Notation} We will use the standard notation to denote the grid points in the finite-difference method. Let $L=|\Omega|$ be the size of the simulation domain, $N_x$ be the number of points used to discretize the domain, and $\Delta x=L/N_x$ be the grid spacing in the spatial domain. We evaluate the value of $\bar{h}$ only at the discrete grid points $i\Delta x$ for $i=1,\dots,N_x$. Furthermore, we will denote $\bar{h}_i(t) = \bar{h}(x=i\Delta x, t)$. 
% With this notation, the central difference is given by
% \begin{equation}
%     \left(\pder[\bar{h}]{x}\right)_i = \frac{\bar{h}_{i+1} - \bar{h}_{i-1}}{2\Delta x} + O(\Delta x^2).
% \end{equation}
It would also be useful to collect these values of $\bar{h}$ into a vector $\bar{\bm{h}} = (\bar{h}_1,\dots,\bar{h}_{N_x})^T$. 
For the temporal domain, we let $\Delta t$ to denote the step size in the temporal domain, and we write $\bar{\bm{h}}^n=\bar{\bm{h}}(t=n\Delta t)$. 

\paragraph{Spatial discretization} We discretize the first, second, and third-order spatial derivatives with the second-order accurate central differencing scheme. Thus, in discrete form, Equation~\eqref{eq:smooth_tfe} becomes 
\begin{equation} \label{eq:discretize_tfe}
    \pder[\bar{\bm{h}}]{t} = -L^{-1}[D_1(\bm{h}\odot\bar{\bm{h}}^2\odot(D_3\bar{\bm{h}}))], \qquad \bm{h} = L\bar{\bm{h}}. 
\end{equation}
Here, $D_1$, $D_2$, and $D_3$ are the standard centred-difference derivative operators of the first, second, and third order, respectively. $L$ is the bi-Helmholtz operator in matrix form and can be constructed as $L=(1-\alpha^2D_2)^2$. 
%
% Here, $D_1$ is the first-order derivative operator, which is a banded tri-diagonal matrix of size $N_x\times N_x$ with entries
% \begin{equation}
%     D_1 = \frac{1}{2\Delta x}
%     \begin{pmatrix}
%     0 & 1 \\
%     -1 & 0 & 1 \\
%     & \ddots & \ddots & \ddots \\
%     && -1 & 0 & 1 \\
%     &&& -1 & 0
%     \end{pmatrix}.
% \end{equation}
% Similarly, $D_3$ is the third-order derivative operator, which is a banded penta-diagonal matrix of the same size with entries
% \begin{equation}
%     D_3 = \frac{1}{2\Delta x^3}
%     \begin{pmatrix}
%     0 & -2 & 1 \\
%     2 & 0 & -2 & 1 \\
%     -1 & 2 & 0 & -2 & 1 \\
%     & \ddots & \ddots & \ddots & \ddots & \ddots \\
%     && -1 & 2 & 0 & -2 & 1 \\
%     &&& -1 & 2 & 0 & -2 \\
%     &&&& -1 & 2 & 0
%     \end{pmatrix}.
% \end{equation}
% Finally, $L$ is the bi-Helmholtz operator in matrix form and can be constructed as $L=(I-\alpha^2D_2)^2$ where $D_2$ is a banded tri-diagonal matrix with entries
% \begin{equation}
%     D_2 = \frac{1}{\Delta x^2}
%     \begin{pmatrix}
%     -2 & 1 \\
%     1 & -2 & 1 \\
%     & \ddots & \ddots & \ddots \\
%     && 1 & -2 & 1 \\
%     &&& 1 & -2
%     \end{pmatrix}.
% \end{equation}
%
In practice, we use a sparse matrix representation to store these matrix operators ($I$, $D_1$, $D_2$, $D_3$, and $L$). %, and since these matrices do not change, we can initialize them at the start of the simulation once. Furthermore, we can pre-decompose the matrix $L$, either using the LU decomposition or the Choleskey decomposition, to speed up the matrix inversion step. 
We also remark that for our purpose of studying the droplet spreading, ideally, we would like to use an infinite boundary condition, which is not possible with the finite difference method. Instead, we initialize the simulation domain with a boundary far away from the initial foot of the droplet to anticipate the spreading of the droplet. For the same reason, the boundary terms of the derivative operators are not especially important, as the values of $\bar{h}$ far away from the droplet core are negligible. 

For convenience, we will denote the right-hand-side of Equation~\eqref{eq:discretize_tfe} by 
\begin{equation}
    \mathcal{C}(\bar{\bm{h}}) = D_1(\bm{h}\odot\bar{\bm{h}}^2\odot(D_3\bar{\bm{h}})), \qquad \bm{h}=L\bar{\bm{h}}.
\end{equation}

\paragraph{Temporal discretization} Using the $\theta$-weighted scheme, Equation~\eqref{eq:discretize_tfe} becomes
\begin{align}
    \frac{\bar{\bm{h}}^{n+1}-\bar{\bm{h}}^n}{\Delta t} &= -\theta L^{-1}\mathcal{C}(\bar{\bm{h}}^{n+1})-(1-\theta)L^{-1}\mathcal{C}(\bar{\bm{h}}^n), \\
    \bar{\bm{h}}^{n+1}+\Delta t \theta L^{-1}\mathcal{C}(\bar{\bm{h}}^{n+1}) &= \bar{\bm{h}}^{n}-\Delta t(1-\theta)L^{-1}\mathcal{C}(\bar{\bm{h}}^n). \label{eq:discretize_cost}
\end{align}
When $\theta=1$, we recover the backward Euler scheme, and when $\theta=1/2$, this is the Crank-Nicolson scheme. Here, the objective is to find $\bar{\bm{h}}^{n+1}$ such that Equation~\eqref{eq:discretize_cost} is satisfied, given $\bar{\bm{h}}^n$.

This can then be turned into a non-linear optimization problem by considering the objective function
\begin{equation} \label{eq:fdm_objective}
    \bm{F}(\bar{\bm{v}}) = \bar{\bm{v}} + \Delta t \theta L^{-1}\mathcal{C}(\bar{\bm{v}}) + F_0(\bar{\bm{h}}^n),
\end{equation}
where $F_0(\bar{\bm{h}}^n)=-\bar{\bm{h}}^{n}+\Delta t(1-\theta)L^{-1}\mathcal{C}(\bar{\bm{h}}^n)$. Thus, if $f(\bar{\bm{v}})$ defined in Equation~\eqref{eq:gn_cost} is very close to zero, then $\bar{\bm{v}}$ is a good approximation for $\bar{\bm{h}}^{n+1}$. In order to employ the Gauss-Newton method discussed in Section~\ref{sec:fdmgn}, we need to find the Jacobian of Equation~\eqref{eq:fdm_objective}. This can be computed by
\begin{align}
    J(\bar{\bm{v}}) &= \nabla_{\bar{\bm{v}}}\bm{F}(\bar{\bm{v}}), \\
    &= \nabla_{\bar{\bm{v}}}(\bar{\bm{v}}+\Delta t\theta L^{-1}\mathcal{C}(\bar{\bm{v}})+F_0(\bar{\bm{h}}^n)), \\
    &= I + \Delta t\theta L^{-1}\nabla_{\bar{\bm{v}}}\mathcal{C}(\bar{\bm{v}}). \label{eq:jgtfe}
\end{align}
Next we compute $\nabla_{\bar{\bm{v}}}\mathcal{C}(\bar{\bm{v}})$ using the pointwise matrix algebra discussed in Section~\ref{sec:fdmpw} giving
\begin{equation} \label{eq:fdm_jacobian}
    \nabla_{\bar{\bm{v}}}\mathcal{C}(\bar{\bm{v}}) = D_1[(\bar{\bm{v}}^2\odot L + 2\bar{\bm{v}}\odot\bm{v}\odot I)\odot D_3\bar{\bm{v}} + \bm{v}\odot\bar{\bm{v}}^2\odot D_3].
\end{equation}
We note that $\nabla_{\bar{\bm{v}}}\mathcal{C}(\bar{\bm{v}})$ is sparse, but if not for $L^{-1}$ in Equation~\eqref{eq:jgtfe}, we would have $J(\bar{\bm{v}})$ be sparse matrix too. Inverting a dense matrix $J(\bar{\bm{v}})$ is computationally costly, and we would like to keep the sparsity nature of the finite difference operators. As such, we add an additional step by factoring $L^{-1}$ from $J(\bar{\bm{v}})$ such that $J(\bar{\bm{v}})=L^{-1}\tilde{J}(\bar{\bm{v}})$ with
\begin{equation}
    \tilde{J}(\bar{\bm{v}}) = L + \Delta t\theta \nabla_{\bar{\bm{v}}}\mathcal{C}(\bar{\bm{v}}).
\end{equation}
Similarly, we can factor $L^{-1}$ from $\bm{F}(\bar{\bm{v}})$ to get 
\begin{equation}
    \tilde{\bm{F}}(\bar{\bm{v}}) = \bm{v} + \Delta t\theta \mathcal{C}(\bar{\bm{v}}) - \bm{h}^n + \Delta t(1-\theta)\mathcal{C}(\bar{\bm{h}}^n).
\end{equation}
Then the equation for the descent direction (for minimizing $\bm{F}(\bar{\bm{v}})$) becomes 
\begin{equation}
    L^{-1}\tilde{J}(\bar{\bm{v}})\delta\bar{\bm{v}} = L^{-1}\tilde{\bm{F}}(\bar{\bm{v}}),
\end{equation}
giving
\begin{equation}
    \delta\bar{\bm{v}} = \tilde{J}(\bar{\bm{v}})^{-1}\tilde{\bm{F}}(\bar{\bm{v}}).
\end{equation}
This step is crucial, especially when we solve the Geometric Thin-Film Equation in higher dimension because the size of $J(\bar{\bm{v}})$ would grow like $O(N^{2d})$, where $d$ is the number of dimensions. 

In \cref{alg:fdm1}, we describe a pseudocode for solving the Geometric Thin-Film Equation using the Newton method. 

\begin{algorithm}[htb]
\caption{Newton method} \label{alg:fdm1}
\begin{algorithmic}
\State $n\gets0$
\State Initialize $I, D_1, D_3, L$
\While{$n\Delta t < T$}
    \State Compute $F_0(\bar{\bm{h}}^n)$
    \State $\bar{\bm{v}} \gets \bar{\bm{h}}^n$ \Comment{Initial guess for $\bar{\bm{h}}^{n+1}$} 
    \While{$f(\bar{\bm{v}}) > \varepsilon$}
        \State Compute $\bm{F}(\bar{\bm{v}})$ using Equation~\eqref{eq:fdm_objective}
        \State Compute $J(\bar{\bm{v}})$ using Equation~\eqref{eq:fdm_jacobian}
        \State $\delta\bar{\bm{v}} \gets -J(\bar{\bm{v}})^{-1}\bm{F}(\bar{\bm{v}})$
        \State Find $\alpha$ with a line search algorithm
        \State $\bar{\bm{v}} \gets \bar{\bm{v}} + \alpha\delta\bar{\bm{v}}$
    \EndWhile
    \State $\bar{\bm{h}}^{n+1} \gets \bar{\bm{v}}$
    \State $n\gets n+1$
\EndWhile
\end{algorithmic}
\end{algorithm}

\subsection{Numerical Results} \label{sec:fdm_result}

To characterize the spreading phenomenon, we solve the Geometric Thin-Film Equation in dimensionless variables. For the initial condition, we use a parabola cap with area $A$ and a base of width $2r_0$ given by
\begin{equation}
    h_0(x) = \begin{dcases}
    \frac{3A}{4r_0^3}(r_0^2-x^2), & \text{if }|x|\leq r_0,\\
    0, &\text{otherwise}.
    \end{dcases}
\end{equation}

The simulations are carried out in a finite spatial domain $x\in[-2,2]$ with a periodic boundary condition (although we emphasize that the boundary condition is not important as long as the droplet does not reach the boundary during the simulation).  Furthermore, we chose $A=1/4$ and $r_0=1/2$ so that the droplet remains in the simulation domain throughout the simulation. The smoothing parameter $\alpha$ should be small but non-zero unless specified otherwise. We have used a value of $\alpha=0.05$ for all of our simulations.  

The numerical calculations indicate that the backward Euler scheme and the Crank-Nicolson scheme produce indistinguishable results by eye: we, therefore, show only results for the Crank-Nicolson scheme. A yet more rigorous comparison between the two schemes is presented in the next subsection. For the high-resolution simulation, we use a grid spacing of $\Delta x=0.02$ and a step size of $\Delta t=0.02$. The Geometric Thin Film Equation is solved until $t=50$ such that the droplet reaches a natural spreading profile.

\begin{figure}[tbh]
	\centering
	\includegraphics[width=0.7\textwidth]{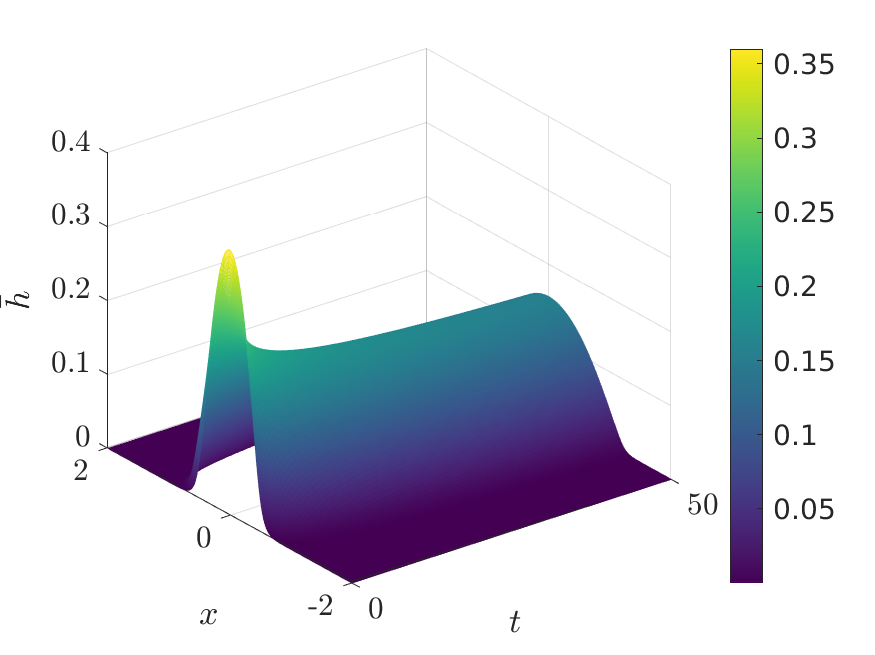}
	\caption{Spacetime plot of the solution of the Geometric Thin-Film Equation. }
	\label{fig:fdm_solution}
\end{figure}

\begin{figure}[htb]
	\centering
	\includegraphics[width=0.7\textwidth]{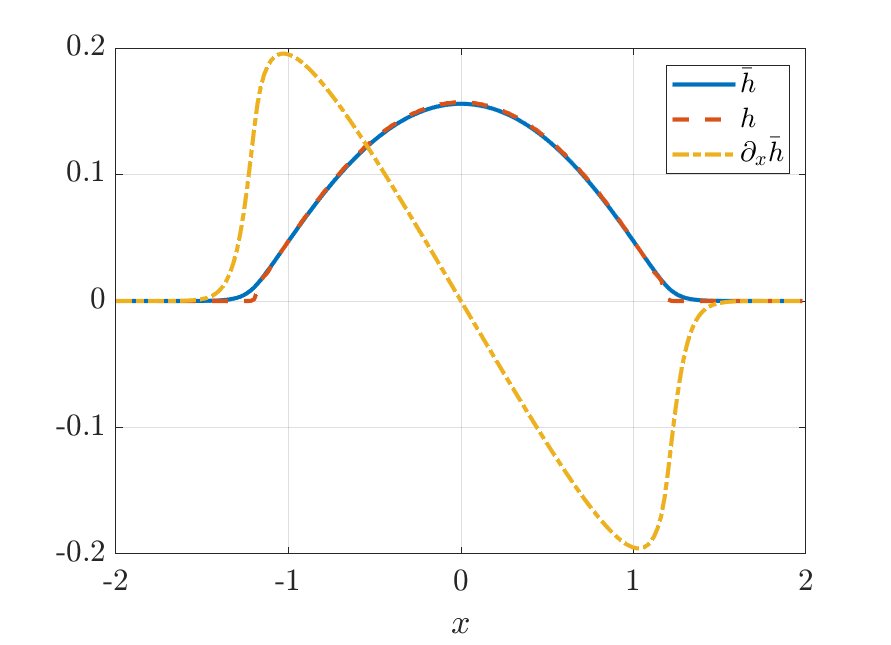}
	\caption{Numerical solution of the Geometric Thin-Film Equation at $t=50$. Here, we plot the filtered surface height $\bar{h}$, the noisy surface height $h$, and the first derivative of $\barh$.}
	\label{fig:sol_theta}
\end{figure}

Figure~\ref{fig:fdm_solution} shows a space-time plot of the smoothened free-surface height $\barh$. From this plot, the width of the droplet increases over time, and the height of the droplet decreases to conserve the volume (mass) of the droplet, demonstrating that the droplet is indeed spreading. Figure ~\ref{fig:sol_theta} shows a snapshot of the solution at $t=50$. We see that the $\barh$ and $\partial_x\barh$ are indeed smooth as a result of the smoothing operation. We also note that the deconvolved free-surface height $h$ is smooth in the droplet core but exhibits a large slope near the foot of the droplet (at $x=\pm1.2$). This is in agreement with the findings in Section~\ref{sec:revise}. 


\begin{figure}[tbh]
	\centering
	\includegraphics[width=0.7\textwidth]{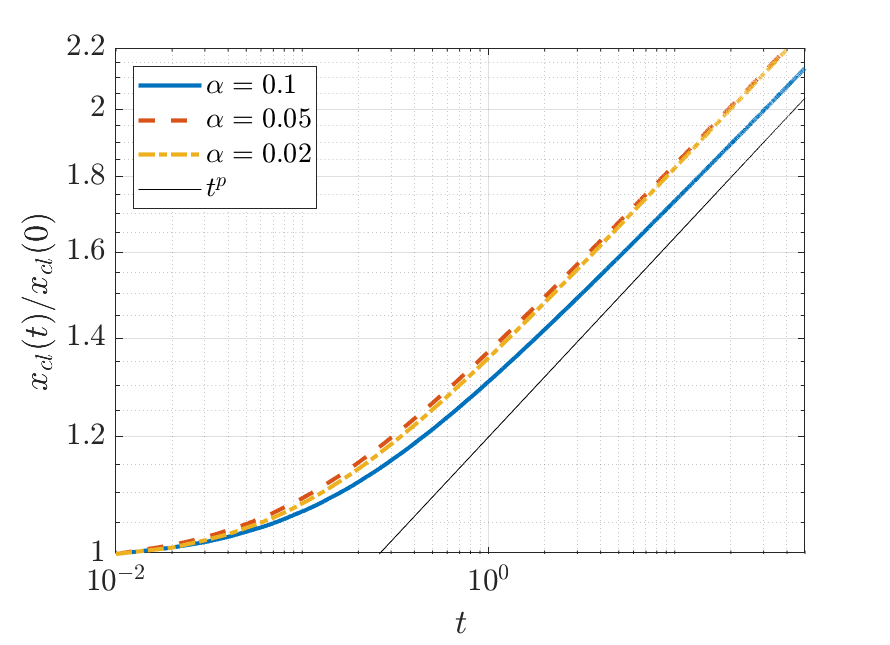}
	\caption{Trajectories of the contact line with various values of $\alpha$. We also plotted $t^p$ (with $p=0.135$) alongside the trajectories, showing that the contact line tends towards $t^p$ as $t\rightarrow\infty$.}
	\label{fig:sensitivity}
\end{figure}

The position of the contact line is obtained by extrapolation: we take $x_{cl}$ to be the $x$-intercept of the tangent line at the inflection point of $\bar{h}(x,t)$. That is, the tangent line at $(x_*(t),\barh(x_*,t))$, where $x_*(t)=\argmin_x\partial_x\barh(x,t)$. This procedure was used in Reference \cite{holm2020gdim} and produces a demarcation between the droplet core and far-field regions, where the droplet profile decays to zero exponentially. In Figure~\ref{fig:sensitivity}, we plot the trajectories of the contact line $x_{cl}(t)$ with different values of $\alpha$. It can be seen that $x_{cl}(t)$ behaves as a power law at late times, with $x_{cl}(t)\sim t^p$ and $p=0.135$. This is very close to the theoretical value $p=1/7$ given by Tanner's Law. Moreover, we see that the effect of varying $\alpha$ on the contact line is small and diminishes as $\alpha$ is decreased. 

\begin{figure}[tbh]
	\centering
	\includegraphics[width=0.7\textwidth]{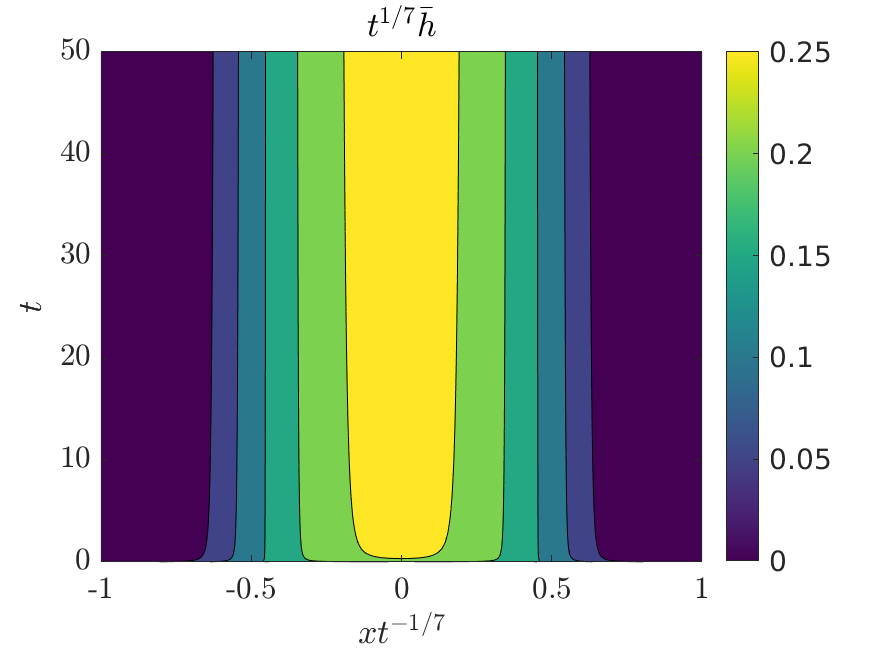}
	\caption{Contour plot of Geometric Thin-Film Equation in similarity variables. The vertical contour lines indicate that the solution profile is time-independent in the similarity variables.}
	\label{fig:similarity_var}
\end{figure}

The finding that the Geometric Thin-Film Equation satisfies Tanner's Law of droplet spreading indicates that the regularized model captures the large-scale physics in the droplet-spreading problem.  In order to demonstrate this even further, we introduce the function $f_\alpha(\eta,t)=t^{1/7}\barh(x,t)$, with $\eta =xt^{-1/7}$.  In Figure \ref{fig:similarity_var}, we produce a space-time plot of $f_\alpha(\eta,t)$ -- this is seen to relax to a constant profile at late times as $t\rightarrow \infty$.  

\begin{figure}[tbh]
	\centering
    \includegraphics[width=0.7\textwidth]{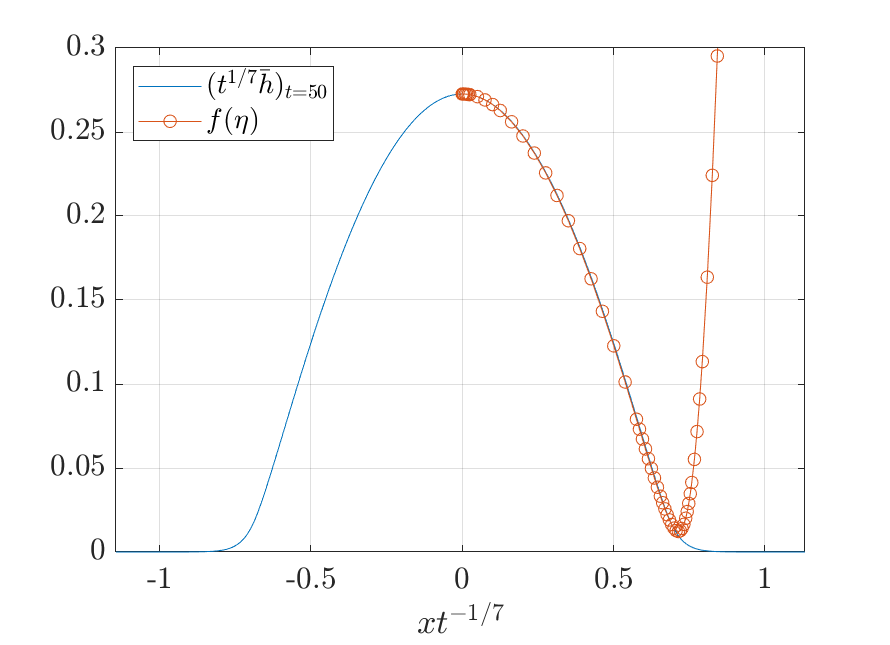}
    \caption{Comparison between $f_\alpha(\eta,t=50)$ and the similarity solution. The two plots agree within the droplet core, but the similarity solution diverges around the droplet foot. }
    \label{fig:similarity_var2}
\end{figure}

Furthermore, the profile of $f_\alpha(\eta,t)$ at fixed $t$ ($t$ large) can be compared with a similarity solution of the unregularized problem, $f^3f'''=\eta f/7$ (\textit{cf.} Equation~\eqref{eq:simtr3}).  This ordinary differential equation is then solved with the shooting method together with appropriate initial conditions~\cite{holm2020gdim}.  The results are shown in Figure~\ref{fig:similarity_var2}. This figure, therefore, shows that the Geometric Thin-Film Equation describes the expected large-scale droplet-spreading physics in the droplet core.  Where the classical Thin-Film Equation breaks down at the contact line, the height profile of the Geometric version decays smoothly to zero. 

% \clearpage
\subsection{Error Analysis} \label{sec:fdm_err}

We analyse the truncation error associated with the finite difference methods. As such, let $\barh$ denote the exact solution of Equation~\eqref{eq:gtfe}, and let $\barh_{\Delta x}$ denote the numerical solution with step size $\Delta x$. We assume that the error $\|\barh-\barh_{\Delta x}\|$ depends smoothly on $\Delta x$, then
\begin{equation}
    \|\barh-\barh_{\Delta x}\| = C\Delta x^p + O(\Delta x^{p+1}),
\end{equation}
for some constant $C$ and $p$. Since $\barh$ is unknown, we instead compute
\begin{equation} \label{eq:varepsilondef0}
\varepsilon(\Delta x) := \|\barh_{\Delta x}-\barh_{\Delta x/2}\|
\end{equation}
By the triangle inequality, 
\begin{align*}
    \varepsilon(\Delta x) &= \|\barh-\barh_{\Delta x}+\barh_{\Delta x/2}-\barh\|, \\
    &\leq \|\barh-\barh_{\Delta x}\|+\|\barh_{\Delta x/2}-\barh\|, \\
    &= C\Delta x^p + C(\Delta x/2)^p + O(\Delta x^{p+1}), \\
    &= C\Delta x^p(1+1/2^p) + O(\Delta x^{p+1}). 
\end{align*}
%
% \begin{equation}
%     \varepsilon(\Delta x) \leq  C\Delta x^p(1-1/2^p) + O(\Delta x^{p+1}).
% \label{eq:varepsilondef1}
% \end{equation}
%
Taking the natural log gives
\begin{equation}
    \log\varepsilon \leq p\log(\Delta x) + \log(C) + \log(1-1/2^p).
\label{eq:varepsilondef2}
\end{equation}
Thus, the rate of convergence (or the order of accuracy) of the numerical method is $p$; $p$ can be computed from the numerical simulations as the slope of the log-log plot between the error $\varepsilon$ and the grid spacing $\Delta x$. We can do the same to compute the temporal rate of convergence $q$ using the formula $\log\varepsilon(\Delta t)\approx q\log(\Delta t)$. 

\begin{figure}[htb]
	\centering
	\includegraphics[width=0.7\textwidth]{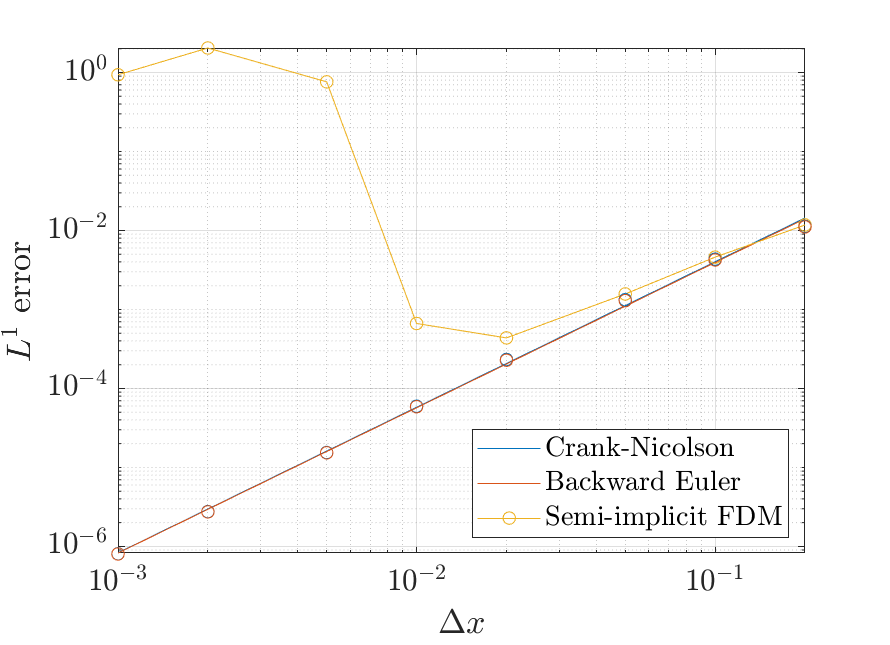}
	\caption{Spatial convergence of the Backward Euler scheme, Crank-Nicolson scheme, and the semi-implicit FDM. The lines for the Backward Euler scheme and the Crank-Nicolson scheme overlap and have the same error. The semi-implicit FDM becomes unstable when $\Delta x<0.01$.}
	\label{fig:conv_newton}
\end{figure}

To compare the stability of the numerical schemes, we restrict the simulation domain to $x\in[-1,1]$ and solve the Geometric Thin-Film Equation until $t=1$. We have also used a fixed step size of $\Delta t=0.02$ across all methods and plot the error of varying values of grid size $\Delta x$. Figure~\ref{fig:conv_newton} shows the rate of convergence of the implicit methods (backward Euler and Crank-Nicolson schemes) and the semi-implicit finite-difference method introduced in \cite{holm2020gdim}. Here we use the $L^1$ norm applied to Equation~\eqref{eq:varepsilondef0} (our choice of the norm is obtained because we will be using $L^1$ norm in our analysis in the following chapters). We observed that the semi-implicit scheme FDM remains stable only when $\Delta x > \Delta t$, whereas the implicit schemes are stable for all values of $\Delta x$. We also see that both implicit schemes make a slope of approximately $1.83$ which indicates a superlinear convergence in space. 

\begin{figure}[htb]
	\centering
	\includegraphics[width=0.7\textwidth]{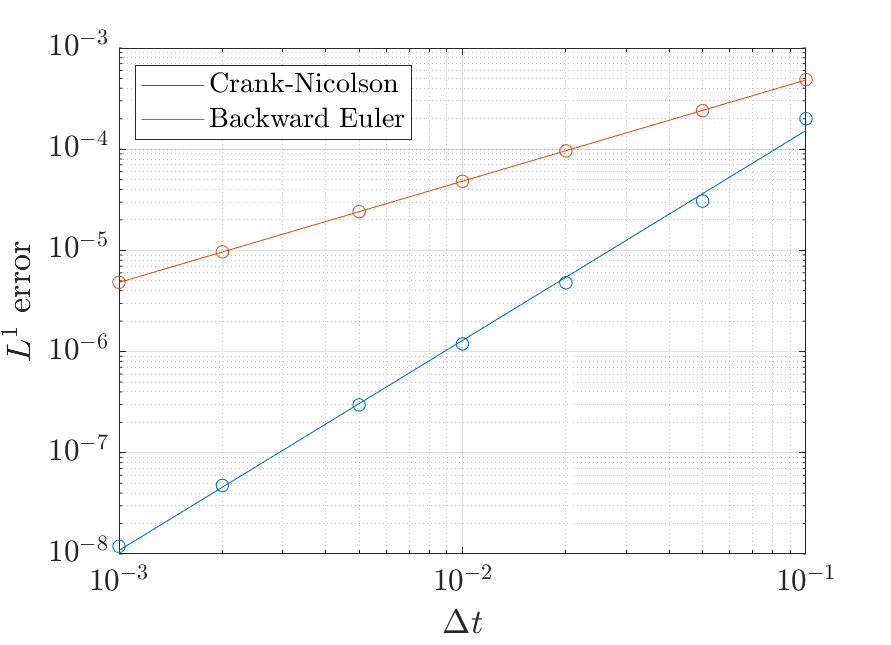}
	\caption{Temporal convergence of the Backward Euler scheme $(\theta=1)$ and the Crank-Nicolson scheme ($\theta=0.5$). We have also plotted the line of best fit (solid lines) through the numerics (hollow circles) to measure each scheme's slope, and thus the convergence rate. }
	\label{fig:conv_dt}
\end{figure}

Figure~\ref{fig:conv_dt} shows the rate of convergence of the Backward Euler scheme ($\theta=1$) and the Crank-Nicolson scheme ($\theta=0.5$) in the temporal domain. Here, we have instead fixed the spatial resolution of $\Delta x=0.04$ and varied the step size $\Delta t$ while keeping all other parameters the same as the spatial convergence plot. The slope of the Backward Euler scheme is $1.001$ indicating that it is order $1$ accurate, whereas the slope of the Crank-Nicolson scheme is $2.075$ suggesting that it is order $2$ accurate in time. 

% \clearpage
\subsection{\markup{Axisymmetric Droplet Spreading}}

The Geometric Thin-Film Equation with two-spatial dimensions is given by
\begin{equation}
    \pder[h]{t} = -\nabla\cdot\left\{h\mob\nabla\nabla^2\bar{h}\right\}, \qquad (1-\alpha^2\nabla^2)^2\bar{h}=h,
\end{equation}
on the domain $\mathbb{R}^2$, $t\geq0$. In Cartesian coordinate, the del operator is given by $\nabla = (\partial_x, \partial_y)$. In 2D, the corresponding Green's function for the bi-Helmholtz operator (using $r=\sqrt{x^2+y^2}$) is given by \cite{bi_helmholtz}
\begin{equation}
    K_2(r) = 
    \begin{dcases}
        \frac{r}{4\pi\alpha^3}\mathcal{K}_1\left(\frac{r}{\alpha}\right), & r>0, \\
        \frac{1}{4\pi\alpha^2}, & r=0.
    \end{dcases}
\end{equation}
where $\mathcal{K}_1$ is the modified Bessel function of the second kind of order 1:
\begin{equation}
    \mathcal{K}_1(r) = \int_0^\infty \mathe^{-r\cosh t}\cosh t\,\mathd t.
\end{equation}

When unperturbed, the most common shape of a sessile droplet is an axisymmetric one. That is, the droplet profile is invariant to rotation around a central $z$-axis. The natural coordinate system for describing the droplet in this case is the cylindrical polar coordinate $(r,\varphi,z)$, where the droplet profile is constant in $\varphi$. In this case, the axisymmetric Geometric Thin-Film Equation becomes 
\begin{align} \label{eq:gtferad}
    \pder[h]{t} &= -\frac{1}{r}\pder{r}\left\{rh\bar{h}^2\pder{r}\left(\frac{1}{r}\pder{r}\left(r\pder[\bar{h}]{r}\right)\right)\right\}, 
    % u &= \left(1-\alpha^2\frac{1}{r}\der{r}\left(r\der{r}\right)\right)^2\bar{u}.
\end{align}
on $r\in(0,\infty)$, $t\geq0$. The boundary conditions at the pole $r=0$ are given by
\begin{equation}
    \partial_rh(0,t) = \partial_{rrr}h(0,t) = 0. 
\end{equation}
We use the same fully-implicit finite-difference scheme to solve this equation. 

\begin{figure}[tbh]
	\centering
    \includegraphics[width=0.7\textwidth]{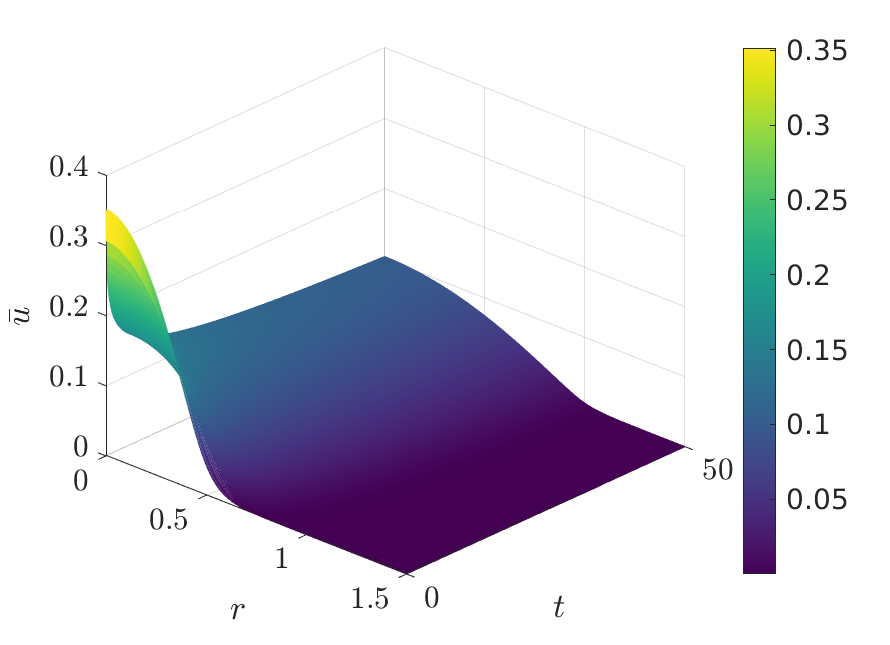}
    \caption{Spacetime plot of the Geometric Thin-Film Equation for 3D radially symmetrical droplet, where $r$ is the radial distance from the centre of the droplet.}
    \label{fig:solution_rad}
\end{figure}

For the initial condition, we use a paraboloid cap with volume $A$ and radius $r_0$:
\begin{equation}
    h_0(r) = \begin{cases}
        \dfrac{2A}{\pi r_0^4}(r_0^2 - r^2), &|r|<r_0 \\ 
        0, &\text{otherwise.}
    \end{cases}
\end{equation}
We set $A=0.15$ and $r_0=0.5$ for the initial condition and $\alpha=0.05$ for the smoothing parameter. The radially symmetrical Geometric Thin-Film Equation is solved using the Crank-Nicolson scheme ($\theta=0.5$) on the domain $r=[0,1.5]$ and until a final time of $t=50$. We use a grid spacing of $\Delta x=0.005$ and step size of $\Delta t=0.05$, the result is shown in Figure~\ref{fig:solution_rad}. In 3D, the spreading rate is slower compared to the 2D counterpart, thus we are allowed to initialize a smaller domain. Although we do not show it here, the solution also converges quadratically as the grid size $\Delta x$ gets smaller. 

\begin{figure}[tbh]
	\centering
    \includegraphics[width=0.7\textwidth]{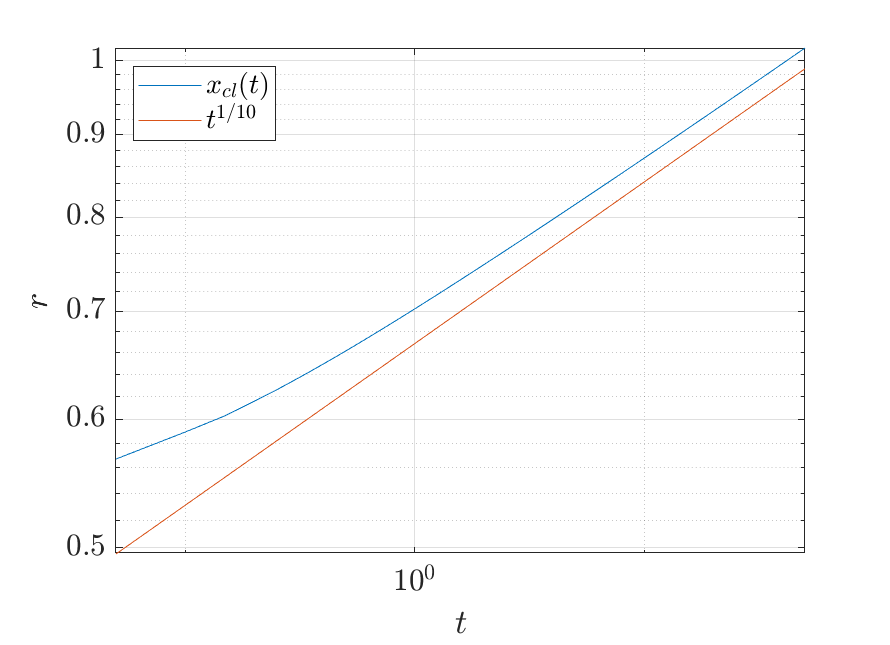}
    \caption{Trajectory of the contact line for the radially symmetrical droplet following a power law behaviour at late time.}
    \label{fig:contact_line_rad}
\end{figure}

We compute the position of the contact line of the droplet in the same way as we did in Section~\ref{sec:fdm_result} with the $x$-intercept of the tangent line at the inflection point of $\bar{u}$. In Figure~\ref{fig:contact_line_rad}, we plot the location of the contact line over time. Same as in the 2D case, the contact line also follows a power law at late time, but with a slower rate of $x_{cl}\sim t^{1/10}$. This agrees with Tanner's law for three-dimensional droplets. 

\clearpage
\section{Particle Method for Gradient Flows} \label{sec:particles}

In this section, we discuss the (deterministic) particle method for the Geometric Thin-Film Equation. The particle method is a numerical method for solving gradient flow PDEs of the form
\begin{equation}
    \pder[\rho]{t} = \nabla\cdot\left(\rho m(\rho)\nabla\frac{\delta E}{\delta\rho}\right).
\end{equation}
It is different from the finite difference method introduced in the previous section in that the particle method is a mesh-free Lagrangian method where a fluid quantity (usually the mass) is discretized into elements called particles, and the location of these particles is then updated at each time step in accordance to the Navier-Stokes Equation or the corresponding PDE. Mathematically, the particle method seeks weak solutions of the PDE as a sum of weighted Dirac delta distributions, also known as particle solution. In particular, we will focus on the solution of the form
\begin{equation} \label{eq:p_sol_wass}
    \rho(x,t) = \sum_{i=1}^N w_i\delta(x-x_i(t)),
\end{equation}
where the location of the particles is described by a system of $N$ coupled ODEs 
\begin{equation} \label{eq:p_odewass}
    \dot{x}_i(t) = -\left(m(\rho)\nabla\frac{\delta E}{\delta\rho}\right)_{x=x_i}, \qquad i=1,\dots,N.
\end{equation}
Here, $N$ is a positive integer corresponding to a truncation of an infinite sum, $w_i\geq0$ are weights to be computed, and $\delta(x)$ is the Dirac delta function. The motivation for seeking out such highly simplified particle solutions is that they make the task of solving the PDE numerically very simple: instead of discretizing a fourth-order parabolic-type PDE in the case of the Geometric Thin-Film Equation, we can solve a set of ordinary differential equations for the delta-function centres $x_i(t)$ using standard time-marching algorithms. This simplifies the numerical computation greatly and is similar to the concept in the Method of Lines. We refer to the weights $w_i$ together with the delta-function centres $x_i(t)$ as the `particles'. 

The particle method is similar conceptually to the smoothed-particle hydrodynamics (SPH) \amend{where the fluid field is discretized into `particles' that represent packets of the fluid masses \cite{wang2016}. 
% They are both Lagrangian methods, which provide better resolution of the fluid interface compared to the standard grid-based method.  
The main difference between the particle method and the SPH is the discretized PDE, where the SPH directly discretizes the Navier-Stokes equations. In contrast, the particle method is designed with the thin-film equation in mind. This provides a performance advantage to using the particle method in the droplet spreading scenario to utilise the lubrication approximation. Furthermore, the particle solutions also provide a framework to establish the regularity of the solutions through Banach space theory, which we explore in \cref{sec:theory_wellpose}. }
%The main difference between the two methods is that the particle method describes a weak solution of the PDE in a mathematically precise sense, which will be discussed in the following sections, whereas the SPH describes an approximate solution. 
%For the same reason, the particle method only applies to problems involving the gradient flow of specific structures, whereas the SPH can be applied more generally. 
\amend{In general, the particle method is a numerical method specific to solving problems with a gradient flow structure \cite{chertock2017}. }
We also note that the particles are referred to as the Dirac delta distributions in the particle method. Meanwhile, the particles are referred to as the smooth kernel functions in the SPH. 

% In practice, we use built-in ODE solver in MATLAB or Python for solving Equation~\eqref{eq:p_odewass}. This requires that the right-hand-side of Equation~\eqref{eq:p_odewass} to be a classical function such that it is well defined when evaluated at $x_i$. This is the motivation for the higher order smoothing discussed in Section~\ref{sec:tfe_higher} and we take the mobility $\mu(h,\bar{h})=\bar{h}^2$. 

The particle method provides a few advantages in investigating the contact-line motion: (i) Since the droplet is compactly supported and the domain of simulation is potentially large (ideally the whole of $\mathbb{R}$), the finite-difference method can be slow as it requires the initialization of a large simulation domain (the mesh) to anticipate for the spreading of the droplet. The particle method can be made sparse to reduce the computational cost by only simulating the location where the fluid is present. Furthermore, the particle method is mesh-free, and the particles will occupy a larger domain as the droplet spread at no cost to the simulation time. (ii) The flow field of the fluid, in particular, the flow at the contact line, needs to be measured separately in the finite difference method, but in the particle method, this is simply the trajectories of the particles. (iii) The particle method has a low numerical diffusion compared to Eulerian methods. This is because the derivative in space can be computed exactly in the particle method. (iv) Much effort was made to develop positivity-preserving numerical algorithms to solve the thin-film equation and equation of the like. The particle solution, however, is both positivity preserving and mass conserving by definition as long as $w_i\geq0$. 

% In Section~\ref{sec:ptheory}, we will derive the particle solution for the Geometric Thin-Film Equation. In Section~\ref{sec:fast_summation}, we discuss a fast summation algorithm for optimizing the evaluation of the ODEs. In section~\ref{sec:pnumerics}, we perform some numerical simulation and compare the particle method with the finite difference method. % Finally, in Section (??), we discuss the possibility of using the particle method for simulating droplet in higher dimension using the fast multipole method. 

\subsection{A Family of Particle Solutions} \label{sec:ptheory}

As stated above, we start by considering solutions of the form
\begin{equation} \label{eq:hNdef}
    h^N(x,t) = \sum_{i=1}^N w_i \delta(x-x_i(t)).
\end{equation}
For a general smoothing kernel $K$, the smoothed surface height is given by
\begin{equation}
    \bar{h}^N(x,t) = K(x)*\bar{h}^N(x,t) = \sum_{i=1}^N w_i K(x-x_i(t)).
\end{equation} 
The weights $w_i$ and the initial position of the particle $x_i(0)$ need to be chosen such that the solution reflects the initial condition $h_0(x)$. In particular, we require the particle solution to converge to the initial condition weakly, 
\begin{equation} \label{eq:compatibity1}
    \lim_{N\rightarrow\infty} \intinf h^N(x,0)\phi(x)\mathd x = \intinf h_0(x)\phi(x)\mathd x,
\end{equation}
where $\phi(x)$ is an arbitrary smooth, integrable test function. In practice, because the droplet is compactly supported, say on $[-L,L]$, we can define 
\begin{equation}
    x_i(0) = x_i^0 = \left(i-\frac{N}{2}\right)\Delta x, \qquad i=1,\dots,N,
\end{equation}
where $\Delta x=2L/N$ and $N$ is the number of particles. We then define $w_i$ to be
\begin{equation}
    w_i = \int_{x_i^0-\Delta x/2}^{x_i^0+\Delta x/2} h_0(x) \mathd x.
\end{equation}
Then Equation~\eqref{eq:compatibity1} is satisfied automatically. In fact, the total weight is fixed relative to $N$ and is equal to the volume (area) of the droplet
\begin{equation}
    \sum_{i=1}^N w_i = \intinf h_0(x) \mathd x.
\end{equation}

% We now define the weak form of the Geometric Thin-Film Equation
% \begin{equation} \label{eq:weak1}
%     \intinf \phi \pder[h]{t} \mathd x + \frac{\surften}{3\vis} \intinf \phi \pder{x}\left(h \barh^2 \pder[^3\barh]{x^3}\right)\mathd x = 0, 
% \end{equation}
% for all test function $\phi$. 

% We substitute Equations~\eqref{eq:hNdef} into Equation~\eqref{eq:weak1}.  Owing to the judicious choice of mobility $\mu=(1/3\vis)\barh^2$ (\textit{cf.} Equation~\eqref{eq:mobility}), no instance of the singular solution $h^N(x,t)$ gets squared (only the smoothened solution $\barh^N$ gets squared). For the first term, we have
% \begin{align}
%     T_1 =& \intinf \phi(x)\pder{t}\left[\sum_{i=1}^N w_i \delta(x-x_i)\right]\mathd x, \\
%     =& \sum_{i=1}^N w_i \der[x_i]{t} \intinf \phi'(x)\delta(x-x_i) \mathd x, \\
%     =& \sum_{i=1}^N w_i \phi'(x_i) \der[x_i]{t}.
% \end{align}
% On the other hand, the second integral in Equation~\eqref{eq:weak1} becomes
% \begin{align}
%     I_2 =& - \intinf \phi'(x) h \barh^2 \pder[^3\barh]{x^3}\mathd x, \\
%     =& - \intinf \phi'(x) \barh^2 \left[\sum_{i=1}^N w_i\delta(x-x_i)\right] \left[\sum_{j=1}^N w_j K'''(x-x_j)\right]\mathd x.
% \end{align}
% If $K=K_2$ is the Green's function of the bi-Helmholtz operator, we interpret $K'''$ to be the weak derivative 
% \begin{align}
%     =& - \sum_{i=1}^N\sum_{\substack{j=1\\j\neq i}}^N w_i w_j \phi'(x_i)\barh^N(x_i,t)^2K'''(x_i-x_j) \nonumber\\
%     &\qquad - \sum_{i=1}^N w_i^2 \intinf \phi'(x)\barh^2 \delta(x-x_i)K'''(x-x_j) \mathd x
% \end{align}

We now multiply both sides of Equation~\eqref{eq:gtfe} by the test function $\phi(x)$, integrate from $x=-\infty$ to $x=\infty$, and apply vanishing boundary conditions at these limits.  We thereby obtain
\begin{equation}
    \left\langle \phi,\pder[h]{t}\right\rangle + \frac{\surften}{3\vis}\left\langle \phi,\pder{x}\left(h \barh^2 \pder[^3\barh]{x^3}\right)\right\rangle=0,
    \label{eq:weak1}
\end{equation}
where $\langle \cdot,\cdot\rangle$ denotes the standard pairing of square-integrable functions with respect to $x$:
\begin{equation}
    \langle f,g\rangle=\int_{-\infty}^\infty f(x) g(x) \,\mathd x,\qquad f,g, \in L^2(\mathbb{R}).
\end{equation}
We substitute Equations~\eqref{eq:hNdef} into Equation~\eqref{eq:weak1}.  Owing to the judicious choice of mobility $\mob=(1/3\vis)\barh^2$ (\textit{cf.} Equation~\eqref{eq:mobility}), no instance of the singular solution $h^N(x,t)$ gets squared (only the smoothened solution $\barh^N$ gets squared). For the first term, we have  
\begin{align}
    \left\langle\phi,\pder[h]{t}\right\rangle =& \sum_{i=1}^N w_i \frac{\mathd x_i}{\mathd t} \langle\phi', \delta(\cdot-x_i)\rangle, \\
    =& \sum_{i=1}^N w_i \phi'(x_i)\frac{\mathd x_i}{\mathd t}.
\end{align}
On the other hand, the second inner product in Equation~\eqref{eq:weak1} becomes
\begin{align}
    \left\langle \phi,\pder{x}\left(h \barh^2 \pder[^3\barh]{x^3}\right)\right\rangle =& - \left\langle\phi',h \barh^2 \pder[^3\barh]{x^3} \right\rangle, \\
    =& -\sum_{i=1}^N w_i \left\langle \phi' , \delta(\cdot-x_i)\barh^2 \pder[^3\barh]{x^3} \right\rangle, \label{eq:singularprob} \\
    =& -\sum_{i=1}^N w_i \phi'(x_i) \left[(\barh^N)^2 \partial_{xxx}\barh^N\right]_{x=x_i}.
\end{align}
Thus, Equation~\eqref{eq:weak1} becomes:
\begin{equation}
    \sum_{i=1}^N w_i\phi_x(x_i)\frac{\mathd x_i}{\mathd t}-\frac{\surften}{3\vis}\sum_{i=1}^N w_i \phi_x(x_i) \left[(\barh^N)^2 \partial_{xxx}\barh^N\right]_{x=x_i}=0,
    \label{eq:weak2}
\end{equation}
where now $[(\barh^N)^2 \partial_{xxx}\barh^N]_{x=x_i}$ is taken to mean
\begin{equation} \label{eq:odevdef}
    \left[\sum_{j=1}^N w_j K(x_i-x_j)\right]^2\left[\sum_{\substack{j=1\\j\neq i}}^N w_j K'''(x_i-x_j)\right].
\end{equation}
Equation~\eqref{eq:weak2} is re-arranged to give
\begin{equation}
    \sum_{i=1}^N w_i\phi_x(x_i)\left[ \frac{\mathd x_i}{\mathd t}-\frac{\surften}{3\vis}\left[(\barh^N)^2 \partial_{xxx}\barh^N\right]_{x=x_i}\right]=0.
    \label{eq:weak3}
\end{equation}
Equation~\eqref{eq:weak3} is true for all test functions $\phi(x)$. Therefore,
\begin{equation}
    \frac{\mathd x_i}{\mathd t}-\frac{\surften}{3\vis}\left[(\barh^N)^2 \partial_{xxx}\barh^N\right]_{x=x_i}=0.
    \label{eq:ode}
\end{equation}
Thus, Equation~\eqref{eq:hNdef}, together with the ordinary differential equations
\begin{equation}
    \frac{\mathd x_i}{\mathd t}=\frac{\surften}{3\vis}\left[(\barh^N)^2 \partial_{xxx}\barh^N\right]_{x=x_i},\qquad i=1,\cdots,N,
    \label{eq:odedefx}
\end{equation}
give a  so-called \textit{particle solution} to the Geometric Thin-Film Equation~\eqref{eq:gtfe}.  The velocity $v_i$ of the $i^{\text{th}}$ particle is identified with the right-hand side in  Equation~\eqref{eq:odedefx}. We can write this more compactly as
\begin{equation}
    \dot{\bm{x}} = \bm{v}(\bm{x}(t)),
\end{equation}
where $\dot{\bm{x}}=(\dot{x}_1,\dots,\dot{x}_N)$ and the dot represents derivative with respect to time, and $\bm{v}(\bm{x})=(v_1(\bm{x}),\dots,v_N(\bm{x}))$ with
\begin{equation} \label{eq:odedefx1}
    v_i(\bm{x})=\frac{\surften}{3\vis}\left[(\barh^N)^2 \partial_{xxx}\barh^N\right]_{x=x_i}.
\end{equation}

We notice that in Equation~\eqref{eq:odedefx1}, evaluation of $K'''$ is required -- this is the rationale for our choice of the bi-Helmholtz kernel $K_2$ as the smoothing kernel in Equation~\eqref{eq:gtfe}.  Using the single Helmholtz kernel $K_1$ would not be sufficient, as $K_1'''$ is singular at the origin.  Furthermore, as the reconstructed interface profile $\barh(x,t)=\sum_{i=1}^Nw_iK(x-x_i(t))$ involves the positive weights $w_i$ and a positive kernel $K\geq 0$, the particle method is positivity-preserving: if $h$ and $\barh$ are initially positive, then they stay positive for all time.  The numerical particle method is manifestly positivity-preserving; this is a key advantage as a numerical method that led to erroneous negative values of $h$ and $\barh$ would produce unphysical results.

We remark that the evaluation of $\partial_{xxx}\bar{h}$ at $x=x_i$ as defined in Equation~\eqref{eq:odevdef} does not include the evaluation of $K_2'''(0)$. This is because $K_2$ is only twice differentiable in the classical sense, and the third derivative is taken as a weak derivative. So the value of $K_2'''$ is only uniquely determined on $\mathbb{R}\setminus\{0\}$. When evaluating the inner product in Equation~\eqref{eq:singularprob}, the $j=i$ term in the summation can be shown to be identically zero by an integration by parts of the bi-Helmholtz operator. 

% A pseudocode (in MATLAB syntax) for solving the Geometric Thin-Film Equation is implemented below, where we have used the built-in \hltexttt{ode45} solver. We have also utilized multi-dimensional arrays and pointwise operator \hltexttt{.*} and \hltexttt{.\^} to vectorize the summation. 
% \lstinputlisting[language=Octave]{code/direct_sum.m}

\subsection{Fast Summation Algorithm} \label{sec:fast_summation}

The particle method requires us to compute the evolution equations for the particle trajectories; these are given in Equation~\eqref{eq:odedefx}.  Thus,  one is required to evaluate
\begin{align}
	\barh^N(x_i) &= \sum_{j=1}^N w_j K_2(x_i-x_j), \label{eq:particle_sum1} \\
	\partial_{xxx}\barh^N(x_i) &= \sum_{\substack{j=1\\j\neq i}}^N w_j K_2'''(x_i-x_j), \label{eq:particle_sum2}
\end{align}
for $i=1,\dots,N$, at each time step. Here, we provide the explicit expression for $K_2$ and $K_2'''$:
\begin{align}
    K_2(x) =& \frac{1}{4\alpha^2} (\alpha + |x|) \mathe^{-|x|/\alpha}, \\
    K_2'''(x) =& \frac{1}{4\alpha^4} \left(2\operatorname{sgn}(x)-\frac{x}{\alpha}\right)\mathe^{-|x|/\alpha}.
\end{align}
Since there are $N$ particles and each particle is influenced by all $N$ particles, a direct evaluation of Equation~\eqref{eq:odedefx} would require an operational count of $O(N^2)$. Now, suppose that the particle relative ordering of the particles remains the same throughout the simulation, that is, $x_i<x_j$ for all $i<j$ for all $t\geq0$, % then we can sweep from left to right to collect the contribution from particles left of $x_i$ and do the same for contribution from particles right of $x_i$. 
then the contribution from particles of either side of $x_i$ becomes a degenerate case of the fast multipole method. A theorem for the no-crossing of particles is presented in \cref{sec:collision}. 

The key idea of the fast summation algorithm is to isolate the source contribution from the evaluation of the summation
\begin{equation} \label{eq:fast}
    \bar{h}^N(x) = \sum_{k=1}^p a_k(x_1,\dots,x_N) \Phi_k(x),
\end{equation}
where $a_k(x_1,\dots,x_N)$ can be evaluated without the knowledge of the target location. 

We start by splitting the summation into two parts,  
\begin{equation}
S_1 = \sum_{j=1}^{i-1}w_jK_2(x_i-x_j), \qquad S_2 = \sum_{j=i+1}^{N}w_jK_2(x_i-x_j).
\end{equation}
Since $x_i-x_j>0$ for $j=1,\dots,i-1$, we can drop the absolute value in the function $K_2$ in $S_1$,
\begin{align}
    S_1 &= \sum_{j=1}^{i-1}w_j(\alpha+x_i-x_j)\mathe^{-(x_i-x_j)/\alpha}, \\
    &= (\alpha+x_i)\mathe^{-x_i/\alpha}\underbrace{\sum_{j=1}^{i-1}w_j\mathe^{x_j/\alpha}}_{a_i} - \mathe^{-x_i/\alpha}\underbrace{\sum_{j=1}^{i-1}w_jx_j\mathe^{x_j/\alpha}}_{b_i}.
\end{align}
Similarly, $x_i-x_j<0$ for $j=i+1,\dots,N$ and we obtain for $S_2$,
\begin{align}
    S_2 &= (\alpha-x_i)\mathe^{x_i/\alpha}\underbrace{\sum_{j=i+1}^Nw_j\mathe^{-x_j/\alpha}}_{c_i} + \mathe^{x_i/\alpha}\underbrace{\sum_{j=i+1}^Nw_jx_j\mathe^{-x_j/\alpha}}_{d_i}.
\end{align}
So Equation~\eqref{eq:particle_sum1} can be expressed as 
\begin{align}
    \barh^N(x_i) &= S_1 + w_iK_2(0) + S_2, \\
    &= \frac{1}{4\alpha^2}(\alpha+x_i)\mathe^{-x_i/\alpha}a_i - \frac{1}{4\alpha^2}\mathe^{-x_i/\alpha}b_i + w_iK(0) \nonumber \\ 
    &\qquad + \frac{1}{4\alpha^2}(\alpha-x_i)\mathe^{x_i/\alpha}c_i + \frac{1}{4\alpha^2}\mathe^{x_i/\alpha}d_i, \\
    &= \frac{1}{4\alpha^2} \left[\mathe^{-x_i/\alpha}((\alpha+x_i)a_i-b_i) + \mathe^{x_i/\alpha}((\alpha-x_i)c_i+d_i)\right] \nonumber \\
    &\qquad + w_iK_2(0).
\end{align}
Note that $a_i, b_i, c_i$, and $d_i$ can be computed at the start of each time step before the evaluation of $\barh^N(x_i)$ in $O(N)$ operations using a recursion formula 
\begin{align}
&
\begin{rcases}
    a_1=0, & \qquad a_i = a_{i-1} + w_{i-1}\mathe^{x_{i-1}/\alpha} \\
    b_1=0, & \qquad b_i = b_{i-1} + w_{i-1}x_{i-1}\mathe^{x_{i-1}/\alpha}
\end{rcases}
\ i=2,\dots,N, \\ 
&
\begin{rcases}
    c_N=0, & \qquad c_i = c_{i+1} + w_{i+1}\mathe^{-x_{i+1}/\alpha} \\
    d_N=0, & \qquad d_i = d_{i+1} + w_{i+1}x_{i+1}\mathe^{-x_{i+1}/\alpha}
\end{rcases}
\ i=N-1,\dots,1.
\end{align}
We follow the same procedure for Equation~\eqref{eq:particle_sum2} to obtain, 
\begin{align}
    \partial_{xxx}\barh^N(x_i) &= \frac{1}{4\alpha^4}\left[\mathe^{-x_i/\alpha}((2\alpha-x_i)a_i+b_i) + \mathe^{x_i/\alpha}(-(2\alpha+x_i)c_i+d_i)\right].
\end{align}
This decomposition reduces the complexity of evaluating the Equation~\eqref{eq:ode} from $O(N^2)$ to $O(N)$.

\subsection{Numerical Results} \label{sec:pnumerics}

We use the same initial condition as discussed in \cref{sec:fdm_result} to compare the particle method to the finite-difference method. 

\begin{figure}[htb]
	\centering
    \includegraphics[width=0.7\textwidth]{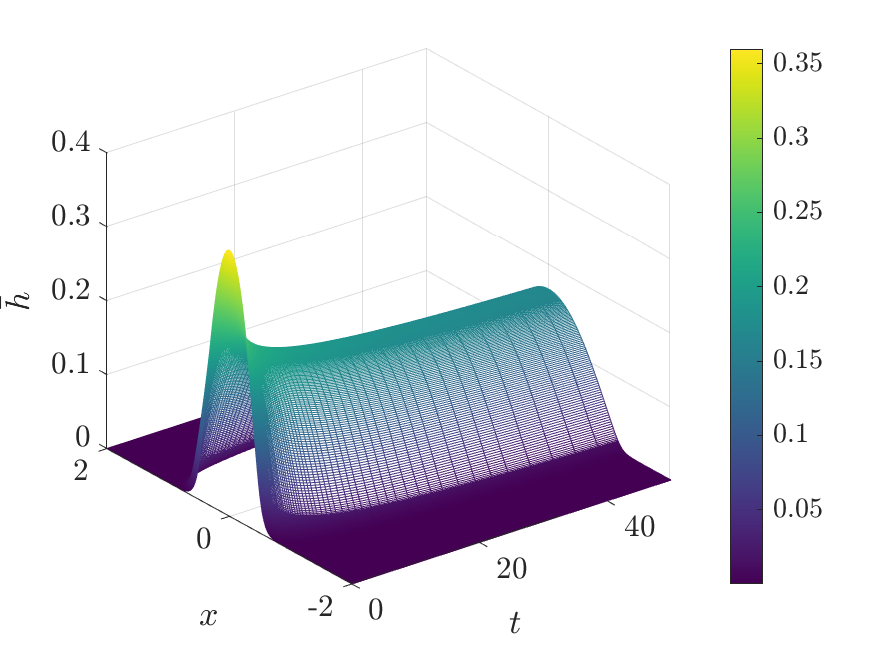}
    \caption{Spacetime plot of the solution of the Geometric Thin-Film Equation using the particle method. \amend{$N=800$, $\alpha=0.05$}.}
    \label{fig:solution1}
\end{figure}

In Figure~\ref{fig:solution1}, we show a space-time plot of the smoothened free-surface height $\bar{h}$ where the spatial grid is evaluated at the particle positions $x_i(t)$ -- effectively, a discretization of $\barh$ on a non-uniform grid. We have used the built-in MATLAB ODE solvers to solve Equation~\eqref{eq:odedefx}, meaning that the particle method inherits the adaptive time stepping of the ODE solver and takes larger time step $\Delta t$ when the solution is sufficiently smooth. 

\begin{figure}[htb]
	\centering
    \includegraphics[width=0.7\textwidth]{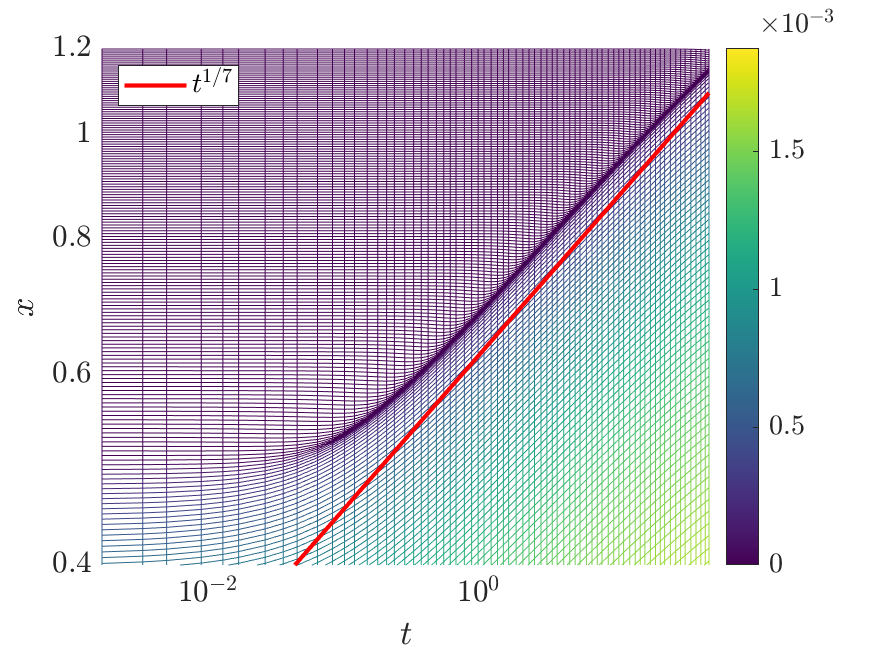}
    \caption{Evolution of the particle trajectories $x_i(t)$. The colours indicate the weight corresponding to each particle $w_i$. The line $t^{1/7}$ is imposed to show that the trajectories follow a power law at late time.}
    \label{fig:trajectory2}
\end{figure}

In Figure~\ref{fig:trajectory2}, we plot the particle trajectories $x_i(t)$ for the solution of the Geometric Thin-Film Equation. From the Equation~\eqref{eq:odedefx1}, we see that the particles within the droplet core will move outwards, while the particles outside the droplet core, where $\barh$ is small, $v_i$ is approximately zero. Hence one can reasonably deduce that the particles will accumulate near the foot of the droplet. When a finite-difference or finite-volume solver is able to execute local grid refinement in the region of interest, this is because of adaptive mesh refinement. Here, the particle method demonstrates a tendency to mimic the effect of adaptive mesh refinement without explicitly implementing adaptive mesh procedures \amend{which would have a computational overhead to re-mesh the domain every few timesteps}. 

A further advantage of the particle method is that it provides a numerical description of the flow field of the contact line by simply following the trajectories of the particles near the contact line. This is also shown in Figure~\ref{fig:trajectory2}, where the flow field around the contact line is found to satisfy Tanner's Law, with $x_i(t)\sim t^{1/7}$. The scaling persists even at late times, indicating that finite-size effects are not important here. Indeed, the particle method is intrinsically free from finite-size effects. Since the kernel function extends to infinity in both directions, the particle method and the ODE solver seamlessly track the particles as they move away from the origin and hence, as the droplet spreads. This is a key benefit of using the particle method. 

\subsection{Error and Performance Comparison}

\begin{figure}[htb]
	\centering
    \includegraphics[width=0.7\textwidth]{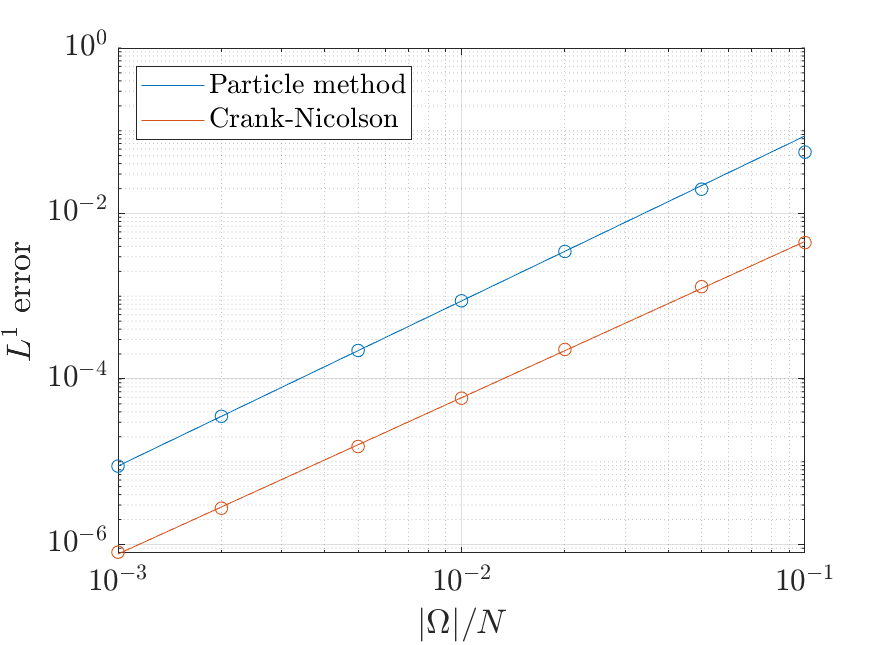}
    \caption{Error of the particle method plotted against the Crank-Nicolson scheme. The $x$-axis is the domain size over the number of particles for the particle method and the number of grid cells in the finite difference Crank-Nicolson scheme.}
    \label{fig:conv_particle}
\end{figure}

Figure~\ref{fig:conv_particle} shows the rate of convergence of the finite-difference method and the particle method. The finite-difference method is implemented with a step size of $\Delta t=0.01$, while the particle method uses the ODE45 solver in MATLAB, hence an adaptive time step. Both methods use the same numerical parameter of $\alpha=0.05$, with final time $T=1$, and periodic boundary condition on the spatial domain $x\in[-1,1]$. From this figure, both the particle method and the standard finite-difference method are estimated to be second-order accurate in the spatial domain, albeit the absolute error is smaller for the finite-difference method.

\begin{figure}[htb]
	\centering
    \includegraphics[width=0.7\textwidth]{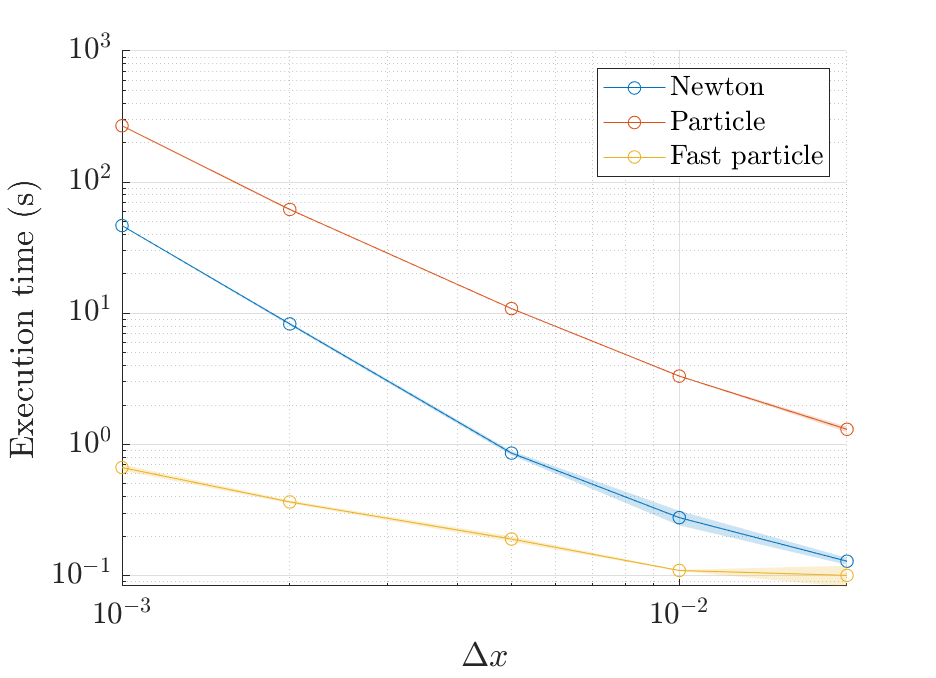}
    \caption{Execution time measured in seconds of various numerical methods and the resolution $\Delta x$ is varied. Lower $\Delta x$ corresponds to higher resolutions. }
    \label{fig:performance}
\end{figure}

Furthermore,  in Figure~\ref{fig:performance}, we evaluate the execution time of the different numerical methods to see if any one method outperforms the rest.   A comparison of the average execution time over ten runs between the finite-difference method, the direct implementation of the particle method (computational complexity $O(N^2)$), and the fast implementation particle method (computational complexity $O(N)$). The numerical parameters used are the same as the ones used in the convergence analysis, and the calculations are performed on an Intel i7-9750H with six hyper-threaded cores. 

\subsection{\markup{Fast Multipole Method for bi-Helmholtz Kernel}}

In two spatial dimensions, the particle solution is given by 
\begin{align}
    h^N(x,y;t) = \sum_{i=1}^N w_i\delta(x-x_i(t),y-y_i(t)), \\ 
    \barh^N(x,y;t) = \sum_{i=1}^N w_iK_2(x-x_i(t),y-y_i(t)),
\end{align}
with $K_2$ given by the Green's function of the bi-Helmholtz operator in two dimensions $(1-\alpha^2\nabla^2)^2$.
The trajectories $\bm{x}_i(t)=(x_i(t),y_i(t))\in\mathbb{R}^2$ satisfy the ODE system
\begin{align}
    % \dot{x}_i = \left[(\bar{h}^N)^2\pder{x}\nabla^2\barh^N\right]_{\bm{x}=\bm{x}_i}, \qquad
    % \dot{y}_i = \left[(\bar{h}^N)^2\pder{y}\nabla^2\barh^N\right]_{\bm{x}=\bm{x}_i}.
    \begin{cases}
        \dot{\bm{x}}_i = \left[(\bar{h}^N)^2\nabla\nabla^2\barh^N\right]_{\bm{x}=\bm{x}_i}, \\
        \bm{x}_i(0) = \bm{x}_i^0.
    \end{cases}
\end{align}
The particle method requires one to evaluate 
\begin{equation} \label{eq:fmmhbar}
    \barh^N(\bm{x}_j) = \sum_{i=1}^N w_iK_2(\bm{x}_j-\bm{x}_i), \qquad j=1,\dots,N.
\end{equation}
A direction evaluation of \cref{eq:fmmhbar} requires $O(N^2)$ computations. In the case of two-dimensional droplets, we introduce a fast summation algorithm to reduce the computations of such summation to $O(N)$. However, in higher dimensions, the assumption of the fast summation algorithm no longer holds. Indeed, while the particles do not collide, the well-ordering assumption no longer applies when the particles are roaming in $\mathbb{R}^2$. While the particle method is still valid, a naive implementation is not scalable. For this reason, we look to the fast multipole method to reduce the cost of evaluating \cref{eq:fmmhbar}. 

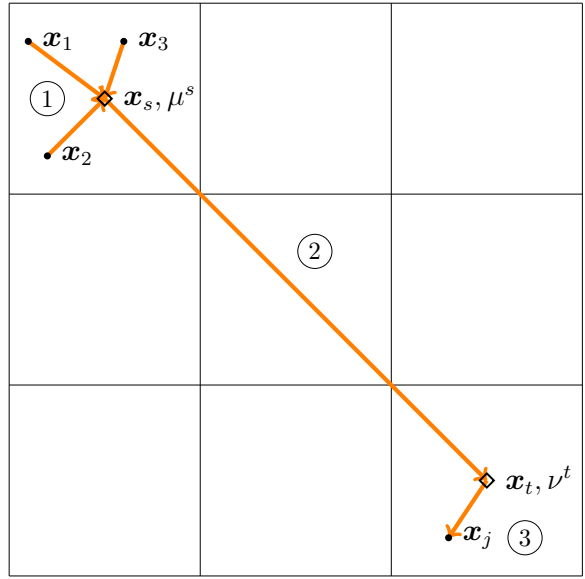
\begin{figure}[tbh]
    \centering
    \begin{tikzpicture}[scale=2.7]
        % \draw (0,3) -- (1,1);
        \foreach \i in {0,...,3} {
            \draw (0,\i) -- (3,\i);
            \draw (\i,0) -- (\i,3);
        }
        % source moments
        \draw [->,line width=0.6mm,orange] (0.1,2.8) -- (0.5,2.5);
        \draw [->,line width=0.6mm,orange] (0.2,2.2) -- (0.5,2.5);
        \draw [->,line width=0.6mm,orange] (0.6,2.8) -- (0.5,2.5);
        \draw (0.1,2.8) node[circle,fill,inner sep=1pt,label=right:$\bm{x}_1$]{};
        \draw (0.2,2.2) node[circle,fill,inner sep=1pt,label=right:$\bm{x}_2$]{};
        \draw (0.6,2.8) node[circle,fill,inner sep=1pt,label=right:$\bm{x}_3$]{};
        \draw [->,line width=0.6mm,orange] (0.5,2.5) -- (2.5,0.5);
        \draw [->,line width=0.6mm,orange] (2.5,0.5) -- (2.3,0.2);
        \draw (0.5,2.5) node[diamond,draw,thick,inner sep=1.5pt,label=right:${\bm{x}_s,\mu^s}$]{};
        \draw (2.5,0.5) node[diamond,draw,thick,inner sep=1.5pt,label=right:${\bm{x}_t,\nu^t}$]{};
        \draw (2.3,0.2) node[circle,fill,inner sep=1pt,label=right:$\bm{x}_j$]{};
        % labels
        \node at (0.2,2.5) [shape=circle,draw,inner sep=2pt] (char) {\footnotesize 1};
        \node at (1.6,1.7) [shape=circle,draw,inner sep=2pt] (char) {\footnotesize 2};
        \node at (2.7,0.2) [shape=circle,draw,inner sep=2pt] (char) {\footnotesize 3};
    \end{tikzpicture}
    \caption{An example of a 9-cell grid and schematics of the fast multipole method. Step {\large \textcircled{\small 1}} computes the source moments, step {\large \textcircled{\small 2}} converts the source moments into the target moments, and step {\large \textcircled{\small 3}} evaluates the summation at the target point using the target moments.}
    \label{fig:fmm}
\end{figure}

The fast multipole method is a class of numerical optimization techniques for approximating $N$-point summations of the form
\begin{equation} \label{eq:fmmsum}
    u(\bm{x}_j) = \sum_{i=1}^N w_i K(\bm{x}_j-\bm{x}_i), \qquad j=1,\dots,N, 
\end{equation}
for some kernel $K$. The fast multipole method is based on four steps \cite{fmm}:
\begin{enumerate}
    \setcounter{enumi}{-1}
    \item Partition the domain into smaller cells. We will denote the set of all cells as $C$, and for each cell $s\in C$, we choose a reference point (usually the centre of the cell) and denote the point as $\bm{x}_s$.
    % \item A far-field expansion that allows far-away particles to be treated as a point. Using this, we can compute a set of moments $\mu^s$ of the source cell with respect to $\bm{x}_s$.
    \item Compute the set of moments $\mu^s$ of the source cell centred at $\bm{x}_s$ using the particles in cell $s$.  
    % \item A local expansion that allows us to translate the set of moments of the source cell $\mu^s$ to a target cell $t$ as another set of moments $\nu^t$. 
    \item Translate the set of moments of the source cell $\mu^s$ into another set of moments $\nu^t$ centred at $\bm{x}_t$ of the target cell. 
    \item Evaluate the summation at targets using the pre-computed moments.
\end{enumerate}
A schematic of the fast multipole method is shown in \Cref{fig:fmm}. 
The reference points $\bm{x}_s$ and $\bm{x}_t$ are anchors for the moments to be computed. The reference points act as local ``hubs'' for an efficient transfer of information. 

We note that the fast multipole method is kernel specific, and the version developed herein is inspired by \amend{a version of the} fast Gauss transform \cite{greengard1998}.
We also distinguish the fast summation algorithm from the fast multipole method since the fast summation algorithm does not require a reference grid. Moreover, the fast summation algorithm is exact, while the fast multipole method gives only an approximation of the summation. 

\begin{figure}[bht]
    \centering
    \begin{tikzpicture} [scale=3]
        \draw [->, line width=0.3mm] (-2,0) -- (2,0) node[right]{$k$};
        % \draw [orange!20,line width=0.6mm] plot[samples=50,domain=-2:2,smooth] (\x,{sin(\x*180)*exp(-abs(\x)^2)});
        \draw [orange,line width=0.6mm] plot[samples=50,domain=-2:2,smooth] (\x,{cos(\x*180)*exp(-abs(\x)^2)});
        \pgfmathsetmacro{\dk}{0.3}
        \foreach \i in {-4,...,4} {
            \pgfmathsetmacro{\k}{\dk*\i}
            \draw ({\k-\dk/2},0) rectangle ({\k+\dk/2},{cos(\k*180)*exp(-abs(\k)^2)});
            \draw (\k,-0.03) -- (\k,0.03) node[below]{\footnotesize $k_{\i}$};
        }
        \draw [<->] ({-\dk/2},1.1) -- ({\dk/2},1.1); 
        \draw (0,1.1) node[above]{$\Delta k$};
        \draw (0.24,0.7) node[circle,fill,inner sep=1pt]{} -- (1,1) node[right]{$\frac{1}{(1+\alpha^2 k^2)^2}\mathe^{\mathi k x}$};
    \end{tikzpicture}
    \caption{Approximating the inverse Fourier transform with Riemann summation. Since the function $1/(1+\alpha^2 k^2)^2$ decays sufficiently fast, the area of strip $n$ is negligible for large $n$.}
    \label{fig:riemann}
\end{figure}
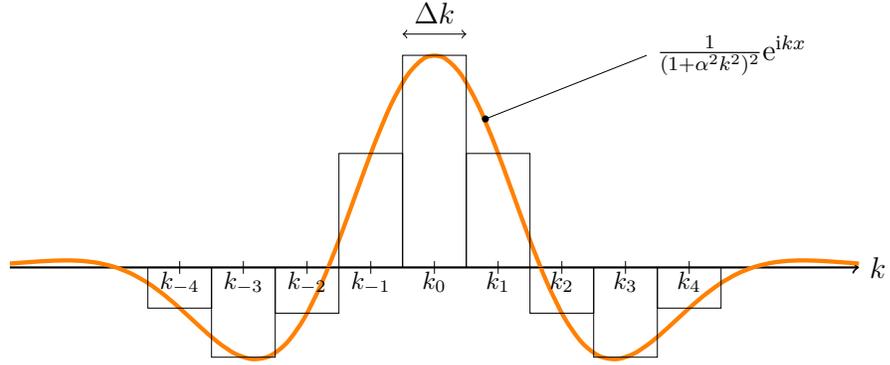

We start by considering the Fourier transform of the bi-Helmholtz kernel in one spatial dimension:
\begin{equation}
    \mathcal{F}[K_2](k) = \frac{1}{(1+\alpha^2 k^2)^2},
\end{equation}
with inverse Fourier transform given by
\begin{equation}
    K_2(x) = \frac{1}{2\pi} \int_{-\infty}^\infty \frac{1}{(1+\alpha^2 k^2)^2} \mathe^{\mathi k x} \,\mathd k.
\end{equation}
We can approximate this integral using a truncated Riemann summation with strips of width $\Delta k$ (as seen in \Cref{fig:riemann}):
\begin{equation}
    K_2(x) \approx \frac{1}{2\pi} \sum_{n=-p}^p \frac{1}{(1+\alpha^2 k_n^2)^2} \mathe^{\mathi k_n x} \,\Delta k, \qquad k_n = n\Delta k.
\end{equation}
This is called the plane wave expansion. 
% As the integrand decays sufficiently fast, we are allowed to take this truncation. A sketch of this approximation is shown in \Cref{fig:riemann}.
% To make the equations more compact, we denote $\Phi_{mn}(\bm{x})=\mathe^{\mathi(k_m x+k_n y)}$. The far-field expansion for a Gaussian in 2D is simply
% \begin{equation}
%     \mathe^{-(\bm{x}-\bm{x}_s+\bm{x}_s-\bm{x}_i)^2} = \frac{\Delta k^2}{4\pi} \sum_{mn} \mathe^{-(k_m^2+k_n^2)/4} \Phi_{mn}(\bm{x}-\bm{x}_s) \Phi_{mn}(\bm{x}_s-\bm{x}_i).
% \end{equation}

In two spatial dimensions, the 2D Fourier transform of $K_2$ is given by
\begin{equation}
    \mathcal{F}[K_2](\bm{k}) = \frac{1}{(1+\alpha^2|\bm{k}|^2)^2},
\end{equation}
and the inverse Fourier transform is 
\begin{equation}
    K_2(\bm{x}) = \frac{1}{4\pi^2} \int_{-\infty}^\infty\int_{-\infty}^\infty \frac{1}{(1+\alpha^2|\bm{k}|^2)^2} \mathe^{\mathi \bm{k}\cdot\bm{x}} \,\mathd\bm{k}.
\end{equation}
We employ the same approximation of the double integral with truncated Riemann summations 
\begin{equation}
    K_2(\bm{x}) \approx \frac{\Delta k^2}{4\pi^2} \sum_{mn} \frac{1}{(1+\alpha^2(k_m^2+k_n^2))^2} \mathe^{\mathi(k_m x+k_n y)},
\end{equation}
where $k_n=n\Delta k, k_m=m\Delta k$, $-p\leq m,n\leq p$. To make the equations more compact, we shall denote $\Phi_{mn}(\bm{x}):=\mathe^{\mathi(k_m x+k_n y)}$. What is remarkable about this expansion is the separability of $\Phi_{mn}$, where $\Phi_{mn}(\bm{x}+\bm{y})=\Phi_{mn}(\bm{x})\Phi_{mn}(\bm{y})$. This allows us to isolate the contributions of $\bm{x}_i$ with respect to a reference point $\bm{x}_s$, 
\begin{align}
    K_2(\bm{x}-\bm{x}_i) &= \frac{\Delta k^2}{4\pi^2} \sum_{mn} \frac{1}{(1+\alpha^2(k_m^2+k_n^2))^2} \Phi_{mn}(\bm{x}-\bm{x}_s+\bm{x}_s-\bm{x}_i), \\
    &= \frac{\Delta k^2}{4\pi^2} \sum_{mn} \frac{1}{(1+\alpha^2(k_m^2+k_n^2))^2} \Phi_{mn}(\bm{x}-\bm{x}_s)\Phi_{mn}(\bm{x}_s-\bm{x}_i),
\end{align}
Using this, the $N$-point summation becomes
\begin{align}
    \barh^N(\bm{x}) &= \sum_{i=1}^N w_i K_2(\bm{x}-\bm{x}_i), \\
    &= \sum_{i=1}^N w_i \frac{\Delta k^2}{4\pi^2} \sum_{mn} \frac{1}{(1+\alpha^2(k_m^2+k_n^2))^2} \Phi_{mn}(\bm{x}-\bm{x}_s)\Phi_{mn}(\bm{x}_s-\bm{x}_i), \\
    &= \sum_{mn} \underbrace{\frac{\Delta k^2}{4\pi^2} \frac{1}{(1+\alpha^2(k_m^2+k_n^2))^2} \sum_{i=1}^N w_i \Phi_{mn}(\bm{x}_s-\bm{x}_i)}_{\mu_{mn}} \Phi_{mn}(\bm{x}-\bm{x}_s). \label{eq:fmm1}
\end{align}
Thus, the moments are given by $\mu_{mn}$. However, instead of summing from $i=1$ to $N$ to compute the moments, we use the partition of the particles generated by the cells. 

This gives us a first strategy for computing the summation:
\begin{enumerate}
    \item For each cell $s\in C$, compute the set of moments of the particles within the cell:
    \begin{equation} \label{eq:fmm_mu}
        \mu_{nm}^s = \frac{\Delta k^2}{4\pi^2} \frac{1}{(1+\alpha^2(k_m^2+k_n^2))^2} \sum\left\{w_i \Phi_{mn}(\bm{x}_s-\bm{x}_i): \text{$\bm{x}_i$ in cell $s$} \right\}.
    \end{equation}
    Since $-p\leq n,m \leq p$, this step requires $O(4p^2 N)$ computations. The moments encode the contributions and distributions of all the particles within the cell. 
    \item Evaluate the summation at each target using the pre-computed moments of nearby cells: 
    \begin{equation} \label{eq:fmm_simple}
        \barh^N(\bm{x}_j) = \sum \left\{ \sum_{mn} \mu_{mn}^s \Phi_{mn}(\bm{x}_j-\bm{x}_s) : s\in C \right\}. 
    \end{equation}
    This is called the far-field expansion as the sources far from the target are grouped together and treated as a point at $\bm{x}_s$. If there are $|C|$ total cells, this step requires $O(4p^2 |C| N)$ computations. 
\end{enumerate}
Therefore, instead of evaluating particle-particle interactions, here we evaluate the particle-cell interactions. As long as $4p^2 |C|<N$, this strategy provides a speedup over direct summation. However, we see that $p$ and $|C|$ are inversely related: if we partition the domain with fewer but larger cells, then we have to compute the higher-order moments for approximation to remain accurate. The optimal choice of $p$ and $|C|$ depends on the desired accuracy and the distributions of the particles. 

Here, the conceptual leap is that the set of moments $\mu^s$ is able to encode all the information of the particles within that cell. We can apply this idea one step further by noting that the information of all the cells can also be encoded as another set of moments $\nu^t$. Thus, \Cref{eq:fmm_simple} can be improved: instead of using $\{\mu^s:s\in C\}$, we only need $\nu^t$ to evaluate the $N$-point summation in cell $t$. From \Cref{eq:fmm1}, we have
\begin{align}
    \barh^N(\bm{x}) &= \sum_{mn} \mu_{mn} \Phi_{mn}(\bm{x}-\bm{x}_t+\bm{x}_t-\bm{x}_s), \\
    &= \sum_{mn} \underbrace{\mu_{mn} \Phi_{mn}(\bm{x}_t-\bm{x}_s)}_{\nu_{mn}} \Phi_{mn}(\bm{x}-\bm{x}_t). 
\end{align}

Thus, the new strategy is then:
\begin{enumerate}
    \item Compute the set of moments $\mu^s$ for each cell as before (using \cref{eq:fmm_mu}). 
    \item For each cell $t\in C$, compute the contributions of all other cells exerting on cell $t$. 
    \begin{equation}
        \nu_{mn}^t = \sum \left\{ \mu_{mn}^s \Phi_{mn}(\bm{x}_t-\bm{x}_s) : s\in C \right\}.
    \end{equation}
    This step is called translation, and can be thought of as the cell-cell interactions. This step requires $O(4p^2|C|^2)$ computations, but Greengard and Sun~\cite{greengard1998} showed that this step can be accelerated to $O(24p^2|C|)$ using a recurrence relation. 
    \item Evaluate the summation at each target using 
    \begin{equation}
        \barh^N(\bm{x}_j) = \sum_{mn} \nu_{mn}^t  \Phi_{mn}(\bm{x}_j-\bm{x}_t),
    \end{equation}
    where $t$ is the cell containing particle $j$. This is called the local expansion as the function is expressed in terms of the local reference point $\bm{x}_t$ of particle $j$. This step requires $O(4p^2 N)$ computations. 
\end{enumerate}
Therefore, the total computational cost using the new strategy is $O(4p^2N) + O(24p^2|C|)$. A speedup over direct summation is achieved when $4p^2,6|C|<N$. In particular, we note that increasing the cell count is cheaper than increasing the order of the moments. 

Finally, to solve the Geometric Thin-Film Equation, the higher order derivatives of $\barh^N$ are also needed. This can be easily computed with
\begin{equation}
    \pder[^{a+b}]{x^a\partial y^b} \barh^N(\bm{x}) = \sum_{mn} (\mathi k_m)^a (\mathi k_n)^b \nu_{mn}^t \Phi_{mn}(\bm{x}-\bm{x}_t),
\end{equation}
for $\bm{x}$ in cell $t$. Thus, obtaining the higher-order derivatives does not require much more effort.

We also remark on a few complications when trying to integrate the fast multipole method into the particle method solver. (i) For droplet spreading, the domain that the particles occupy increases over time. If the size of the cells is fixed, this would result in more cells (increased $|C|$) and a greater number of moments needed to be computed. (ii) The Fourier transform of $K_2$ has longer tails as $\alpha$ gets smaller, and one would need to compute more terms of the Riemann summation (increasing $p$) to maintain the accuracy. This is not ideal in terms of the G-DIM framework as $\alpha$ is taken to be a small parameter. Therefore, while the fast multipole method provides a significant speedup to the particle method, the efficiency of the method depends greatly on the problem at hand.

\section{\amend{Quantitative Comparison with Existing Models}}

\begin{figure}[htb]
    \centering
    \begin{tikzpicture}
        % interface
        \draw[line width=0.7mm,orange] (0,0) arc[start angle=210, end angle=270, x radius=3cm, y radius=2cm] -- (5,-1);
        \draw[line width=0.7mm,orange] (0,0) -- (3,3) arc[start angle=210, end angle=270, x radius=3cm, y radius=2cm] -- (8,2);
        \draw[line width=0.3mm,orange] (1.5,1.5) arc[start angle=210, end angle=270, x radius=3cm, y radius=2cm] -- (6.5,0.5);
        \draw[line width=0.3mm,orange] (2.3,-1) -- (5.3,2);
        % slope
        \draw[line width=0.6mm] (-1,2) -- (1,-2) (2,5) -- (3,3);
        \draw[dashed,line width=0.6mm] (3,3) -- (4,1);
        \draw[dashed] (1,-2) -- (0,-2);
        \draw (0.5,-2) arc[start angle=180, end angle=120, x radius=0.5cm, y radius=0.5cm] (0.4,-1.6) node[]{$\varphi$};
        \draw[->] (-0.5,0.5) -- (0,-0.5) node[pos=0.7,left]{$U_0$};
        \draw[->] (5,4) -- (5,3) node[pos=0.5,right]{$g$};
        % axes
        % \draw[->,line width=0.4mm] (-1,4.5) -- (-1,3.5) node[below]{$x$};
        % \draw[->,line width=0.4mm] (-1,4.5) -- (0,4.5) node[right]{$y$};
    \end{tikzpicture}
    \caption{Schematic of the forced wetting setup. }
    \label{fig:forced_wetting_sketch}
\end{figure}
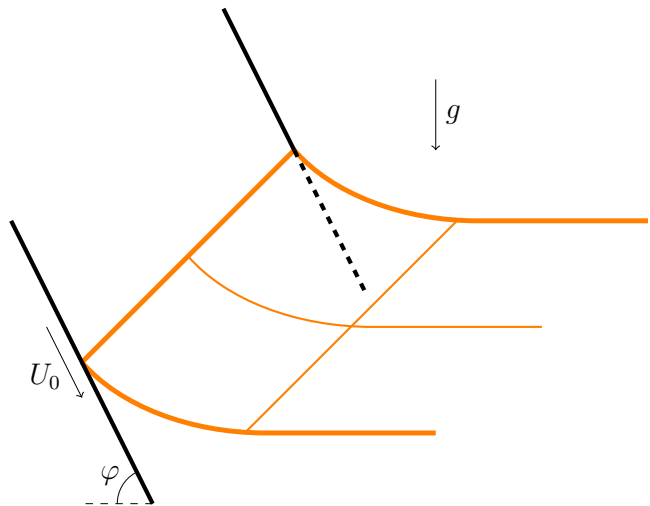

\amend{In this section, we shall discuss quantitative comparisons of the Geometric Diffuse-Interface model with the major existing regularization models in the literature. In order to discuss a meaningful equilibrium profile near the contact line, we consider a forced wetting setup where the substrate is plunged into the liquid with a speed $U$ at an angle $\varphi$. A schematic of the setup is shown in Figure \ref{fig:forced_wetting_sketch}. } 

\subsection{\amend{Derivations}}

At hydrostatic equilibrium, the interface height is given by \cite{eggers2004characteristic,de1985wetting}
\begin{equation} \label{eq:fwdef}
    \frac{3\Ca}{h^2} = \frac{p'}{\gamma}\left(3\frac{\lambda}{h}+1\right). 
\end{equation}

\paragraph{No regularization} 
If we take the slip coefficient $\lambda$ to be zero and the dynamic pressure to be $p = -\gamma h''$, Equation \eqref{eq:fwdef} becomes
\begin{equation} 
    \frac{3\Ca}{h^2} = -h''',
\end{equation}
which is valid far away from the contact line (regardless of the regularization model). We can then rescale the solution with $h(x)=\phi(\xi)$ with $\xi=\Ca^{1/3}x$ to get
\begin{equation}
    \phi''' = -\frac{3}{\phi^2}. \label{eq:fwtanner}
\end{equation} 
This equation was originally studied by Tanner \cite{tanner1979}, in which they chose to use the boundary conditions
\begin{equation} 
    \phi(0) = 1, \qquad \phi'(0) = 0, \qquad \phi''(\infty) = 0. 
\end{equation}
The boundary condition $\phi''(\infty) = 0$ corresponds to a simple wedge in the macroscopic limit. 

\paragraph{Slip-length model} 
For the slip-length model ($\lambda\neq0, p = -\gamma h''$), Equation \eqref{eq:fwdef} becomes
\begin{equation} \label{eq:fwslip}
    \frac{3\Ca}{h^2} = -h'''\left(3\frac{\lambda}{h}+1\right). 
\end{equation}
Using the scalling $h(x)=\lambda\phi(\xi)$ with $\xi=\lambda^{-1}\Ca^{1/3}x$
\begin{equation} \label{eq:fwslip2}
    \phi''' = -\frac{3}{\phi^2 + 3\phi}.
\end{equation}
We wish to impose the boundary conditions
\begin{equation}
    \phi(0) = 0, \qquad \phi'(0) = 0, \qquad \phi''(\infty) = 0. 
\end{equation}
However, $\phi=0$ poses a problem for the purpose of numerical integration. Instead, we use an asymptotic solution valid near the contact line \cite{hocking1992rival}
\begin{equation}
    \phi(\xi) \sim \sqrt{\frac{8}{3}}\xi\exp\left(\frac{\ln\xi}{2}\right), \qquad \xi\rightarrow0.
\end{equation}

\paragraph{Precursor-film model} 
For the precursor-film model ($\lambda=0$, $p = -\gamma h'' + \Phi'$), Equation \eqref{eq:fwdef} becomes
\begin{equation} \label{eq:fwprecursor}
    \frac{3\Ca}{h^2} = -h''' + 3\tilde{\mathcal{A}}\frac{h'}{h^4}, 
\end{equation}
where we have used an attractive van der Waals potential $\Phi(x)=-\tilde{\mathcal{A}}h^{-3}$ with $\mathcal{A}\varepsilon^3$. We scale the solution by $h(x)=\tilde{\mathcal{A}}^{1/2}\Ca^{-1/3}\phi(\xi)$ with $\xi=\tilde{\mathcal{A}}^{-1/2}\Ca^{2/3}x$, giving
\begin{equation} \label{eq:fwprecursor2}
    \phi''' = -\frac{3}{\phi^2} + \frac{3\phi'}{\phi^4}. 
\end{equation}
This is to be matched with a ``maximal'' film solution \cite{de1985wetting} given by
\begin{equation}
    \phi(\xi) \sim -\frac{1}{\xi} + A\exp(\frac{\xi^3}{\sqrt{3}}), \qquad \xi\rightarrow-\infty,
\end{equation}
where $A$ is a constant to be determined. We choose $A$ such that $\phi''(\infty)=0$ is satisfied. 

\paragraph{Geometric diffuse-interface model} For the geometric diffuse-interface model ($\lambda=0, p=-\gamma\bar{h}''$), Equation \eqref{eq:fwdef} becomes
\begin{equation} \label{eq:fwgdim}
    \frac{3\Ca}{h^2} = -\bar{h}'''. 
\end{equation}
We use the transformation $\bar{h}(x)=\alpha\Ca^{1/3}\phi(\xi)$ with $\xi=\alpha^{-1}x$, then
\begin{equation}
    h(x) = (1-\alpha^2\partial_{xx})^2\bar{h}(x) = \alpha\Ca^{1/3}(1-\partial_{\xi\xi})^2\phi(\xi), 
\end{equation}
giving
\begin{equation} \label{eq:fwgdim2}
    \phi''' = -\frac{3}{(\phi-2\phi''+\phi'''')^2} \approx -\frac{3}{\phi^2-4\phi\phi''+4\phi''^2}.
\end{equation}
Following the maximal-film solution, we match the solution with the kernel $K_2$ where
\begin{equation}
    \phi(\xi) \sim A\exp(-|\xi|), \qquad \xi\rightarrow-\infty. 
\end{equation}
Again, $A$ is chosen such that $\phi''(\infty)=0$. 

\begin{figure}[tbh]
    \centering
    \includegraphics[width=0.7\textwidth]{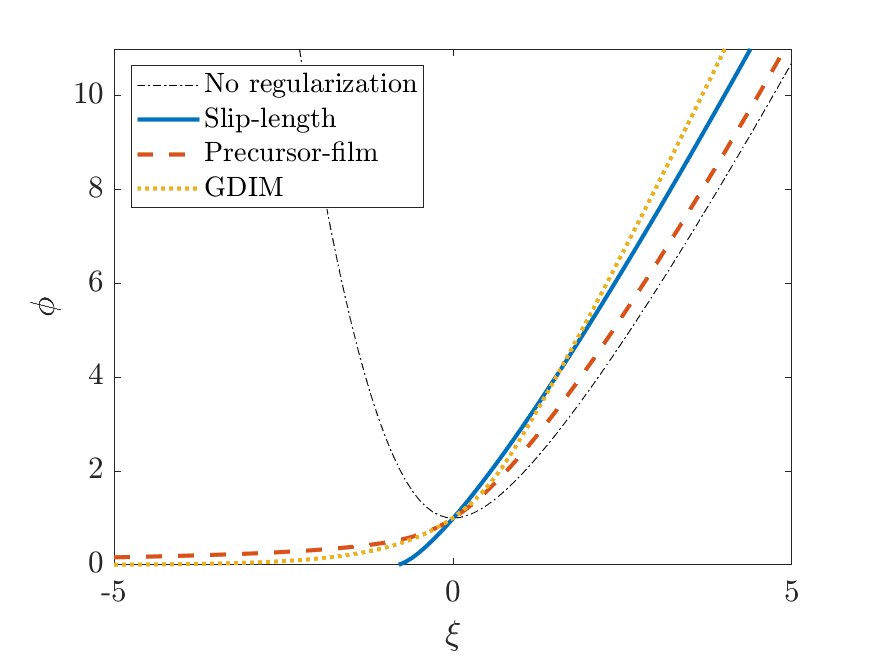}
    \caption{Solution of the forced-wetting film for each regularization model. }
    \label{fig:forced_wetting_sol}
\end{figure}

We integrate Equations (\ref{eq:fwtanner}, \ref{eq:fwslip2}, \ref{eq:fwprecursor2}, \ref{eq:fwgdim2}) numerically using a shooting method with MATLAB's \texttt{ode89} until $\xi=10^4$ where the boundary condition $\phi''(\xi)=0$ is imposed. The solutions are plotted in Figure~\ref{fig:forced_wetting_sol}. Since all the ODEs are implicit in $\xi$, we can translate the solutions by $\xi\mapsto\xi+\xi_0$. We chose $\xi_0$ to align the solutions such that $\phi(0)=1$ for a comparison. 

\subsection{\amend{Asymptotic Solution}}

The asymptotic behaviour of the solution at large $\xi$ is given by \cite{eggers2004characteristic,de1985wetting}
\begin{equation} \label{eq:fwasym}
    \phi(\xi) \sim 3^{2/3}\xi\ln^{1/3}(b\xi), \qquad \xi\rightarrow\infty, 
\end{equation}
where $b$ is a constant to be determined depending on the regularization model. The results of fitting Equation~\eqref{eq:fwasym} to the solution for each model is shown in Table~\ref{tab:fw}. 

\begin{table}[bht]
	\centering
		\caption{Fitting parameter for each model. The values for the slip-length and precursor-film models agree with \cite{hocking1992rival} and \cite{de1985wetting}, respectively. }
		\begin{tabular}{|c|c|}
		\hline
		model & $b$ \\
		\hline
		\hline
        Slip-length & 0.61\\
        Precursor-film & 0.48\\
        GDIM & 1.71\\
		\hline	
		\end{tabular}
		\label{tab:fw}
\end{table}

\begin{figure}[tbh]
    \centering
    \includegraphics[width=0.7\textwidth]{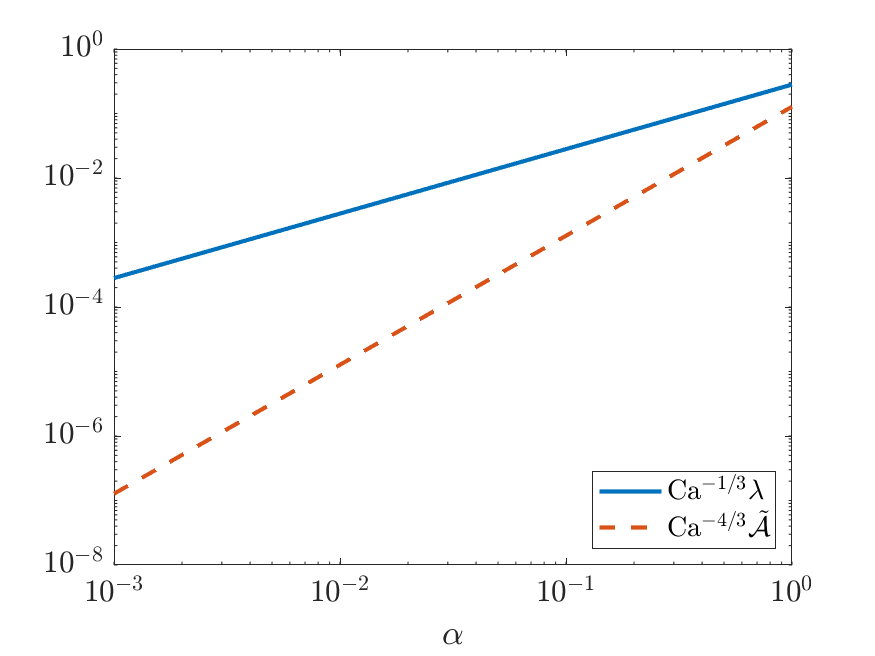}
    \caption{Mapping between regularization parameters.} 
    \label{fig:params}
\end{figure}

Since the solution is moving relative to the substrate at a constant speed $U$, the microscopic length is given by \cite{eggers2004characteristic}
\begin{equation}
    U\tilde{t} = L = \frac{x}{b\xi}.
\end{equation}
If we require that the microscopic length of each model be the same such that the models give comparable spreading rate, then one obtains relationships between the regularization parameters. In particular, we have 
\begin{equation}
    \Ca^{-1/3}\lambda = \frac{0.61}{1.71}\alpha, \qquad \Ca^{-4/3}\tilde{\mathcal{A}} = \left(\frac{0.48}{1.71}\alpha\right)^2.  
\end{equation}
The relationship is plotted in Figure~\ref{fig:params}. 

\section{\markup{Conclusion}}

In summary, we have introduced a regularized thin-film equation where the missing small-scale physics is not modelled but parametrised instead. The idea of the model is that $h(x,t)$ provides an incomplete description of the droplet evolution but nonetheless contains important physical information.  A refined description of the interface is then obtained via a smoothened interface profile $\barh=K_2*h$.  The missing physics is encoded either as a fixed level of uncertainty between $h$ and $\barh$ (model A), or such that the uncertainty in the model description occurs below a lengthscale $\alpha$ (model B).  These two models are equivalent, although model B is preferred for computational simplicity. % Overall, the model evolves to minimize the interfacial energy (area) while keeping the difference between $h$ and $\barh$ as small as possible.

% The model, as formulated, envisages that $h(x,t)$ is an incomplete description of the interface profile (with missing small-scale physics).  The missing physics is encoded either as a fixed level of uncertainty between $h$ and $\barh$ (model A), or such that the uncertainty in the model description occurs below a lengthscale $\alpha$ (model B).  These two models are equivalent, although model B is preferred for computational simplicity.
%
% The model is concerned with physics on the droplet scale, which is typically on the millimetre or micrometre scale.  The model does not explicitly include the intermolecular forces between the droplet and the substrate, which occur typically on the nanometre scale~\cite{de1985wetting}.  Therefore, the uncertainty in the model can be said to occur on the nanometre scale, with the value of $\alpha$ chosen accordingly.  However, our calculations  (e.g. Section~\ref{sec:fdm_result}) show that a much larger value of $\alpha$ can be chosen without affecting the large-scale droplet spreading.

Two numerical schemes were developed -- a finite-difference method and a particle method. The finite-difference solver uses an iterative Gauss-Newton and line search method for optimization to obtain the solution for each time step. The Gauss-Newton method converges quadratically compared to the gradient descent method and usually only requires 2 or 3 iterations to achieve a tolerance of $10^{-9}$. We have compared our solver to the semi-implicit finite difference method used in Ref~\cite{holm2020gdim} and show that our solver is unconditionally stable. That is, the step size $\Delta t$ is insensitive to the change in $\Delta x$ for the solution to converge. We have also shown that the Newton method is second-order accurate in space, first-order accurate in time with the Backward Euler scheme, and second-order accurate in time with the Crank-Nicolson scheme. Using the finite difference solver, we showed that the Geometric Thin-Film Equation is able to reproduce spreading. The spreading rate of the droplet tends towards $t^{1/7}$ in 2D and $t^{1/10}$ in 3D, in agreement with Tanner's law. One downside of the finite-difference solver is that computing and inverting the Jacobian matrix is slow, especially in higher dimensions. 

The second numerical scheme we developed is a particle method solver, which provides a fast and robust mesh-free method for solving gradient flow problems. The flow field at the contact line can be inferred from the trajectory of the particles. Furthermore, the particle method is positive preserving and mass conserving. We numerically showed that the particle method is second-order accurate in space, comparable to the finite-difference method. The convergence rate with respect to $t$ depends on the choice of the time-marching algorithm. When a no-crossing assumption is satisfied, the fast summation algorithm reduces the computational cost of each time step to $O(N)$ as supposed to the $O(N^2)$ for the finite-difference method. In higher dimensions, the fast multipole method can be used instead to speed up the particle method. These optimization techniques could reduce the time necessary for large-scale simulations of thin-film flow. In the next chapter, we classify the regularity of the particle solution and the convergence of the particle method.

\chapter{Existence and Uniqueness Theory for the GTFE}
\label{sec:theory_wellpose}

\section{\markup{Overview}}

We wish to classify the solutions of the Geometric Thin-Film Equation. A partial differential equation (PDE) given an initial condition is deemed well-posed if a unique solution exists and the solution varies smoothly in a neighbourhood of the initial condition. In this chapter, we will address the global existence and uniqueness of the solution of the Geometric Thin-Film Equation for a wide range of initial values.  % Furthermore, we show that the solution of the Geometric Thin-Film Equation exists for all time. 
The results of this chapter stem from joint works of \'O N\'araigh, Pang, and Smith (\'OPS) \cite{smith,smith2} and will be noted as such accordingly. The continuity of solutions with respect to the initial condition is the subject of ongoing work. 

The existence proof relies heavily on the particle solution inherent to the Geometric Thin-Film Equation. The particle solution of the Geometric Thin-Film Equation is restated in \Cref{sec:intro_particle}. \Cref{sec:wellposelit} summarises related work and highlights the similarities with the Camassa-Holm Equation, which also admits particle solutions. In \Cref{sec:texist}, we prove the global existence and characterize the regularity of the solution of the Geometric Thin-Film Equation. This is done in a few steps. Starting with the finite-dimensional system of first-order ODEs given by the particle solution, the global existence and uniqueness of the particle solution can be established using standard dynamical system theory. However, the particle solution is merely an approximation of the original problem. To obtain an exact solution, we will show that by taking successively better approximation of the initial condition, the particle solution converges to an exact solution of the original problem. Furthermore, the converged solution inherits the global existence property from the particle solution. The convergence of the particle solution also validates the particle method solver introduced previously as a numerical method for solving the Geometric Thin-Film Equation. 

% In \Cref{sec:intro_particle}, we briefly review the particle solution of the Geometric Thin-Film Equation. \Cref{sec:wellposelit} goes to summarize related work. In particular, the similarity with the Camassa-Holm Equation which also admits particle solutions are highlighted. \Cref{sec:texist} 

In order to establish the uniqueness of the solution of the Geometric Thin-Film Equation, we have to consider the solution as a pushforward of measure with respect to a family of curves satisfying an ODE. This, in turn, allows us to write the solution in terms of the initial measure, and the curve that satisfies the ODE can be shown to be unique. This alternate framework is discussed in \Cref{sec:pushforward}. 

\section{Background}
\label{sec:intro_particle}

For the sake of simplicity, we focus on the complete wetting model with surface tension only in one spatial dimension
\begin{equation} \label{eq:gtfe_again}
    \partial_th = -\partial_x\left(h\bar{h}^2\partial_{xxx}\bar{h}\right), \qquad \left(1-\alpha^2\partial_{xx}\right)\bar{h}=h,
\end{equation}
on the domain $(x,t)\in\mathbb{R}\times[0,\infty)$ subjected to initial condition $h(x,0)=h_0(x)$. For the rest of this chapter, we will also drop the $\surften/3\vis$ prefactor from the thin-film equation, as it does not affect the analysis. 

One of the key properties of the proposed regularization is that Equation~\eqref{eq:gtfe_again} admits weak solutions of the form
\begin{subequations} \label{eq:hdefall}
\begin{equation}
    h^N(x,t) = \sum_{i=1}^N w_i \delta(x-x_i(t)).
    \label{eq:hNdef2}
\end{equation}
Equation~\eqref{eq:hNdef2} describes a sum of $N$ point particles, each associated with a weight $w_i$ and position $x_i(t)$. For their resemblance to classical particles, the point masses are referred to as particles and Equation~\eqref{eq:hNdef2} as a particle solution. 
%
% Substituting Equation~\eqref{eq:hNdef2} into a weak form of Equation~\eqref{eq:hdefall} yields a system of $N$ first order ODEs given by
The equation of motion for each particle satisfies the following N-coupled autonomous differential equation
\begin{equation}
    \label{eq:ivp1}
    \begin{cases}
        \dot{x}_i(t) = v_i(\bm{x}(t)) = \left[\bar{h}^2\partial_{xxx}\bar{h}^N\right]_{x=x_i(t)}, \\
        x_i(0) = x_i^0,
    \end{cases} 
    \qquad i = 1,\dots,N.
\end{equation}
where we have used a compact notation for
\begin{align}
    \bar{h}^N(x_i(t),t) &= \sum_{j=1}^N w_j K(x_i(t)-x_j(t)), \\
    \partial_{xxx}\bar{h}^N(x_i(t),t) &= \sum_{\substack{j=1\\j\neq i}}^N w_j K'''(x_i(t)-x_j(t)).
\end{align}
\end{subequations}
We note that $K$ is only twice differentiable in the classical sense and that $K'''$ is not defined at the origin.  However, one of the key results in this chapter is that a solution that avoids this discontinuous point at $t=0$ stays away from this discontinuous point for all later times.  %and the third derivative is taken in the weak sense.

We are left with the freedom to choose the initial particle positions $x_i^0$ and the particle weights $w_i$, but a natural choice is to ensure the particle solution is ``close'' to the initial condition given in Equation~\eqref{eq:hdefall}. To be more precise, $(w_i,x_i^0)_{i=1}^N$ is chosen such that for all test functions $\phi(x)$, we have
\begin{equation}
    \lim_{N\rightarrow\infty} \int_{-\infty}^\infty h^N(x,0)\phi(x) \mathd x = \int_{-\infty}^\infty h_0(x)\phi(x) \mathd x.
\end{equation}
Furthermore, since $h_0(x)\geq0$ and $\|h_0\|_1<\infty$, we shall also have $w_i\geq0$ and $W:=\sum_{i=1}^N w_i<\infty$. These choices will be justified in later sections. The choice of $(x_i^0,w_i)_{i=1}^N$ is discussed later in the chapter.

% For the rest of this chapter, we will also drop the $\surften/3\vis$ prefactor in the thin-film equation, as it does not affect the analysis. We gather the relevant equations in one place. The particle solution of the Geometric Thin-Film Equation is
% \begin{align}
%     h^N(x,t) = \sum_{i=1}^N w_i \delta(x-x_i(t)), \label{eq:pweakstart} \\
%     \barh^N(x,t) = \sum_{i=1}^N w_i K(x-x_i(t)),
% \end{align}
% where $K=K_2$ is the bi-Helmholtz kernel, and $x_i$ satisfies the initial value problem
% \begin{equation}
%     \label{eq:ivp1}
%     \begin{cases}
%         \dot{x}_i(t) = v_i(\bm{x}(t)) = \left[\bar{h}^2\partial_{xxx}\bar{h}^N\right]_{x=x_i(t)}, \\
%         x_i(0) = x_i^0,
%     \end{cases} 
%     \qquad i = 1,\dots,N. 
% \end{equation}
% We are left with a freedom to choose the initial particle positions $x_i^0$ and the particle weights $w_i$ but a natural choice is to ensure the particle solution is ``close" to the initial condition. To be more precise, $(w_i,x_i^0)_{i=1}^N$ is chosen such that for all test function $\phi(x)$, we have
% \begin{equation}
%     \lim_{N\rightarrow\infty} \int_{-\infty}^\infty h^N(x,0)\phi(x) \mathd x = \int_{-\infty}^\infty h_0(x)\phi(x) \mathd x.
% \end{equation}
% Furthermore, since $h_0(x)\geq0$ and $\|h_0\|_1\leq\infty$ , we shall also have $w_i\geq0$ and $A=\sum_{i=1}^N w_i<\infty$. These choices will be important in later sections. 

\subsection{Relevant Literature}
\label{sec:wellposelit}

Particle solutions are a key feature of the Camassa--Holm (CH) equation, a partial differential equation for water waves.  In this context, the particles are referred to as `peakons' and are analogous to solitons found in other water-wave models.  Peakons are weak solutions of the CH equation; the peakon positions and momenta satisfy a set of ODEs with a Hamiltonian structure.  As such, the authors of Reference~\cite{chertock2012a} have shown the convergence of particle solutions to a weak solution of the CH equation. The main theoretical tool used here is Helly's selection theorem.  \'OPS~\cite{smith} have generalized this approach using a metric Arzel\'a-Ascoli compactness theorem. 
Other continuum models also admit particle solutions.  For instance, in \cite{laurent2007}, the author analysed the interval of existence of an aggregation equation and provided an ``acceptability'' condition for the kernel function for which the solution exhibits finite time blow-up behaviour.   Our approach to the present problem is similar to these existing works. %We will show that our results agree with the condition.

The Geometric Thin-Film equation~\eqref{eq:hdefall} has already been characterized in the previous chapter.  In particular, we used the particle solution as a basis for the particle method to numerically solve the Geometric Thin-Film equation. We have demonstrated using numerical means that the particle method converges. Using standard dynamical system theory and mathematical analysis, the uniqueness and convergence of the particle method can be established. Furthermore, we introduced a fast summation algorithm to reduce the numerical cost of evaluating the system of ODEs in Equation~\eqref{eq:ivp1} from $O(N^2)$ to $O(N)$. The algorithm relies on the assumption that the relative ordering of the particles is preserved throughout the interval of the existence of the solution. In other words, crossings (or collisions) are not permitted between the particles.  Indeed, one of the main aims of this chapter is to prove the no-crossing theorem rigorously.

We emphasize that other models which admit particle solutions do possess a no-crossing theorem (e.g. the CH equation~\cite{camassa2006, chertock2012b}).  In the case of the CH equation, the no-crossing theorem is proved using arguments about the conservation of total momentum \cite{camassa2006}. Their argument relies on the underlying Hamiltonian structure of the CH equation.  However, there is no such Hamiltonian structure in the present case and a new approach to proving the no-crossing theorem must be developed. We instead introduce a new geometrical argument on the manifold in which the particle collisions occur to prove a no-crossing theorem for the particle solution of the Geometric Thin-Film Equation.

\section{Existence} \label{sec:texist}

\subsection{Particle Solutions of the Geometrical Thin-Film Equation} \label{sec:weak}

We start by defining the weak solution of the Geometric Thin-Film Equation and show that the particle solution given in Equation~\eqref{eq:hdefall} is indeed a weak solution of the Geometric Thin-Film Equation. 

\begin{definition}
Let $\bar{h}\in C(\mathbb{R}^+;W^{3,1}_{loc}(\mathbb{R}))$, $h=(1-\alpha^2\partial_{xx})^2\bar{h}$. We say that the pair $(h,\bar{h})$ is a weak solution of the Geometric Thin-Film Equation if it satisfies
\begin{multline}
    \int_{-\infty}^\infty (h\phi)_{t=0}\,\mathd x + \int_0^\infty \int_{-\infty}^\infty \bar{h} (1-\alpha^2\partial_{xx})^2\phi_t\,\mathd x\,\mathd t + \int_0^\infty \int_{-\infty}^\infty \bar{h}^3 \partial_{xxx}\bar{h} \phi_x\,\mathd x\,\mathd t \\
    -2\alpha^2 \int_0^\infty \int_{-\infty}^\infty \bar{h}^2 \partial_{xx}\bar{h} \partial_{xxx}\bar{h} \phi_x\,\mathd x\,\mathd t - \alpha^4 \int_0^\infty \int_{-\infty}^\infty \bar{h} \partial_{x}\bar{h} (\partial_{xxx}\bar{h})^2 \phi_x\,\mathd x\,\mathd t \\
    -\frac{1}{2}\alpha^4 \int_0^\infty \int_{-\infty}^\infty (\bar{h} \partial_{xxx}\bar{h})^2 \phi_{xx}\,\mathd x\,\mathd t = 0,
    \label{eq:weakdef}
\end{multline}
for all test functions  $\phi\in C_c^\infty (\mathbb{R}\times \mathbb{R}^+)$.
\end{definition}

\begin{theorem} \label{thm:pweaksol}
The pair of functions $(h^N,\bar{h}^N)$ defined in Equation~\eqref{eq:hdefall} is a weak solution of the Geometric Thin-Film Equation for all $N\in\mathbb{N}$.
\end{theorem}

\begin{proof}
We compute the following terms (omitting the superscript $N$): The first term in Equation~\eqref{eq:weakdef} gives
\begin{align}
    T_1 = \sum_{i=1}^N w_i \int_{-\infty}^\infty \delta(x-x_i^0) \phi(x,0) \,\mathd x\,\mathd t = \sum_{i=1}^N w_i \phi(x_i^0, 0). 
\end{align}
Since $\langle \bar{h}, (1+\alpha^2\partial_{xx})^2\phi_t \rangle = \langle h, \phi_t \rangle$ the second term becomes
\begin{align}
    T_2 = \sum_{i=1}^N w_i \int_0^\infty\int_{-\infty}^\infty \delta(x-x_i(t))\phi_t(x,t) \,\mathd x\,\mathd t = \sum_{i=1}^N w_i \int_0^\infty \phi_t(x_i(t),t) \,\mathd t.
\end{align}
Using the identity
\begin{equation}
    \frac{\mathd}{\mathd t}\phi(x_i(t), t) = \dot{x}_i(t)\phi_x(x_i(t),t) + \phi_t(x_i(t),t),
\end{equation}
$T_2$ then becomes
\begin{align}
    T_2 = -\sum_{i=1}^N w_i \phi(x_i^0, 0) - \sum_{i=1}^N w_i \int_0^\infty \dot{x}_i(t) \phi_x(x_i(t),t) \,\mathd t.
\end{align}
Now consider the inner product % $\langle\cdot,\cdot\rangle := \langle\cdot,\cdot\rangle_{L^2(\mathbb{R})}$
\begin{align}
    \langle h, \bar{h}^2 \partial_{xxx}\bar{h} \phi_{x} \rangle &= \langle \bar{h}, (1-\alpha^2\partial_{xx})^2 (\bar{h}^2 \partial_{xxx}\bar{h} \phi_{x}) \rangle, \\
    &= \langle \bar{h}, \bar{h}^2 \partial_{xxx}\bar{h} \phi_{x} \rangle - 2\alpha^2 \langle \bar{h}, \partial_{xx}(\bar{h}^2 \partial_{xxx}\bar{h} \phi_{x}) \rangle \nonumber \\
    &\qquad + \alpha^4 \langle \bar{h}, \partial_{xxxx}(\bar{h}^2 \partial_{xxx}\bar{h} \phi_{x}) \rangle, \\
    &= \langle \bar{h}, \bar{h}^2 \partial_{xxx}\bar{h} \phi_{x} \rangle - 2\alpha^2 \langle \partial_{xx}\bar{h}, \bar{h}^2 \partial_{xxx}\bar{h} \phi_{x} \rangle \nonumber \\
    &\qquad - \alpha^4 \underbrace{\langle \partial_{xxx}\bar{h}, \partial_{x}(\bar{h}^2 \partial_{xxx}\bar{h} \phi_{x}) \rangle}_{I}. \label{eq:weakI1}
\end{align}
Expand $I$ we get
\begin{align}
    I &= 2\langle \partial_{xxx}\bar{h}, \bar{h}\partial_{x}\bar{h} \partial_{xxx}\bar{h} \phi_{x} \rangle + \overbrace{\langle \partial_{xxx}\bar{h}, \bar{h}^2 \partial_{xxxx}\bar{h} \phi_{x} \rangle}^{-I} \nonumber \\
    &\qquad + \langle \partial_{xxx}\bar{h}, \bar{h}^2 \partial_{xxx}\bar{h} \phi_{xx} \rangle, \\
    2I &= 2\langle \partial_{xxx}\bar{h}, \bar{h}\partial_{x}\bar{h} \partial_{xxx}\bar{h} \phi_{x} \rangle + \langle \partial_{xxx}\bar{h}, \bar{h}^2 \partial_{xxx}\bar{h} \phi_{xx} \rangle.
\end{align}
So Expression~\eqref{eq:weakI1} becomes
\begin{align}
    \langle h, \bar{h}^2 \partial_{xxx}\bar{h} \phi_{x} \rangle &= \langle \bar{h}, \bar{h}^2 \partial_{xxx}\bar{h} \phi_{x} \rangle - 2\alpha^2 \langle \partial_{xx}\bar{h}, \bar{h}^2 \partial_{xxx}\bar{h} \phi_{x} \rangle \nonumber \\
    &\qquad - \alpha^4 \langle \partial_{xxx}\bar{h}, \bar{h}\partial_{x}\bar{h} \partial_{xxx}\bar{h} \phi_{x} \rangle - \frac{1}{2}\alpha^4 \langle \partial_{xxx}\bar{h}, \bar{h}^2 \partial_{xxx}\bar{h} \phi_{xx} \rangle .
\end{align}
Thus the rest of the terms in Equation~\eqref{eq:weakdef} is precisely 
\begin{align}
    T_3 &= \int_0^\infty\int_{-\infty}^\infty h\bar{h}^2\partial_{xxx}\bar{h}\phi_x \,\mathd x\,\mathd t, \\
    &= \sum_{i=1}^N w_i \int_0^\infty\int_{-\infty}^\infty \delta(x-x_i(t))\bar{h}^2\partial_{xxx}\bar{h}\phi_x \,\mathd x\,\mathd t, \\
    &= \sum_{i=1}^N w_i \int_0^\infty \left(\bar{h}^2\partial_{xxx}\bar{h}\phi_x\right)_{x=x_i(t)} \,\mathd t.
\end{align}
Putting these terms together gives
\begin{align}
    T_1 + T_2 + T_3 &= -\sum_{i=1}^N w_i \int_0^\infty  \phi_x(x_i(t),t)\left[\dot{x}_i(t) - \left(\bar{h}^2\partial_{xxx}\bar{h}\right)_{x=x_i(t)}\right] \,\mathd t.
\end{align}
Which is identically zero by Equation~\eqref{eq:ivp1}. So $(h,\bar{h})$ is a weak solution of Equation~\eqref{eq:gtfe_again}.
\end{proof}

In terms of the particle solution, the regularized free energy is given by
\begin{align}
    \lagrange[h^N] &= \tfrac{1}{2}\surften\intinf \partial_xh^N\,\partial_x\barh^N\,\mathd x, \\
    &= -\tfrac{1}{2}\surften\sum_{i=1}^N\sum_{j=1}^N w_iw_j \intinf \delta(x-x_i)K''(x-x_j)\,\mathd x, \\
    &= -\tfrac{1}{2}\surften\sum_{i=1}^N\sum_{j=1}^N w_iw_j K''(x_i-x_j).
\end{align}
Furthermore, the energy of the particle solution is strictly decreasing over time
\begin{lemma}
    Let $h^N$ be the particle solution satisfying Equation~\eqref{eq:hdefall} with $w_i\geq0$ for all $i=1,\dots,N$, and $w_i>0$ for some $i$, then
    \begin{equation}
        \der{t}\ell[h^N] < 0.
    \end{equation}
\end{lemma}
\begin{proof}
We compute the quantity
\begin{align}
    \der{t}\ell[h^N] &= -\tfrac{1}{2}\surften\sum_{i=1}^N\sum_{j=1}^N w_iw_j K'''(x_i-x_j)(\dot{x}_i-\dot{x}_j), \\
    &= -\tfrac{1}{2}\surften\sum_{i=1}^N\sum_{j=1}^N w_iw_jK'''(x_i-x_j)\dot{x}_i \nonumber \\ 
    &\qquad +\tfrac{1}{2}\surften\sum_{i=1}^N\sum_{j=1}^N w_iw_jK'''(x_i-x_j)\dot{x}_j. \label{eq:particle_energy}
\end{align}
Swapping the indicies of the second summand ($i\leftrightarrow j$) gives
\begin{equation}
    \der{t}\ell[h^N] = -\tfrac{1}{2}\surften\sum_{i=1}^N\sum_{j=1}^N w_iw_j\dot{x}_i\bigg(K'''(x_i-x_j) - K'''(x_j-x_i)\bigg).
\end{equation}
Since $K'''$ is odd, we get
\begin{align}
    \der{t}\ell[h^N] &= -\surften\sum_{i=1}^N\sum_{j=1}^N w_iw_j\dot{x}_i K'''(x_i-x_j), \\
    &= -\surften\sum_{i=1}^N w_i\dot{x}_i \sum_{j=1}^N w_j K'''(x_i-x_j), \\
    &= -\surften\sum_{i=1}^N w_i\dot{x}_i \partial_{xxx}\barh^N(x_i).
\end{align}
Finally, substituting Equation~\eqref{eq:ivp1} gives
\begin{equation}
    \der{t}\ell[h^N] = -\surften\sum_{i=1}^N w_i[\barh^N(x_i)\partial_{xxx}\barh^N(x_i)]^2.
\end{equation}
Since $w_i\geq0$ for all $i=1,\dots,N$, and not all zero, we conclude that the energy is strictly decreasing
\begin{equation}
    \der{t}\ell[h^N] < 0. 
\end{equation}
So the particle solution preserves the gradient flow structure and maintains the monotonic decreasing of the free energy. 
\end{proof}

\subsection{Global Existence of Particle Solutions and a No Crossing Theorem}\label{sec:collision}

In this section, we show that the system of ODEs~\eqref{eq:ivp1} has a unique, globally defined solution -- for suitable initial conditions.  The strategy of the proof is to show that particles that are initially separated stay separated (do not cross).  This ensures that the trajectories avoid regions where the function on the right-hand side of the ODE system is non-Lipschitz.  Hence, a unique, globally-defined solution is guaranteed.  The key insight which enables us to complete the proof is to demonstrate that the set  $D=\{\bm{x}\in\mathbb{R}^N: x_i<x_j, i<j\}$ is a trapping region for Equation~\eqref{eq:ivp1}

We start with the following definitions:
\begin{definition} [Crossing]
Let $U\subseteq \mathbb{R}^N$, and let $\bm{v}:U\rightarrow \mathbb{R}^N$. Consider the initial value problem
\begin{equation}
    \label{eq:auto}
    \begin{cases}
        \dot{\bm{x}}(t) = \bm{v}(\bm{x}(t)) \qquad t\geq0,\\
        \bm{x}(0) = \bm{x}_0.
    \end{cases}
\end{equation}
We say that Equation~\eqref{eq:auto} has no-crossing in finite time if the solution exists for all $t\geq0$ and 
\begin{equation}
    x_i(t)\neq x_j(t) \qquad \text{for all $i\neq j$ and $t\geq0$}.
\end{equation}
\end{definition}
\noindent Let $\Delta_{ij}=\{\bm{x}\in\mathbb{R}^N: x_i=x_j\}$ and define $\Delta=\bigcup_{i\neq j}\Delta_{ij}$. Thus, $\Delta$ is the set where crossings occur between two or more variables/particles. We note that the solution of an autonomous system has no crossing if and only if $\bm{x}(t)\notin\Delta$ for all $t\geq0$.
\begin{definition} [Trapping region]
Let $U\subseteq\mathbb{R}^N$ be a domain with $\partial U$ piecewise smooth, and let $\bm{n}(\bm{x})$ be the inward-pointing normal to $\partial U$ defined at regular points of $\partial U$. Let $\bm{v}: U\rightarrow\mathbb{R}^N$ be bounded and Lipschitz. We say that $U$ is a trapping region for $\dot{\bm{x}}=\bm{v}(\bm{x})$ if for all $\bm{x}$ regular points of $\partial U$, 
\begin{equation}
    \label{eq:trapping_condition}
    \bm{n}(\bm{x})\cdot\bm{v}(\bm{x}) > 0.
\end{equation}
Note that here we define $\bm{v}(\bm{x})$ for $\bm{x}\in\partial U$ by
\begin{equation}
    \bm{v}(\bm{x}) := \lim_{n\rightarrow\infty}\bm{v}(\bm{x}_n), 
\end{equation}
for any sequence $(\bm{x}_n)\subseteq U$ with $\bm{x}_n\rightarrow\bm{x}$. This limit exists because of the Lipschitz assumption of $\bm{v}$ on U. 
\end{definition}
\noindent We will also use the following theorem from which we take from Reference~\cite{schaeffer2016}:
\begin{theorem}[Sufficient condition for global existence \cite{schaeffer2016}]
\label{thm:schaeffer}
Let $U\in\mathbb{R}^N$ be a domain with a piecewise smooth boundary. Let $\bm{v}:U\rightarrow\mathbb{R}^N$ be bounded and Lipschitz on U. If $U$ is a trapping region for $\dot{\bm{x}}=\bm{v}(\bm{x})$ and $\bm{x}(0)\in U$ then the solution exists for all $t\geq0$ and the solution is contained in $U$. 
\end{theorem}

\begin{corollary}[No-crossing theorem]
\label{thm:cross}
Consider again the assumption of Theorem \ref{thm:schaeffer}. If $U\cap\Delta=\emptyset$, then for any initial data $\bm{x}(0)\in U$, the system has no crossing in finite time. 
\end{corollary}
\begin{proof}
% By Theorem \ref{thm:schaeffer}, for all initial data $\bm{x}(0)\in U$, the solution exist for all $t\geq0$ and $\bm{x}(t)\in U$. Since $U\cap\Delta=\emptyset$, we have that $\bm{x}(t)\notin\Delta$ for all $t\geq0$. Thus the system has no crossing for all $t\geq0$.  
The proof follows from the fact that (the image of) $\bm{x}(t)$ is contained in $U$.
\end{proof}

\begin{lemma}
Let $D=\{\bm{x}\in\mathbb{R}^N: x_i<x_j, i<j\}$, and let $g:(0,\infty)\rightarrow\mathbb{R}$ and $f:(-\infty,0)\rightarrow\mathbb{R}$ be Lipschitz continuous. Let $\bm{v}:D\rightarrow\mathbb{R}^N$ be defined by
\begin{equation}
    v_i(\bm{x}) = \sum_{j=1}^{i-1} w_j f(x_i-x_j) + \sum_{j=i+1}^{N} w_j g(x_i-x_j).
\end{equation}
Then $\bm{v}$ is Lipschitz on $D$.
\end{lemma}
\begin{proof}
Let $\bm{x},\bm{y}\in D$, 
\begin{align}
    \|\bm{v}(\bm{x}) - \bm{v}(\bm{y})\|_1 &= \sum_{i=1}^N |v_i(\bm{x}) - v_i(\bm{y})|, \\
    &\leq \sum_{i=1}^N \sum_{j=1}^{i-1} w_j|f(x_i-x_j)-f(y_i-y_j)| \nonumber \\
    &\qquad + \sum_{i=1}^N \sum_{j=i+1}^{N} w_j|g(x_i-x_j)-g(y_i-y_j)|.
\end{align}
Since $g$ and $f$ are Lipschitz on their respective domains, there exists $L>0$ such that
\begin{align}
    \|\bm{v}(\bm{x}) - \bm{v}(\bm{y})\|_1 &\leq \sum_{i=1}^N \sum_{\substack{j=1\\j\neq i}}^N w_jL|x_i-x_j-y_i+y_j|, \\ 
    &\leq \sum_{i=1}^N \sum_{\substack{j=1\\j\neq i}}^N w_jL(|x_i-y_i|+|x_j-y_j|), \\ 
    &\leq \sum_{i=1}^N \sum_{\substack{j=1\\j\neq i}}^N w_j2L\|\bm{x}-\bm{y}\|_1, \\
    &\leq 2LNW\|\bm{x}-\bm{y}\|_1.
\end{align}
\end{proof}

\begin{example}\label{ex:lip}
The velocity field defined by
\begin{equation}
    v_i(\bm{x}) = \partial_{xxx}\bar{h}^N|_{x=x_i} = \sum_{\substack{j=1\\ j\neq i}}^N w_j K'''(x_i-x_j).
\end{equation}
is Lipschitz on $D$ because $K'''$ can be written as 
\begin{equation}
    K'''(x) = f(x)\chi_{(-\infty,0)}(x) + g(x)\chi_{(0,\infty)}(x),
\end{equation}
for some Lipschitz function $g$ and $f$. It follows that since $\bar{h}^N|_{x=x_i}$ and $\partial_{xxx}\bar{h}^N|_{x=x_i}$ is bounded and Lipschitz on $D$, the velocity field defined in Equation~\eqref{eq:ivp1} are bounded and Lipschitz on $D$.
\end{example}

% \begin{lemma}
% Let $\Omega\subseteq\mathbb{R}^N$, and let $u,v:\Omega\rightarrow\mathbb{R}$ be bounded and Lipschitz continuous. Then the map $x\mapsto u(x)v(x)$ is bounded and Lipschitz continuous on $\Omega$.
% \end{lemma}
% %
% \begin{proof}
% Let $|u(x)|,|v(x)|\leq M$ for all $x\in\Omega$ and $L=\max\{\lip(u),\lip(v)\}$. Let $x,y\in\Omega$
% \begin{align}
%     |u(x)v(x) - u(y)v(y)| &\leq |u(x)v(x) - u(x)v(y)| + |u(x)v(y) - u(y)v(y)| \\
%     &= |u(x)||v(x)-v(y)| + |v(x)||u(x)-u(y)| \\
%     &\leq 2ML|x-y|
% \end{align}
% \end{proof}

\begin{figure}
    \centering
    \begin{tikzpicture}
        \draw[line width=0.5mm] (4,-4) -- (-1,1) node[anchor=east] {$\Delta_{i'j'}$};
        \draw[line width=0.5mm] (-4,-4) -- (1,1) node[anchor=west] {$\Delta_{ij}$};
        \node at (-3.2,-3.8) {$D$};
        \foreach \x in {0,0.5,1,1.5,2,2.5,3}{
            \foreach \y in {0,0.5,1,1.5,2,2.5,3}{
                \pgfmathtruncatemacro{\z}{\x+\y} 
                \pgfmathtruncatemacro{\fac}{(\x-\y)}
                \pgfmathtruncatemacro{\nl}{\z*(\z-2.7)}
                \ifnum\z<4
                    \draw[->, black!40] (\x-\y,-\x-\y) -- (\x-\y+0.1*\nl*\fac,-\x-\y-0.5);
                \fi
            }
        }
        \draw[line width=0.5mm, ->] (2.5,-2.5) -- (2.5-0.7,-2.5-0.7) node[anchor=north] {$\bm{n}_{i'j'}$};
        \draw[line width=0.5mm, ->] (-2.5,-2.5) -- (-2.5+0.7,-2.5-0.7) node[anchor=north] {$\bm{n}_{ij}$};
    \end{tikzpicture}
    \caption{Sketch of the trapping region and the velocity field near the non-smooth manifold. We show that the velocity field is smooth in $D$ and points inwards on $\partial D$.}
    \label{fig:traping_region}
\end{figure}
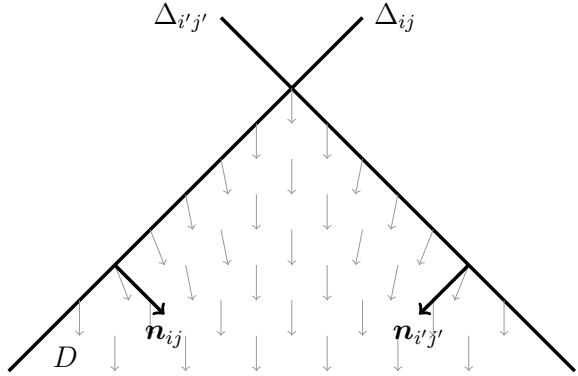

\begin{theorem}[Global existence of particle solution]
Consider the initial value problem defined in Equation~\eqref{eq:ivp1}. If the initial condition satisfies
\begin{align}
    x_i^0 < x_j^0 \qquad \text{for all $i<j$},
\end{align}
then there exists a unique global solution to the initial value problem. Furthermore, the system has no crossing in finite time.
\end{theorem}
\begin{proof}
We show that $D=\{\bm{x}\in\mathbb{R}^N: x_i<x_j, i<j\}$ is a trapping region for the system. A sketch of the proof is given in \Cref{fig:traping_region}.
% First, notice that $\bm{v}$ defined in Equation~\eqref{eq:ivp1} is bounded and Lipschitz on $D$. 
From Example~\ref{ex:lip}, $\bm{v}$ defined in Equation~\eqref{eq:ivp1} is bounded and Lipschitz on $D$. 
Let $\Gamma$ be the set of regular points of $\partial D$ (i.e. where the normal vector is well defined) and let $\bm{n}_{ij} \in\mathbb{R}^N,i\neq j$ with $-1$ for the $i$-th component, $+1$ for the $j$-th component, and $0$ everywhere else. Note that for all $\bm{x}\in\Gamma$, $\bm{x}\in\Delta_{i,i+1}$ for some unique $i=1,\dots,N-1$. So the inward pointing normal vector at $\bm{x}\in\Gamma\cap\Delta_{i,i+1}$ is given by $\bm{n}_{i,i+1}$ and Equation~\eqref{eq:trapping_condition} corresponds to 
\begin{equation}
    \label{eq:trapping_condition2}
    -v_i(\bm{x}) + v_{i+1}(\bm{x}) > 0 \qquad \forall\bm{x}\in\Gamma\cap\Delta_{i,i+1}.
\end{equation}
for all $i=1,\dots,N-1$. To shorten the equation, fix $\bm{x}\in D$ and define $f:\mathbb{R}\rightarrow\mathbb{R}$ by 
\begin{equation}
    f(x) = \bar{h}^N(x)^2 = \sum_{i,j=1}^N w_iw_jK(x-x_i)K(x-x_j), 
\end{equation}
then $f$ is continuous and positive on $\mathbb{R}$. We compute the quantity
\begin{align}
    &-v_i(\bm{x}) + v_{i+1}(\bm{x}) \nonumber \\
    % &\qquad= \sum_{j=1}^N w_j [f(x_{i+1})K'''(x_{i+1}-x_j)-f(x_i) K'''(x_i-x_j)], \\
    &\qquad= \sum_{\substack{j=1\\j\neq i+1}}^N w_jf(x_{i+1})K'''(x_{i+1}-x_j) - \sum_{\substack{j=1\\j\neq i}}^N w_jf(x_i)K'''(x_i-x_j), \\
    &\qquad= \sum_{j=1}^{i-1} w_j [f(x_{i+1})K'''(x_{i+1}-x_j)-f(x_i) K'''(x_i-x_j)] \nonumber \\
    &\qquad\qquad +w_if(x_{i+1})K'''(x_{i+1}-x_i) - w_{i+1}f(x_i)K'''(x_i-x_{i+1}) \nonumber \\
    &\qquad\qquad\qquad +\sum_{j=i+2}^N w_j [f(x_{i+1})K'''(x_{i+1}-x_j)-f(x_i) K'''(x_i-x_j)].
\end{align}
To evaluate at $\bm{x}\in\partial D\cap\Delta_{i,i+1}$, we take $d(\bm{x},\Delta_{i,i+1})\rightarrow0$, equivalently $x_{i+1}-x_i\downarrow0$. Because $f$ is continuous on $\mathbb{R}$ and $K'''$ is continuous when restricted on $(-\infty,0)$ and $(0,\infty)$ separately, we have
\begin{equation}
    \lim_{x_{i+1}\rightarrow x_i} [f(x_i) K'''(x_i-x_j)-f(x_{i+1})K'''(x_{i+1}-x_j)] = 0,
\end{equation}
for $j=1,\dots,i-1,i+2,\dots,N$. So
\begin{align}
    \lim_{d(\bm{x},\Delta_{i,i+1})\rightarrow0} [-v_i(\bm{x}) + v_{i+1}(\bm{x})] = w_if(x_i)K'''(0_+) - w_{i+1}f(x_i)K'''(0_-).
\end{align}
Given that $K'''(0_+)>0$ and $K'''(0_-)<0$, we conclude
\begin{equation}
    -v_i(\bm{x}) + v_{i+1}(\bm{x}) > 0 \qquad \forall\bm{x}\in\Gamma\cap\Delta_{i,i+1}.
\end{equation}
Therefore $D$ is a trapping region for Equation~\eqref{eq:ivp1}. By Theorem \ref{thm:schaeffer} and Corollary \ref{thm:cross}, the solution of Equation~\eqref{eq:ivp1} exists globally for all initial data $\bm{x}_0\in D$ and has no crossing in finite time. 
\end{proof}

For velocity fields that are Lipschitz continuous on the whole of $\mathbb{R}^N$, a sufficient test for the no-crossing property can be established. The theorem covers a large family of autonomous ODE 
systems, including ones discussed in \cite{carrillo2019}. % Hence, the proof develop within could  % Thus, one would be able to apply a fast summation algorithm to greatly improve the performance of the particle method. 

We denote the orbit of the solution $\bm{x}\in C^1(I;\mathbb{R}^N)$ of Equation~\eqref{eq:auto} passes through a point $\bm{x}_0$ by 
\begin{equation}
    \gamma(\bm{x}_0) = \{\bm{x}(t)\in\mathbb{R}^N: t\in I, \bm{x}(0)=\bm{x}_0\}.
\end{equation}

\begin{lemma} [\cite{strogatz1994}] \label{thm:intersect}
Suppose that for all $\bm{x}_0\in U$, there exist a unique $\bm{x}\in C^1(I;\mathbb{R}^n)$ satisfying Equation~\eqref{eq:auto}. Then either two orbits do not have any common points or are the same set.  
\end{lemma}

Lemma \ref{thm:intersect} is a standard result in dynamical system theory to ensure the uniqueness of solution for systems with a smooth velocity field. 

\begin{lemma}\label{thm:nocoll1}
Suppose that for all $\bm{x}_0\in U$, there exist a unique $\bm{x}\in C^1(I;\mathbb{R}^N)$ satisfying Equation~\eqref{eq:auto}. Let $\gamma(\bm{x}_0)$ be the orbit of Equation~\eqref{eq:auto} that passes through the point $\bm{x}_0$. Then the following statements are equivalent:
\begin{enumerate}[label=(\roman*)]
    \item Equation~\eqref{eq:auto} has no collision in $t\in I$.
    \item For all $\bm{x}_0\in\Delta$, $\gamma(\bm{x}_0)\subset\Delta$.
\end{enumerate}
\end{lemma}
\begin{proof}
Suppose (i) is true. Let $\bm{x}_0\in\Delta$ and let $\bm{y}_0\in U\setminus\Delta$. By assumption, $\bm{x}_0\notin\gamma(\bm{y}_0)$, so by Lemma \ref{thm:intersect}, we have that $\bm{y}_0\notin\gamma(\bm{x}_0)$ for all $\bm{y}_0\in U\setminus\Delta$. Thus $\gamma(\bm{x}_0)\subset \Delta$.

Conversely, suppose (ii) is true. Let $\bm{x}_0\in U\setminus\Delta$ and let $\bm{y}_0\in\Delta$. By assumption, $\bm{x}_0 \notin \gamma(\bm{y}_0)$, so by Lemma \ref{thm:intersect}, we have that $\bm{y}_0 \notin \gamma(\bm{x}_0)$ for all $\bm{y}_0\in\Delta$. Thus $\bm{x}(t)\in\Delta^c$ for all $t\in I$.
\end{proof}

\begin{theorem} [A sufficient condition for no collision] \label{thm:nocoll2}
Let $\bm{v}:\mathbb{R}^N\rightarrow\mathbb{R}^N$ be Lipschitz continuous, and suppose
\begin{align}
    x_i = x_j \implies v_i(\bm{x}) = v_j(\bm{x}), \qquad \forall i,j
\end{align}
Then the initial value problem \eqref{eq:auto} has no collision for all $t\in\mathbb{R}$. 
\end{theorem}
\begin{proof}
By the Picard-Lindel\"of theorem, for all $\bm{x}_0\in\mathbb{R}^n$, there exist a unique solution $\bm{x} \in C^1(\mathbb{R};\mathbb{R}^n)$ satisfying Equation~\eqref{eq:auto}. 
% Let $\gamma(\bm{x}_0)$ denote the orbit of Equation~\eqref{eq:auto2}. 
Suppose $x_i^0=x_j^0$ for some $i\neq j$. Then $\bm{x}_0\in\Delta$ and by assumption, $x_i$ and $x_j$ satisfy the same differential equation, so we have
\begin{align}
    \dot{x}_i - \dot{x}_j = v_i(\bm{x}) - v_j(\bm{x}) = 0.
\end{align}
Thus $x_i(t)=x_j(t)$ for all $t\in\mathbb{R}$ and $\gamma(\bm{x}_0)\subset\Delta$. By Lemma~\ref{thm:nocoll1}, we conclude that the system has no collision for all time.
\end{proof}

\begin{remark}
The set $\Delta$ partitions $\mathbb{R}^N$ into $N!$ regions, and any solution starting in one region is not allowed to move to another region as that would require the intersection of solutions. Furthermore, if the velocity field is bounded, then each of the $N!$ partitions is a trapping region. 
\end{remark}

\subsection{Convergence of Particle Solutions and Existence of a Weak Solution}

In this section, we show that the family of particle solution $(h^N)$ converges and classifies the regularity of the limiting function. Then we conclude by showing that the limiting function is a weak solution of Equation~\eqref{eq:hdefall}. 
We start with the following definition.

\begin{definition}[$w^*$-lower semicontinuity]
    A metric $d$ is $w^*$-lower semicontinuous on $B_{X^*}$ if it is a lower semicontinuous function on the product space $(B_{X^*},w^*)\times(B_{X^*},w^*)$; equivalently, given $f,g\in B_{X^*}$ and $\varepsilon>0$, there exist $w^*$-open $U\ni f$ and $V\ni g$ such that
    \begin{equation}
        d(f',g') > d(f,g) - \varepsilon, \qquad \text{whenever $f'\in U$ and $g'\in V$.}
    \end{equation}
\end{definition}

The convergence result can be established using the Arzel\'a-Ascoli theorem, which is a minor variant of a previous result taken from \cite{ambrosio2005}:
\begin{theorem}\label{thm:arzela}(A metric Arzel\`a-Ascoli theorem \cite{ambrosio2005})
Let $(X,\tau)$ be a sequentially compact Hausdorff topological space, and let $d$ be a $\tau$-lower semicontinuous metric on $X$. Let $f_n:\mathbb{R}^+\rightarrow X$ such that
\begin{equation}
    \limsup_{n\rightarrow\infty} d(f_n(s),f_n(t)) \leq L|s-t| \qquad \forall s,t\in\mathbb{R}^+,
\end{equation}
for some $L\geq0$. Then there exists a subsequence of $(f_n)$, labelled in the same way, and $f:\mathbb{R}^+\rightarrow X$ such that
\begin{equation}
    f_n(t) \xrightarrow{\tau} f(t) \qquad \forall t\in\mathbb{R}^+,
\end{equation}
and $f$ is $d$-Lipschitz with Lipschitz constant $L$.
\end{theorem}
For our purpose, we set $(X,\tau)=(B_{\mathcal{M}^+(\mathbb{R})},w^*)$ where $\mathcal{M}^+(\mathbb{R})$ is the space of positive Radon measures equipped with the usual variational norm $\|\cdot\|_1$, $B_{\mathcal{M}^+(\mathbb{R})}=\{m\in\mathcal{M}^+(\mathbb{R}):\|m\|_1\leq1\}$, and $w^*$ denotes the weak-$^*$ topology with respect to $C_0(\mathbb{R})^*$. As $C_0(\mathbb{R})^*$ is separable, $(X,\tau)$ is metrizable and compact, hence sequentially compact. Consider the set of Lipschitz functions
\begin{equation}
    A = \{f\in B_{C_0(\mathbb{R})}: \lip(f)\leq1\},
\end{equation}
and a norm $\opnorm{\cdot}$ on $\mathcal{M}^+(\mathbb{R})$ defined by
\begin{equation}
    \opnorm{m} = \sup_{f\in A} m(f) < \infty.
\end{equation}
For all $f\in A$, the map $f:\mathcal{M}(\mathbb{R})\rightarrow\mathbb{R}$ defined by $m\mapsto m(f)$ is a real-valued $w^*$-continuous linear functional on $\mathcal{M}(\mathbb{R})$ (i.e. $f\in\mathcal{M}(\mathbb{R})^*\equiv C_0(\mathbb{R})^{**}$).
As $\opnorm{\cdot}$ is the pointwise supremum of a family of $w^*$-continuous functions, $\opnorm{\cdot}$ is $w^*$-lower semicontinuous on $\mathcal{M}^+(\mathbb{R})$. 
% Indeed, for all $f\in A$, we have $f\in \mathcal{M}(\mathbb{R})^*\equiv C_0(\mathbb{R})^{**}$. Since $\|f\|_{\infty}\leq 1$, $\opnorm{m}<\infty$. Therefore the map $\opnorm{\cdot}$ is $w^*$
%
It follows that $d(m,m'):=\opnorm{m-m'}$ defines a $w^*$-lower semicontinuous metric on $B_{\mathcal{M}^+(\mathbb{R})}$. For more details on this norm, we refer the reader to \'OPS~\cite[\nopp Section 3]{smith}. %for the justification of our choice of norm. 
Finally, a function $f:\mathbb{R}\rightarrow\mathbb{R}$ having finite pointwise bounded variation shall be called a BV function and the Banach space of (equivalence classes of) integrable BV functions is denoted $\operatorname{BV}(\mathbb{R})$.

From here on, $\delta$ denotes the Dirac measure instead of a distribution, and we assume by rescaling that $W\leq1$. Thus, the particle solution are $h^N:\mathbb{R}^+ \rightarrow B_{\mathcal{M}^+(\mathbb{R})}$ given by
\begin{equation}
    h^N(t) = \sum_{i=1}^N w_i\delta_{x_i(t)}.
\end{equation}
Define the space of functions $\mathcal{X}=\{h:\mathbb{R}^+ \rightarrow B_{\mathcal{M}^+(\mathbb{R})}: \text{$h$ is $d$-continuous}\}$. In the next proposition, we show that the family of particle solutions satisfy the assumption of Theorem \ref{thm:arzela}. This result echoes \'OPS~\cite[\nopp Proposition 3.1]{smith}. 

\begin{proposition}
We have $h^N\in \mathcal{X}$ for all $N\in\mathbb{N}$. Moreover, 
\[\sup_{N\in\mathbb{N}}\lip(h^N)<\infty,\] 
with respect to $d$. 
\end{proposition}
\begin{proof}
Let $s,t\geq0$, and let $f\in A$. Then 
\begin{align}
    |(h^N(t) - h^N(s))(f)| &\leq \sum_{i=1}^N w_i |(\delta_{x_i(t)}-\delta_{x_i(s)})(f)|, \\
    &= \sum_{i=1}^N w_i |f(x_i(t))-f(x_i(s))|, \\
    &\leq \sum_{i=1}^N w_i |x_i(t)-x_i(s)|. \label{eq:equi1}
\end{align}
On the other hand, we have the following bounds for the bi-Helmholtz kernel:
\begin{equation}
    \label{eq:kernel_bound}
    0 \leq |\bar{h}^N(x)| \leq W\|K\|_\infty = \frac{1}{4\alpha}, \qquad 0 \leq |\partial_{xxx}\bar{h}^N(x)| \leq W\|K'''\|_\infty = \frac{1}{2\alpha^4},
\end{equation}
for all $x\in\mathbb{R}$. By the Mean-Value Theorem, % there exists $\tau\in(t,s)$ such that
\begin{align}
    % \dot{x}_i(\tau) &= \frac{x_i(t)-x_i(s)}{t-s} \\
    % \left[(\bar{h}^N)^2\partial_{xxx}\bar{h}^N\right]_{x=x_i(\tau)} &= \frac{x_i(t)-x_i(s)}{t-s} \\
    |x_i(t)-x_i(s)| &\leq \left\|(\bar{h}^N)^2\partial_{xxx}\bar{h}^N\right\|_\infty|t-s| \leq \frac{1}{32\alpha^6}|t-s|.
\end{align}
Thus
\begin{equation}
    d(h^N(t),h^N(s)) \leq |(h^N(t) - h^N(s))(f)| \leq \frac{1}{32\alpha^6}|t-s|,
\end{equation}
giving $\sup_{N\in\mathbb{N}}\lip(h^N)<\infty$ with respect to $d$.
\end{proof}

By Theorem \ref{thm:arzela}, there exists a subsequence of $(h^N)$, labelled in the same way, and $h\in \mathcal{X}$ such that 
\begin{equation}
    h^N(t) \xrightarrow{w^*} h(t) \qquad \forall t\geq0.
\end{equation}
% Using the compactness of $B^+_{\mathcal{M}(\mathbb{R})}$ with respect to the $w^*$-topology, and because $d$ is a $w^*$-lower semicontinuous metric on this set, we can extend the convergence result to all $t\geq0$. In fact, using \cite[\nopp Proposition 5.5]{smith}, the convergence result can be extended to the entire sequence of particle solutions. 
% \commentkp{uniqueness of the solution of the PDE is required for the whole sequence of the singular solution to converge.}

To show that $\bar{h}(t):=K*h(t)\in H^3(\mathbb{R})$ for all $t\geq0$, we define a bounded linear map $T:H^3(\mathbb{R})\rightarrow C_0(\mathbb{R})$ by
\begin{align}
    (Tu)(x) &= \langle K(\cdot-x)|u \rangle_{H^3(\mathbb{R})} = \sum_{i=0}^3\int_{-\infty}^\infty K^{(i)}(y-x)u^{(i)}(y)\,\mathd y
\end{align}
Since $K\in H^3(\mathbb{R})$, and is even, by \'OPS~\cite[\nopp Proposition 2.2]{smith}, $T$ is well defined and the dual operator $T^*:\mathcal{M}^+(\mathbb{R})\rightarrow H^3(\mathbb{R})$ is given by
\begin{equation}
    T^*m = K*m.
\end{equation}
Thus $\bar{h}(t)\in H^3(\mathbb{R})$ for all $t\geq0$. Also note that since $K\in W^{3,1}(\mathbb{R})$ and $K'''\in \operatorname{BV}(\mathbb{R})$, and $h$ is $d$-Lipschitz, by \'OPS~\cite[\nopp Proposition 4.4]{smith}, $\bar{h}\in C_b^{0,\frac{1}{2}}(\mathbb{R}^+;H^3(\mathbb{R}))$. 

% For the rest of the section, we show that $h$ and $\bar{h}$ are weak solution of Equation~\eqref{eq:hdef}. 

% \begin{lemma}\label{thm:conv1}
% Let $\phi\in C_c^\infty(\mathbb{R}^+\times\mathbb{R})$. We have
% \begin{equation}
%     \lim_{N\rightarrow\infty} \int_\infty^\infty \partial_{x}^k\bar{h}^N(t,x)\phi(t,x) \,\mathd x = \int_\infty^\infty \partial_{x}^k\bar{h}(t,x)\phi(t,x) \,\mathd x.
% \end{equation}
% for all $k=0,1,2,3$.
% \end{lemma}

\begin{theorem}\label{thm:conv}
$(h$, $\bar{h})$ defined above is a weak solution of Equation~\eqref{eq:gtfe_again}. %satisfies Equation~\eqref{eq:gtfe_again}.
\end{theorem}
We require one last lemma in order to prove Lemma \ref{thm:conv}. 

% \begin{lemma}\label{thm:conv1}
% Let $\Omega=\mathbb{R}^+\times\mathbb{R}$. Let $p:C(\mathbb{R}^+;H^3(\mathbb{R}))\rightarrow C(\mathbb{R}^+;L^2(\mathbb{R}))$ be defined by
% \begin{equation}
%     p(u(t,x)) = \Pi_{n=0}^3 (\partial_x^n u(t,x))^{l_n} \qquad l_n\in\mathbb{N}.
% \end{equation}
% Then 
% \begin{equation}
%     \lim_{N\rightarrow\infty} \int_0^\infty \int_{-\infty}^\infty (p(\bar{h}^N(t,x)) - p(\bar{h}(t,x)))\phi(t,x) \,\mathd x\,\mathd t = 0,
% \end{equation}
% for all $\phi\in C_c^\infty(\Omega)$.
% \end{lemma}
% %
% \begin{proof}
% % Note that $\partial_x^k\bar{h}$ and $\partial_x^k\bar{h}^N$, $k\leq3$, $N\in\mathbb{N}$, are uniformly boundedn on $\Omega$. By \cite[\nopp Proposition 5.2]{smith}, we have $p(\bar{h}^N) \rightarrow p(\bar{h})$ in $L^1_{loc}(\Omega)$. The result follows from the fact that $\operatorname{supp}(\phi)\subset\subset\Omega$. 
% We note that as $h,h^N\in\mathcal{X}$ and $K^{(n)}, n\leq3$, is bounded, $\partial_x^n\bar{h}$ and $\partial_x^n\bar{h}^N, n\leq3$ are uniformly bounded on $\Omega$. 
% \end{proof}

\begin{lemma} \label{thm:conv1}
Let $(u^N),(v^N),u,v\in L^\infty(\Omega)$ be uniformly bounded with respect to $\|\cdot\|_\infty$ 
% such that
% \begin{equation}
%     \|u^N\|_\infty, \|v^N\|_\infty, \|u\|_\infty, \|v\|_\infty \leq \infty
% \end{equation}
and let $u^N\rightarrow u$ and $v^N\rightarrow v$ in $L^1_{loc}(\Omega)$. Then $u^Nv^N\rightarrow uv$ in $L^1_{loc}(\Omega)$. 
\end{lemma}

\begin{proof}
Let $U\subset\subset\Omega$, then
\begin{align}
    &\int_U |u^N(x)v^N(x) - u(x)v(x)| \,\mathd x \nonumber \\
    &\leq \int_U |u^N(x)v^N(x) - u^N(x)v(x)| \,\mathd x + \int_U |u^N(x)v(x) - u(x)v(x)| \,\mathd x, \\
    &\leq \|u^N\|_\infty \int_U |v^N(x) - v(x)| \,\mathd x + \|v\|_\infty \int_U |u^N(x) - u(x)| \,\mathd x \rightarrow 0.
\end{align}
\end{proof}

\begin{proof}[Proof of Lemma \ref{thm:conv}]
We show that $(h,\bar{h})$ satisfy Equation~\eqref{eq:gtfe_again}. The First term follows directly from the fact that $h^N(0)\rightarrow h_0$ in $(\mathcal{M}(\mathbb{R}),w^*)$ and $\phi|_{t=0}$ is in the predual space $C_0(\mathbb{R})$. %The rest of the terms in Equation~\eqref{eq:gtfe_again} follows from \cite[\nopp Proposition 5.2]{smith} and Lemma \ref{thm:conv1}.
From \'OPS~\cite[\nopp Proposition 5.2]{smith}, we have that $\partial_x^k\bar{h}^N\rightarrow\partial_x^k\bar{h}$ in $L^1_{loc}(\mathbb{R})$,for $k=0,1,2,3$, so the linear term converges as $N\rightarrow\infty$. Furthermore, since for all $k=0,1,2,3$, the functions $(\partial_x^k\bar{h}^N),\partial_x^k\bar{h}\in L^\infty(\mathbb{R})$ are uniformly bounded with respect to $\|\cdot\|_\infty$, the nonlinear terms in Equation~\eqref{eq:gtfe_again} also converges as $N\rightarrow\infty$. 
\end{proof}

\section{\markup{Solution as Pushforward of Measure}}
\label{sec:pushforward}

The above work has been generalized by \'OPS \cite{smith2} in a collaborative effort to establish the existence and uniqueness of the solution of the Geometric Thin-Film Equation. We include a discussion of these results in this section for completeness. As such, the following section should be treated as a literature review. 

\begin{figure}
    \centering
    \begin{tikzpicture} [scale=2]
        \draw [<-, line width=0.3mm] (0,1) node[above] {$x$} -- (0,-1);
        \draw [->, line width=0.3mm] (0,-1) -- (3.5,-1) node[right] {$t$};
        % straight
        \draw (0.75,0) node[above] {$\delta_0$};
        \draw [line width=0.6mm,orange] (0,0) -- (1.5,0);
        % decay
        \draw [orange, line width=0.6mm, dashed] plot[domain=0:1.5,smooth] (\x+1.5,{0.3*ln(4*\x+1)});
        \draw [orange, line width=0.6mm, dashed] plot[domain=0:1.5,smooth] (\x+1.5,{-0.3*ln(4*\x+1)});
        \draw (3.3,0.6) node {$\tfrac{1}{2}\delta_{x_1(t)}$};
        \draw (3.3,-0.6) node {$\tfrac{1}{2}\delta_{x_2(t)}$};
        % slope 
        \draw [->] (1.5,0) -- (1.8,0.33) node[above]{$\dot{x}_1(t_0)$};
        \fill (1.5,0) circle [radius=0.3mm];
        \draw [dashed] (1.5,0) -- (1.5,-1) node[below] {$t_0$};
    \end{tikzpicture}
    \caption{Spontaneous decay of particles is prohibited when $h$ is defined by a pushforward. The splitting velocity $\dot{x}_1(t_0)$ is bounded below by $\|K\|_{\infty}^2\|K'''\|_\infty/4$.}
    \label{fig:sketch_splitting}
\end{figure}
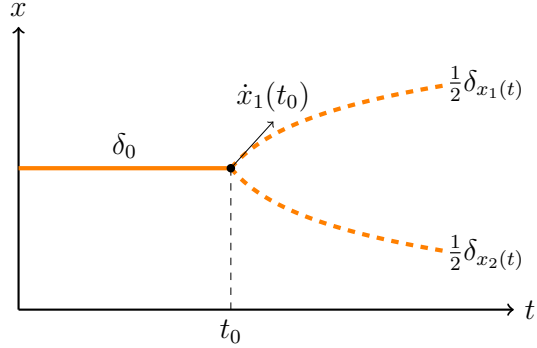

A generalization is necessary as establishing the uniqueness and continuity with respect to the initial condition of the solution of the Geometric Thin-Film equation is not possible in the space $C(\mathbb{R}^+,\mathcal{M}^+(\mathbb{R}))$. In \'OPS \cite{smith2}, a quantitative no-crossing theorem is established, which gives a lower bound for the minimum particle separation:
\begin{equation} \label{eq:qnocross}
    |x_i(t)-x_j(t)|\geq\frac{\|K\|_{\infty}^2\|K'''\|_{\infty}\max\{w_i^3,w_j^3\}}{A}(1-\mathe^{-At}),
\end{equation}
for some constant $A>0$ depending only on $K$. Consider the initial condition $\mu_1\in \mathcal{M}^+(\mathbb{R})$ given by $\mu_1=\delta_0$. Since there is no interaction between particles, we deduce that the solution is given by $h_1(t)=\delta_0$ for all $t\geq0$. Now consider another initial condition $\mu_2\in \mathcal{M}^+(\mathbb{R})$ with $\mu_2 = \tfrac{1}{2}(\delta_{-\varepsilon} + \delta_\varepsilon)$ with $\varepsilon>0$. In particular, $\mu_2\xrightarrow{w^*}\mu_1$ as $\varepsilon\rightarrow 0$. By the convergence of the particle method, we expect the solution $h_2(t)$ to coincide with $h_1(t)$ as $\varepsilon\rightarrow 0$. The particle method suggests a solution of the form
\begin{equation}
    h_2(t) = \frac{1}{2}(\delta_{x_1(t)} + \delta_{x_2(t)})
\end{equation}
with $x_1(0)=-\varepsilon$ and $x_2(t)=\varepsilon$. From \cref{eq:qnocross}, the separation between the two particles is bounded below by
\begin{equation}
    |x_1(t)-x_2(t)| \geq \frac{\|K\|_{\infty}^2\|K'''\|_{\infty}}{8A}(1-\mathe^{-At}),
\end{equation}
which is independent of $\varepsilon$. Thus $h_2(t)\neq h_1(t)$ for all $\varepsilon$, leading to two solutions satisfying the same initial condition. In this example, the function $h_1$ follows a single path while $h_2$ immediately splits into two paths (e.g. \Cref{fig:sketch_splitting}). The wider class of weak solutions admits all kinds of path-splitting. 

For the above reason, we have to restrict ourselves to a narrower class of solutions to rule out path-splitting solutions. 
Let $\mu\in \mathcal{M}(\mathbb{R}^+)\setminus\{0\}$, $\|\mu\|_1\leq1$, we define the support of $\mu$ as
\begin{equation}
    M := \supp \mu = \{x\in\mathbb{R}:|\mu|(U)>0 \text{ whenever $U\ni x$ is open}\}.
\end{equation}
That is, $M$ is the smallest closed set such that $\mu(\mathbb{R}\setminus M)=0$. Next, we consider the vector space $X_M$ of locally bounded functions $c:M \times \R^+\rightarrow\R$, such that $c(\cdot,t)$ is Borel measurable for all $t \in \R^+$ and $c(x,\cdot)$ is $C^1(\mathbb{R}^+)$ for all $x \in M$. We use $c$ to define potential solutions $h(t)$ of \cref{eq:weakdef} as the pushforward of the initial measure 
\begin{equation}
    h(t)(x) = (c(t,\cdot)_*\mu)(x) = \mu(c(t,\cdot)^{-1}(x)).
\end{equation}
In order to satisfy the initial condition $h(0)=\mu$ of the PDE, we require $c(0,x) = x, x \in M$. In this framework, the no-crossing condition is equivalent to saying $c(t,\cdot)$ is strictly increasing on $M$. For this reason, we define
\begin{equation}
    I_M = \{c\in X_M: c(t,\cdot) \text{ is strictly increasing on $M$ for all $t\in\mathbb{R}^+$} \}.
\end{equation}
We show that given an initial measure $\mu$, there exists a unique curve $c\in I_M$ such that the corresponding $h$ is a weak solution of the Geometric Thin-Film equation. 

Note that $\bar{h}(t,\cdot)$ can be expressed in terms of $c(t,\cdot)$. In particular, we observe the identity
\begin{align}
    \bar{h}(t,c(t,x)) &= \int_{-\infty}^\infty K(c(t,x)-z)\,\mathd h(t)(z), \\
    &= \int_{-\infty}^\infty K(c(t,x)-c(t,z))\,\mathd \mu(z).
\end{align}
To make this dependence on $c$ explicit, we will write $H(c,t,x)=\bar{h}(t,c(t,x))$. This allows us to describe the solution in terms of the initial condition. 

\begin{theorem} \label{thm:pfweaksol}
    Let $c\in I$ satisfies $c(0,x)=x,x\in M$, and suppose that for $(t,x)\in(0,\infty)\times M$. If $c$ satisfies the ODE
    \begin{equation} \label{eq:pfode}
        \partial_t c(t,x) = H(c,t,x)^2\int_{z\neq x}K'''(c(t,x)-c(t,z))\,\mathd\mu(z),
    \end{equation}
    then the corresponding map $h$ is a solution of \cref{eq:weakdef}. 
\end{theorem}
This is the generalization of \Cref{thm:pweaksol}. In particular, if we take $\mu$ to be a weighted sum of Dirac measures $\mu = \sum_{i=1}^N w_i\delta_{x_i^0}$, then \cref{eq:pfode} is equivalent to the finite-dimensional ODE system given in \cref{eq:ivp1}. Furthermore, the trajectory of the particles is given by $c(t,x_i)=x_i(t)$. % For this reason, the proof of \cref{thm:pfweaksol} and \cref{thm:pweaksol} are fundamentally similar, which we omit here. The proof is given in \cite{smith2} in full. 

To construct a solution $c$, the proof again makes use of the particle solutions as they exhibit properties such as global existence and continuity with respect to time. A sequence of approximating measures $(\mu_n) \subseteq \mathcal{M}^+(\R)\setminus\{0\}$, each supported on $F_n$ is constructed. The particle method then gives a sequence of curves $c_n\in I_{F_n}$ corresponding to the $\mu_n$. A limiting procedure is used to obtain a solution $c\in I_M$ of \cref{eq:pfode} corresponding to $\mu$. However, the main difference between this and \Cref{sec:texist} is that special care needs to be taken for the choice of $F_n$. In particular, the sequence of particle solutions needs to be constructed around the set of `problematic' points labelled $P$ such that
\begin{equation}
    P\subseteq F := \bigcup_{n\geq1} F_n.
\end{equation}
Furthermore, since $F_n$ is finite (because the particle solution is a finite sum of Dirac measures) while $M$ may be uncountable, to ensure the convergence, we require that $F$ is dense in $M$. One such choice is to define $F=P\cup\mathbb{Q}$, which ensures the density of $F$ in $\mathbb{R}$. In \'OPS \cite{smith2}, the set of problematic points are identified, which are points whose singletons are atoms of $\mu$:
\begin{equation}
    P = \{x\in M:\mu\{x\}>0\}.
\end{equation}
Moreover, $P$ is at most countably infinite. Otherwise, $(F_n)$ would not be able to cover $P$. This also covers the basis of constructing numerical solutions and the necessary conditions for convergence. By defining the solution in terms of pushforward of measure, path-splitting solutions are no longer allowed, which renders the solution unique. 

% Equation \eqref{eq:pfode} allows us to define a map $\Phi:I_M\rightarrow X_M$, by 
% \begin{equation}
%     (\Phi c)(t,x) = x + \int_0^t H(c,s,x)^2 \int_{z\neq x} K'''(c(s,x)-c(s,z)) \,\mathd\mu(x) \,\mathd s. 
% \end{equation}
% Then it can be shown that if $c,c'\in I_M$, with $c(0,x)=c'(0,x)=x$ for all $x\in M$ and satisfying the ODE \cref{eq:pfode}. Then $c=c'$. Hence establishing the global uniqueness of the Geometric Thin-Film Equation in $I_M$. 

\begin{figure}[htb]
    \centering
    \includegraphics[width=0.8\textwidth]{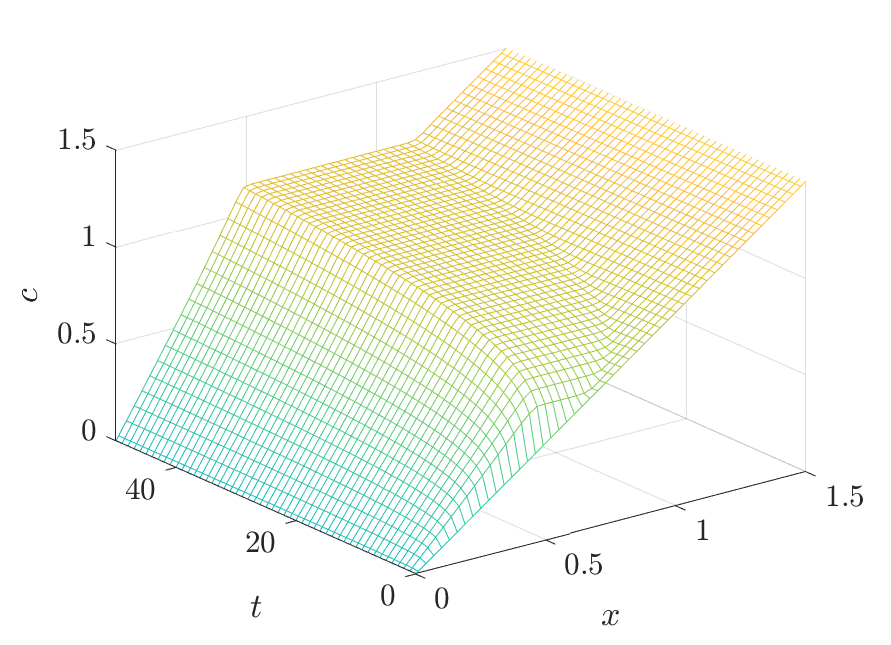}
    \caption{The solution of the Geometric Thin-Film Equation in terms of $c(t,x)$. }
    \label{fig:pfsolution}
\end{figure}

In \Cref{fig:pfsolution}, the corresponding curve $c(t,x)$ for the droplet spreading problem is plotted. Only the positive half of the solution is plotted. By the symmetry of the initial condition, $c(\cdot,x)$ is an odd function in $x$. Furthermore, since the initial condition is non-atomic, $c$ is continuous and unique solution of the Geometric Thin-Film Equation.  

% \begin{figure}
%     \centering
%         \begin{tikzpicture}[scale=1.2,fill=gray!50]
%         % right hand
%         \scope
%         \clip (-2,-2) rectangle (2,2)
%               (0,0) circle (1);
%         \fill (1,0) circle (1);
%         \endscope
%         % outline
%         \draw (0,0) circle (1) (-1,0)  node [text=black,left] {$P$}
%               (1,0) circle (1) (2,0)  node [text=black,right] {$Q$}
%               (-2,-2) rectangle (3,2) node [text=black,right] {$\mathbb{R}$};
%         \end{tikzpicture}
%     \caption{Caption}
%     \label{fig:enter-label}
% \end{figure}

\section{\markup{Conclusion}}

In this chapter, the existence and uniqueness of the solutions of the Geometric Thin-Film Equation with respect to a class of Radon measure initial condition is established: Given an initial condition $h_0\in\mathcal{M}^+(\mathbb{R})$, we proved that there exists a global unique solution to the Geometric Thin-Film equation. Furthermore, we showed that the solution is $\tfrac{1}{2}$-H\:older continuous in time and the smoothened solution $\barh(\cdot,t)\in H^3(\mathbb{R})$. These results are an important first step in justifying the use of the Geometric Thin-Film equation as a model for describing physical processes such as droplet spreading. The convergence of the particle solutions combined with the uniqueness results provide the necessary rigour for the numerical exposition of the Geometrical Thin-Film Equation given in the previous chapter.

We see that the particle solution underpins the existence theorem, giving another justification for studying particle solutions and the rationale of seeking particle-admitting regularization of the thin-film equation. The particle solution transforms the PDE into a system of finite dimensional ODEs that describes the equation of motion of the particles, and standard dynamical system theory is used to characterize the solution. Notably, by proving a Lipschitz condition and a trapping region on $D=\{\bm{x}\in\mathbb{R}^N:x_i<x_j,i<j\}$ we established the uniqueness and global existence of the particle solution. We emphasise that the global existence property suggests that the contact-line singularity has been alleviated with the regularization. The trapping region also ensures that the ordering of the particles remains the same for all $t\geq0$, thus satisfying the assumption of the fast summation algorithm. The quantitative lower bound of the separation between particles given in \Cref{eq:qnocross} also has implications on the density and distribution of the particles. This could serve as a starting point for finding the optimal parameters of the fast multipole method. 
% The resulting fast particle method then provides a fast and robust numerical method for solving the Geometric Thin-Film Equation. 

The convergence of the particle method is established using a metric Arzel\`a-Ascoli theorem, guaranteeing a solution of the Geometric Thin-Film equation. In general, the solution is non-unique. However, by considering solutions induced via pushforward of the initial measure by curves in $I_M$, the solution can be shown to be unique.

\chapter{Partial Wetting}
\label{sec:partial}

\section{Overview}

In this chapter, we extend the Geometric Thin-Film Equation to the case of partial wetting, where a droplet on a substrate spreads initially before assuming an equilibrium shape.  This requires the inclusion of an extra, stabilizing term, to the Geometric Thin-Film Equation.  We derive this additional term in \Cref{sec:partial_theory}. An alternate nondimensionalization with respect to the droplet volume and the equilibrium contact angle is introduced.  In \Cref{sec:equilib_sol}, we construct an analytical solution for the equilibrium droplet shape and study the behaviour of the solution when the parameters are changed.  In \Cref{sec:partial_num}, we use both the finite-difference method and the particle method to simulate transient droplet spreading, up to the point where the droplet assumes its equilibrium shape. Additional remarks are given for incorporating the partial wetting terms into the solver. Finally, a conclusion is given in \Cref{sec:partial_conclusion}. 

\section{Theoretical Formulation}
\label{sec:partial_theory}

The starting-point for the theoretical formulation is to consider an unregularized description of the droplet, with $h(x)$ as the droplet profile.  Then, the unregularized energy associated with a droplet of radius $r$ is:
\begin{equation}
    \Energy[h]=\gamma_{lg}\int_{-r}^r \sqrt{1+ h_x^2}\,\mathd x+2r\,\gamma_{sl}+\gamma_{sg}\left(S-2r\right),
    \label{eq:physics}
\end{equation}
Here, $\gamma_{lg}$ is the surface tension between the gas and the liquid droplet (previously referred to as $\gamma$), $\gamma_{sl}$ is the surface tension between the liquid and the substrate, and $\gamma_{sg}$ is the surface tension between the gas and the substrate; $S$ is an arbitrary lengthscale denoting the extent of the system in the lateral direction.  In the longwave limit, $\sqrt{1+h_x^2}$ is expanded as $1+(1/2)h_x^2$, and Equation~\eqref{eq:physics} becomes:
\begin{equation}
    \Energy[h]=\tfrac{1}{2}\gamma_{lg}\int_{-r}^r h_x^2\,\mathd x+2r\left(\gamma_{lg}+\gamma_{sl}-\gamma_{sg}\right)+\mathrm{Const.}
    \label{eq:physics1}
\end{equation}
The three surface-tension coefficients are related via the Laplace-Young condition,
\begin{equation}
    \gamma_{sl}+\gamma_{lg}\cos\theta_{eq}-\gamma_{sg}=0,
\end{equation}
where $\theta_{eq}$ is the equilibrium contact angle.  Thus, Equation~\eqref{eq:physics1} can be re-written as
\begin{equation}
    \Energy[h]=\tfrac{1}{2}\gamma_{lg}\int_{-r}^r h_x^2\,\mathd x+2r\gamma_{lg}\left(1-\cos\theta_{eq}\right)+\mathrm{Const.}
    \label{eq:physics2}
\end{equation}
Now, using the asymmetrical regularization of the geometric diffuse-interface method introduced previously, we propose herein a regularized energy,
\begin{equation}
    \Energy[h]=\tfrac{1}{2}\gamma_{lg}\int_{-\infty}^\infty h_x\barh_x \mathd x+w\gamma_{lg}\left(1-\cos\theta_{eq}\right)+\text{Const.},
    \label{eq:physics3}
\end{equation}
where $w$ is an estimate of the size of the droplet footprint, based on the interfacial profile $h$, and on the smoothened interfacial profile, $\barh=K_2*h$. 

\begin{figure}
    \centering
    \begin{tikzpicture}
        \fill [black!10] (0,0) rectangle (10,-0.6);
        % axes
        \draw [->, line width=0.3mm] (0,0) -- (10,0);
        \draw [->, line width=0.3mm] (0,0) -- (0,3);
        \draw (10,0) node[right] {$x$};
        \draw (0,3) node[above] {$z$};
        % length scale
        % \draw [<->, line width=0.2mm] (0,-0.1) -- (2*3.14,-0.1);
        % \draw (2*3.14,0) node[below] {$\lambda_0$};
        % \draw (0,1.8) node[left] {$h_0$};
        % interface
        \draw[orange,line width=0.7mm] plot[domain=0:6.28,smooth] (\x,{1.8*(1-(\x/6.28)^2)});
        \draw [orange,line width=0.7mm] (0,0) -- (6.28,0);
        \draw (3.5,1.8) node {$h(x,t)$};
        \draw (1.2,2.2) node {gas};
        \draw (1.2,0.8) node {liquid};
        \draw (1.2,-0.3) node {solid};
        % contact angle
        \draw [->, line width=0.3mm] (6.28,0) -- (5.3,0.65) node[above] {$\gamma_{lg}$};
        \draw [line width=0.3mm] (6.28-0.7,0) arc (180:147:0.7);
        \draw (5.5,0.4) node[left] {$\theta_{eq}$};
        % contact line velocity
        \draw [->, line width=0.3mm] (6.28,0.02) -- (7.8,0.02) node[above] {$\gamma_{sg}$};
        \draw [->, line width=0.3mm] (6.28,0.02) -- (5.2,0.02) node[below] {$\gamma_{sl}$};
        \fill [orange] (6.28,0) circle (0.08);
        \draw (6.28,0) node[below] {$r$};
    \end{tikzpicture}    
    \caption{Schematic of }
    \label{fig:sketch_partial}
\end{figure}
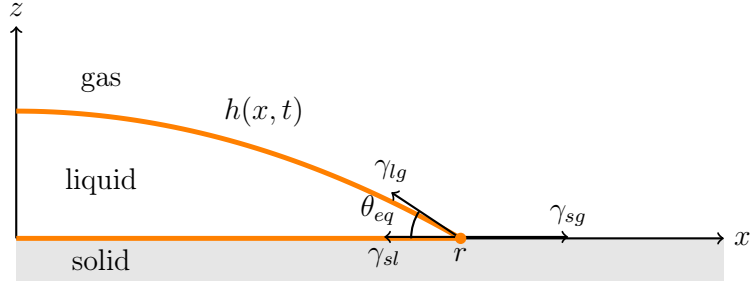

We estimate the size of the droplet footprint as
\begin{equation}
    w(t)=c\frac{\|\barh\|_1^2}{\langle h,\barh\rangle }=c\frac{A^2}{\langle h,\barh\rangle},
    \label{eq:wapprox}
\end{equation}
where $A$ is the constant droplet volume, $A=\int_{-\infty}^\infty h(x,t)\mathd x=\int_{-\infty}^\infty \barh(x,t)\mathd x$, the inner product paring $\langle h,\barh\rangle(t) = \int_{-\infty}^\infty h(x,t)\barh(x,t)\,\mathd x$, and $c$ is an $O(1)$ parameter to be determined.
The estimate in Equation~\eqref{eq:wapprox} is dimensionally correct, but also yields good agreements with some model droplet profiles: for instance, if $h$ were a parabolic cap with the form $h(x)=\max\{0,(3A/4r)[1-(x/r)^2]\}$, then we would have (by direct calculation) the pairing $\langle h,\barh\rangle=3A^2/5r + O(\alpha^2)$, where $\alpha\ll1$ is the smoothing parameter, and
\[
\frac{\|\barh\|_1^2}{\langle h,\barh\rangle}=\frac{5}{6}(2r) + O(\alpha^2),
\]
i.e. a width proportional to the droplet footprint $2r$, with a constant of proportionality $6/5 + O(\alpha^2)$ close to one.  
%
%Similarly, if $h$ were a Gaussian, with $h(x)=(A_0/\sqrt{\pi}a)\mathe^{-x^2/a^2}$, then we would have
% %
% \[
% \frac{\|\barh\|_1^2}{\langle h,\barh\rangle}=\frac{A_0^2}{\langle h,\barh\rangle}=(TBC)\,a,
% \]
% %
% and again, the ratio of the norms-squared is proportional to the `width' of the function $a$, with a constant of proportionality near one.  
%
%
Thus, the regularized energy becomes:
\begin{equation}
\Energy[h]=\frac{1}{2}\gamma\int_{-\infty}^\infty h_x \barh_x \,\mathd x+ \underbrace{\gamma \chi \frac{A^2}{\langle h,\barh\rangle}}_{E_w[h]}.
\label{eq:E_partial}
\end{equation}
where $\chi=c(1-\cos\theta_{eq})$ is an $O(1)$ constant, which will be selected \textit{a priori} in what follows; we also use $\gamma$ instead of $\gamma_{lg}$, for consistency with the previous sections.  Finally, the constant term in the energy has been dropped in Equation~\eqref{eq:E_partial} because only energy differences are important for the purpose of deriving evolution equations.

To derive the evolution equation associated with Equation~\eqref{eq:E_partial}, we use the framework of geometric gradient-flow introduced previously.  Thus, the generalized force associated with Equation~\eqref{eq:E_partial} is:
\begin{equation}
    f=-\frac{\partial}{\partial x}\frac{\delta \Energy}{\delta h}. 
\end{equation}
The functional derivative of the partial wetting term can be computed from the first principle. We start by perturbing the energy functional by a small amount $\delta h$, 
\begin{align}
    E_w[h+\delta h] &= \gamma\chi A^2 \left(\int_{-\infty}^\infty (h+\delta h)(\barh + \bar{\delta h})\,\mathd x\right)^{-1}, \\ 
    &= \gamma\chi A^2 \left(\int_{-\infty}^\infty (h\barh+\barh\delta h + h\bar{\delta h} + \delta h\bar{\delta h})\,\mathd x\right)^{-1}.
\end{align}
Since $\delta h$ is small, second-order terms with respect to $\delta h$ are dropped. Also note that $\langle h,\bar{\delta h}\rangle = \langle \barh,\delta h\rangle$, we get
\begin{align}
    E_w[h+\delta h] &= \gamma\chi A^2\left(\langle h,\barh\rangle + 2\langle\barh,\delta h\rangle\right)^{-1}, \\
    &= \gamma\chi \frac{A^2}{\langle h,\barh\rangle}\left(1 + 2\frac{\langle\barh,\delta h\rangle}{\langle h,\barh\rangle}\right)^{-1}.
\end{align}
Using the binomial approximation, this becomes
\begin{align}
    E_w[h+\delta h] &= \gamma\chi \frac{A^2}{\langle h,\barh\rangle}\left(1 - 2\frac{\langle\barh,\delta h\rangle}{\langle h,\barh\rangle} + O(\delta h^2)\right), \\
    &= E_w[h] - 2\gamma\chi \frac{A^2}{\langle h,\barh\rangle^2}\langle\barh,\delta h\rangle.
\end{align}
Hence, the energy gradient of the partial wetting term is given by
\begin{equation}
    \frac{\delta E_w}{\delta h} = -2\gamma\chi\frac{A^2}{\langle h,\barh\rangle^2}\barh. 
\end{equation}
We again take the mobility to be $\mob=(1/3\vis)\barh^2$.  Thus, 
\[
f=-\frac{\partial}{\partial x}\left(-\gamma\partial_{xx}\barh-2\gamma \chi\frac{A^2}{\langle h,\barh\rangle^2}\barh\right),
\]
and the mass conservation evolution equation for $h$ is 
% %
% \[
% \frac{\partial h}{\partial t}+\frac{\partial}{\partial x}\left(hU\right)=0,
% \]
% hence
% %
\begin{equation}
\frac{\partial h}{\partial t}=\frac{\partial}{\partial x}\left\{h\mob\frac{\partial}{\partial x}\left(-\gamma\barh_{xx}-2\gamma \chi \frac{A^2}{\langle h,\barh\rangle^2}\barh\right)\right\}.
\label{eq:particle_partial_dim}
\end{equation}

\subsection{Nondimensionalization}

We introduce a different nondimensionalization with respect to the equilibrium contact angle $\theta_{eq}>0$ and the droplet volume $A<\infty$. Such that under the transformation, we have
\begin{equation}
    \tilde{A} = \int_{-\infty}^\infty \tilde{h} \,\mathd\tilde{x} = 1, \qquad \tilde{\theta}_{eq} = -\pder[\tilde{h}]{\tilde{x}}(r,\infty) = 1.
\end{equation} 
This is achieved by considering the following rescaling
\begin{equation} \label{eq:partial_rescale1}
    \tilde{x} = x\sqrt{\frac{\theta_{eq}}{A}}, \qquad \tilde{h} = \frac{h}{\sqrt{A\theta_{eq}}}, \qquad \tilde{t} = \frac{t}{t_0},
\end{equation}
for some timescale $t_0$ to be determined. In this context, the ratio between the two lengthscale is precisely $\epsilon=\theta_{eq}$. We require $\theta_{eq}\ll 1$ for the lubrication approximation to apply. We also note that under this limit, $\chi = c(1-\cos\theta_{eq}) \approx c\theta_{eq}^2/2$. So $\chi$ transform like $\theta_{eq}^2$, giving 
\begin{equation} \label{eq:partial_rescale2}
    \tilde{\chi} = \frac{\chi}{\epsilon^2} = \frac{c}{2}.
\end{equation}

Substituting \cref{eq:partial_rescale1,eq:partial_rescale2} into \cref{eq:particle_partial_dim} and dropping the tildes, we get
\begin{equation}
    \frac{\sqrt{A\theta_{eq}}}{t_0}\pder[h]{t} = -\frac{\gamma}{3\mu}\epsilon^4\left\{h\barh^2 \pder{x}\left(\barh_{xx}+2\chi\frac{\barh}{\langle h,\barh\rangle^2}\right) \right\}.
\end{equation}
Thus, we are also motivated to scale the time by
\begin{equation}
    t_0 = \frac{3\mu\sqrt{A\theta_{eq}}}{\gamma \epsilon^4},
\end{equation}
finally giving
\begin{equation}
\frac{\partial h}{\partial t}=-\frac{\partial}{\partial x}\left\{h\barh^2\frac{\partial}{\partial x}\left(\barh_{xx}+2\chi \frac{\barh}{\langle h,\barh\rangle^2}\right)\right\}.
\label{eq:particle_partial}
\end{equation}
% Also, in dimensionless variables, $\int_{-\infty}^\infty h(x,t)\mathd x=1$.
% %
% %
% Also in this context, the ratio $\epsilon=h_0/\lambda_0$ is precisely $\tan\theta_{eq}$; strictly speaking therefore, $\theta_{eq}$ should be small, for inertial effects to be negligible, and hence, for the lubrication theory underlying Equation~\eqref{eq:particle_partial} to be valid.

\section{Equilibrium Solution}
\label{sec:equilib_sol}

Equation~\eqref{eq:particle_partial} has an equilibrium solution with $\partial h/\partial t=0$.  In this limiting case, Equation~\eqref{eq:particle_partial} reduces to
\begin{equation}
%h\barh^2 \Phi*\left(\partial_{xx} h+\xi^2 h\right)=0,
h\barh^2\partial_x\left(\partial_{xx}\barh+\xi^2\barh\right)=0,
\label{eq:eqm1}
\end{equation}
where $\xi^2$ is a positive constant,
\begin{equation}
\xi^2=\frac{2\chi}{\langle h,\barh\rangle^2}.
\label{eq:eqm2}
\end{equation}
We seek solution of Equation~\eqref{eq:eqm1} using an Ansatz $\bar{h}(x)=\mathe^{\lambda x}$ and recalling that $h=(1-\alpha^2\partial_{xx})^2\barh$, the characteristic equation of Equation~\eqref{eq:eqm1} is given by
\begin{equation}
    (1-\alpha^2\lambda^2)^2\lambda(\lambda^2+\xi^2) = 0.
\end{equation}
Since $\alpha,\xi>0$, the roots of the characteristic polynomial are
\begin{equation}
    \lambda = \pm\frac{1}{\alpha} \text{ (repeated roots) }, \qquad \text{and} \qquad \lambda=0, \qquad \text{and} \qquad \lambda = \pm\mathi\xi.
\end{equation}
Thus, the general solution of Equation~\eqref{eq:eqm1} is given by
\begin{multline}
    \bar{h}(x) = B_1\cos(\xi x) + B_2\sin(\xi x) + C_1\mathe^{x/\alpha} + C_2\mathe^{-x/\alpha} \\+ D_1|x|\mathe^{x/\alpha} + D_2|x|\mathe^{-x/\alpha} + \mathrm{Const.} 
\end{multline}
For a droplet solution, we require that the solution to be even, bounded, has a finite volume, and differentiable at $x=0$. Therefore, we deduced that the solution has the from
\begin{equation}
\barh(x)=\begin{cases} B_1\cos(\xi x)+B_2. & |x|<r,\\
                  C_1 \mathe^{-|x|/\alpha}+C_2|x|\mathe^{-|x|/\alpha}, & |x|>r,\end{cases}
\label{eq:hbar_solution}
\end{equation} 
%
%the choice of coefficients in Equation~\eqref{eq:h_solution} means that $\int_{-\infty}^\infty h\mathd x=1$.
which is parametrized by  $\xi$, and by a radius $r$. Correspondingly,
\begin{equation}
h(x)=\begin{cases} B_1(1+\alpha^2\xi^2)^2\cos(\xi x)+B_2, & |x|<r,\\
                          0, & |x|>r.\end{cases}
\label{eq:h_solution}
\end{equation} 
Here, $B_1,B_2,C_1$, and $C_2$ are constants of integration.  Equation~\eqref{eq:hbar_solution}--\eqref{eq:h_solution} are then subjected to the following boundary conditions:
\begin{equation}
\begin{gathered}
    A = \intinf h\,\mathd x = \intinf \barh\,\mathd x, \\
    \text{$\barh$, $\barh'$, and $\barh''$ continuous at $r$.}  
\end{gathered} \label{eq:partial_bc}
\end{equation}
These conditions then determine the value for $B_1$, $B_2$, $C_1$, $C_2$, and $r$. 

\begin{proposition}
    If $\bar{h}$ as given in Equation~\eqref{eq:hbar_solution} satisfies the boundary conditions given in Equation~\eqref{eq:partial_bc}, then $\barh\in C^3(\mathbb{R})$. That is, we obtain the condition ``$\barh'''$ continuous at $r$'' without explicitly imposing the condition. 
\end{proposition}
\begin{proof}
We note that the conditions
\begin{enumerate}
    \item Mass conservation for $h$,
    \begin{equation} \label{eq:partial_bc1}
        \frac{1}{2}A = B_1(1+\alpha^2\xi^2)^2\frac{1}{\xi}\sin(\xi r) + B_2r,
    \end{equation}
    \item Mass conservation for $\barh$,
    \begin{equation} \label{eq:partial_bc2}
        \frac{1}{2}A = B_1\frac{1}{\xi}\sin(\xi r) + B_2r + \alpha(C_1+rC_2+\alpha C_2)\exp\left(-\frac{r}{\alpha}\right).
    \end{equation} 
    \item $\barh'$ continuous at $r$,
    \begin{equation} \label{eq:partial_bc3}
        -B_1\xi\sin(\xi r) = -\frac{1}{\alpha}(C_1-\alpha C_2+rC_2)\exp\left(-\frac{r}{\alpha}\right).
    \end{equation}
    \item $\barh'''$ continuous at $r$,
    \begin{equation} \label{eq:partial_bc4}
        B_1\xi^3\sin(\xi r) = -\frac{1}{\alpha^3}(C_1-3\alpha C_2+rC_2)\exp\left(-\frac{r}{\alpha}\right).
    \end{equation}
\end{enumerate}
are linearly \textit{dependent}. Indeed, we see that $\eqref{eq:partial_bc1} - \eqref{eq:partial_bc2} - 2\alpha^2\eqref{eq:partial_bc3}$ gives Equation~$\eqref{eq:partial_bc4}$. Therefore, choosing three out of the four boundary conditions \cref{eq:partial_bc1,eq:partial_bc2,eq:partial_bc3,eq:partial_bc4} yields the same system of equations.   
\end{proof}

% We note that the boundary condition ``$\barh'''$ continuous at $r$'' is linearly \textit{dependent} to the above set of boundary conditions, thus by imposing Equation~\eqref{eq:partial_bc}, we get that $\barh\in C^3(\mathbb{R})$ automatically. 

Furthermore, from the set of boundary conditions given in Equation~\eqref{eq:partial_bc}, we can obtain a relation between $r$ and $\xi$. For this, we list the other two conditions
\begin{enumerate}
    \setcounter{enumi}{4}
    \item $\barh$ continuous at $r$,
    \begin{equation} \label{eq:partial_bc5}
        B_1\cos(\xi r) + B_2 = (C_1+rC_2)\exp\left(-\frac{r}{\alpha}\right).
    \end{equation}
    \item $\barh''$ continuous at $r$,
    \begin{equation} \label{eq:partial_bc6}
        -B_1\xi^2\cos(\xi r) = \frac{1}{\alpha^2}(C_1-2\alpha C_2+rC_2)\exp\left(-\frac{r}{\alpha}\right).
    \end{equation}
\end{enumerate}
We start by eliminating $B_1$ and $B_2$ by taking $(\eqref{eq:partial_bc1} - \eqref{eq:partial_bc2})\div\eqref{eq:partial_bc3}$ and $\eqref{eq:partial_bc3}\div\eqref{eq:partial_bc5}$, leading to two linearly independent equations
\begin{align}
2 + \alpha^2\xi^2 &= \frac{C_1+rC_2+\alpha C_2}{C_1+rC_2-\alpha C_2}, \\
\frac{1}{\xi}\tan(\xi r) &= -\frac{1}{\alpha}\frac{C_1+rC_2-\alpha C_2}{C_1+rC_2-2\alpha C_2}.
\end{align}
In matrix form, these are
\begin{equation}
    \begin{pmatrix}
        1+\alpha^2\xi^2 & r(1+\alpha^2\xi^2)-\alpha(3+\alpha^2\xi^2) \\
        \tan(\xi r)+\alpha\xi & r(\tan(\xi r)+\alpha\xi)-\alpha(2\tan(\xi r)+\alpha\xi) 
        \end{pmatrix}
    \begin{pmatrix}
    C_1 \\
    C_2
    \end{pmatrix} = 0.
\end{equation}
For a non-trivial solution, we require the determinant of the matrix to be zero, giving 
\begin{align}
    (1+\alpha^2\xi^2)(2\tan(\xi r)+\alpha\xi) - (3+\alpha^2\xi^2)(\tan(\xi r)+\alpha\xi) = 0, 
\end{align}
which expended becomes
\begin{equation}
    -\tan(\xi r) - 2\alpha\xi + \alpha^2\xi^2\tan(\xi r) = 0.
\end{equation}
So the root finding condition for $r$ is
\begin{equation}
    \tan(\xi r) = -\frac{2\alpha\xi}{1-\alpha^2\xi^2}.
    \label{eq:r_val}
\end{equation}
Equation~\eqref{eq:r_val} gives $r$ as a function of $\xi$.  However, $\xi$ is not arbitrary, but is instead fixed by its own root-finding condition $\xi^2=2\chi/\langle h,\barh\rangle^2$ with
\begin{equation}
	\langle h,\bar{h} \rangle = r\tilde{B}_1B_1 + 2rB_2^2 + \frac{\tilde{B}_1B_1 + 4\tilde{B}_1B_2 + 4B_1B_2}{2\xi}\sin(\xi r),
\end{equation}
where $\tilde{B}_1 = B_1(1+\alpha^2\xi^2)^2$.
Although this procedure is somewhat involved, the point remains: the Geometric Thin-Film Equation with partial wetting admits an analytical equilibrium solution in terms of elementary functions (via Equations~\eqref{eq:hbar_solution}--\eqref{eq:h_solution}).  Furthermore, the elementary solution~\eqref{eq:hbar_solution} for $\barh(x)$ coincides with the expression for a parabolic-cap droplet in the core region $x\rightarrow 0$,
\[
\barh(x)\approx \left(B_1+B_2\right)-\tfrac{1}{2}B_1\xi^2x^2,\qquad x\rightarrow 0.
\]
It is noted, however, that this expression is valid only in the core region $x\rightarrow 0$.  In particular, the analytical solution $\barh(x)$ does not converge to a parabolic-cap profile in the limit as $\alpha\rightarrow 0$; rather, $\barh(x)$ maintains its cosine shape in this limit.  Therefore, in order to reproduce the parabolic-cap profile in the limit as $\alpha\rightarrow 0$, it would be necessary to make a more judicious choice for an estimate of the width of the droplet footprint in Equation~\eqref{eq:E_partial}. 

Equation~\eqref{eq:r_val} also lets us approximate the transition region $r$. By expanding the right hand side of Equation~\eqref{eq:r_val} in terms of $\alpha\ll1$, we get
\begin{align}
    \tan(\xi r) &= -2\alpha\xi + O(\alpha^3), \\
    \xi r &= -\tan^{-1}(2\alpha\xi + O(\alpha^3)). \label{eq:partial_arctan}
\end{align}
Using the series expansion 
\begin{equation}
    \tan^{-1}(x) = x - \frac{x^3}{3} + \frac{x^5}{5} - \frac{x^7}{7} + \dots, \qquad \text{for } -1<x<1,
\end{equation}
Equation~\eqref{eq:partial_arctan} becomes
\begin{align}
    \xi r &= -2\alpha\xi + O(\alpha^3) + n\pi, \\
    r &= n\frac{\pi}{\xi} - 2\alpha + O(\alpha^3),
\end{align}
where $n\in\mathbb{Z}$. We are interested in $r$ satisfying $0<r\leq\pi/\xi$. This gives only one solution for $r$
\begin{equation}
    r=\frac{\pi}{\xi} - 2\alpha + O(\alpha^3).
\end{equation}
At $x=r$, we have
\begin{align}
    \bar{h}'(r) &= -B_1\xi\sin(\xi r), \\
    &= -B_1\xi\sin(\pi-2\alpha\xi+O(\alpha^3)), \\
    &= -2B_1\xi^2\alpha + O(\alpha^3). 
\end{align}
So the droplet foot is proportionate to $\alpha$ or $\xi^2$. These calculations show that the equilibrium solution is completely determined by $\xi$ and $\alpha$. 

\begin{figure}[thb]
	\centering
		\includegraphics[width=0.7\textwidth]{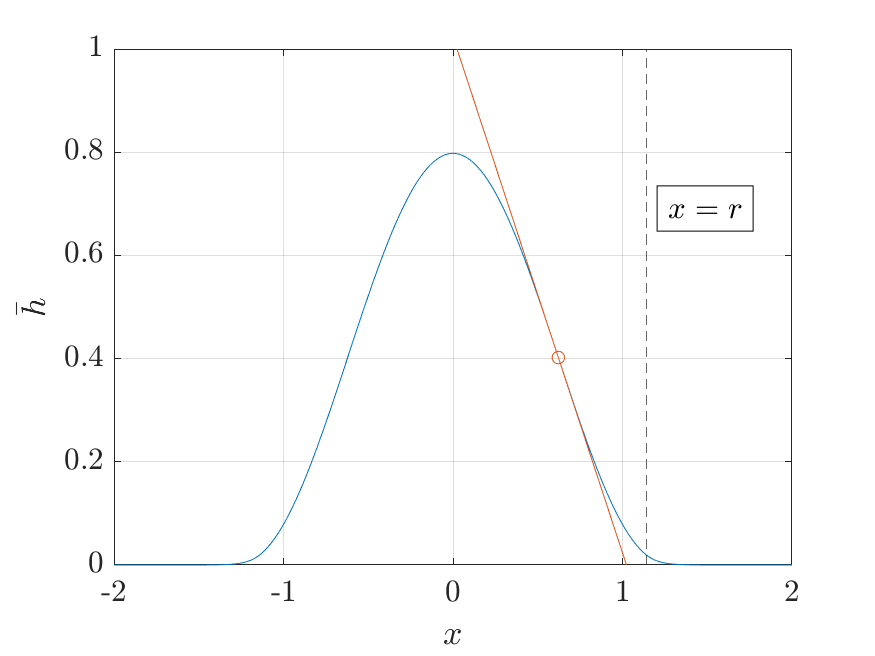}
		\caption{Equilibrium droplet profile $\barh(x)$ in dimensionless variables ($\alpha=0.05$). The red line is the tangent of the droplet profile at $x=\mathrm{argmax}_x[-\partial_{x}\barh(x)]$. The transition threshold $r$ from the droplet core to the exponentially decaying tail is marked with a dashed line.}
	\label{fig:eqm}
\end{figure}

The equilibrium (apparent) contact angle is now computed as:
\[
\theta_{eq}=-\epsilon \partial_{x}\barh(x=x_{ref}),
\]
where  $\barh_x$ is expressed in dimensionless variables and $x_{ref}>0$ is a reference point.  Since $\epsilon=\theta_{eq}$ in the chosen dimensionless variables, we set
\[
1=-\partial\barh(x=x_{ref}),
\]
We choose the reference point $x_{ref}$ to be the positive value of $x$ which maximizes $|\partial_x\barh|$, thus $x_{ref}=\pi/(2\xi)$.  Thus, we require $B_1\xi=1$.  But $B_1$ and $\xi$ depend parametrically on $\chi$, hence we require $B_1(\chi)\xi(\chi)=1$.  This, therefore, fixes the model parameter $\chi$ as a global constant, $\chi\approx 1.1602$ (for filter width $\alpha=0.05$).  The resulting equilibrium droplet profile is shown in Figure~\ref{fig:eqm}.  Values of $\chi$ for different values of $\alpha$ are given in Table~\ref{tab:opt}.

\begin{table}[htb]
	\centering
		\caption{Optimum value of $\chi$ for different values of filter width $\alpha$ such that $B_1\xi=1$.}
		\begin{tabular}{|c|c|}
		\hline
		$\alpha$& $\chi$ \\
		\hline
		\hline
		0.01   & 1.1264\\
		0.02   & 1.1306\\
		0.05   & 1.1602\\
		\hline	
		\end{tabular}
		\label{tab:opt}
\end{table}

\section{Time Dependent Solution}
\label{sec:partial_num}

\subsection{Particle Method} 
With the additional potential term, the particle evolution for Equation~\eqref{eq:particle_partial} now becomes
\begin{equation}
    \der[x_i]{t} = \left[\mob\pder{x}\left(\partial_{xx}\barh + 2\chi\frac{A^2}{\langle h,\barh\rangle^2}\barh\right)\right]_{x=x_i}, \qquad i=1,\dots,N. 
\end{equation}
In terms of the particle solution, the inner product pairing is given by
\begin{align}
    \langle h^N,\barh^N\rangle =& \sum_{i=1}^N \sum_{j=1}^N w_iw_j\intinf \delta(x-x_i) K(x-x_j)\, \mathd x, \\
    =& \sum_{i=1}^N \sum_{j=1}^N w_iw_j K(x_i-x_j), \\
    =& \sum_{i=1}^N w_i\barh^N(x_i),
\end{align}
which only requires $O(N)$ operations to compute once $\barh^N(x_i)$'s are obtained using the fast summation algorithm. This justifies the use of the pairing of $h$ with $\barh$ in Equation~\eqref{eq:particle_partial}. If we have use $\langle\barh,\barh\rangle = \|\barh\|_2$ instead, then one would have to compute
\begin{equation}
    \|\barh^N\|_2 = \sum_{i=1}^N \sum_{j=1}^N w_iw_j (K*K)(x_i-x_j),
\end{equation}
and evaluate $K*K$. We choose $\langle h,\barh\rangle$ for the simplicity. Finally, we also apply the fast summation algorithm to compute $\partial_x\barh^N$ and get 
\begin{equation}
    \partial_x\barh^N(x_i) = \frac{1}{4\alpha^3}[e^{-x_i/\alpha}(xa_i-b_i)+e^{x_i/\alpha}(xc_i-d_i)],
\end{equation}
where
\begin{gather*}
    a_i = \sum_{j=1}^{i-1} w_j\mathe^{x_j/\alpha}, \qquad b_i = \sum_{j=1}^{i-1} w_jx_j\mathe^{x_j/\alpha}, \\
    c_i = \sum_{j=i+1}^{N} w_j\mathe^{-x_j/\alpha}, \qquad d_i = \sum_{j=i+1}^{N} w_jx_j\mathe^{-x_j/\alpha}.
\end{gather*}
Thus, we see that Equation~\eqref{eq:particle_partial} is still compatible with the fast summation algorithm. 

\subsection{Finite-Difference Method} 
For the Newton method, we modify the objective function $\bm{F}(\bar{\bm{v}})$ by adding the potential term explicitly giving
\begin{align}
    \bm{F}(\bar{\bm{v}}) &= \bar{\bm{v}} + \Delta t \theta \mathcal{C}(\bar{\bm{v}}) + F_1(\bar{\bm{h}}^n), \\
    F_1(\bar{\bm{h}}^n) &= F_0(\bar{\bm{h}}^n)+\Delta t(1-\theta)\xi^2D_1\bar{\bm{h}}^n.
\end{align}
The inner product is computed using the finite difference approximation $\langle h,\barh\rangle \approx \ell = (\lambda_0\bar{\bm{h}})^{T}\bar{\bm{h}}\Delta x$, with $\Delta x$ being the grid spacing. Since $F_1(\bar{\bm{h}}^n)$ does not depends on $\bar{\bm{v}}$, the Jacobian $J(\bar{\bm{v}})$ remains unchanged. 

\subsection{Numerical Results}

\begin{figure}[tbh]
        \centering
        \includegraphics[width=0.7\textwidth]{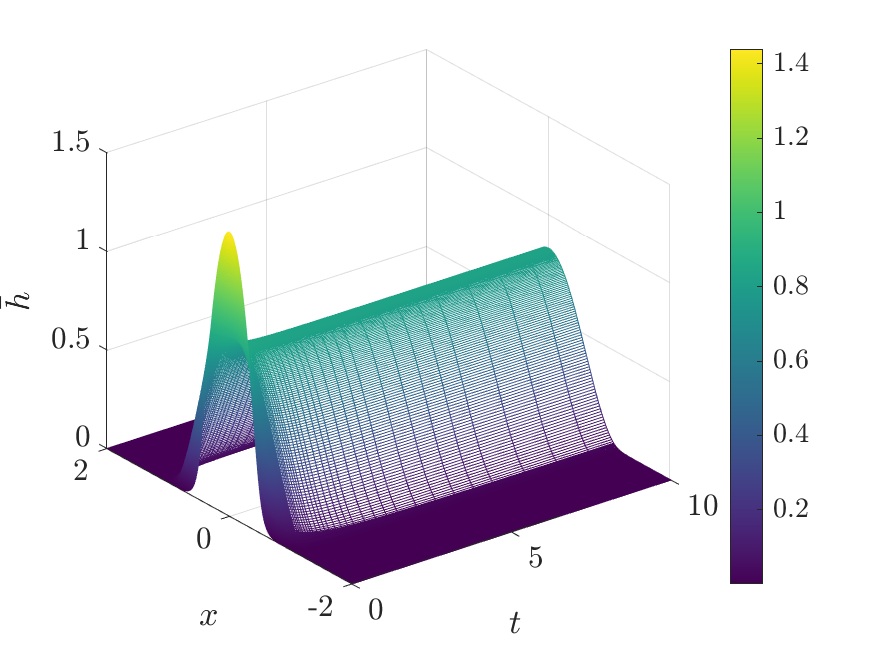}
    \caption{Space-time plot of $\barh(x,t)$ showing the spreading of the droplet for the partial-wetting case. The equilibrium droplet profile is stable over time. }
    \label{fig:eq_solution1}
\end{figure}

\begin{figure}[tbh]
        \centering
		\includegraphics[width=0.7\textwidth]{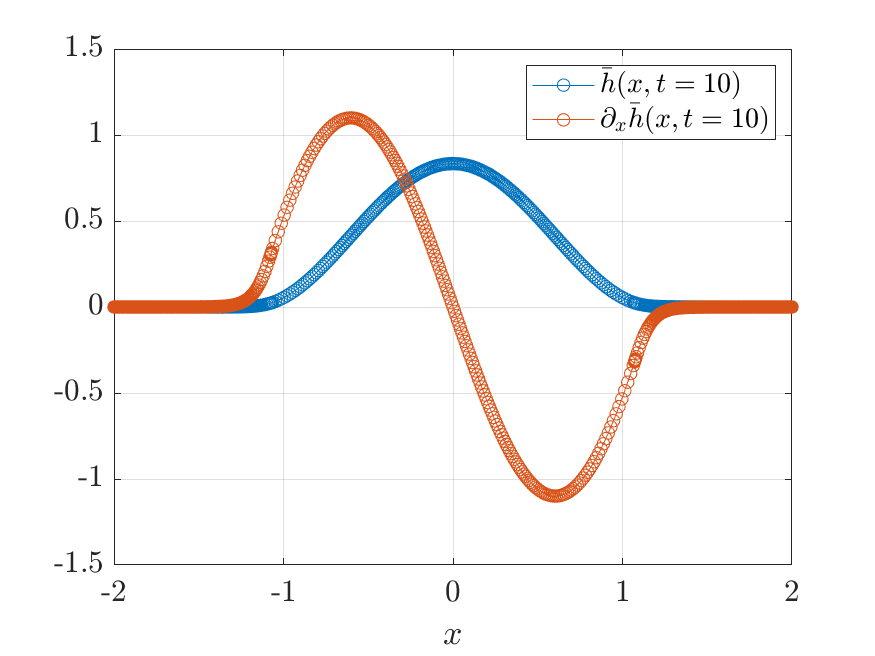}
    \caption{Droplet shape $\barh$ and the slope $\partial_x\barh$ at $t=10$. The derivative of the droplet profile shows that the equilibrium solution does not exhibit sharp changes in space even at the contact region. }
    \label{fig:eq_solution2}
\end{figure}

For the simulations of partial wetting, we again use the initial condition
\[
h_0(x)=
\begin{dcases}
    \frac{3}{4r_0^3}(r_0^2-x^2),&|x|<r_0,\\
    0,&|x|>r_0.
\end{dcases}
\]
with $r_0=0.5$, and $\int_{-\infty}^\infty h_0(x)\,\mathd x=1$.  We also take $\alpha=0.05$.  We use both the particle method and the finite-difference method: the results are the same in each case, thus, we only show the results of the particle method.  In Figure~\ref{fig:eq_solution1}, we show a space-time plot of the solution for the partial wetting case up to $t=10$. Figure~\ref{fig:eq_solution2} shows a snapshot of the droplet profile at $t=10$. We see that the transient simulation agrees with the analytical equilibrium solution obtained in \Cref{sec:equilib_sol}. Furthermore, the solution has a constant and finite volume throughout the whole duration of the simulation. 

\begin{figure}[tbh]
	\centering
		\includegraphics[width=0.7\textwidth]{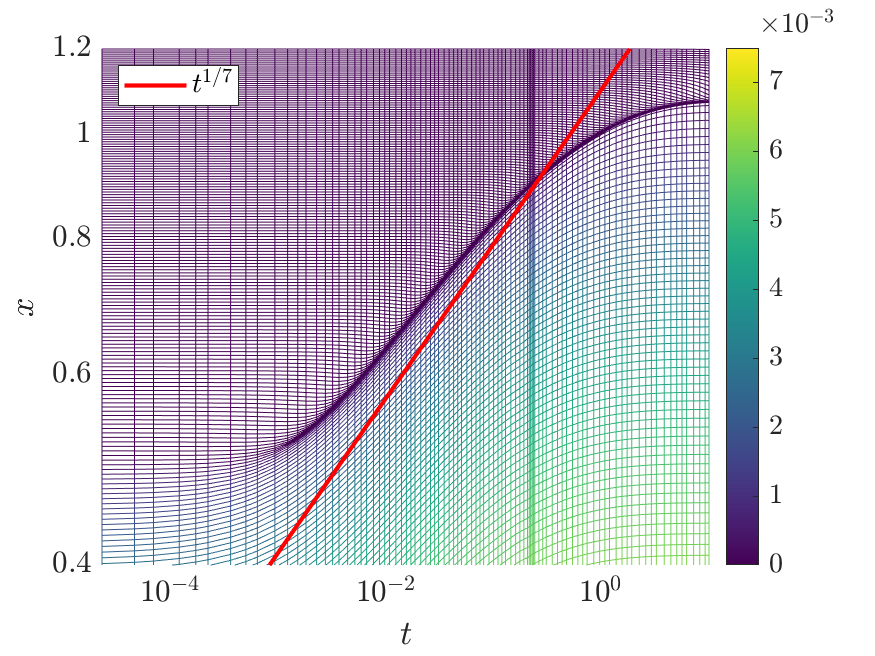}
		\caption{Spacetime plot showing the evolution of the particle trajectories (partial wetting).  Logarithmic scale on both axes. The trajectories depart from the power law around $t=1$. }
	\label{fig:eq_trajectory}
\end{figure}

Figure~\ref{fig:eq_trajectory} is based exclusively on the particle method: here, we show a log-log plot of the particle trajectories.  At intermediate times, the particle trajectories are parallel to the path $x=t^{1/7}$ before attaining a steady state at late times.  Thus, in the case of partial wetting, the system obeys Tanner's law at intermediate times. Until at late times, the partial wetting stabilizes the droplet, and the droplet assumes its equilibrium shape.

\begin{figure}[tbh]
	\centering
		\includegraphics[width=0.7\textwidth]{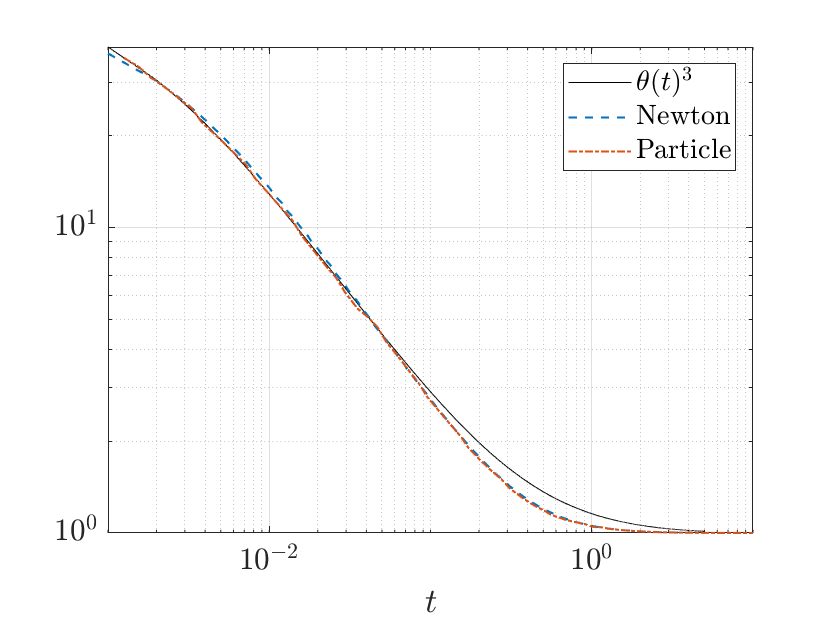}
		\caption{Plot of $\theta_{cl}(t)^3$ (solid line) and $1+c\dot{x}_{cl}\log(x_{cl}/d)$ (dashed and dot-dashed) as a function of time showing the agreement between GDIM and the Cox--Voinov theory for droplet spreading in the case of partial wetting.  Here, the dynamic contact angle $\theta_{cl}(t)$ and the contact-line position $x_{cl}(t)$ are defined operationally as in the text.  The values of $c$ and $d$ are chosen to optimize the fit between the two curves.}
	\label{fig:cox_voinov2}
\end{figure}

The foregoing statement that the droplet spreading obeys Tanner's Law at intermediate times until the onset of equilibrium also applies to the Cox--Voinov Law~\cite{bonn2009wetting} of droplet spreading, given by
\begin{equation}
    \theta_{cl}(t)^3=\theta_{eq}^3 + c\dot{x}_{cl}\log\left(\frac{x_{cl}}{d}\right).
\end{equation}
Here, $\theta_{cl}(t)$ is the dynamic contact angle, which is obtained from the slope of the interface profile $h(x,t)$ at some appropriate location $x$, and $c$ and $d$ are constants.  In order to validate the applicability of the Cox--Voinov law to the droplet spreading within the framework of the Geometric Thin-Film Equation, we operationally define the contact angle as $\theta_{cl}(t)=\max_x [-\partial_x\barh(x,t)])$.  The tangent line to $\barh(x,t)$ at $\mathrm{argmax}_x[-\partial_x\barh(x,t)]$ is constructed, and the contact line is then taken to be the intersection of this tangent line with the $x$-axis. This is the same procedure as in \Cref{sec:fdm}. A plot of $\theta_{cl}(t)^3$ constructed in this way is shown in Figure~\ref{fig:cox_voinov2}.  Shown also is a plot 
$1+c\dot{x}_{cl}\log(x_{cl}/d)$ -- here, $\theta_{eq}=1$, and $c$ and $d$ are best-fit constants.  Overall, there is good agreement between the two curves at intermediate times -- consistently with Tanner's law.  There is some disagreement at late times, however, this may be expected, in view of the somewhat imprecise operational definition of $\theta_{cl}(t)$ and $x_{cl}$.

\subsection{Error and Performance Analysis}

\begin{figure}[htb]
\centering
\includegraphics[width=0.7\linewidth]{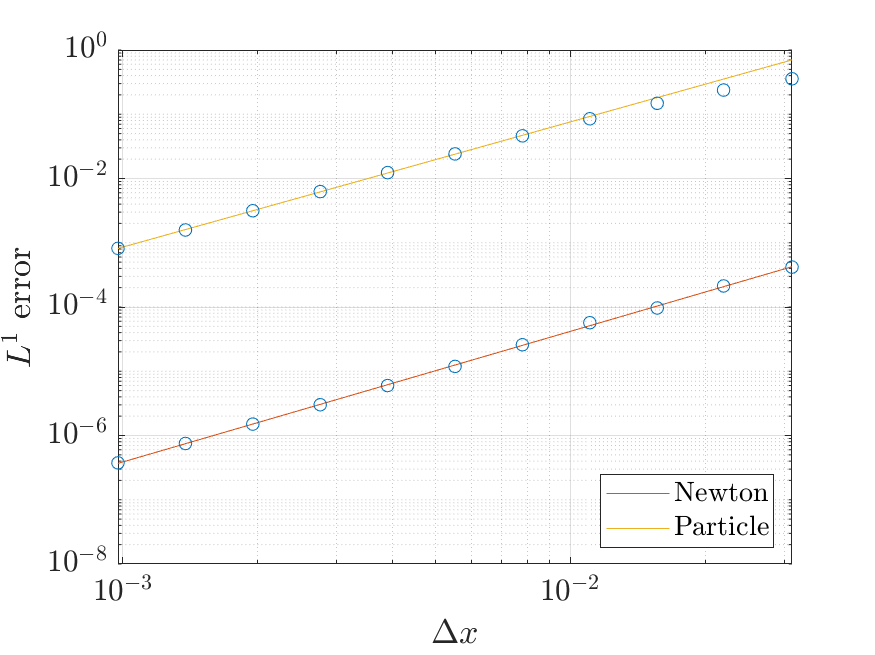}
    \caption{Convergence plot of the finite-difference method and the particle method for the complete wetting case. Both lines have a slope of $2.0$ on the log-log plot. }
    \label{fig:eq_convergence}
\end{figure}

We carry out a rigorous error analysis of both the finite-difference method and the particle method for the case of partial wetting.  Because an analytical equilibrium solution exists, we analyze the results of the numerical simulations at such a late time (specifically, $t=100$), when the numerical solutions are close to the equilibrium state.  In this case, the equation for the rate of convergence of the numerical method is simply
\begin{equation}
    \log\|\barh-\barh_{\Delta x}\|_1 = p\log(\Delta x),
\end{equation}
where $\barh$ denotes the analytical equilibrium solution, and $\barh_{\Delta x}$ denotes the numerical equilibrium solution (or what amounts to the same, the numerical solution at $t=100$).
Figure~\ref{fig:eq_convergence} shows the convergence rate for the finite-difference and particle methods. Both methods have $\alpha=0.05$ and were performed on the spatial domain $x\in[-2,2]$ for various spatial resolutions $\Delta x$.  We have used the MATLAB solver for the particle method. Again, we observed both the finite-difference and particle methods to be second-order accurate in the spatial domain. 

Although the fast-particle method is less accurate for fixed $N$, the fast-particle method is more efficient.  Therefore, for fixed execution time (but varying $N$), the fast particle method and the finite-difference method achieve a similar level of accuracy, albeit the finite-difference method still has the edge.   However, beyond this standard performance analysis, the particle method still retains a key benefit, namely that it is free from finite-size effects.

%This further suggests that Equation~\eqref{eq:hbar_solution} and subsequent boundary conditions are indeed an equilibrium solution of Equation~\eqref{eq:particle_partial_dim}. 
%

\begin{figure}[htb]
    \centering
    \includegraphics[width=0.7\linewidth]{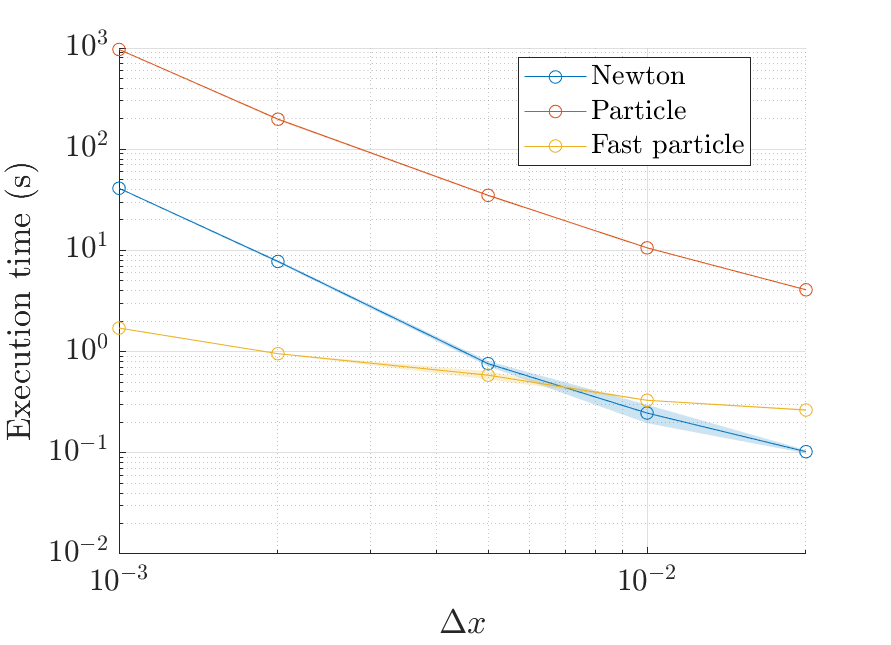}
    \caption{Performance of the finite-difference method, the direct implementation of the particle method, and the fast implementation of the particle method (partial wetting).}
    \label{fig:perf_eq}
\end{figure}

Finally, we have looked at the performance of the different numerical methods (fully-implicit finite-difference method, particle method, and `fast' particle method): the results are similar to what was observed in the case of complete wetting (Figure~\ref{fig:perf_eq}).

\section{\markup{Conclusion}}
\label{sec:partial_conclusion}

In this chapter, we model partially wetting droplets by including an additional surface tension term that arises from the interaction between the liquid at the base of the droplet and the solid substrate. The additional solid-liquid surface tension force opposes the tendency of the droplet to spread, leading to an equilibrium state of nonzero equilibrium contact angle. More specifically, while the liquid-gas interface evolves to minimize the surface curvature via spreading, the solid-liquid interface seeks to minimize the surface area, preventing the droplet footprint of the droplet from increasing indefinitely. 

The associated energy for partial wetting is proportional to the droplet footprint. However, in the geometric diffuse-interface method framework, the contact line contains uncertainty, and the base of the droplet is not well defined. For this reason, the size of the droplet footprint has to be estimated. We estimate this quantity as the ratio between the $L^1$-norm and the $L^2$-norm of the solution. Intuitively, this is the ratio between the area (zeroth moment) and the first moment. Lastly, for the Geometric Thin-Film Equation to admit particle solution, the $L^2$-norm is replaced with the pairing $\langle h,\barh\rangle$. 

Interestingly, the partial wetting model has an equilibrium solution of finite volume, expressed in terms of elementary functions, in contrast to the precursor-film model. The equilibrium solution is characterized completely, and we proved that the smoothened surface height $\barh\in C^3(\mathbb{R})$ has the desirable regularity. A limitation of our model is that the model fails to recover the parabolic-cap profile in the limit as $\alpha\rightarrow 0$, as one might expect for the classical thin-film equation. 
% It would be necessary to make a more judicious choice for estimate of the width of the droplet footprint. 
An alternative model is necessary for this purpose. This is left as future work. 

The time-dependent solution is obtained numerically using numerical methods developed in Chapter~\ref{sec:theory}. Notably, we see that the Geometric Thin-Film Equation still admits particle solutions, and the fast summation optimization is applicable. We observed a slower spreading rate of our model during late time when compared to the Cox--Voinov theory. Since the exact equilibrium solution is available, the absolute error of the numerical schemes can be computed. Even though the error of the particle method solver is much larger than the finite-difference solver, the particle method remains faster to compute and offers compelling reasons for its use, including the mass-conserving and positivity-preserving properties.

% \subfile{chapters/3d}

\chapter{Conclusion}

In this chapter, we summarise all the results presented in this thesis. A short discussion on potential future work that builds on this thesis is presented at the end. 

% \section{Summary}

In this thesis, we studied the stability of the internal Marangoni current of a point-heated droplet using classical thin-film theory. 
% A model in the lubrication theory is introduced for describing a thin droplet on an inhomogeneously heated substrate. 
The equilibrium solutions were shown to be stable to small perturbations when the heating is perfectly centred. We proposed that the mechanism for the twin-vortices formation is a slight deviation of the heat source and demonstrated this using two independent numerical methods. The complications regarding the classical droplet spreading models were discussed. 
To address this, we formulated the Geometric Thin-Film Equation to model the moving contact line using the geometric diffuse-interface method. 
% The geometric diffuse-interface method is used to regularize the energy functional in order to alleviate the contact line singularity. 
An improved finite-difference solver with better numerical stability and accuracy is introduced. Using the finite-difference solver, we demonstrated that our model reproduces spreading, and the spreading rates agree with Tanner's law in two and three spatial dimensions. 
We showed how the gradient-flow structure of the Geometric Thin-Film Equation leads to the existence of the particle solutions and a particle method solver. The advantages of the particle method over the finite-difference method were highlighted. Moreover, we introduced a fast summation algorithm and a fast multipole method to improve the efficiency of the particle method solver. The particle solutions also allow us to prove the existence theorem for the Geometric Thin-Film Equation over a wide class of initial conditions. We also showed that the solution is continuous for all time and has the desired regularity in both space and time. 
Finally, we modelled the partial wetting phenomena by incorporating the surface tension between the base of the droplet and the solid substrate into the Geometric Thin-Film Equation. We are able to obtain a droplet-shaped equilibrium solution analytically. The transient simulation shows that our model has a slightly weaker stabilizing force when compared to the Cox--Voinov law. 

% \section{Future Work}

Based on the work presented in this thesis, there are a number of ways to further expand on the research. In \Cref{sec:point_heating}, we concluded that the axisymmetric Marangoni current is stable to small perturbations of the droplet interface, but we see that the twin vortical flow can be induced by a very small asymmetry in the system. A promising direction is to include a description of the fluid evaporation in the model, where the evaporation flux could potentially initiate the symmetry breaking of droplets. We have also observed rich dynamics related to point-heated droplets, including rupturing and thermotaxis, which are worthy of investigation. 

This thesis is a preliminary study of using the Geometric Thin-Film Equation to describe droplet spreading. Much work is still needed to validate the model. This includes a quantitative comparison between the solution of the Geometric Thin-Film Equation with other classical models (slip-length and precursor-film model) as well as experimental data. To do this, a complete description of partial wetting is needed. The partial wetting model introduced in \Cref{sec:partial} is incomplete. Ideally, the equilibrium solution of the Geometric Thin-Film Equation should coincide with the classical thin-film equation as the regularization parameters $\alpha$ tends to zero. While the complete wetting model satisfies this condition, the partial wetting model does not. Some care is also needed when formulating the partial wetting term if one wishes to utilize the particle method. 

Finally, in terms of the numerical schemes developed in the thesis, there is a wide range of particle-solution admitting partial differential equations that could potentially benefit from the optimization discussed here. While the fast multipole method can speed up the $N$-point summations for the particle method, the computational complexity of the algorithm depends on the width of the kernel. In particular, as $\alpha$ decreases, the cell count $|C|$ required increases. Unless the number of particles is very large and the particles are densely packed, the fast particle method might not be faster than the finite-difference method. An improved fast multipole method that decouples the kernel width with the cell size would greatly improve the efficiency of the particle method.

\appendix

\chapter{Matched Asymptotic Expansion}

We consider the following thin-film equation with slip length 
\begin{equation} \label{eq:tfeslip}
    \pder[h]{t} + \pder{x}\left\{h^2(h+\beta)\pder[^3h]{x^3}\right\} = 0.
\end{equation}
For compact and symmetrical droplet solution with the boundary condition given by
\begin{subequations} \label{eq:maebc}
\begin{align}
    \partial_xh(0,t) &= 0, \\ 
    h(x_{cl}(t),t) &=0, \\
    \partial_xh(x_{cl}(t),t) &= -\theta_{eq}, \\
    2\int_0^{x_{cl}(t)}h(x,t) \,\mathd x &= V,
\end{align}
where $\theta_{eq},V>0$ is the equilibrium contact angle and the droplet volume respectively. Hocking \cite{hocking1983spreading,hocking1982the} proposed a two boundary layer matched asymptotic expansion to obtain the solution near the contact line, which is used to augment the similarity solution. We note that this derivation is not readily available and wish to address it here. 
\end{subequations}

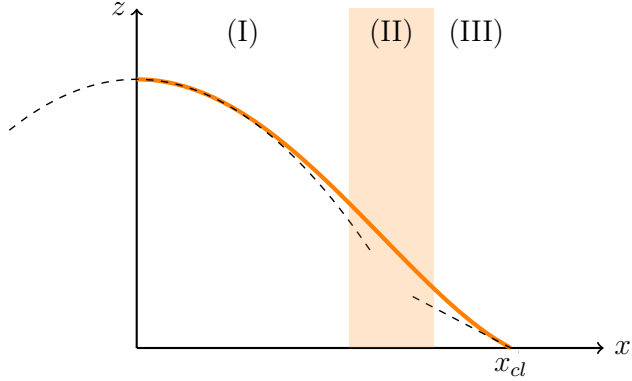
\begin{figure}[ht]
    \centering
    \begin{tikzpicture}[x=6cm, y=4cm]
        % axis
        \draw [->, line width=0.3mm] (0,0) -- (0,1.2) node[left] {$z$};
        \draw [->, line width=0.3mm] (0,0) -- (1.1,0) node[right] {$x$};
        % profile
        \draw[orange, line width=0.6mm] plot[domain=0:0.8811,smooth] (\x,{(\x^2-1)^2-0.05});
        \draw (0.8811,0) node[below] {$x_{cl}$};
        % approx
        \draw[black, dashed, line width=0.2mm] plot[domain=-0.3:0.55,smooth] (\x,{-2*abs(\x)^2+1-0.05});
        \draw[black, dashed, line width=0.2mm] plot[domain=0.65:0.9,smooth] (\x,{-0.788*(\x-0.881)});
        % region
        \fill [orange, fill opacity=0.2] (0.5,0) rectangle (0.7,1.2);
        % \fill [blue, fill opacity=0.2] (0.7,0) rectangle (0.9,1.2);
        \draw (0.25,1.2) node[below] {(I)}; 
        \draw (0.6,1.2) node[below] {(II)}; 
        \draw (0.8,1.2) node[below] {(III)}; 
    \end{tikzpicture} 
    \caption{Region I, II, and III corresponding to the outer, intermediate, and inner regions are solved separately with appropriate approximation. }
\end{figure}

Assuming the spreading rate is small, we expand $\dot{x}_{cl}$ as a power series of $\varepsilon$, where $\beta\ll\varepsilon\ll1$ is an intermediate parameter to be determined, 
\begin{equation}
    \dot{x}_{cl} = \varepsilon u_1 + \varepsilon^2u_2 + O(\varepsilon^3).
\end{equation}
By the chain rule, we have
\begin{equation}
    \pder[h]{t} = \dot{x}_{cl}\pder[h]{x_{cl}}. 
\end{equation}

\paragraph{Region I} At the outer region where $x-x_{cl}=O(1)$, we expand $h=h_0+\varepsilon h_1+O(\varepsilon^2)$. Then the linearised \cref{eq:tfeslip} becomes
\begin{multline}
    \varepsilon u_1\pder[h_0]{x_{cl}} + \pder{x}\left(h_0^2(h_0+\beta)\pder[^3h_0]{x^3}\right) \\ + \varepsilon\pder{x}\left((3h_0^2h_1+2\beta h_0h_1)\pder[^3h_0]{x^3}\right) 
    + \varepsilon\pder{x}\left(h_0^2(h_0+\beta)\pder[^3h_1]{x^3}\right) = 0.
\end{multline}
If $\beta\sim O(\varepsilon^2)$, then we can dropped the $\beta$ terms 
\begin{align} \label{eq:match1}
    \varepsilon u_1\pder[h_0]{x_{cl}} + \pder{x}\left(h_0^3\pder[^3h_0]{x^3}\right) + 3\varepsilon\pder{x}\left(h_0^2h_1\pder[^3h_0]{x^3}\right) + \varepsilon\pder{x}\left(h_0^3\pder[^3h_1]{x^3}\right) = 0.
\end{align}
Matching the $\varepsilon$ terms in \cref{eq:match1} gives 
\begin{align}
    \varepsilon^0 \text{-term:} &\qquad \pder{x}\left(h_0^3\pder[^3h_0]{x^3}\right) = 0, \label{eq:outer1} \\
    \varepsilon^1 \text{-term:} &\qquad u_1\pder[h_0]{x_{cl}} + \pder{x}\left(h_0^3\pder[^3h_1]{x^3}\right) = 0, \label{eq:outer2}
\end{align}
which allow us to solve for $h_0$ and $h_1$. Subjecting to the boundary condition in \eqref{eq:maebc} the solutions are given by
\begin{align}
    h_0 &= \frac{1}{2x_{cl}^3}(x_{cl}^2-x^2), \\
    h_1 &= u_1 x_{cl}^4\bigg((x_{cl}+x)\ln(x_{cl}+x) + (x_{cl}-x)\ln(x_{cl}-x) \nonumber \\
    &\hspace{150pt} - 2x_{cl}\ln(2x_{cl}) + \frac{3}{2x_{cl}}(x_{cl}^2-x^2)\bigg).
\end{align}
Note that the first order approximation in the bulk of the droplet for the low spreading regime is a parabola. Next we take the inner limit $x-x_{cl}=\delta\rightarrow0$. For $\delta$ small, $h_0$ becomes
\begin{equation}
    h_0 = \frac{1}{2x_{cl}^3}(2x_{cl}-\delta)\delta = \frac{\delta}{x_{cl}^2} + O(\delta^2).
\end{equation}
and $h_1$ becomes
\begin{align}
    h_1 &= u_1 x_{cl}^4\bigg((2x_{cl}-\delta)\ln(2x_{cl}-\delta) + \delta\ln\delta \nonumber\\
    &\hspace{150pt} - 2x_{cl}\ln(2 x_{cl}) + 3\delta-\frac{3}{2x_{cl}}\delta^2\bigg), \\
    &= u_1x_{cl}^4\delta\left(\ln\frac{\delta}{2x_{cl}} + 2\right) + O(\delta^2).
\end{align}
So the inner limit of the outer solution is given by
\begin{align}
    h_{out} \sim \frac{x-x_{cl}}{x_{cl}^2} + \varepsilon u_1x_{cl}^4(x-x_{cl})\left(\ln\frac{x-x_{cl}}{2a}+2\right), \label{eq:outer_sol}
\end{align}
as $x-x_{cl}\rightarrow0$.

\paragraph{Region III} For the inner region where $x-x_{cl}=O(\beta)$, we rescale $x$ and $h$ such that
\begin{align}
    x=x_{cl}-\beta X, \qquad h(x,x_{cl}) = \beta H(X,x_{cl})
\end{align}
Then Equation~\eqref{eq:tfeslip} becomes
\begin{align} \label{eq:tfeslip2}
    \dot{x}_{cl}\pder[H]{X} + \pder{X}\left(H^2(H+1)\pder[^3H]{X^3}\right) &= 0,
\end{align}
with boundary conditions
\begin{align}
    H(0,t) &= 0, \qquad \pder[H]{X}(0,t) = \theta_{eq}.
\end{align}
Next, we expand $H$ in terms of $\varepsilon$ such that $H=H_0+\varepsilon H_1+O(\varepsilon^2)$. With this, the linearized \cref{eq:tfeslip2} is given by
\begin{multline}
    \varepsilon u_1\pder[H_0]{X} + \pder{X}\left(H_0^2(H_0+1)\pder[^3H_0]{X^3}\right)
    + \varepsilon\pder{X}\left(H_0^2(H_0+1)\pder[^3H_1]{X^3}\right)
    % +2\varepsilon\pder{X}\left(H_0H_1(H_0+1)\pder[^3H_0]{X^3}\right) + \varepsilon\pder{X}\left(H_0^2H_1\pder[^3H_0]{X^3}\right) 
    = 0.
\end{multline}
Matching the $\varepsilon$ terms gives
\begin{align}
    \varepsilon^0 \text{-term:} &\qquad \pder{X}\left(H_0^2(H_0+1)\pder[^3H_0]{X^3}\right) = 0, \label{eq:outereq1} \\
    \varepsilon^1 \text{-term:} &\qquad u_1\pder[H_0]{X} + \pder{X}\left(H_0^2(H_0+1)\pder[^3H_1]{X^3}\right) = 0. \label{eq:outereq2}
\end{align}
Solving \cref{eq:outereq1,eq:outereq2} gives 
\begin{equation}
    H = \theta_{eq} X + \frac{1}{2}\varepsilon u_1\left((X+1)^2\ln(X+1) - X^2\ln X - X\right).
\end{equation}
As $X\rightarrow\infty$, the outer limit of the inner solution is 
\begin{align} \label{eq:inner_sol}
    H \sim \theta_{eq}X + \varepsilon u_1 X\ln X.
\end{align}
Equivalently,
\begin{equation}
    h_{in}\sim -\theta_{eq}(x-x_{cl}) + \varepsilon u_1(x_{cl}-x)\ln\frac{x_{cl}-x}{\beta},
\end{equation}
away from the contact line. So $h$ is approximately linear at the contact line with some small logarithmic correction. To determine the contact line velocity, \cref{eq:outer_sol} and \cref{eq:inner_sol} are matched togather through an intermediate region.

\paragraph{Region II} For the intermediate region $x=O(\varepsilon)$, we use the transformation
\begin{equation}
    y = \varepsilon\ln X, \qquad Q(y)=\frac{H}{X}
\end{equation}
Substituting $Q$ into \cref{eq:tfeslip} we obtain 
\begin{align}
    \dot{x}_{cl} + Q(Q+e^{-y/\varepsilon})\left(\varepsilon^3\pder[^3Q]{y^3}-\varepsilon\pder[Q]{y}\right) &= 0,
\end{align}
and the linearization is given by
\begin{align} \label{eq:match3}
    u_1+\varepsilon u_2 - Q^2\pder[Q]{y} = 0.
\end{align}
Integrating once with respect to $y$ gives 
\begin{align}
    Q^3 &= C_0 + 3u_1y + \varepsilon(C_1+3u_2y),
\end{align}
where $C_0$ and $C_1$ are constants of integration. To match the intermediate solution with the inner solution, we expend in terms of the inner variables
\begin{align}
    Q^3 &= C_0 + \varepsilon C_1 + 3(u_1 + \varepsilon u_2)\varepsilon\ln X, \\
    % &= C_0 + \varepsilon(3u_1\ln X C_1) + O(\epsilon^2), \\
    % &= C_0\left(1 + \frac{1}{C_0}\varepsilon(3u_1\ln X+C_1) + \dots\right) \\
    Q &= C_0^{1/3} + \frac{1}{3C_0^{2/3}}\varepsilon(3u_1\ln X+C_1) + O(\varepsilon^2).
\end{align}
So we obtain the intermediate solution
\begin{align}
    H = C_0^{1/3}X + \varepsilon\frac{1}{C_0^{2/3}} u_1X\ln X+\varepsilon\frac{C_1}{3C_0^{2/3}}X. \label{eq:intermediate_sol} 
\end{align}
Comparing Equation~\eqref{eq:intermediate_sol} with Equation~\eqref{eq:outer_sol} gives
\begin{equation}
    C_0 = \theta_{eq}^3, \qquad C_1 = 0.
\end{equation}
So the updated intermediate solution becomes
\begin{align}
    % Q^3 &= \theta_{eq}^3 + 3u_1 + 3\varepsilon u_2y \\
    % &= (\theta_{eq}^3 + 3u_1)\left(1 + \varepsilon\frac{3u_2 y}{\theta_{eq}^3+3u_1}\right) \\
    % Q &= (\theta_{eq}^3 + 3u_1)^{1/3}\left(1 + \varepsilon\frac{u_2 y}{\theta_{eq}^3+3u_1} + \dots\right) \\
    Q &= (\theta_{eq}^3 + 3u_1)^{1/3} + \varepsilon\frac{u_2 y}{(\theta_{eq}^3+3u_1)^{2/3}} + O(\varepsilon^2). \label{eq:intermediate_sol2}
\end{align}
To match \cref{eq:intermediate_sol2} with the outer solution, we write \cref{eq:intermediate_sol2} in terms of the original variables using the transformation $y=1+\varepsilon\ln(x_{cl}-x)$ and $Q=h/(x_{cl}-x)$ to get
\begin{align}
    % \frac{h}{a-x} &= (\theta_{eq}^3 + 3u_1)^{1/3} +\varepsilon\frac{u_2+\varepsilon u_2\ln(a-x)}{(\theta_{eq}^3+3u_1)^{2/3}} \\
    h &= (x_{cl}-x)\left((\theta_{eq}^3 + 3u_1)^{1/3} +\varepsilon\frac{u_2+u_1\ln(x_{cl}-x)}{(\theta_{eq}^3+3u_1)^{2/3}}\right). \label{eq:intermediate_sol3}
\end{align}
Finally, matching Equation~\eqref{eq:intermediate_sol3} with Equation~\eqref{eq:outer_sol} gives
\begin{align}
    \varepsilon^0 \text{-term:} &\qquad (x_{cl}-x)(\theta_{eq}^3+3u_1)^{1/3} = \frac{x_{cl}-x}{x_{cl}^2} \\
    &\qquad \theta_{eq}^3 +3u_1 = x_{cl}^{-6} \\
    \varepsilon^1 \text{-term:} &\qquad \frac{u_2+u_1\ln(x_{cl}-x)}{(x_{cl}^{-6})^{2/3}}=u_1x_{cl}^4\left(\ln\frac{x_{cl}-x}{2x_{cl}}+2\right) \\
    &\qquad u_2 = -u_1\ln(2x_{cl}) + 2u_1
\end{align}
So the equation of motion of the contact line is
\begin{align}
    \dot{x_{cl}} &= \varepsilon(u_1+\varepsilon u_2), \\
    % &= \varepsilon u_1(1 + \varepsilon(2-\ln(2x_{cl}))) \\
    &= \frac{1}{3}\varepsilon(x_{cl}^{-6}-\theta_{eq}^3)(1+\varepsilon(2-\ln(2x_{cl}))).
\end{align}

\chapter{Derivation of the Smoothing Kernels}

\paragraph{Bi-Helmholtz Kernel}

The bi-Helmholtz kernel $K_2$ is defined as the Green's function of the problem 
\begin{equation} \label{eq:a_bihelm}
    (1-\alpha^2\partial_{xx})^2K_2=\delta.
\end{equation}

We introduce the Fourier transform defined by
\begin{equation}
    \mathcal{F}[f](k) = \hat{f}_k = \int_{-\infty}^\infty f(x)\mathe^{-\mathi kx}\,\mathd x,
\end{equation}
with inverse transform
\begin{equation}
    \mathcal{F}^{-1}[\hat{f}_k](x) =f(x)= \frac{1}{2\pi}\int_{-\infty}^\infty \hat{f}_k \mathe^{\mathi kx}\,\mathd k.
\end{equation}
The Fourier transform of $\delta$ is given by
\begin{equation}
    \mathcal{F}[\delta](k) = \int_{-\infty}^\infty \delta(x)\mathe^{-\mathi kx}\,\mathd x = \mathe^{-\mathi k 0} = 1.
\end{equation}
On the other hand, the derivative can be computed by observing
\begin{align}
    K_2'(x) &= \der{x} \left(\frac{1}{2\pi}\int_{-\infty}^\infty \hat{K}_2 \mathe^{\mathi kx}\,\mathd k\right) \\
    &= \frac{1}{2\pi}\int_{-\infty}^\infty \mathi k\hat{K}_2 \mathe^{\mathi kx}\,\mathd k = \mathcal{F}^{-1}[\mathi k \hat{K}_2](x).
\end{align}
Thus, we obtain the identity
\begin{equation}
    \mathcal{F}[K_2'](k) = \mathi k \hat{K}_2. 
\end{equation}

By taking the Fourier transform of \cref{eq:a_bihelm}, we obtained
\begin{equation}
    \hat{K}_2 = \frac{1}{(1+\alpha^2k^2)^2} = \frac{1}{\alpha^4(k-k_1)^2(k-k_2)^2},
\end{equation}
where 
\begin{equation}
    k_1 = \frac{\mathi}{\alpha}, \qquad k_2 = -\frac{\mathi}{\alpha}, \qquad \in\mathbb{C},
\end{equation}
are the complex roots of $(1+\alpha^2k^2)^2$. 

To obtain $K_2$, we take the inverse Fourier transform of $\hat{K}_2$ 
\begin{equation}
    K_2(x) = \frac{1}{2\pi}\int_{-\infty}^\infty \hat{K}_2 \mathe^{\mathi kx}\,\mathd k.
\end{equation}
By the Cauchy's residue theorem, 
\begin{equation}
    K_2(x) = 
    \begin{dcases}
    \mathi \res_{k=k_1} \hat{K}_2 \mathe^{\mathi kx}, & \text{for $x>0$}, \\
    -\mathi \res_{k=k_2} \hat{K}_2 \mathe^{\mathi kx}, & \text{for $x<0$}. \\
    \end{dcases}
\end{equation}

The residue can be computed using the following Lemma from complex analysis
\begin{lemma} \label{lem:complex}
    If $a$ is a pole of order $m$ at $f$ and $f$ is written in the form
    \begin{equation}
        f(z) = \frac{1}{(z-a)^m}g(z),
    \end{equation}
    for a function $g$ analytic in some $\{z\in\mathbb{C}:|z-a|<r\}$, where $g(a)\neq0$, then
    \begin{equation}
        \res_{z=a}f(z) = \frac{1}{(m-1)!}g^{(m-1)}(a).
    \end{equation}
\end{lemma}

When $x>0$,
\begin{align*}
    K_2(x) &= \mathi \res_{k=k_1} \hat{K}_2 \mathe^{\mathi kx}, \\
    &= \mathi \res_{k=k_1} \frac{1}{(k-k_1)^2}\frac{\mathe^{\mathi k x}}{\alpha^4(k-k_2)^2}, \\
    &= \mathi \der{k} \left[\frac{\mathe^{\mathi k x}}{\alpha^4(k-k_2)^2}\right]_{k=k_1},  \tag{\cref{lem:complex}} \\
    &= \frac{\mathi}{\alpha^4} \left[\frac{\mathi x(k-k_2)-2}{(k-k_2)^3}\mathe^{\mathi k x}\right]_{k=k_1}, \\
    &= \frac{\mathi}{\alpha^4} \frac{\mathi x(2\mathi/\alpha)-2}{(2\mathi/\alpha)^3}\mathe^{\mathi k_1 x}, \\
    &= \frac{1}{4\alpha^2}(x+\alpha)\mathe^{-x/\alpha}.
\end{align*}
Similarly, when $x<0$, we get
\begin{equation}
    K_2(x) = \frac{1}{4\alpha^2}(-x+\alpha)\mathe^{x/\alpha}.
\end{equation}
Therefore the bi-Helmholtz kernel is given by
\begin{equation}
    K_2(x) = \frac{1}{4\alpha^2}(|x|+\alpha)\mathe^{-|x|/\alpha}.
\end{equation}

\paragraph{Extended Helmholtz Kernel}

The extended Helmholtz kernel refers to the Green's function of the following problem
\begin{equation}\label{eq:a_exhelm}
    (1-\alpha^2\partial_{xx}+\tfrac{1}{2}\alpha^4\partial_{xxxx})K_3 = \delta.  
\end{equation}
The Fourier transform of \cref{eq:a_exhelm} is
\begin{equation}
    \hat{K}_3 = \frac{1}{1+\alpha^2k^2+\tfrac{1}{2}\alpha^4k^4} = \cfrac{2}{\alpha^4\left(\cfrac{2}{\alpha^4} + \cfrac{2k^2}{\alpha^2} + k^4\right)}.
\end{equation}
The denominator is zero when
\begin{equation}
    k^2 = \cfrac{-\cfrac{2}{\alpha^2}\pm\sqrt{\cfrac{4}{\alpha^4}-\cfrac{8}{\alpha^4}}}{2} = \frac{1}{\alpha^2}(-1\pm \mathi).
\end{equation}
In polar coordinates, this is equivalent to
\begin{equation}
    k^2 = 
        \frac{2^{1/2}}{\alpha^2}\mathe^{3\pi\mathi/4}, \qquad \text{or} \qquad
        k^2 = \frac{2^{1/2}}{\alpha^2}\mathe^{5\pi\mathi/4}.
\end{equation}
Taking the square root yields the four roots 
\begin{align}
    k_1 = \frac{2^{1/4}}{\alpha}\mathe^{3\pi\mathi/8}, \qquad 
    k_2 = \frac{2^{1/4}}{\alpha}\mathe^{5\pi\mathi/8}, \qquad 
    k_3 = \frac{2^{1/4}}{\alpha}\mathe^{11\pi\mathi/8}, \qquad 
    k_4 = \frac{2^{1/4}}{\alpha}\mathe^{13\pi\mathi/8}.
\end{align}
Note that this can be written as
\begin{gather}
    k_1 = A+B\mathi, \qquad k_2 = -A+B\mathi, \qquad k_3 = -A-B\mathi, \qquad k_4 = A-B\mathi, \\
    A = \frac{\sqrt{2-\sqrt{2}}}{2^{3/4}\alpha}, \qquad B=\frac{\sqrt{2+\sqrt{2}}}{2^{3/4}\alpha}.
\end{gather}
Thus, we can write
\begin{equation}
    \hat{K}_3 = \frac{2}{\alpha^4(k-k_1)(k-k_2)(k-k_3)(k-k_4)}.
\end{equation}
Taking the inverse transform gives
\begin{equation}
    K_3(x) = \begin{dcases}
    \mathi \res_{k=k_1} \hat{K}_3 \mathe^{\mathi kx} + \mathi \res_{k=k_2} \hat{K}_3 \mathe^{\mathi kx}, & \text{for $x>0$}, \\
    -\mathi \res_{k=k_3} \hat{K}_3 \mathe^{\mathi kx} -\mathi \res_{k=k_4} \hat{K}_3 \mathe^{\mathi kx}, & \text{for $x<0$}. \\
    \end{dcases}
\end{equation}

We compute one of the residue
\begin{align}
    \res_{k=k_1} \hat{K}_3\mathe^{\mathi kx} &= \frac{2\mathe^{\mathi k_1 x}}{\alpha^4(k_1-k_2)(k_1-k_3)(k_1-k_4)}, \\
    &= \frac{2\mathe^{\mathi Ax-Bx}}{8\alpha^4A(A+B\mathi)B\mathi}, \\
    &= \frac{(-1+\mathi)\sqrt{-1+\mathi}}{4\alpha}\mathe^{\mathi Ax-Bx}, \\
    &= \frac{1}{4}\mathe^{-Bx}(-1+\mathi)(A+\mathi B)\left(\cos(Ax)+\mathi\sin(Ax)\right), \\
    &= \frac{1}{4}\mathe^{-Bx}\big[-(A+B)\cos(Ax) - (A-B)\sin(Ax) \nonumber\\
    &\qquad\qquad\qquad -(A+B)\mathi\sin(Ax) + (A-B)\mathi\cos(Ax)\big].
\end{align}
Similarly, 
\begin{align}
    \res_{k=k_2} \hat{K}_3\mathe^{\mathi kx} 
    &= \frac{1}{4}\mathe^{-Bx}\big[(A+B)\cos(Ax) + (A-B)\sin(Ax) \nonumber\\
    &\qquad\qquad\qquad -(A+B)\mathi\sin(Ax) + (A-B)\mathi\cos(Ax)\big].
\end{align}
Therefore, for $x>0$,
\begin{align}
    K_3(x) &= \mathi \res_{k=k_1} \hat{K}_3 \mathe^{\mathi kx} + \mathi \res_{k=k_2} \hat{K}_3 \mathe^{\mathi kx}, \\
    &= \frac{1}{2}\mathe^{-Bx}\big[(A+B)\sin(Ax) - (A-B)\cos(Ax)\big].
\end{align}
We can compute $K_3(x)$ for $x<0$ in a similar way. Thus, the kernel is given by
\begin{equation}
    K_3(x) = \frac{1}{2}\mathe^{-|B|x}\big[(A+B)\sin(A|x|) - (A-B)\cos(Ax)\big],
\end{equation}
which is not non-negative on $\mathbb{R}$. 

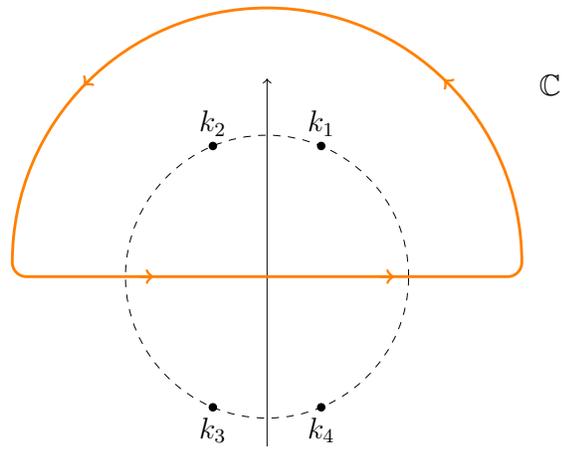
\begin{figure}[tbh]
    \centering
    \begin{tikzpicture} [scale=2]
        \draw (2,1.5) node[below] {$\mathbb{C}$};
        \draw [->] (0,-1.2) -- (0,1.4);
        % \draw [->] (-2.2,0) -- (2.2,0);
        % roots of unity
        \draw [dashed] (0,0) circle (1);
        \fill ({cos(3*180/8)},{sin(3*180/8)}) circle (0.03) node[above] {$k_1$};
        \fill ({cos(5*180/8)},{sin(5*180/8)}) circle (0.03) node[above] {$k_2$};
        \fill ({cos(11*180/8)},{sin(11*180/8)}) circle (0.03) node[below] {$k_3$};
        \fill ({cos(13*180/8)},{sin(13*180/8)}) circle (0.03) node[below] {$k_4$};
        % \draw [dashed] ({1.3*cos(11*180/8)},{1.3*sin(11*180/8)}) -- ({1.3*cos(3*180/8)},{1.3*sin(3*180/8)});
        % \draw [dashed] ({1.3*cos(5*180/8)},{1.3*sin(5*180/8)}) -- ({1.3*cos(13*180/8)},{1.3*sin(13*180/8)});
        % integration path
        \draw [orange, line width=0.4mm] (-1.7,0) -- (1.7,0) arc (270:360:0.1) arc (0:180:1.8) arc (180:270:0.1);
        \draw [->,orange, line width=0.4mm] (-0.9,0) -- (-0.8,0);
        \draw [<-,orange, line width=0.4mm] (0.9,0) -- (0.8,0);
        \draw [->,orange, line width=0.4mm] (1.273,1.373) arc (45:46:1.8);
        \draw [->,orange, line width=0.4mm] (-1.273,1.373) arc (135:136:1.8);
    \end{tikzpicture}
    \caption{The poles of $\hat{K}_3$ correspond to four of the eighth roots of unity. Orange marked the clockwise path for evaluating the integral over the real line. By taking the limit of the radius of the path to infinity, the contribution from the upper semicircle vanishes and we are left with the integral from $-\infty$ to $\infty$.  }
    \label{fig:cauchy_path}
\end{figure}

\newpage
\addcontentsline{toc}{chapter}{References}
\printbibliography
% \bibliographystyle{unsrt}
% \bibliography{thesis}

\end{document}